%% file: ms.tex
\documentclass[manuscript]{aastex}

\def\arcs{$^{\prime\prime}$}
\def\arcss{\rlap{.}$^{\prime\prime}$}

\begin{document}


\title{Pattern Corotation Radii from Potential-Density Phase-Shifts for
153 OSUBGS Sample Galaxies}


\author{Ronald J. Buta}\affil{Department of Physics and Astronomy,
University of Alabama, Tuscaloosa, AL 35487, USA}

\author{Xiaolei Zhang}\affil{Department of Physics and Astronomy, George Mason
University, Fairfax, VA 22030, USA}



\begin{abstract}
The potential-density phase-shift method is an effective new tool for
investigating the structure and evolution of galaxies. In this paper,
we apply the method to 153 galaxies in the Ohio State University Bright
Galaxy Survey (OSUBGS) to study the general relationship between
pattern corotation radii and the morphology of spiral galaxies. The
analysis is based on near-infrared $H$-band images that have been
deprojected and decomposed assuming a spherical bulge. We find that
multiple pattern speeds are common in disk galaxies. By selecting those
corotation radii close to or slightly larger than the bar radius as
being the bar corotation (CR) radius, we find that the average and
standard deviation of the ratio $\cal R$ = $r(CR)/r(bar)$, is
1.20$\pm$0.52 for 101 galaxies having well-defined bars. There is an
indication that this ratio depends weakly on galaxy type in the sense
that the average ranges from 1.03$\pm$0.37 for 65 galaxies of type Sbc
and earlier, to 1.50$\pm$0.63 for 36 galaxies of type Sc and later. Our
bar corotation radii are on average smaller than those estimated from
single-pattern-speed numerical simulations, most likely because these
simulations tend to find the pattern speed which generates a density
response in the gas that best matches the morphology of the outer
spiral structure.  Although we find CR radii in most of the sample
galaxies that satisfy conventional ideas about the extent of bars, we
also consider the alternative interpretation that in many cases the bar
CR is actually inside the bar and that the bar ends close to its
outer Lindblad resonance instead of its CR.  These ``super-fast" bars
are the most controversial finding from our study.  We see evidence in
the phase-shift distributions for ongoing decoupling of patterns, which
hints at the formation pathways of nested patterns, and which in turn
further hints at the longevity of the density wave patterns in
galaxies. We also examine how uncertainties in the orientation
parameters of galaxies and in the shapes of bulges affect our results.

\end{abstract}




\section{Introduction}

There is considerable evidence at this time that internal secular
evolution is an important physical process in disk galaxies. Kormendy \&
Kennicutt (2004) have argued that such evolution accounts for the
disk-like ``pseudobulges" seen in late-type galaxies, and evidence of
bulges built through secular evolution has been found to be present
across the entire Hubble-de Vaucouleurs sequence (Laurikainen et al.
2007). Secular evolution can also explain the transformation of galaxy
morphological features other than bulges along this sequence.  An
important engine for the evolution in many cases appears to be the
presence of bars, ovals, and spiral structure in disk galaxies. The
ability of a bar or oval to drive gas to the central regions of
galaxies is often argued as the dominant process involved (Kormendy \&
Kennicutt 2004), although gas accretion alone is not adequate
to produce the significant mass-flow rate needed to transform a
galaxy's Hubble type from late to early, nor will it be able to produce
the observed stellar content of most bulges (Zhang 2003).

The rate at which a pattern rotates (the ``pattern speed") is an
important feature in secular evolutionary processes. The pattern speed,
together with the rotation curve of the parent galaxy, determines the
location of all wave-particle resonances, which in turn influences how
a wave interacts with the basic state of a galactic disk.\footnote{The
term basic state refers to the axisymmetric disk when the density-wave
modal density distribution is subtracted.} It had been argued by
Lynden-Bell \& Kalnajs (1972, hereafter LBK) that the wave/basic-state
interaction which loads and unloads the angular momentum onto the
density wave train for its outward transport can only happen at
wave-particle resonances for a quasi-steady wave pattern.  This angular
momentum transport is believed to be necessary for the generation of
spirals,\footnote{Thus, the reason for the title of the original LBK
paper ``On the Generating Mechanism of the Spiral Structure.''}, but
wave/basic-state interaction across the entire galactic disk, and the
role of this wide-spread interaction to the maintenance of a
quasi-steady spiral structure and to the long-term evolution of the
basic state, were not addressed in LBK.

Zhang (1996, 1998, 1999) showed that stars must also be involved in the
secular evolutionary process of a galaxy and furthermore should be a
major part of the mass redistribution process which grows bulges and
transforms galaxy morphologies. To enable the participation of stars,
which were traditionally thought to be adiabatic and non-dissipative,
required a significant breakthrough in our understanding of the physical
processes occurring in so-called self-organized large scale structures
of which spirals and bars are examples.  The new secular dynamical 
evolution mechanism discovered in Zhang (1996, 1998, 1999) operates 
mainly through spontaneously formed quasi-steady spiral and bar {\it modes} which,
when interacting with the basic state of the galactic disk, slowly
transforms a galaxy's morphology over its lifetime due to a collective dissipation
process\footnote{The ``collective dissipation" process in a galaxy
possessing a spontaneously formed density wave pattern leads to the ``graininess" effect of
the stellar potential, which breaks the conservation of the Jacobi
integral or the adiabatic condition of a single stellar orbit, and
enables the secular orbital decay or increase needed for secular
evolution. Since this process involves the coordinated motion of all
the individual stars on the disk, it could not be regarded as a kind of
``broadening of resonances" effect due to fluctuating wave amplitudes
because the latter effect is due to the
single orbit's {\em passive} response to an {\em applied}, smooth, large-scale
perturbation potential.  Other evidences that these two points of view
are incompatible are given later in the Discussion section of the
current paper.} enabled by an azimuthal phase-shift between
the density-wave modal density distribution and the potential
distribution implied by that density. In spite of the importance of the
potential-density phase-shift for studies of galactic evolution, the
topic is not discussed even in the latest edition of {\it Galactic
Dynamics} (Binney \& Tremaine 2008). S. Tremaine (private
communication, 2008) stated that the reason for this omission is that
the Zhang papers argue that there must be wave/basic-state interaction
all across the disk of a spiral galaxy, and not just at the resonances,
which is contrary to the widely-accepted claim of the LBK paper.
Without this wide-spread angular momentum exchange between the basic
state and the wave,
however, it is unlikely there could be the continuous mass flow and
basic state change needed to gradually transform galaxy morphologies
across the Hubble time\footnote{Note that what LBK derive, and the
Binney and Tremaine book (equation [6.56] or Appendix J) insists on, is
that the outward angular momentum flux is a constant across the disk
for a steady wave, and this conclusion is also consistent with (in fact
equivalent to) their other claim that the angular momentum exchange
between the wave and the basic state happens only at the wave-particle
resonances.  This means the wave does not pick up angular momentum from
most of the disk stars {\em en route} of its outward transport, and
therefore the stars over most of the disk area cannot lose/gain their
angular momentum and accrete/excrete.  These conclusions of LBK and
Binney and Tremaine were all derived for a steady {\it wave train}
under the tightly wrapped WKBJ approximation.  Zhang (1998) on the
other hand has shown that the {\em total} (gravitational {\em plus} advective) 
angular momentum flux for a spontaneously-formed, open spiral {\em mode} is NOT a
constant, but rather of the shape of a bell peaked at corotation. 
Thus the gradient of this flux, which is proportional to the angular
momentum exchange rate of the wave with the basic state, is no longer
zero across most of the area of the disk, which enables galaxy-wide mass accretion and excretion. 
Evidence for the bell-shaped angular momentum flux was also seen in the
Gnedin, Goodman \& Frei (1995) calculation for M100, though there only
the gravitational angular momentum flux contribution (or gravitational torque couple)
was calculated and not the advective contribution, so the shape of the total
angular momentum flux could not be inferred.  Plus, since these authors deliberately 
chose not to adopt any dynamical model for the interpretation of their 
results, Gnedin et al. (1995) made no comments (and apparently had no
realization) of the significance of this peak.  We have found recently,  
however, that the radial location of this peak, at about 135-140" (see
Figure 7 of Gnedin et al. 1995), coincides almost exactly with that
found by Zhang \& Buta (2007) for the outer CR location of this galaxy, (see Figure 8,
left frame of Zhang \& Buta 2007).  This near coincidence of our CR
location derived using the phase shift method, and the peak of the
bell-shaped gravitational torque couple derived independently by
Gnedin et al. is a strong support for the arguments presented in Zhang
(1998), which was also confirmed by the N-body simulations presented there.}.

In a previous paper (Zhang \& Buta 2007, hereafter paper I; see also
Zhang 1996), we showed that the theoretically-expected
potential-density phase-shifts for spiral/bar modes can be measured
from near-infrared images of galaxies; furthermore, the corotation
radii determined through the zero-crossings of the radial distribution
of phase-shift are arrived at without need of the measurement of actual
kinematics. The idea is that the sign of the phase-shift tells us in
what sense angular momentum is exchanged between a density wave mode
and the basic state of a galaxy, and the sign of the phase-shift has to
change across corotation because the sense of angular momentum density
itself changes sign across corotation. Without the mutual consistency
of these two sets of signs, a spiral/bar mode will not be able to
spontaneously emerge out of an originally featureless disk, as shown in
Zhang (1998).  We confirmed in paper I that if a single steady pattern
with a well-defined pattern speed actually exists in a galaxy disk,
then the radius where the phase-shift changes from positive to negative
appears to be the location of that pattern's corotation radius. If
instead multiple patterns with differing pattern speeds exist in a
single disk galaxy, then multiple positive-to-negative (P/N) crossings
would be seen marking the locations of corotation radii for the
individual modes, and the radii where patterns with different pattern
speeds decouple would be seen as negative-to-positive (N/P) crossings
-- once again using the sign of angular momentum density argument. N/P
crossings are further discussed in section 2.1.

In practice, we have to exercise considerable judgment in interpreting
phase-shift distributions. Our working assumption is that we are
dealing with quasi-steady wave modes. If the modes are not steady
(e.g., out of equilibrium, transient, or too young and unsettled) the
method will not give a reliable corotation radius. Nevertheless, we 
found that we can identify those galaxies for which our working assumption 
is likely to be incorrect, through signatures of the lack of pattern and 
phase shift coherence.  Pattern coherence and phase-shift-plot coherence seem to be
well correlated, and both of them usually tell the same story of
whether the pattern is likely to be steady or not, or else in the
process of decoupling.  Also, less steady patterns should introduce
more noise into the phase-shift plots.

In this paper, we apply the phase-shift method to a much larger sample
of galaxies than in paper I: the Ohio State University Bright Galaxy
Survey (OSUBGS; Eskridge et al. 2002). This survey includes all RC3 (de
Vaucouleurs et al. 1991) galaxies having a type in the range S0/a to Sm (or
stage index 0$\leq T \leq$9), total blue magnitude $B_T$$\leq$12.0,
blue light isophotal diameter $D_{25}$ $\leq$ 6\rlap{.}$^{\prime}$5,
and declination in the range $-80^{\circ} < \delta < +50^{\circ}$. The
whole OSUBGS has 205 galaxies, but our analysis is based on a subset of
153 OSUBGS galaxies having inclinations less than 65$^o$ (Buta,
Laurikainen, \& Salo 2004). Several close pairs were excluded from our
analysis. The images we used were obtained with the 1.65$\mu$m
$H$-band filter, and were deprojected by Laurikainen et al. (2004)
assuming the bulges are spherical. Our goals are to examine the
statistics of the density wave patterns in normal, massive galaxies,
and to get a general picture of the role these patterns are actually
playing on the morphological evolution of the disks of the galaxies. We
also seek to demonstrate the robustness of the phase-shift method by
examining the impact of deprojection uncertainties on the phase-shift
distributions, and by correlating the coherence of the phase-shift
distribution with the coherence of the density wave pattern itself.

\section{Comments on the Physical Basis of the Method and the Calculation Procedure}

A fair questioning of the validity of the potential-density phase-shift 
method is that it is used to determine a partially kinematic feature, i.e. 
the corotation radii, in the absence of any direct kinematic data. The reason 
this is possible is because there are two kinds of phase-shifts: one is
predicted by the Poisson equation after integrating the density, while
the other is predicted by the equations of motion. {\it For a
self-sustained mode, these two have to give a phase-shift distribution
across the galactic radii that
agree}, i.e., the density distribution and the velocity field
distribution have to achieve global self-consistency to lead to a single
phase-shift distribution. This means that
neither the density distribution nor the velocity field of a mode can
be of an arbitrary shape.  It is why we can predict a partially kinematic
quantity such as corotation radii from the density field alone, because {\it
in physical galaxies the requirement of global self-consistency has
presumably already been satisfied.}

The explicit demonstration that stellar orbital motion for a spiral
mode requires that the phase-shift changes sign at corotation is from
Zhang (1996, second Appendix). There the orbital solution itself 
diverges at exactly the corotation radius due to the passive-orbit
resonance condition. The fact that for collective modes the phase
shift actually goes through zero-crossing at corotation, as well as
changes sign, is demonstrated in Zhang (1998), where the relation of
the phase-shift to the flux of total angular momentum across the disk is
established. Since the angular momentum flux is of the shape of a bell
with the top of the bell at corotation, and the sine of the phase-shift
is shown to be proportional to the angular momentum flux gradient
(derivative), then the phase-shift distribution both changes sign and
goes through zero at CR because this is exactly the behavior of the
derivative of the bell-shaped angular momentum flux. 

Kalnajs (1971) showed that for a pure logarithmic spiral, the
phase-shift is actually a constant across all disk radii -- a
well-known result of the potential theory which has been reviewed in
the book by Snow (1952).  The fact that even though the potential
theory predicts a constant phase-shift value whereas a spiral {\it
mode} has a variable phase-shift, changing sign and crossing zero at
CR, is because the modal density distribution is not that of a single
logarithmic spiral of a constant trend of radial density falloff. The
modal shape is carefully balanced globally during the self-organization
process to allow all the required properties, including phase-shift
distributions, to be satisfied simultaneously. In fact, Zhang (1998)
showed that a phase-shift distribution that has such a changing shape
across the disk is the only possible one which allows the mode to
spontaneously emerge. No other phase-shift distribution would be
compatible with the mode creating itself from nothingness.


The phase-shift for any galaxy is defined by equation (3) in paper 1,
which was first obtained in Zhang (1996, 1998),
and this definition involves the integral of the product of the density
and the derivative of the gravitational potential. Our analysis begins
with a deprojected near-infrared image sampled into a square array with
dimension chosen a power of 2. This array is most often assumed to be
proportional to the density distribution for a constant mass-to-light
ratio (we have verified in Paper 1 that a variable mass-to-light ratio
does not significantly change the corotation radii determined by the
phase-shift method).  The density array is then used to calculate a
gravitational potential array using an approach similar to that of
Quillen, Frogel, \& Gonz\'alez (1994).  The phase-shift distribution
$\phi_o$ versus galactic radius is derived using these density and
potential square arrays re-sampled onto a polar grid, and the graph is analyzed
for P/N crossings. As noted in paper I, the phase-shift does not
``know" the sense of rotation of a galaxy, but we deduce this by noting
the sense of winding of the spiral pattern, S or Z, assuming the arms
are trailing (this is generally a safe assumption, but exceptions to
this rule do exist, such as the case NGC 4622 treated in Paper I).  
Furthermore, in practice we limit the phase-shift
analysis to $r < r_o(25) \equiv D_o(25)/2$, where $D_o(25)$ is the
extinction-corrected isophotal diameter at the surface brightness
$\mu_B$=25.0 mag arcsec$^{-2}$ from RC3.

The appearance of a phase-shift distribution can be smooth or noisy. A
noisy distribution does {\it not} necessarily result from a poor or
noisy image, but often appears to indicate the presence of unsteady
patterns. At large radii, image noise can also contribute. Ideally, we
visualize a smooth curve through the distribution. To evaluate the
significance of all suspected CR radii, we overplot circles on the
deprojected near-IR (and, if available, optical) images. We deduce
which radii are likely to be real CRs by comparing actual density wave
patterns like bars, ovals, and different spirals on the images with the
P/N crossing radii on the phase-shift plot. In cases of potential
multiple pattern speeds, we sometimes see in the phase-shift plots
evidence for ongoing pattern decoupling which doesn't have a clear P/N
crossing (or only a weak/shallow crossing) but which shows some form of
``bumpy'' feature, the location of which corresponds to an actual
morphological feature on the image.

\begin{figure}
\figurenum{1}
\plotone{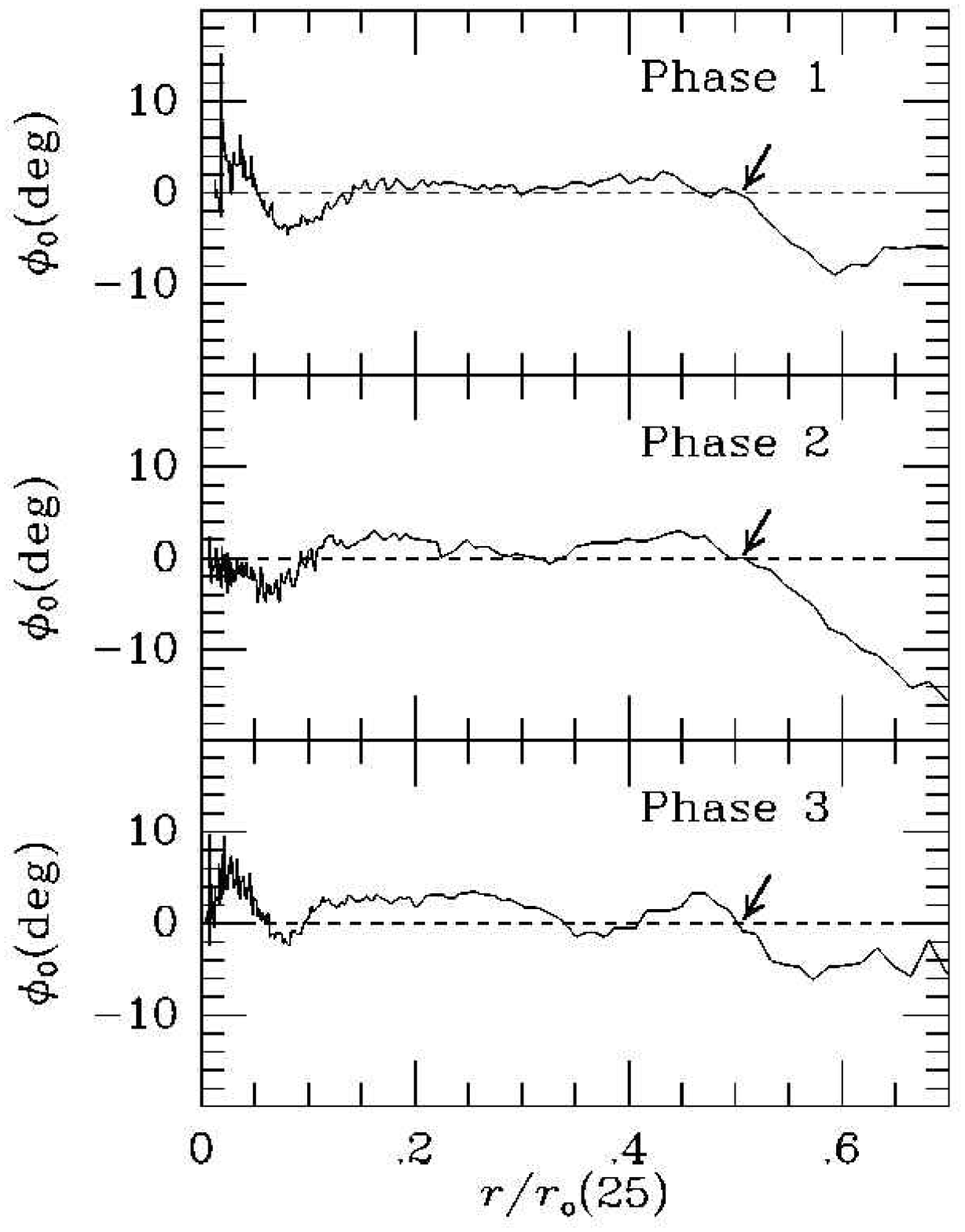}
\caption{}
\label{schematics} 
\end{figure}

The reason we might see ``near-crossings" for some features rather than
full crossings is that decoupling is likely to be a gradual process
that occurs in stages. We do not expect a strong P/N crossing until a
pattern has completely distinguished itself. In fact, we suggest that
the formation of multiple zero crossings in the phase-shift
distribution can be understood in terms of an evolutionary sequence,
such as is shown in Figure~\ref{schematics}. In Phase 1, a bar and
spiral share the same pattern speed and have a single CR which has been
arbitrarily set at $r/r_o(25)$=0.5 (arrow). Except for a small inner
pattern, the bar extends from $r/r_o(25)$$\approx$0.1 to the beginning
of the spiral at 0.5. Over time, the inner part of the bar begins to
decouple from the spiral, as shown in Phase 2. This leads to a
near-crossing at $r/r_o(25)$$\approx$0.3-0.4. Eventually, the inner
part of the bar decouples more completely from the spiral and a
well-defined crossing is seen (Phase 3), again at
$r/r_o(25)$$\approx$0.3-0.4. The three curves in
Figure~\ref{schematics} are based on NGC 7479 (Phase 1), NGC 613 (Phase
2), and NGC 4593 (Phase 3). The curves have all been scaled to give the
main CR at $r/r_o(25)$=0.5. Besides the possibility depicted here,
other pathways for the formation of nested patterns are
possible.

\subsection{N/P Crossings and Mode-Decoupling}

In the case of nested patterns, we have argued that N/P crossings
indicate where an inner pattern and an outer pattern decouple, i.e.,
where they start to have different pattern speeds.  For some galaxies,
there is evidence that these N/P crossings could extend as far as the
outer Lindblad resonance (OLR) of the inner mode, although this has not
been found to be universal. For a bar and a spiral, this seems at odds
with conventional thoughts on mode couplings, where the inner Lindblad
resonance (ILR) of an outer spiral mode coincides with the CR of an
inner bar mode (e.g., Massett \& Tagger 1987), although Rautiainen \&
Salo (1999) have shown that other matchings of resonance locations of
the inner and outer modes (such as CR-inner 4:1 or OLR-ILR) are
possible. The reason an N/P crossing could be at a location outside the
CR of the inner mode and possibly extend all the way to its OLR is that
the complete mode usually has negative angular momentum density inside
CR, and positive angular momentum density outside CR until the OLR is
reached.  A positive phase-shift distribution supports the growth of
the portion of the mode inside CR, and a negative phase-shift
distribution supports the growth of the portion of the mode outside CR
(both these phase-shift senses allow the torquing of the density by the
potential in the right direction to remove/inject angular momentum
from/to the mode to enhance its growth).  The negative portion of the
phase-shift distribution thus definitely belongs to the inner mode and
not the outer one, and it ought to have the same pattern speed as the
inner one. Whether it extends all the way to the OLR of the mode or not
is found to vary for individual galaxies, and this should be studied in detail
in galaxies where kinematic data as well as good images are both available.
But in principle, if a mode sustains itself between its ILR and OLR,
the phase-shift distribution should be one-positive/one-negative hump
and the corresponding radial angular momentum flux is shaped like a
bell.

\section{Phase-Shift Distributions}

We have examined the phase-shift distributions for 153 OSUBGS
galaxies (see Eskridge et al. 2002 for details about the survey) based
on the publicly-available $H$-band (1.65$\mu$m) images. The deprojected
versions of these images were kindly made available to us by E.
Laurikainen (see Laurikainen et al. 2004). The images were prepared by
cleaning the originals of foreground and background objects, assessing
the sky background subtraction, fitting ellipses to optical isophotes
where possible and matching these fits to the $H$-band, using
two-dimensional decomposition to characterize the bulges and derive
disk radial scale-lengths, and finally deprojecting the galaxies using
mean orientation parameters from the ellipse fits and assuming a
spherical shape for the bulge. The latter assumption is incorrect in
many galaxies and when applied causes light to be oversubtracted from
the minor axis.  This leads to what we refer to as ``decomposition
pinch" or ``spherical bulge pinch"
in the inner regions of the deprojected images. We found that
this has some effect on our analysis, especially for the most inclined
galaxies in the sample. To minimize its impact, we use what we call an
``average bulge image", that is, an image based on an average of the
deprojected image we get assuming a spherical bulge and the
deprojected image we get assuming the bulge is as flat as the disk.
Most bulges have a flattening intermediate between these two extremes,
and this approach provides a better approximation than either extreme
does. This is discussed in more detail in section 5.2.

For each galaxy, we illustrate the phase-shift distribution versus the
normalized radius, $r/r_o$(25), as given in the left-hand-side plots of
Figures 2.1 to 2.153.  On each plot, slanted arrows indicate phase-shift
zero crossings we believe to be significant. There are some zero
crossings that we believe to be due to image noise, or are otherwise
insignificant due to the non-steady nature of the patterns. These we
have not compiled. Also, almost half the plots show negative phase
differences just outside a crossing having
$r$$\approx$0$^{\prime\prime}$, with no corresponding inner positive
phase differences, possibly indicating nuclear patterns unresolved
by our current images/analyses. We have also not compiled most of these. The
significant zero crossings are taken to be corotation radii and all are
compiled in Table 1 in arcseconds. For some galaxies, only a single
crossing is found, while for others we can identify as many as five
likely significant crossings. In general, we gave a code to each
crossing as a judgment of its significance: code 3 crossings are taken
to be most significant. Code 1 crossings are less secure, and we tended
to ignore any crossings close to the outer $r_o$ unless we could identify an
actual pattern at such radii. Table 1 also summarizes some of the
scatter in crossing positions. Sharp crossings have no error indicated,
while for noisy crossings, the number under the column ``err" is a
standard deviation of several close crossings. The reason we cannot
assign a general error bar for our approach (even for sharp crossings)
is the difficulty of judging (quantitatively) how self-consistent and
how steady a given mode is. We can only judge by eye how coherent the
pattern and the phase-shift plot are, and how well the features
correspond to one another.  However, one trend we do find is that for
very straight bars, because of the smallness of the resulting phase
shift values, the absolute values of the determined CR locations
are more prone to deprojection error than those for more open or
skewed patterns.

Table 1 also includes judgments of the bar corotation radii, indicated
by an asterisk following the value. In some cases, an alternative
interpretation is indicated by the value in boldface. These choices are
discussed further in section 5.1.

To the right of each phase-shift plot, we show the deprojected $H$-band
image of the galaxy with red circles overlaid showing how the
corotation radii relate to features seen in the galaxies. Code 3 radii
are shown by full red circles while code 1 radii are shown by dotted
circles. In several cases, a significant N/P crossing is indicated by a
green circle. In many cases, we can see clear relationships to observed
patterns, such as bars or different spiral patterns. These are
discussed individually for each galaxy.

Almost all of our phase-shift distributions are calculated using the
Laurikainen et al. (2004) deprojected images, Fourier-smoothed by 21
terms to reduce some of the image noise (Buta et al. 2005). The phase
shift calculations are made using a numerical procedure originally
developed in Zhang (1996), and adapted to the current project during
the work on Paper I.

\section{Description of Individual Galaxies}

{\it Note}: radii discussed are given in arcseconds, but to facilitate
comparisons with the phase-shift plots, the normalized radius
is also given in parentheses.

\begin{figure}                                                                  
\figurenum{2.1}
\plotone{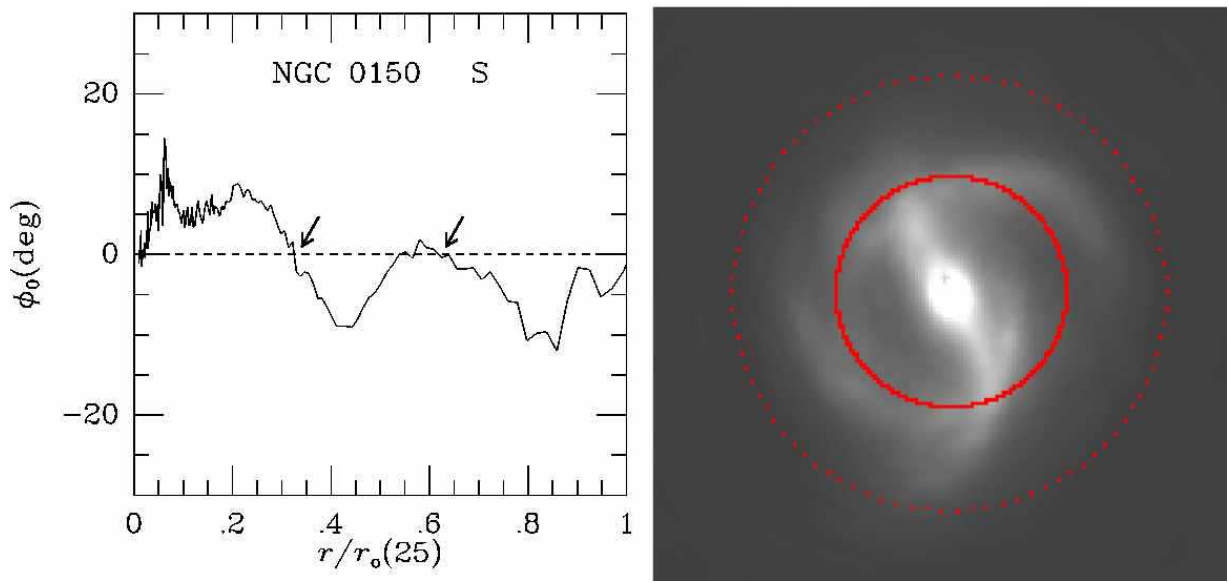}
\vspace{2.0truecm}                                                              
\caption{(left) Phase-shift distribution and (right) 21-term                    
Fourier-smoothed $H$-band image with overlays at CR positions (red              
circles) for NGC 150. In the phase-shift plot, the radius is normalized         
to the ``face-on," extinction-corrected isophotal radius,                       
$r_o(25)=D_o/2$, at $\mu_B$=25.00 mag arcsec$^{-2}$, from RC3. The              
arrows indicate corotation radii from major positive-to-negative                
crossings. These radii are listed in arcseconds in Table 1. In the              
image, the overlaid circles show the arrowed radii.                             
Solid red circles indicate likely more reliable crossings than                  
dotted circles.} \label{ngc0150} \end{figure}

NGC 150 (Figure~\ref{ngc0150}) - A galaxy with a well-defined bar and
spiral pattern.  The best defined P/N crossing is at $r$=38\arcs\ (0.33,
solid red circle), which mostly encircles the bar ends. A second
crossing at $r$=72\arcs\ (0.62, dotted red circle) is less certain and
could signify a decoupling of the main spiral from the bar. Also, at
$r$=13\arcs\ (0.11), the inflection in the positive phase-shift
indicates the decoupling of an inner pattern (the bar-like feature in
the image) in progress.  We show in section 5.2 that the ``spike" seen
at $r/r_o$=0.06 is likely due to decomposition pinch (see section 3).

\begin{figure}                                                                  
\figurenum{2.2}
\plotone{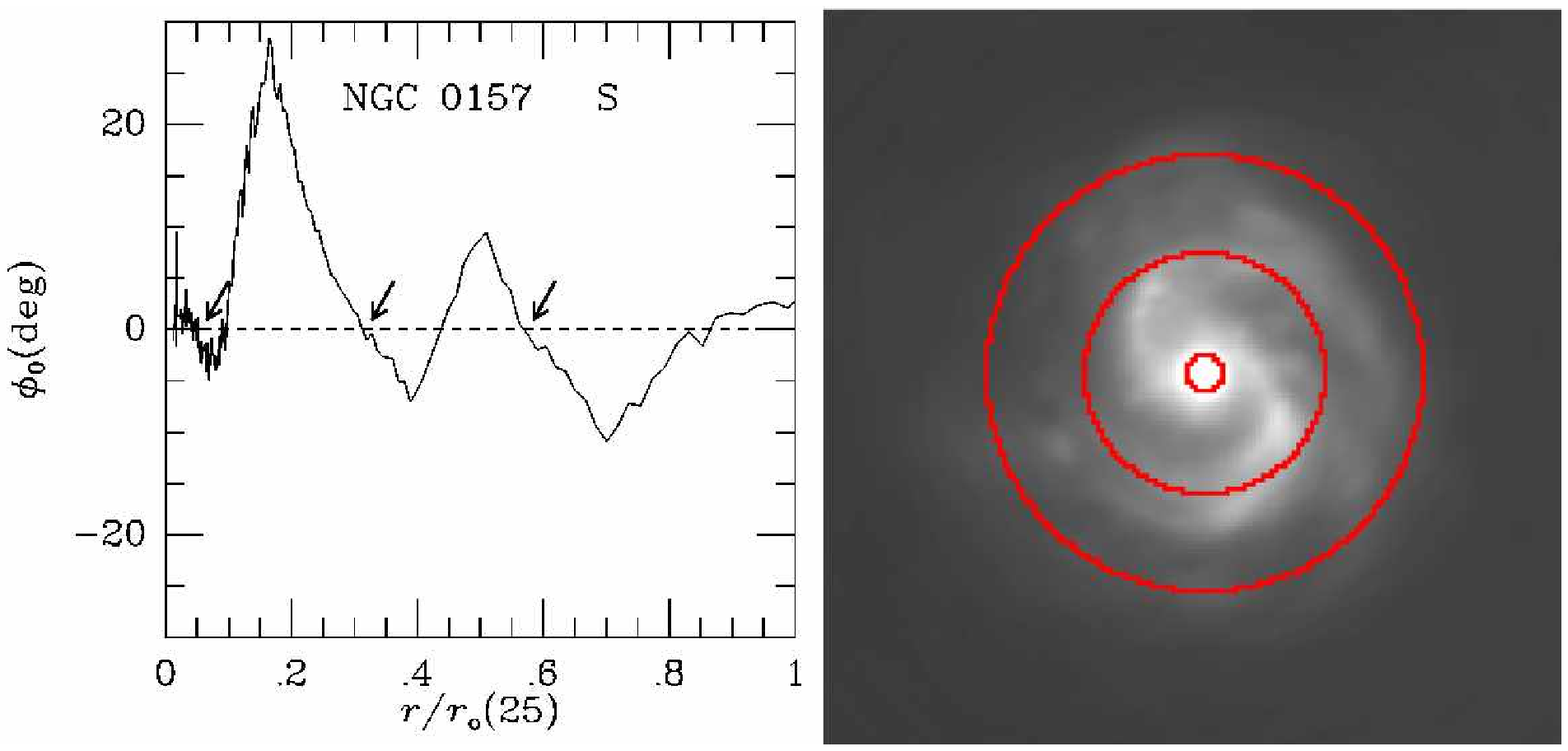}
\vspace{2.0truecm}                                                              
\caption{Same as Figure 2.1 for NGC 157.}                                         
\label{ngc0157}                                                                 
\end{figure}

NGC 157 (Figure~\ref{ngc0157})- The phase-shift distribution 
shows three well-defined P/N crossings,
suggesting all patterns are fully decoupled and have reached quasi-steady state.
The CR for the main inner pattern is at $r$=40\arcs\ (0.31) and for the outer pattern
is at $r$=73\arcs\ (0.57). The graph in Figure~\ref{fourier_plots} shows that
these radii each correspond to a peak in the amplitude of the $m$=2
component of the $H$-band light distribution. For comparison, the dynamical
model of Sempere \& Rozas (1997) gave a single corotation radius of 50\arcs,
intermediate between our two values.

\begin{figure}                                                                  
\figurenum{2.3}
\plotone{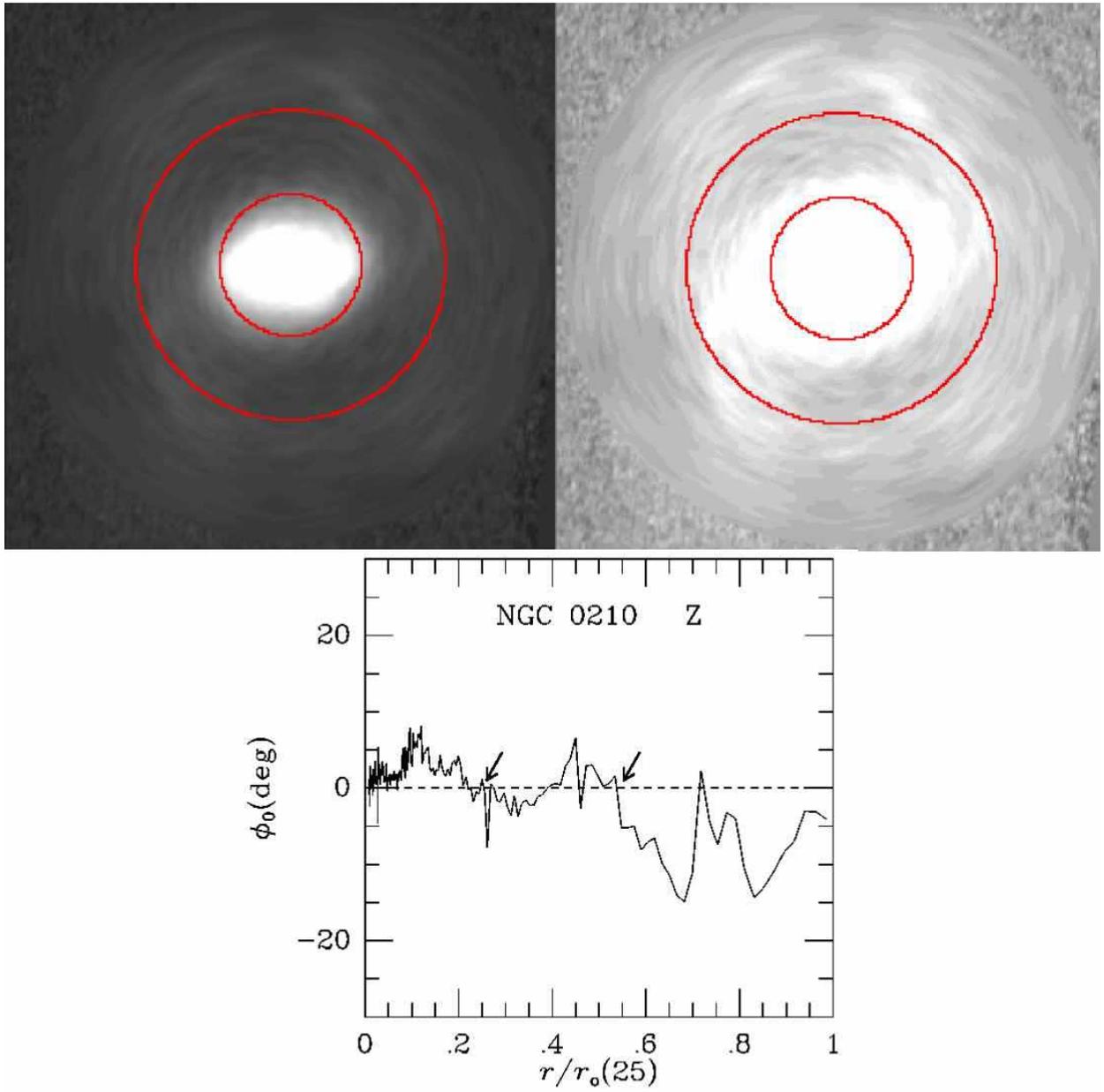}
\vspace{2.0truecm}                                                              
\caption{Same as Figure 2.1 for NGC 210.}                                         
\label{ngc0210}                                                                 
\end{figure}

NGC 210 (Figure~\ref{ngc0210})- Classified in the de Vaucouleurs
Atlas of Galaxies (deVA, Buta et al. 2007) as 
(R$_2^{\prime}$)SAB(s)b, the implication
is that NGC 210 has an OLR subclass feature in its outer regions. In the
$H$-band, the outer pseudoring is mainly an R$_1$ component. The 
phase-shift distribution is complicated but we can identify several features.
First, there is 
a likely decoupling inner pattern at $r$=6\arcss 4 (0.04). In fact, in blue light
there is a nuclear feature of approximately this size (Buta \& Crocker 1993).
The main nonaxisymmetric
feature in NGC 210 is a large oval which is associated with
two crossings: one at 
37\arcs\ $\pm$4\arcs\ (0.25), and a second crossing at 81\arcs\ (0.54). There is
isophote twisting in this feature, which could be the result of two pattern
speeds or else continuous pattern shearing. The larger crossing
appears to circle through the lower density regions perpendicular to the
oval axis, and from this CR two spiral arms appear to be emanating.
This seems to support the idea of an OLR subclass feature since the outer
pseudoring lies outside the larger CR. 
Neither crossing is cleanly-defined, which might be an indication of
residual on-going decoupling and pattern shearing.

\begin{figure}                                                                 
\figurenum{2.4}
\plotone{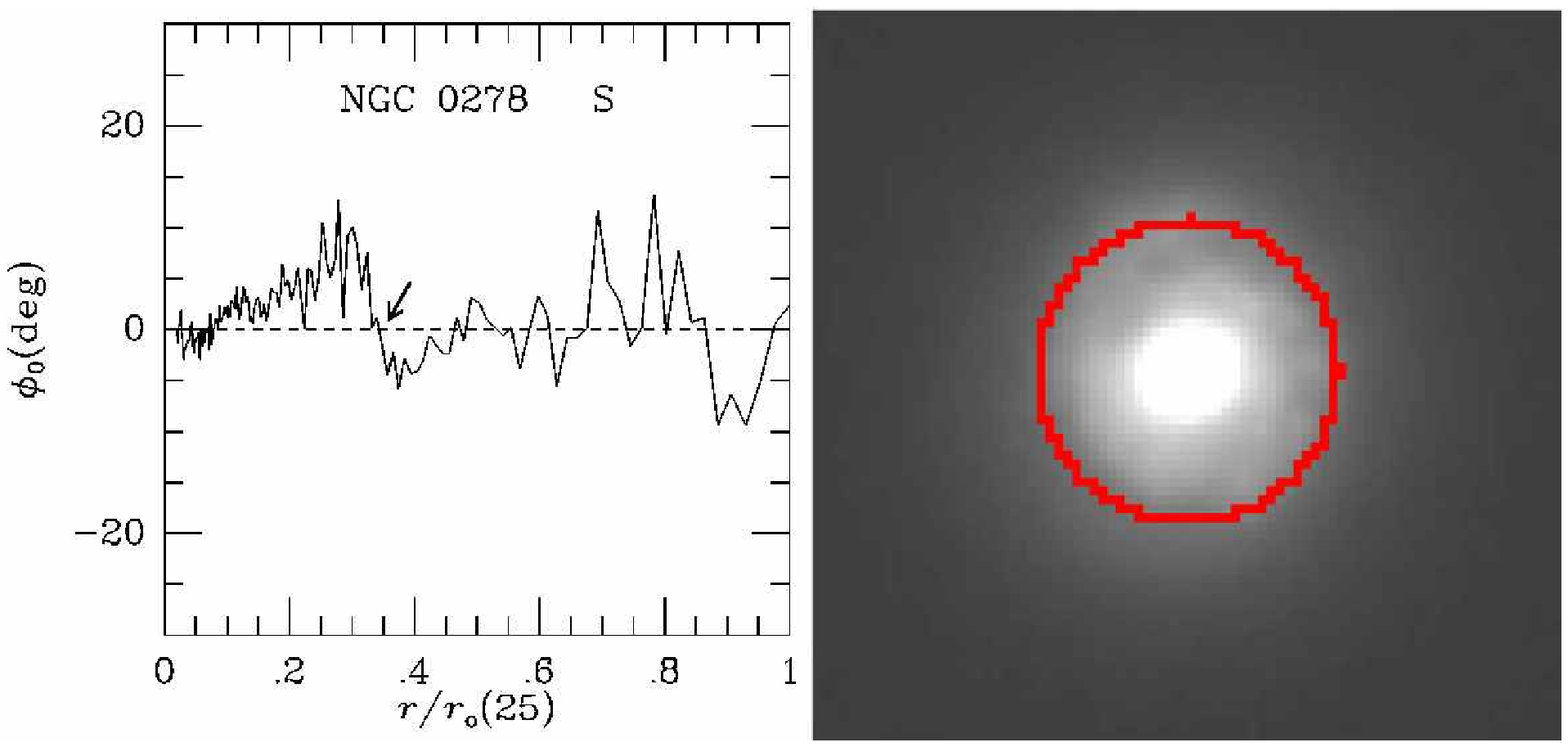}
 \vspace{2.0truecm}                                                             
\caption{Same as Figure 2.1 for NGC 0278}                                         
\label{ngc0278}                                                                 
 \end{figure}

NGC 278 (Figure~\ref{ngc0278}) - This high surface brightness galaxy has an inner spiral
pattern that has been interpreted as a nuclear ring (Knapen et al. 2004). We
find a well-defined P/N crossing at $r$=25\arcs\ (0.34) that just encompasses
this feature.

\begin{figure}                                                                  
\figurenum{2.5}
\plotone{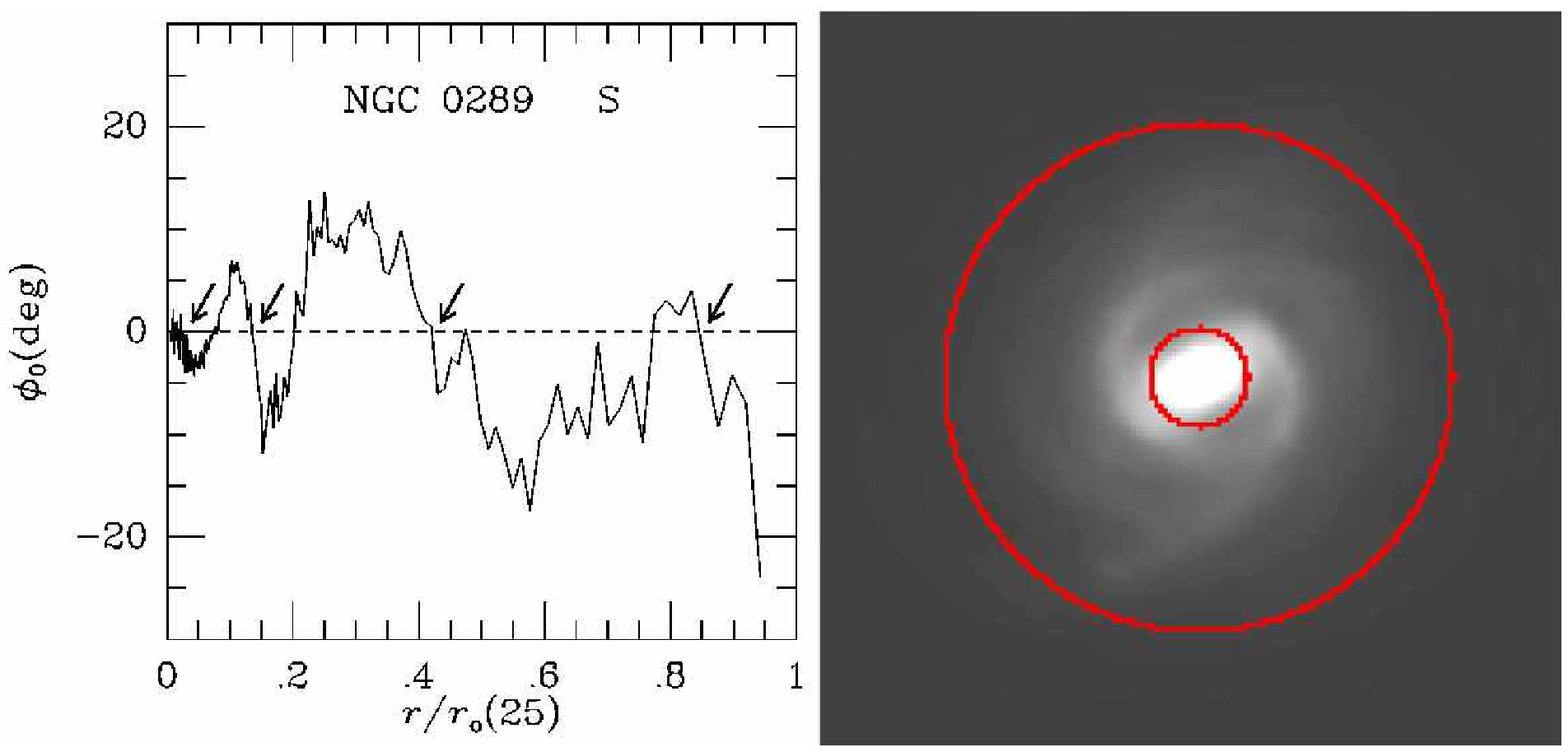}
\vspace{2.0truecm}                                                              
\caption{Same as Figure 2.1 for NGC 289.                                          
Only CR$_2$ and CR$_3$ from Table 1 are                                         
are shown overlaid on the image.}                                               
\label{ngc0289}                                                                 
\end{figure}

NGC 289 (Figure~\ref{ngc0289}) - The complicated phase-shift plot shows several P/N crossings.
The first, at $r$$\approx$4\arcs\ (0.03), is ill-defined, but we believe it is
real because there is a prominent N/P crossing at $r$$\approx$12\arcs\ (0.08).
The overlay image shows that the crossing at $r$=21\arcs\ (0.13) encircles the
ends of an inner oval. Another major crossing near $r$=66\arcs\ (0.42)
signifies a decoupled outer spiral pattern. This galaxy has extensive
spiral structure at large radii (deVA), which could account for the
outermost crossing near $r_o$. 

Rautiainen et al. (2008; hereafter RSL08; see also Rautiainen et al. 2005, hereafter
RSL05) obtained a fairly good representation of the inner
spiral structure of NGC 289 for a pattern speed placing $r(CR)$ at 61\arcs\
(0.39), which is very close to our third CR. Their estimated bar radius of
23\arcss 8 (0.15) is very close to our second CR. Thus, it is likely that
Rautiainen et al. have derived the spiral pattern speed and not the bar
pattern speed.

\begin{figure}                                                                  
\figurenum{2.6}
\plotone{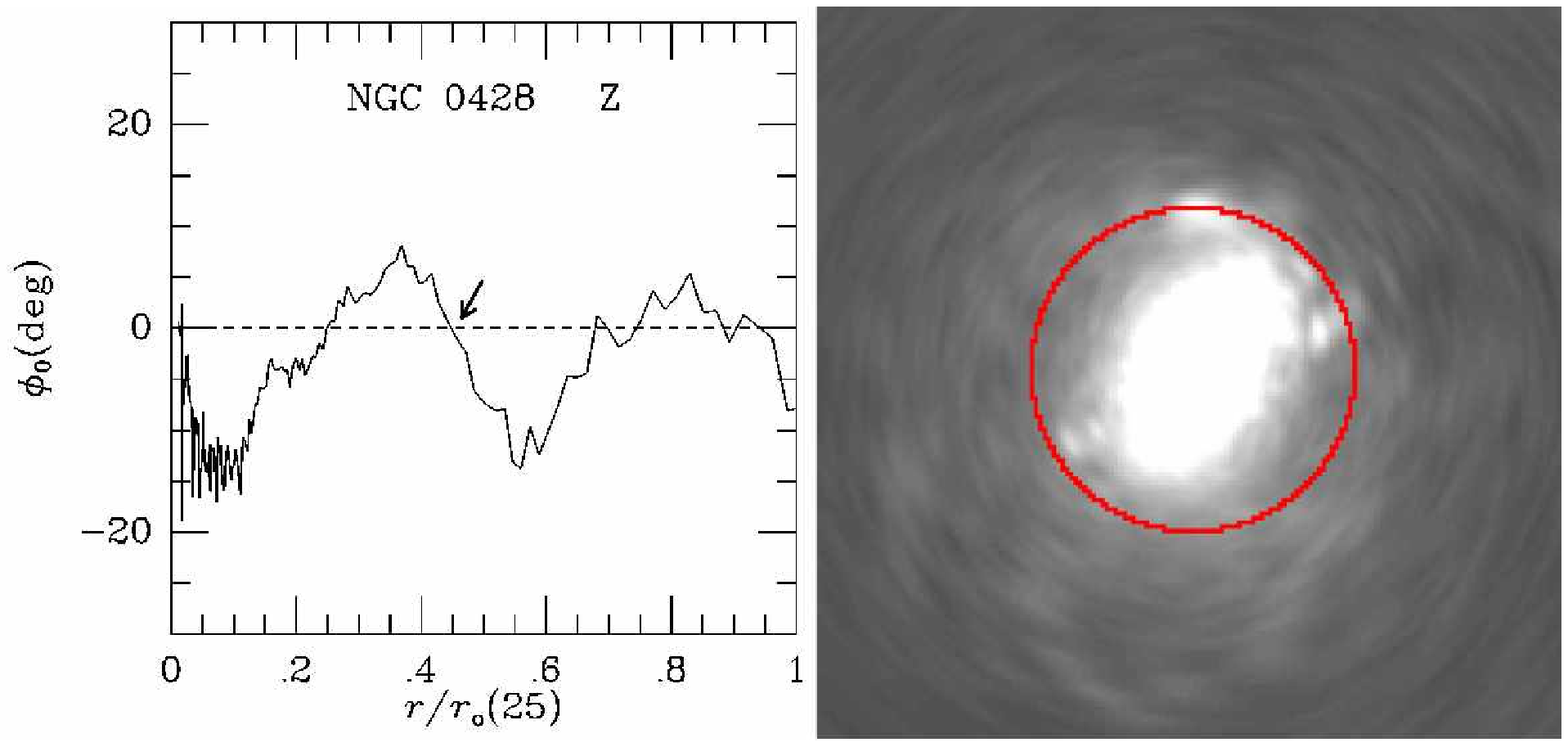}
\vspace{2.0truecm}                                                              
\caption{Same as Figure 2.1 for NGC 428.}                                         
\label{ngc0428}                                                                 
\end{figure}

NGC 428 (Figure~\ref{ngc0428}) - A very late-type galaxy whose sense of unwinding is 
uncertain but which appears to be ``Z" in optical images. The phase-shift
distribution shows a well-defined crossing at $r$=55\arcs\ (0.45) that circles
around the ends of an oval bar feature.

\begin{figure}                                                                  
\figurenum{2.7}
\plotone{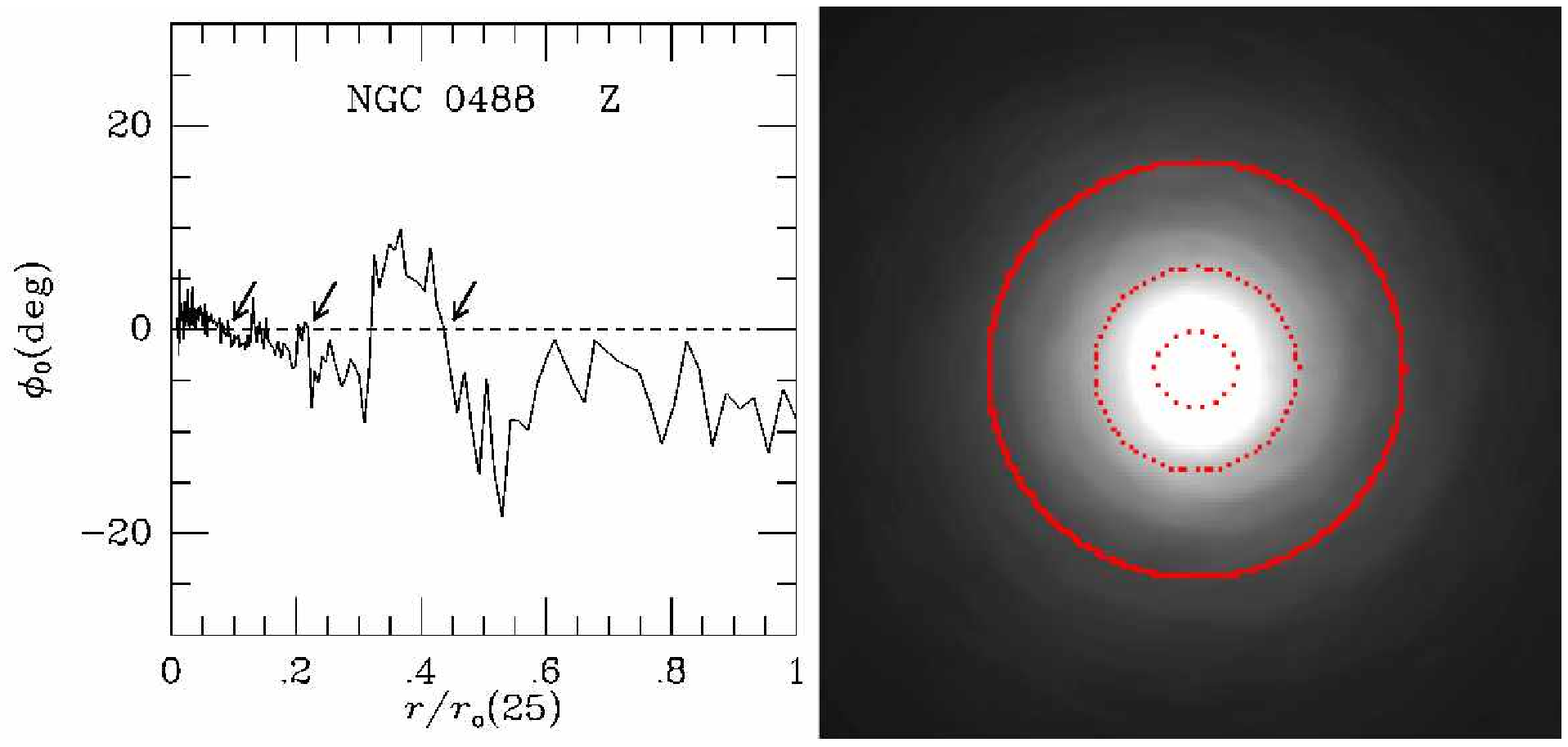}
\vspace{2.0truecm}                                                              
\caption{Same as Figure 2.1 for NGC 488.}                                         
\label{ngc0488}                                                                 
\end{figure}

NGC 488 (Figure~\ref{ngc0488}) - 
there is a well-defined crossing at $r$=70\arcs\ (0.43) that
could be associated with a tightly-wrapped spiral. The crossing at
$r$=34\arcs\ (0.21) could be associated with a weak inner oval. There
is also the evidence of a decoupled nuclear pattern.  Other crossings
are of uncertain significance and could be an indication of on-going
pattern decoupling.

\begin{figure}                                                                  
\figurenum{2.8}
\plotone{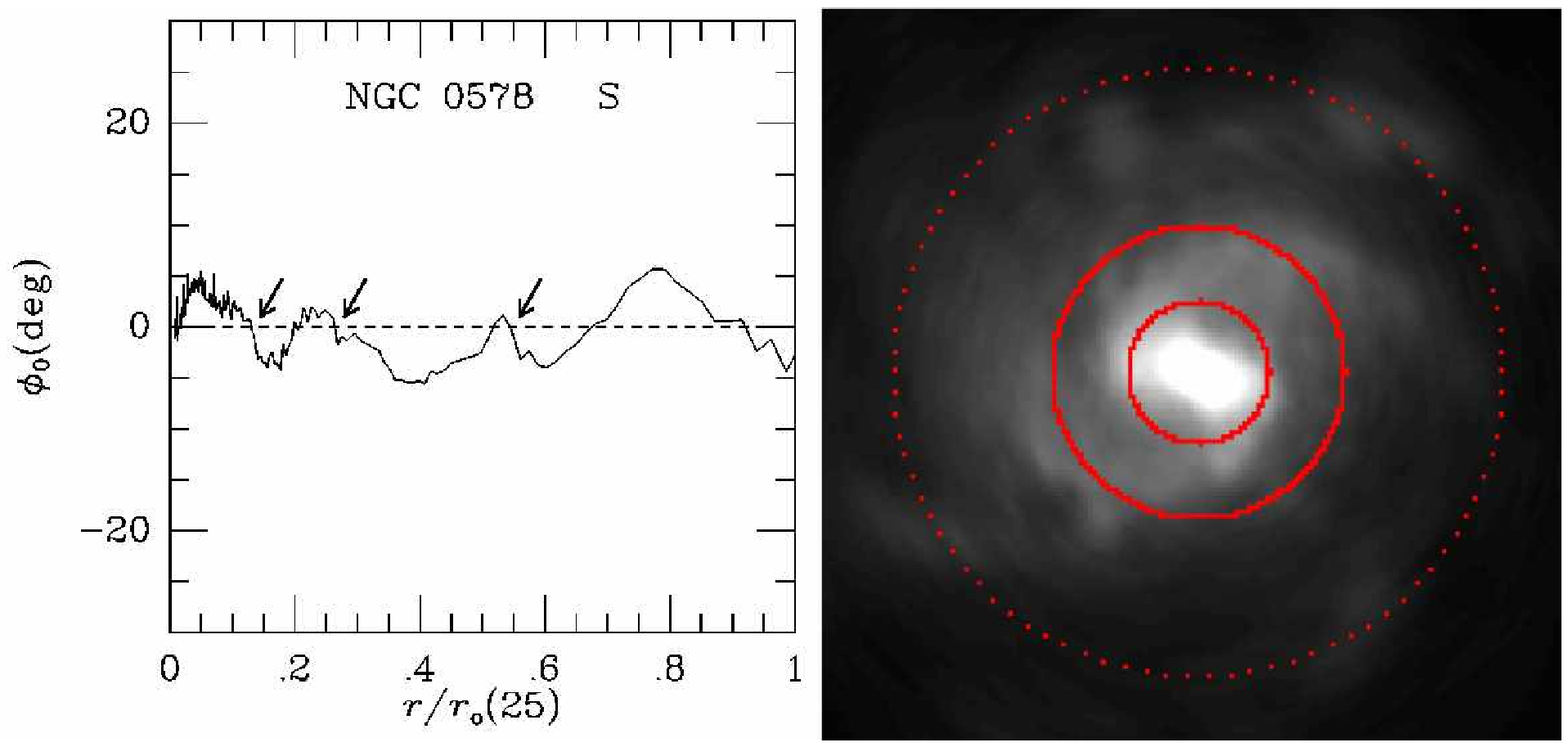}
\vspace{2.0truecm}                                                              
\caption{Same as Figure 2.1 for NGC 578.}                                         
\label{ngc0578}                                                                 
\end{figure}

NGC 578 (Figure~\ref{ngc0578}) - The two innermost P/N crossings for this galaxy are most likely
to be significant. The first, at $r$=19\arcs\ (0.13), encircles just beyond the ends
of the bar of the galaxy. The second, at $r$=39\arcs\ (0.27), is associated with
a bright inner spiral. The crossings at $r$=80\arcs\ (0.54) and 135\arcs\ (0.92)
are uncertain because
they occur either near the RC3 isophotal radius or are associated with an
indistinct pattern of large patches.  

The RSL08 survey gave $r(CR)$=78\arcss 9 (0.54), which
corresponds almost exactly with our third CR. Their bar radius of 23\arcs\ 
(0.16) is close to our first CR.

\begin{figure}                                                                  
\figurenum{2.9}
\plotone{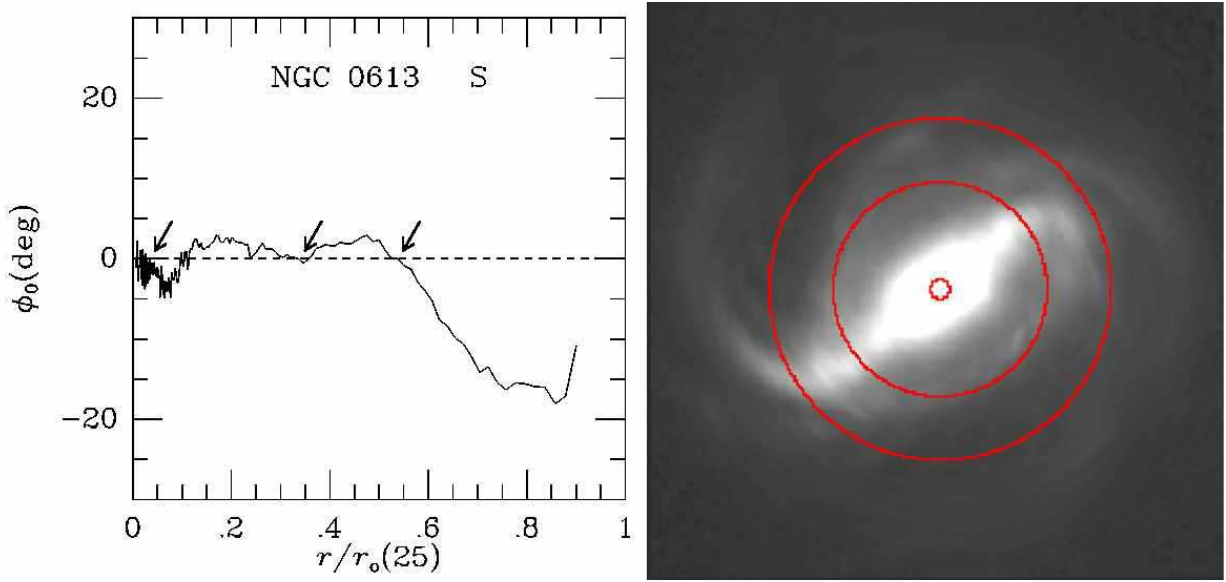}
\vspace{2.0truecm}                                                              
\caption{Same as Figure 2.1 for NGC 613.}                                         
\label{ngc0613}                                                                 
\end{figure}

NGC 613 (Figure~\ref{ngc0613}) - We believe this galaxy is a strong
case of a pattern evolving from a bar-driven spiral to a decoupled
inner bar and outer spiral. The phase-shift plot shows first of all
that there is a pattern in the central regions that is almost decoupled
from the outer regions. In the near-IR image, there is in fact a small
central bar nearly aligned with the main bar in this region. This
feature is also interpreted as a nuclear ring by Mazzuca et al. (2008.)
Well outside this area are two more crossings, the one at
$r$=55\arcs\ (0.33) being shallow while the one at $r$=88\arcs\ (0.53)
is well-defined. The shallowness of crossing 2 is what suggests on-going
decoupling. The $r$=55\arcs\ crossing encircles the inner part of the
apparent bar, while the $r$=88\arcs\ crossing encircles the whole bar
and part of the main spiral pattern. We suspect the galaxy started out
as a pure spiral, and that the inner part of the bar could be speeding
up and decoupling from the spiral. Other evidence in support of the
decoupling hypothesis for this galaxy include the spur spiral arms
between the two CRs which are disconnected from the outer arms.  Since
the different segments of the pattern may not be steady, the pattern
speeds inferred from the CR radii for the two patterns should be
understood as average values.

RSL08 were unable to get a very good fit to the structure
of this galaxy with a single pattern speed. Their value of $r(CR)$=126\arcs\
(0.76) is much larger than any of our CR radii. 

\begin{figure}                                                                  
\figurenum{2.10}
\plotone{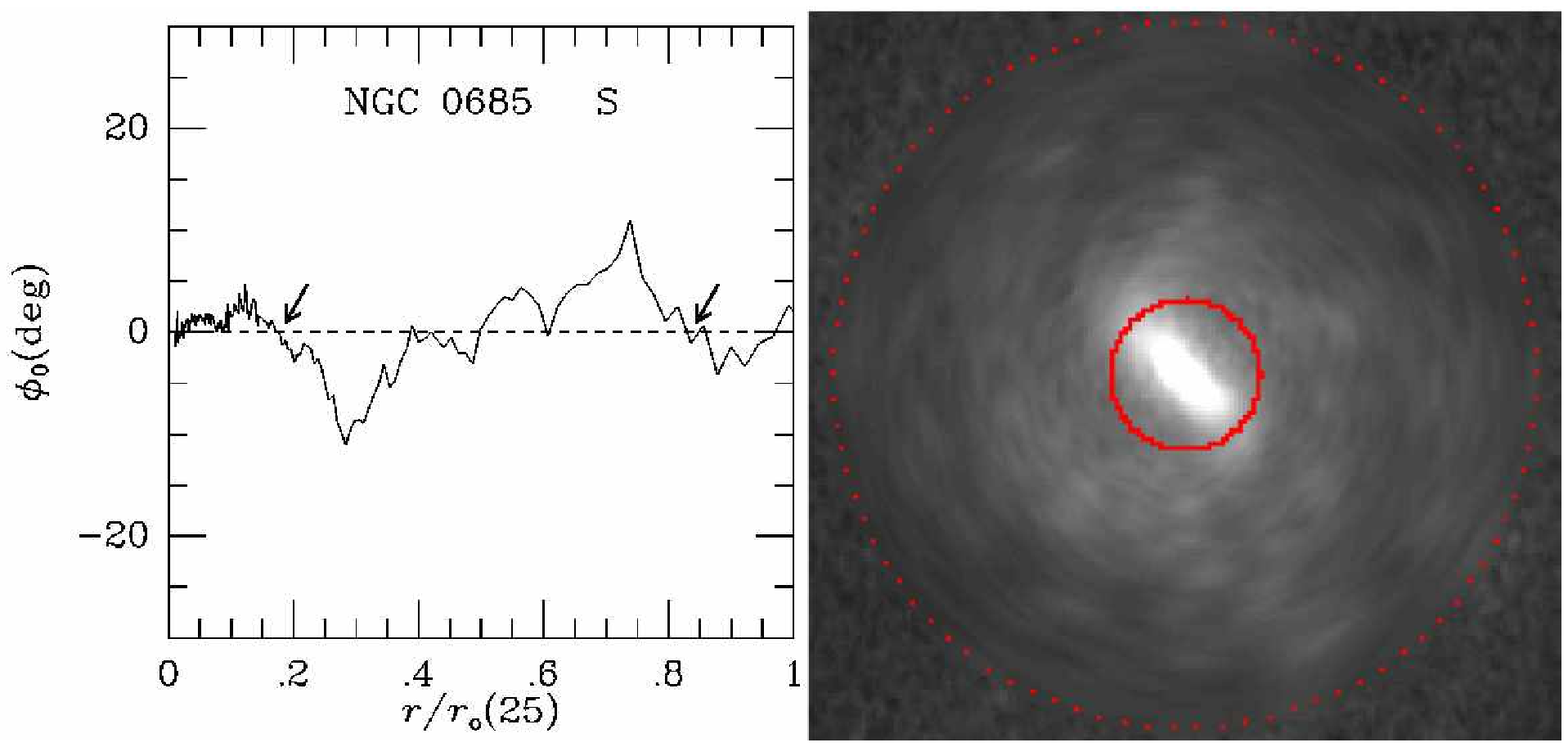}
\vspace{2.0truecm}                                                              
\caption{Same as Figure 2.1 for NGC 685.}                                         
\label{ngc0685}                                                                 
\end{figure}

NGC 685 (Figure~\ref{ngc0685}) - 
The main crossing at $r$=19\arcs\ (0.17) appears to circle the ends of the
bar. The pattern is ill-defined beyond the bar ends and none of the 
crossings in the outer region may be significant. 

\begin{figure}                                                                  
\figurenum{2.11}
\plotone{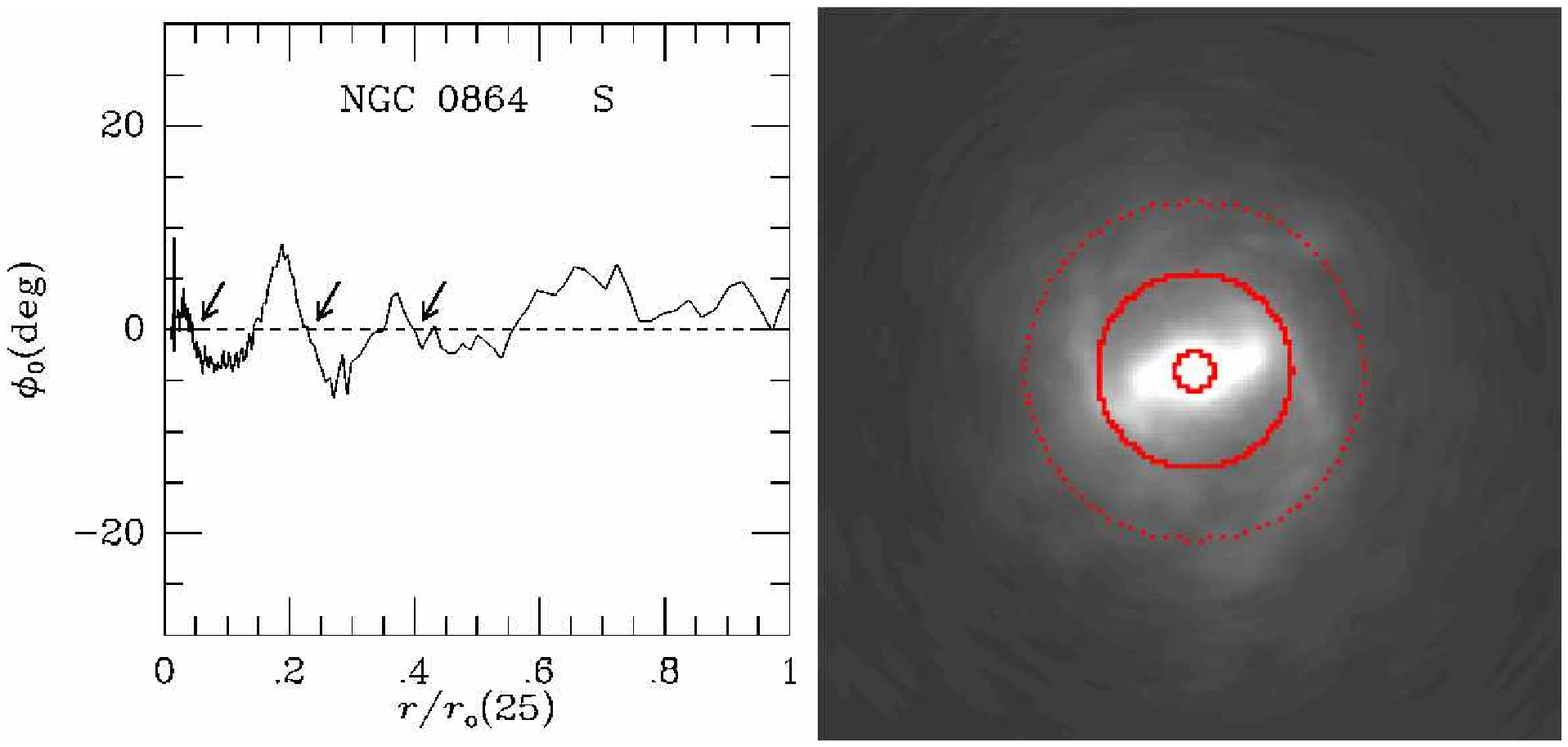}
\vspace{2.0truecm}                                                              
\caption{Same as Figure 2.1 for NGC 864.}                                         
\label{ngc0864}                                                                 
\end{figure}

NGC 864 (Figure~\ref{ngc0864}) - The galaxy shows three well-defined
P/N crossings within $r_o$.  The innermost circle lies within the bar
while the next crossing circles slightly outside the bar ends. The
third crossing lies within the inner spiral arms. All three could be
fully decoupled patterns each with a distinct pattern speed. Although
we have selected the second CR in Table 1 as the bar CR (asterisk), the
negative phase-shifts in the bar region suggest that the inner crossing
could be the bar CR.

\begin{figure}                                                                  
\figurenum{2.12}
\plotone{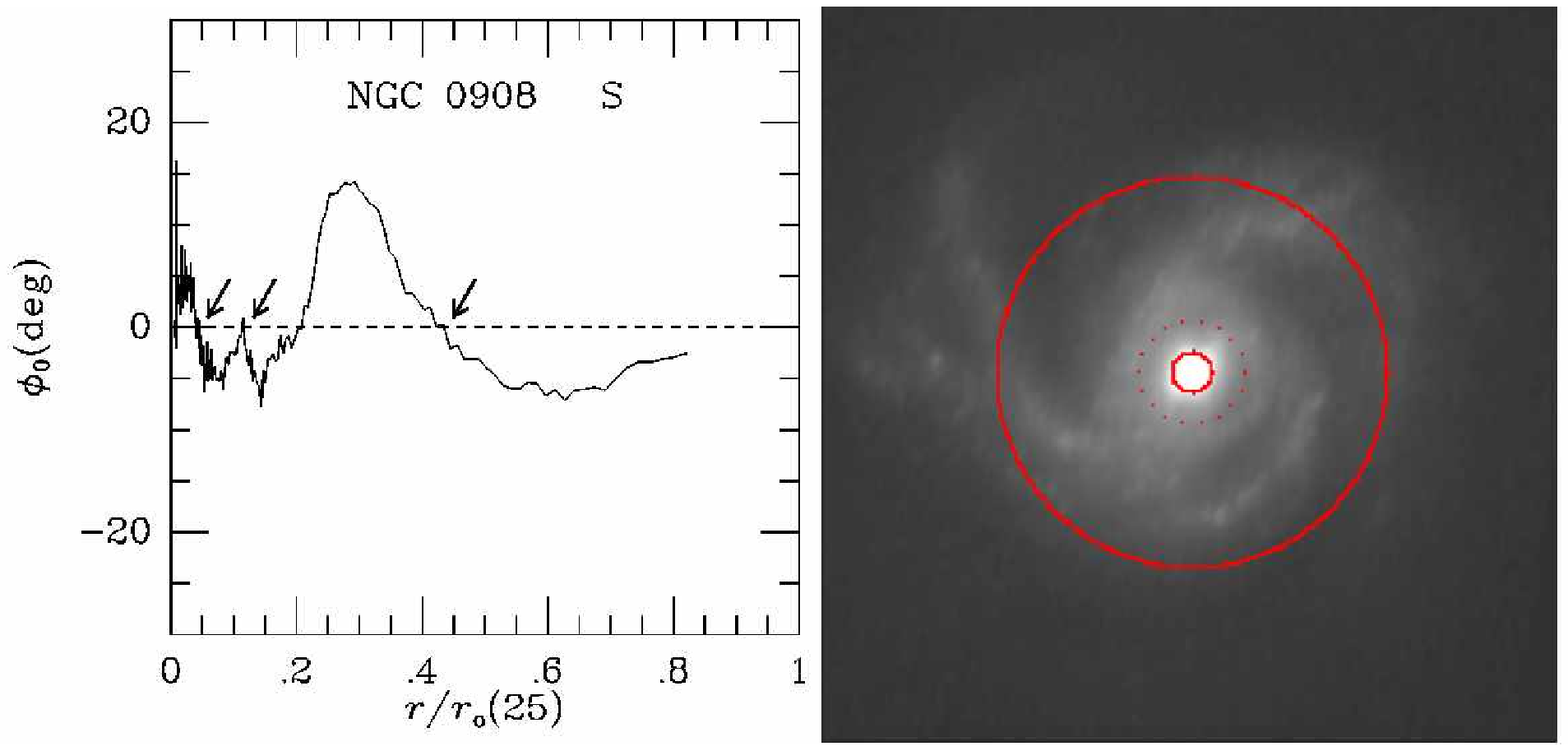}
\vspace{2.0truecm}                                                              
\caption{Same as Figure 2.1 for NGC 908}                                          
\label{ngc0908}                                                                 
\end{figure}

NGC 908 (Figure~\ref{ngc0908}) - The assumption of a perfectly spherical
bulge in this case leads to strong decomposition pinch. Hence,
we use the average bulge image (section 3)
for our analysis. The phase-shift distribution for this
grand-design spiral shows one strong outer P/N crossing and two possibly
significant inner crossings that are not necessarily artifacts of 
residual pinch.  In particular, the second crossing,
if significant, could indicate another on-going
decoupling pattern. The main crossing at $r$=79\arcs\ (0.44)
lies in the middle of the spiral pattern and just inside the peak
of the $m$=2 Fourier intensity amplitudes (Figure~\ref{fourier_plots}).

\begin{figure}                                                                  
\figurenum{2.13}
\plotone{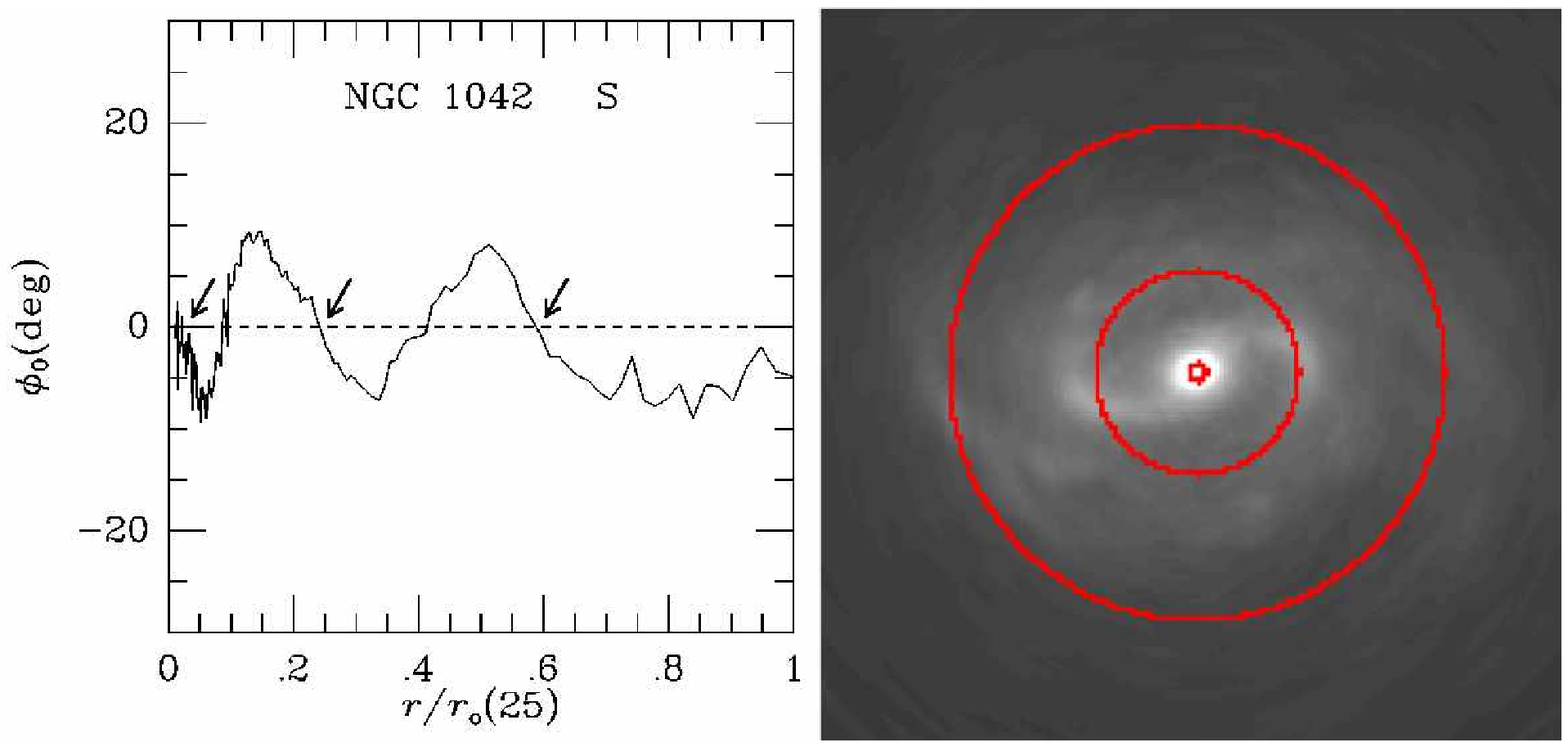}
\vspace{2.0truecm}                                                              
\caption{Same as Figure 2.1 for NGC 1042.}                                        
\label{ngc1042}                                                                 
\end{figure}

NGC 1042 (Figure~\ref{ngc1042}) - The strong N/P crossing at $r$$\approx$14\arcs\ (0.10) suggests there
is an inner pattern, but the P/N crossing associated with it is not very well defined.
The two larger crossings are much better defined. The innermost of these
encircles around the very open inner spiral, which is almost barlike. The
third circle lies within the outer arms. 

\begin{figure}                                                                  
\figurenum{2.14}
\plotone{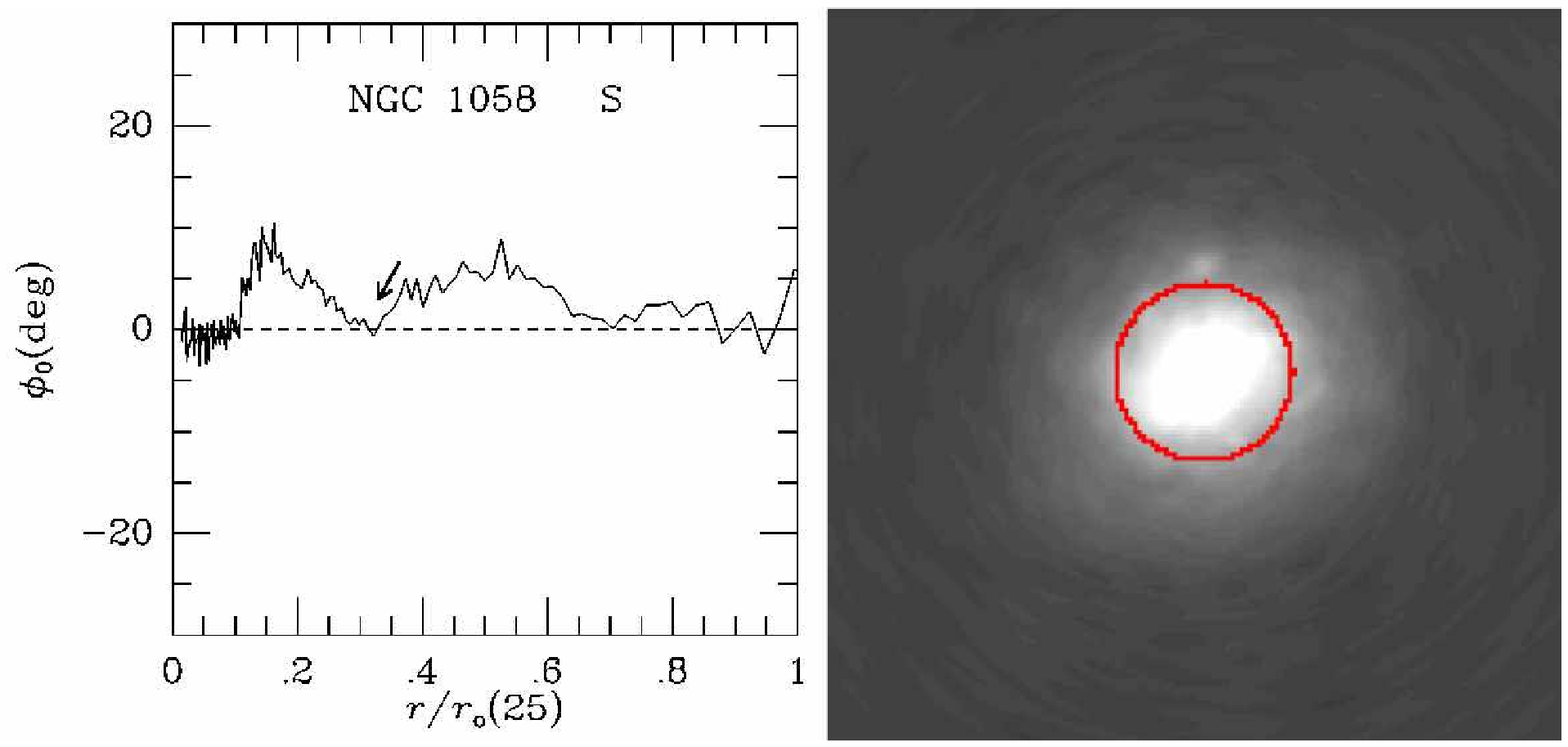}
\vspace{2.0truecm}                                                              
\caption{Same as Figure 2.1 for NGC 1058.}                                        
\label{ngc1058}                                                                 
\end{figure}

NGC 1058 (Figure~\ref{ngc1058}) - The shallow crossing at $r$=30\arcs\ (0.31)
encircles the ends of a weak inner oval.

\begin{figure}                                                                  
\figurenum{2.15}
\plotone{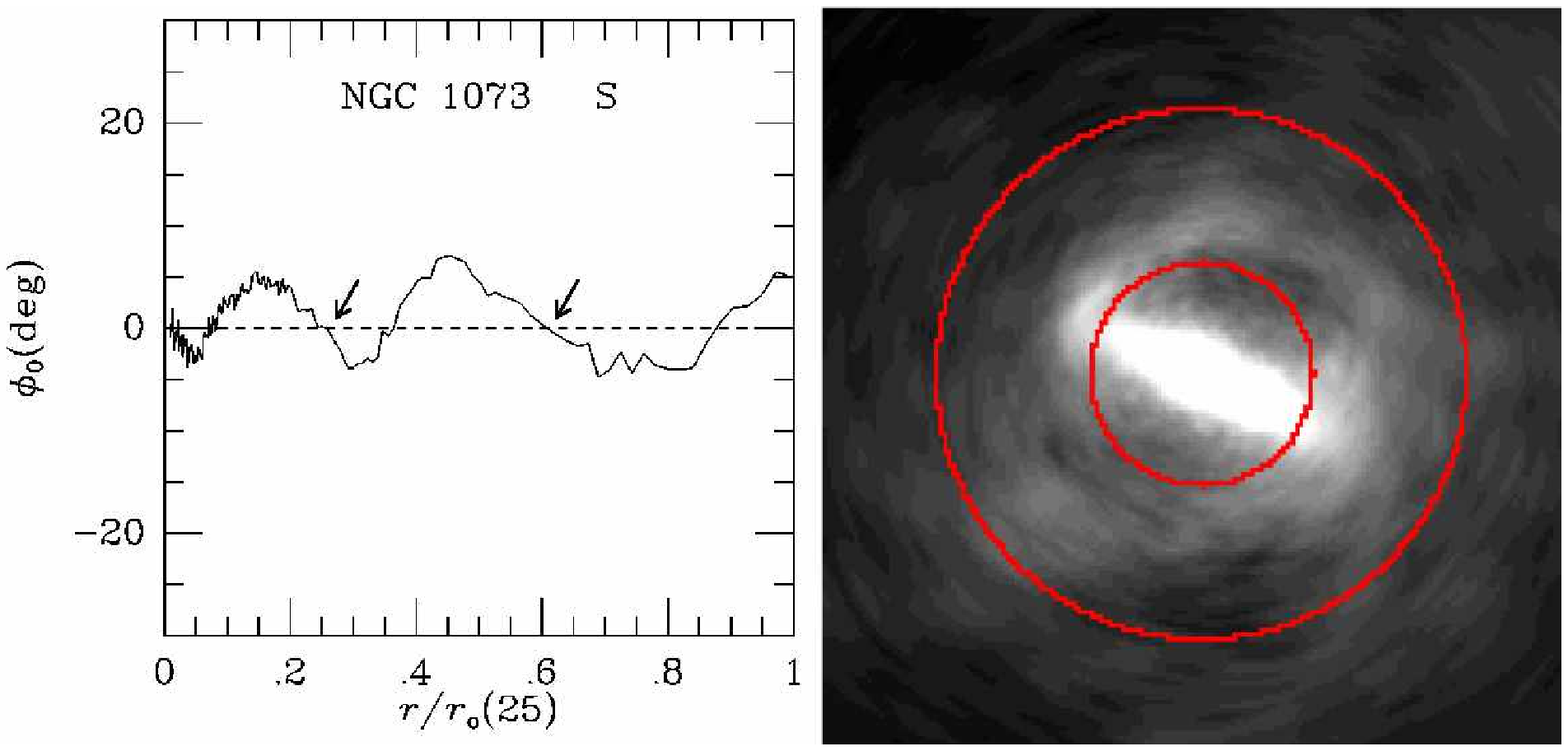}
\vspace{2.0truecm}                                                              
\caption{Same as Figure 2.1 for NGC 1073.}                                        
\label{ngc1073}                                                                 
\end{figure}

NGC 1073 (Figure~\ref{ngc1073}) - The phase-shift distribution for this
galaxy, already described in paper 1, shows two well-defined crossings.
The one at $r$=38\arcs\ (0.26) lies just inside the ends of the prominent
bar. In fact, the bar ends very near the radius of a strong N/P
crossing, suggesting that the bar extends to near its OLR. The crossing
at $r$=89\arcs\ (0.61) lies within the outer spiral arms. 

Interestingly, the best-fitting simulation of RSL08 gave
$r(CR)$=48\arcss 7 (0.33), intermediate between our two CRs but closer
to the inner one. Consistent with our results, these authors get $r(CR)/r(bar)$
$<$ 1. The best-fitting simulation, however, fails to represent the spiral
structure well. 

\begin{figure}                                                                  
\figurenum{2.16}
\plotone{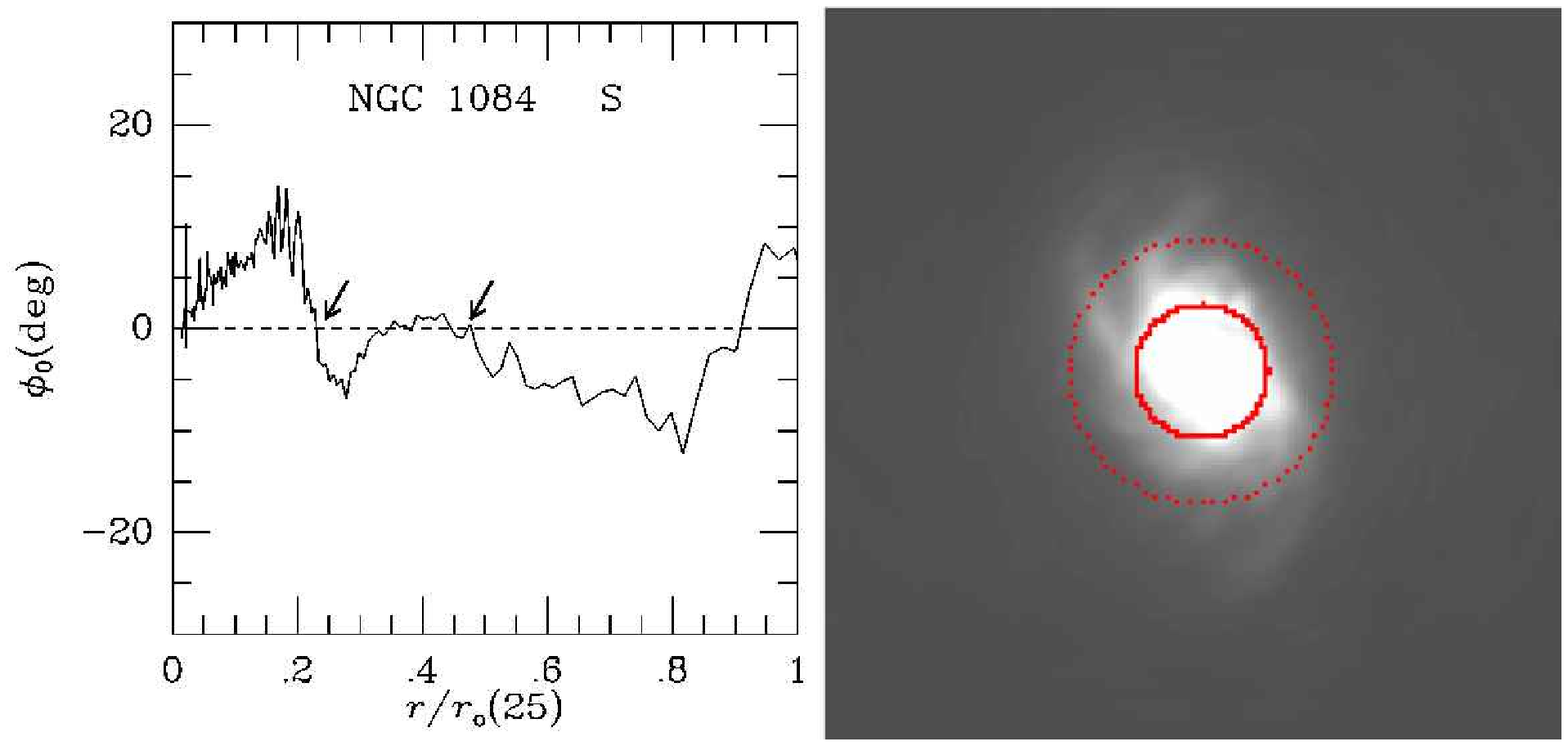}
\vspace{2.0truecm}                                                              
\caption{Same as Figure 2.1 for NGC 1084.}                                        
\label{ngc1084}                                                                 
\end{figure}

NGC 1084 (Figure~\ref{ngc1084}) - Both crossings seem to be associated with the
bright inner structure. The strongest crossing at $r$=22\arcs\ (0.23)
may be associated with an inner oval.

\begin{figure}                                                                  
\figurenum{2.17}
\plotone{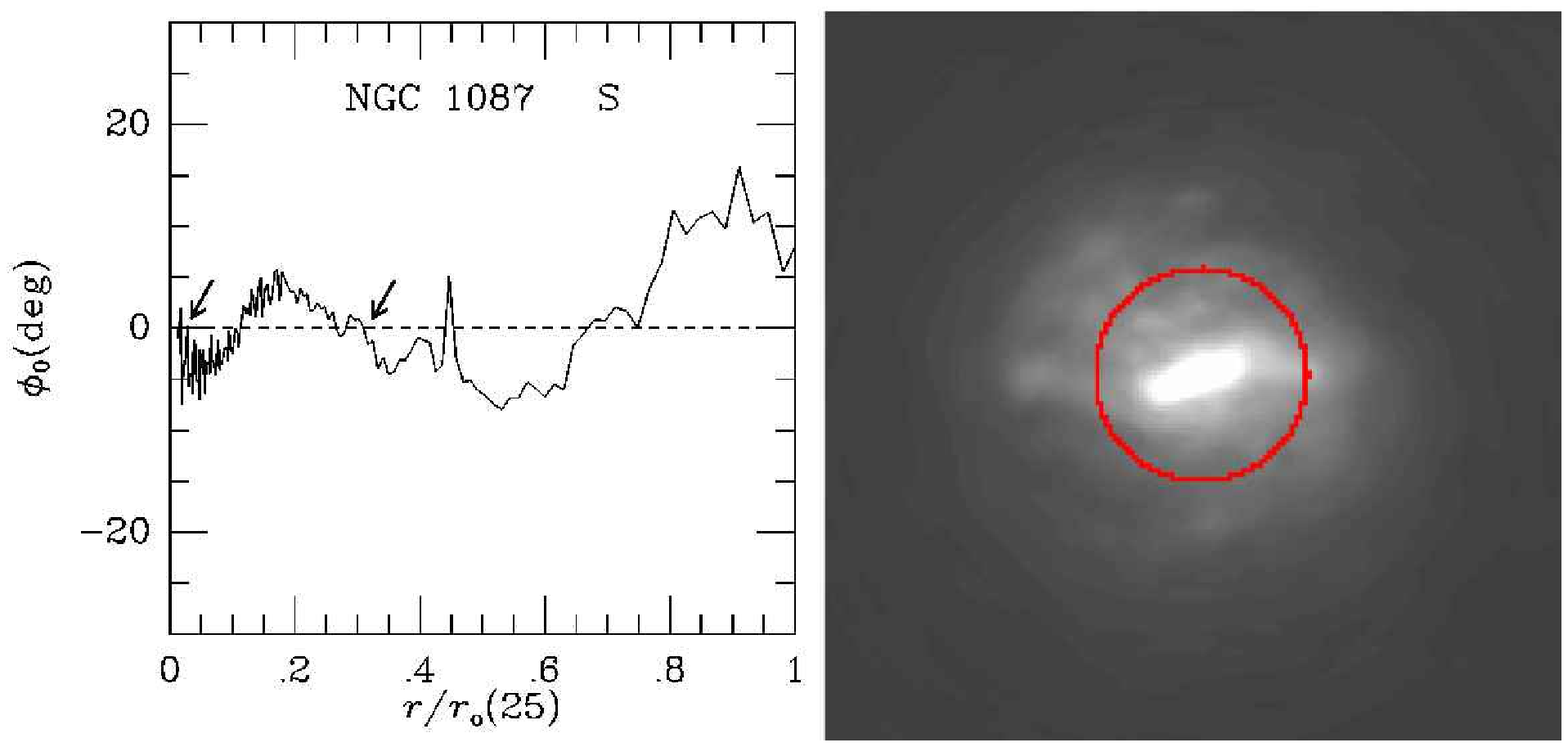}
\vspace{2.0truecm}                                                              
\caption{Same as Figure 2.1 for NGC 1087. Only CR$_2$ from                        
Table 1 is shown overlaid on the image.}                                        
\label{ngc1087}                                                                 
\end{figure}

NGC 1087 (Figure~\ref{ngc1087}) - The main P/N crossing at $r$=35\arcs\ (0.31)
lies well beyond the ends of the prominent (but asymmetric) bar.  There is the
indication of a nuclear
bar but the crossing is not well defined.

\begin{figure}                                                                  
\figurenum{2.18}
\plotone{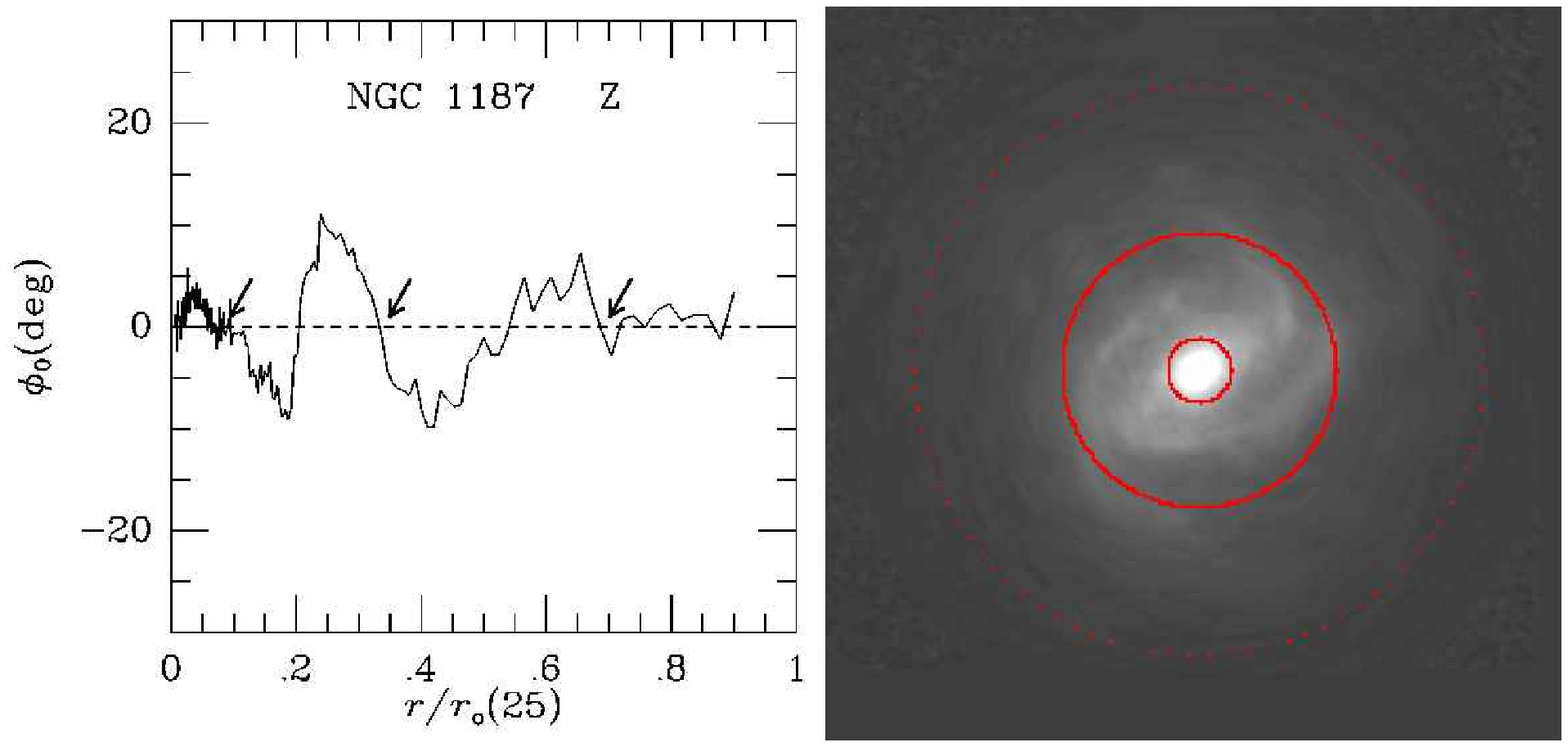}
\vspace{2.0truecm}                                                              
\caption{Same as Figure 2.1 for NGC 1187.}                                        
\label{ngc1187}                                                                 
\end{figure}

NGC 1187 (Figure~\ref{ngc1187}) - The main crossing at
$r$=55\arcs\ (0.33) lies within the bright inner spiral pattern, while
the crossing near $r_o$ may be associated with the outer spiral
pattern. The innermost crossing lies within the weak-looking bar, and
encircles an inner oval. This suggests that CR$_1$ in Table 1 is the
actual corotation radius for the bar of this galaxy.

The best-fitting RSL08 simulation of this galaxy gave
$r(CR)$=74\arcss 1 (0.45), intermediate between our two larger CR radii. 
Although the simulation reproduces the inner arms fairly well, the outer
spiral is likely too complicated to be described by a single pattern 
speed.

\begin{figure}                                                                  
\figurenum{2.19}
\plotone{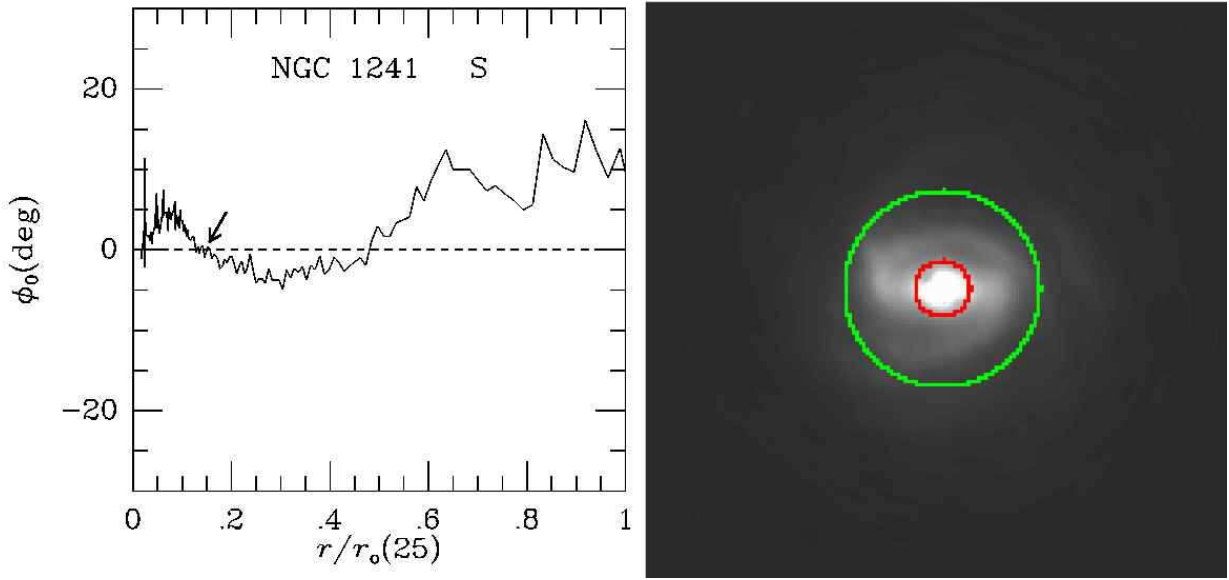}
\vspace{2.0truecm}                                                              
\caption{Same as Figure 2.1 for NGC 1241. The green circle                        
shows the well-defined N/P crossing near                                        
$r/r_o(25)$$\approx$0.5.}                                                       
\label{ngc1241}                                                                 
\end{figure}

NGC 1241 (Figure~\ref{ngc1241}) - This bright barred spiral shows only
one significant P/N crossing within $r_o$. This CR lies within the ends
of the bar, and appears associated with a small inner pattern. The
prominent N/P crossing is shown as the green circle in
Figure~\ref{ngc1241}. indicating that the entire pattern may end at close to its
OLR. There is, however, some indication of a local peak in the negative
portion of the phase-shift plot near $r/r_o = 0.24$, which, if real, could signify
an ongoing mode-decoupling process.  Furthermore, there is the indication 
of some decomposition pinch in the central regions.

An RSL08  simulation which does not represent the spiral 
structure of this galaxy well gives $r(CR)$ = 41\arcss 5 (0.47). This
value is surprisingly close to the N/P crossing we find. 

\begin{figure}                                                                  
\figurenum{2.20}
\plotone{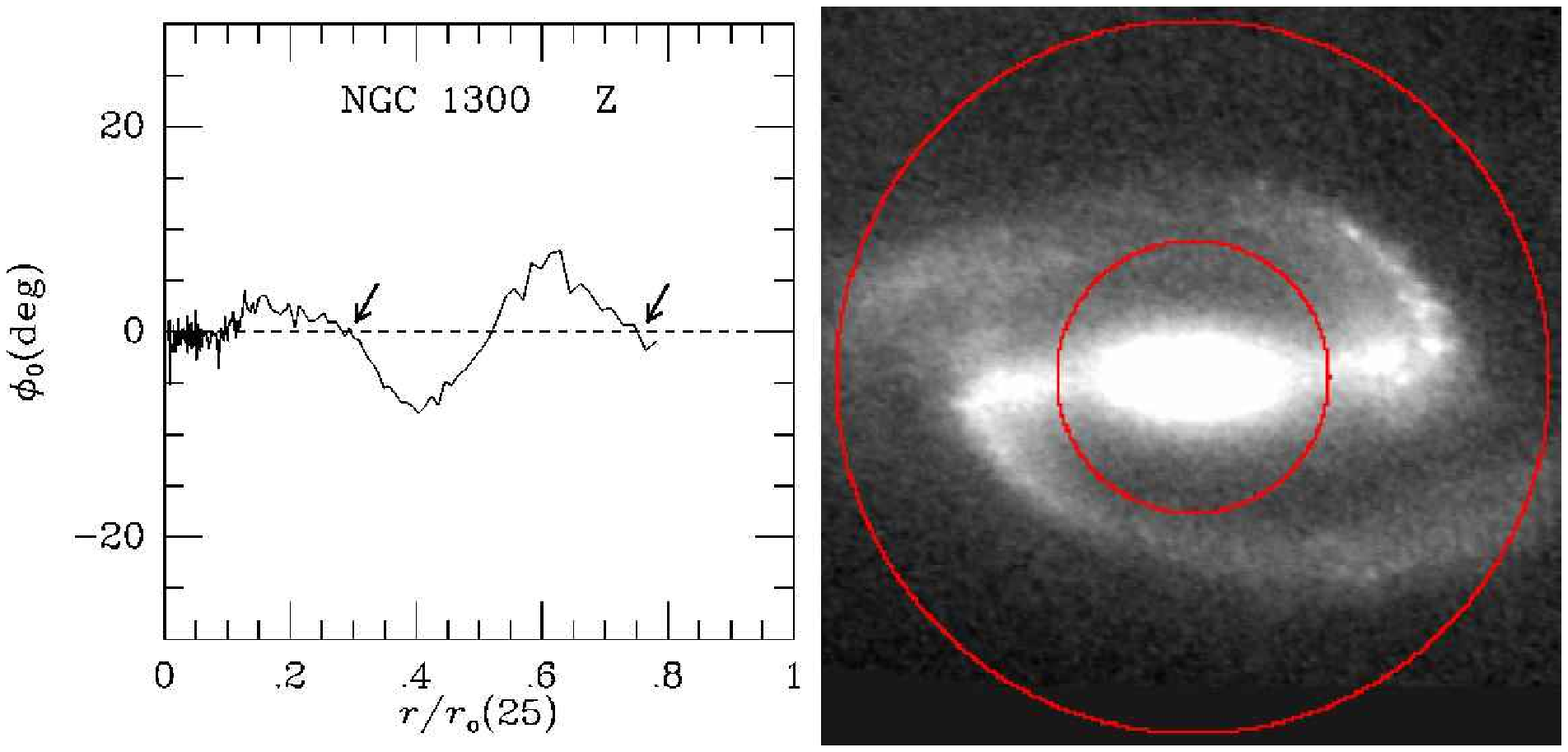}
\vspace{2.0truecm}                                                              
\caption{Same as Figure 2.1 for NGC 1300.}                                        
\label{ngc1300}                                                                 
\end{figure}

NGC 1300 (Figure~\ref{ngc1300}) - The galaxy fits tightly into the image
field, and the phase-shift analysis suggests two significant crossings,
one at $r$=55\arcs\ (0.29) lying just inside the bright ansae of the primary
bar, and one at $r$=142\arcs\ (0.75) lying in the bright spiral arms. The 
image is displayed to emphasize how the inner CR encircles a part of the
bar that is slightly twisted relative to the ansae. The bar appears to
extend well beyond its CR into the negative-to-positive crossing, which
should be close to the OLR of the bar pattern. The shocked arm segment
near the OLR radius could be a kind of OLR ring for the inner mode,
and the shock itself might be related to the shearing of material due to the
sharply-varying pattern speeds for the two modes.

\begin{figure}                                                                  
\figurenum{2.21}
\plotone{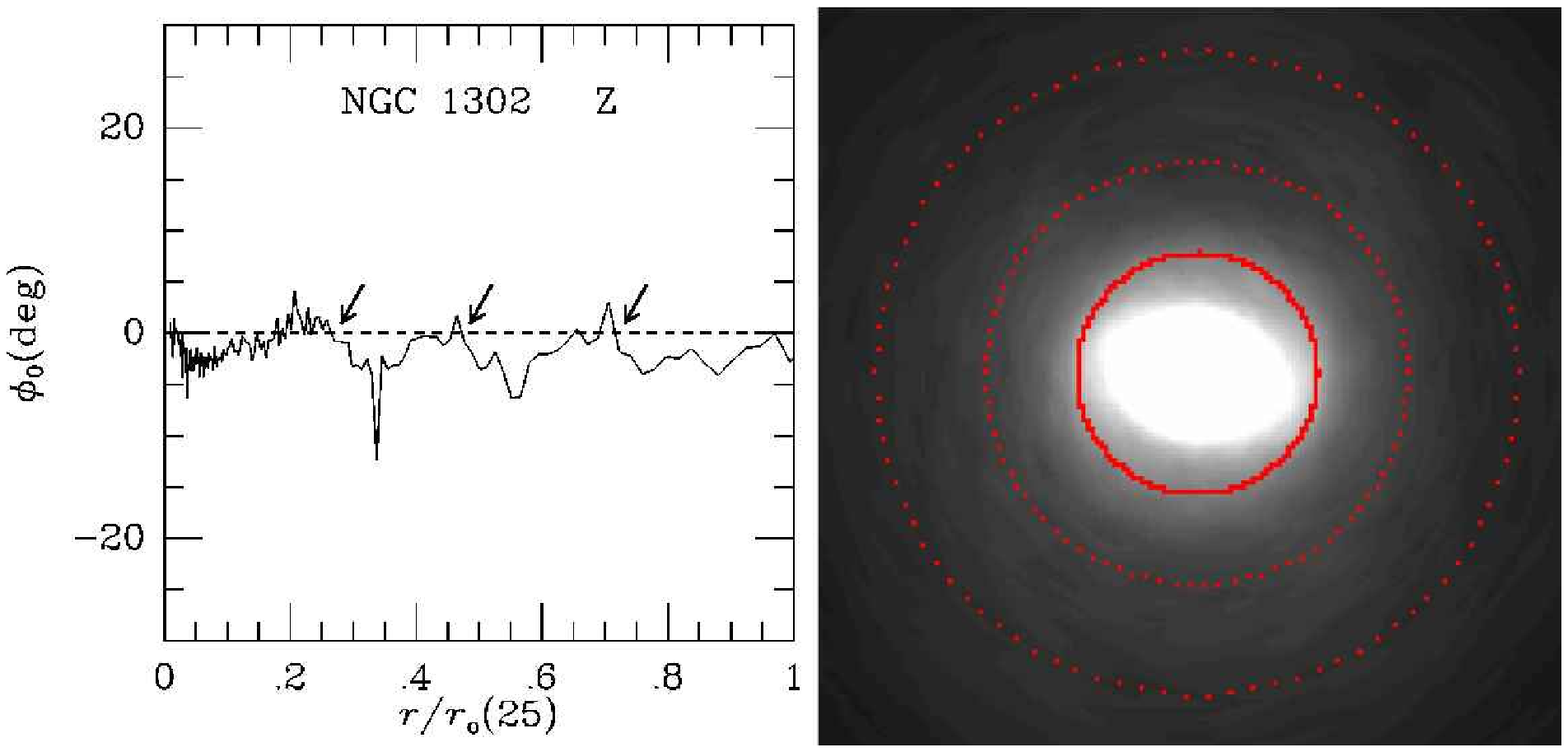}
\vspace{2.0truecm}                                                              
\caption{Same as Figure 2.1 for NGC 1302.}                                        
\label{ngc1302}                                                                 
\end{figure}

NGC 1302 (Figure~\ref{ngc1302}) - The main crossing at $r$=31\arcs\ (0.27)
encircles the ends of
a strong inner oval. Two outer crossings are weak but there is strong
optical structure in this outer region. 

The best-fitting simulation of RSL08 gave $r(CR)$ = 50\arcss 4 (0.43),
which is very close to our second CR at 55\arcs\ (0.47). The bar radius
of 30\arcss 7 (0.26) estimated by RSL08 is almost identical to our first CR.

\begin{figure}                                                                  
\figurenum{2.22}
\plotone{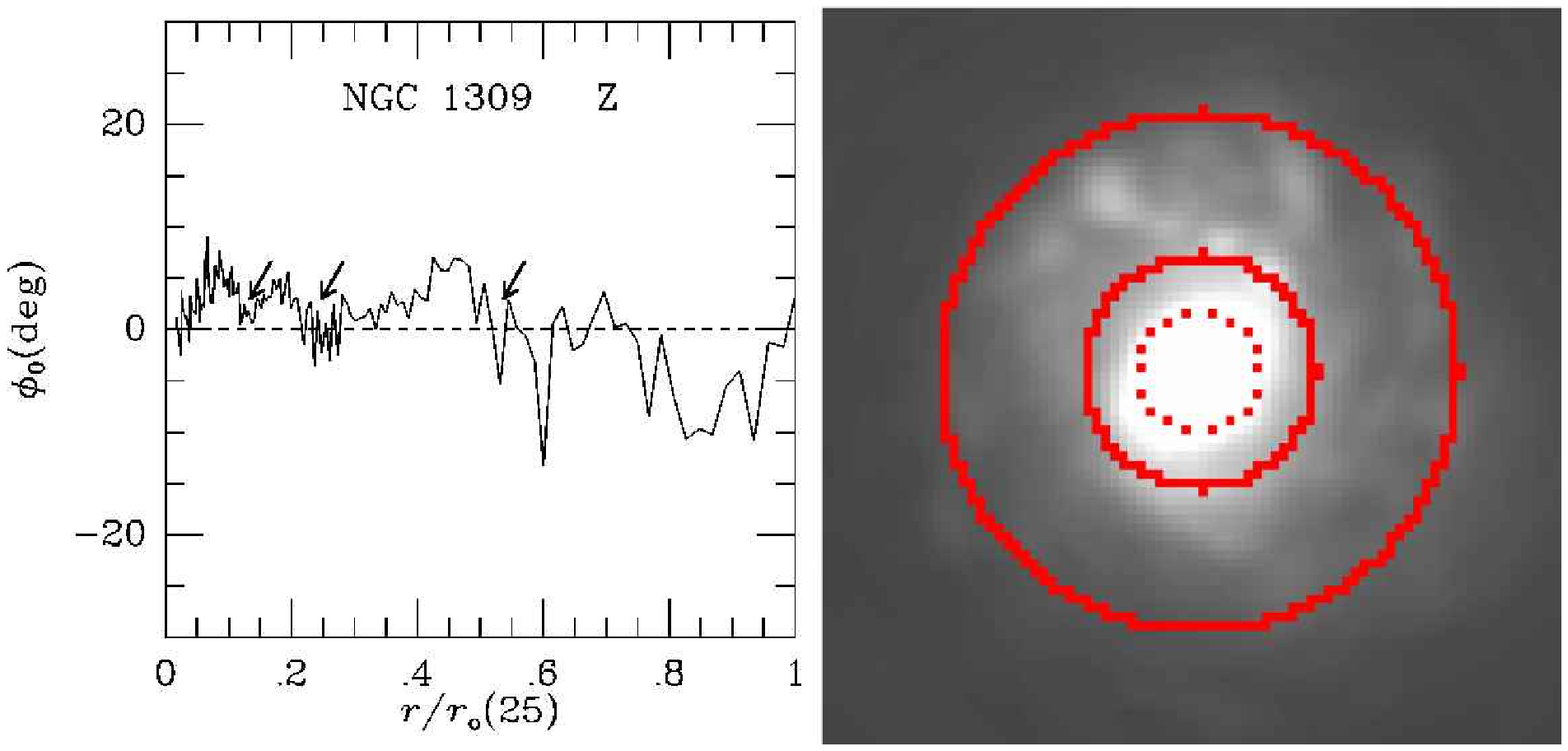}
\vspace{2.0truecm}                                                              
\caption{Same as Figure 2.1 for NGC 1309.}                                        
\label{ngc1309}                                                                 
\end{figure}

NGC 1309 (Figure~\ref{ngc1309})- The crossing at $r$=15\arcs\ (0.23)
appears to encircle an inner oval.
The crossing at $r$=48\arcs\ (0.73) is of uncertain significance.  
There is also a near crossing at 8\arcs\ (0.12) that could indicate a
decoupling pattern in the bulge.

\begin{figure}                                                                  
\figurenum{2.23}
\plotone{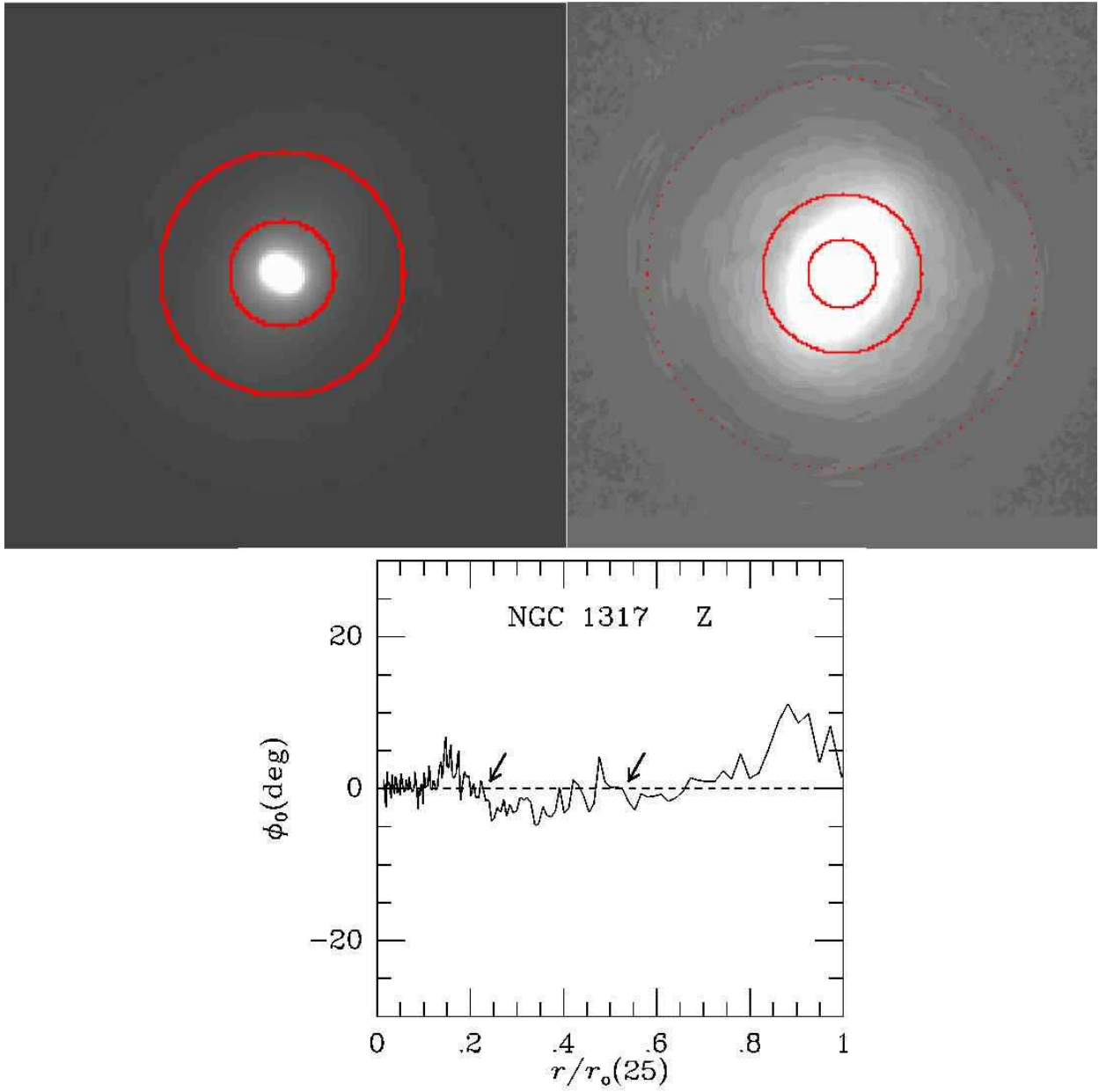}
\vspace{2.0truecm}                                                              
\caption{Same as Figure 2.1 for NGC 1317. The upper left                          
image has CR$_1$ and CR$_2$ from Table 1 overlaid as solid                      
circles. These are also shown at upper right, where CR$_3$                      
is included as a dotted circle.}                                                
\label{ngc1317}                                                                 
\end{figure}

NGC 1317 (Figure~\ref{ngc1317}) - The P/N crossing at $r$=18\arcs\ (0.22)
is inside the main outer
oval and circles at about twice the radius of the prominent nuclear
bar which is aligned perpendicular to the primary oval. The nuclear
ring surrounding this bar has dimensions 26\arcs $\times$ 20\arcs\
and is also aligned perpendicular to the bar. 
The second crossing at $r$=43\arcs\ (0.52) circles just inside the ends
of the main oval. A third crossing at $r$=105\arcs\ (1.27) is of uncertain 
significance but could be associated with an outer spiral.

The simulation of RSL08 gave $r(CR)$ = 54\arcss 5 (0.66), outside our
second CR but still well inside the radius of the primary oval, which
has $r(bar)$ = 61\arcss 1 (0.74), also according to RSL08. Thus,
we are in agreement with RSL08 that NGC 1317 has $\cal R$ $<$ 1.

\begin{figure}                                                                 
\figurenum{2.24}
\plotone{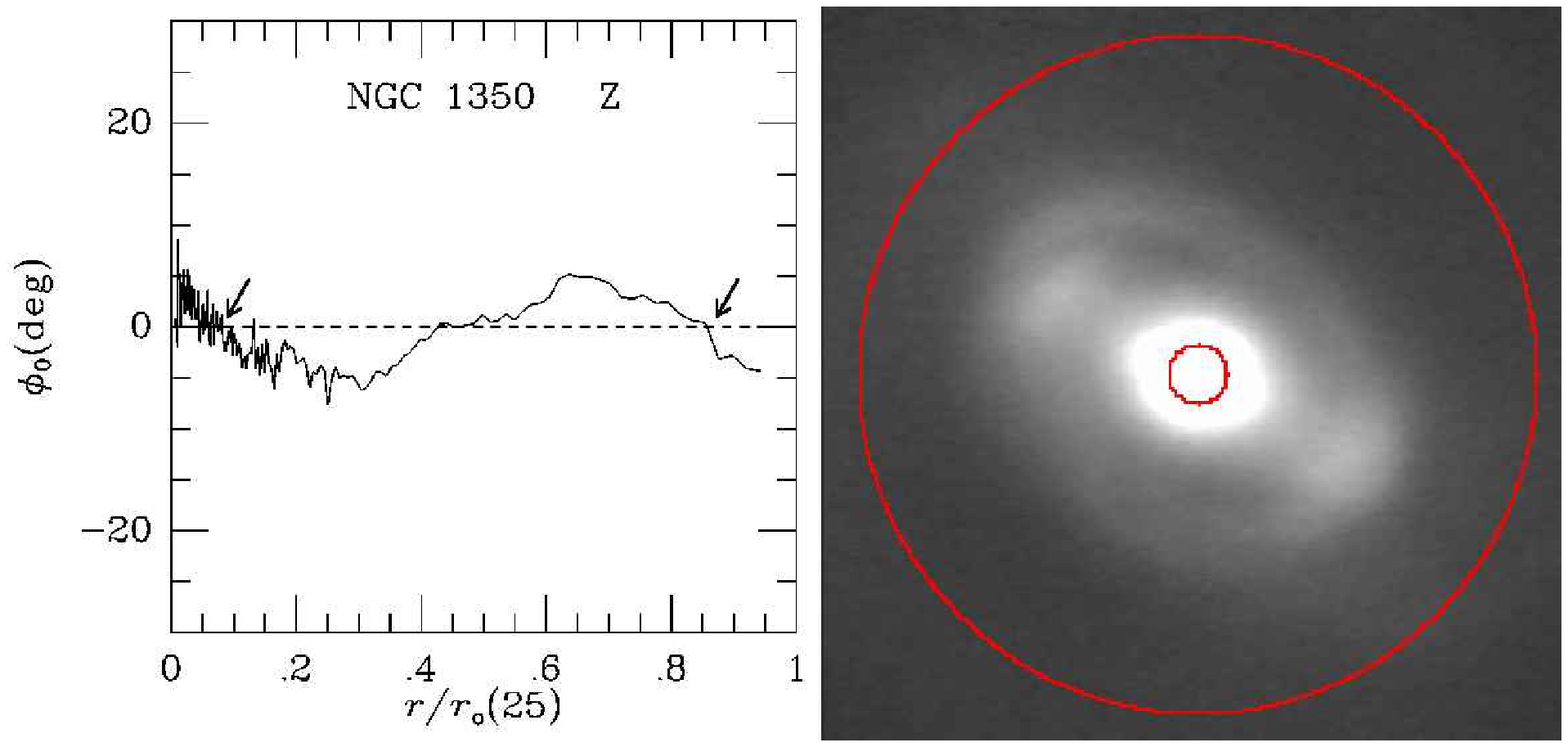}
 \vspace{2.0truecm}                                                             
\caption{Same as Figure 2.1 for NGC 1350}                                         
\label{ngc1350}                                                                 
 \end{figure}

NGC 1350 (Figure~\ref{ngc1350}) - Our analysis for this galaxy is based
on an average bulge image to minimize the strong pinch provided by the
spherical bulge assumption. The phase-shift distribution shows a
significant crossing at $r$=135\arcs\ (0.86), well outside the bar ends
but not beyond the galaxy's well-defined outer R$_1^{\prime}$
pseudoring. This CR instead encompasses an extended oval, much larger
than the apparent ansae-type bar. The prominent N/P crossing near
$r/r_o(25)$ passes within these ansae. The noisy crossing at
$r$=12\arcs\ (0.08) also suggests an inner pattern is present,
and at least the inner oval part of the bar could belong to that
crossing.

\begin{figure}                                                                  
\figurenum{2.25}
\plotone{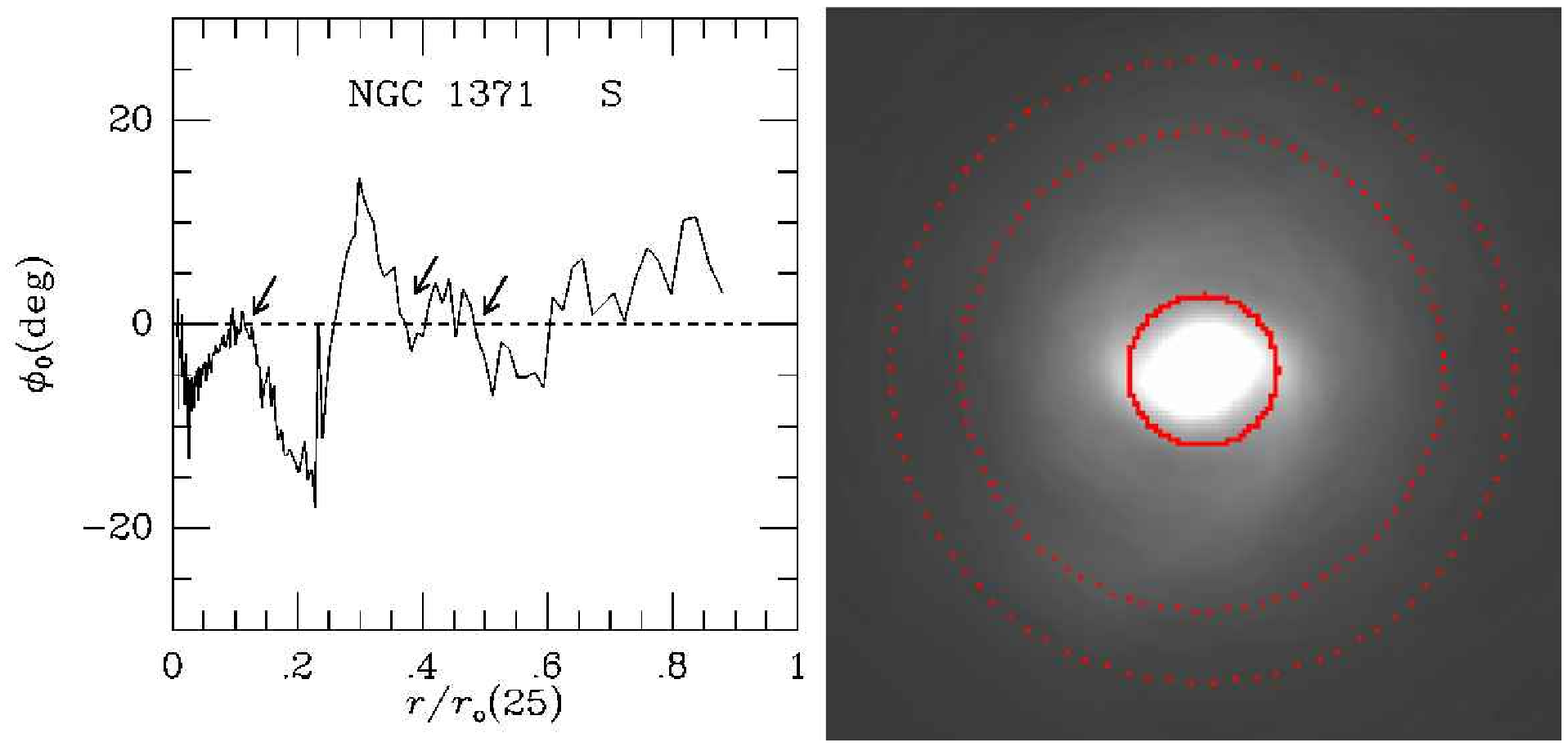}
\vspace{2.0truecm}                                                              
\caption{Same as Figure 2.1 for NGC 1371.}                                        
\label{ngc1371}                                                                 
\end{figure}

NGC 1371 (Figure~\ref{ngc1371}) - The crossing at $r$=19\arcs\ (0.11)
appears to encircle an inner oval. This oval shows isophote twisting
further out and makes the bar appear to spill over the CR circle. The
second crossing is near the outer edge of a fainter spiral pattern,
although the noisiness of the second and the third crossings could imply
a continuous shearing spiral pattern in this region.

\begin{figure}                                                                  
\figurenum{2.26}
\plotone{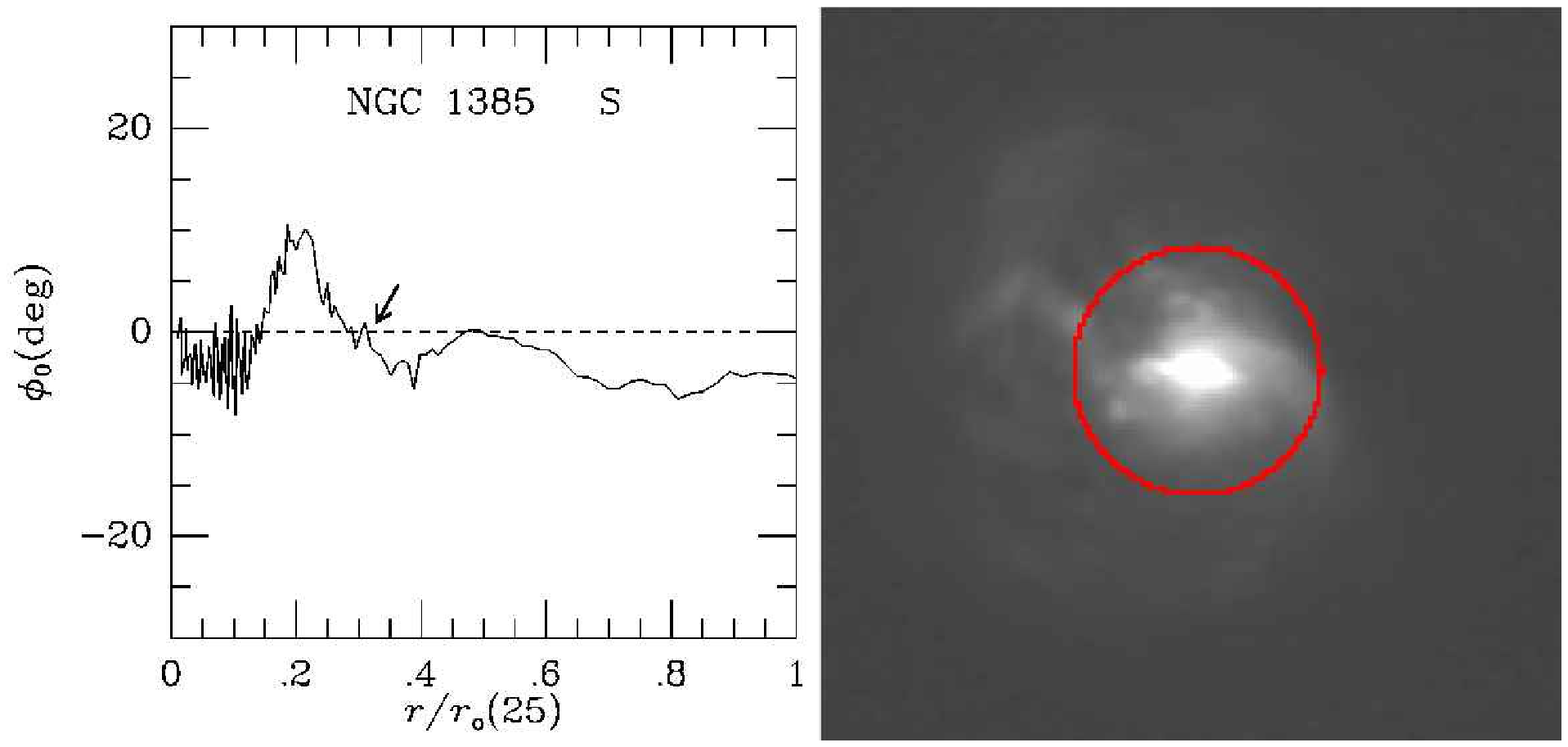}
\vspace{2.0truecm}                                                              
\caption{Same as Figure 2.1 for NGC 1385.}                                        
\label{ngc1385}                                                                 
\end{figure}

NGC 1385 (Figure~\ref{ngc1385}) - The most significant crossing, at $r$=32\arcs, (0.31) circles well
beyond the ends of the galaxy's bar. There may be crossings within 5\arcs\ 
(0.05) and at $\approx$51\arcs\ (0.50), but these are uncertain, especially the latter which
occurs in a region of irregular and likely unsteady structure.

\begin{figure}                                                                  
\figurenum{2.27}
\plotone{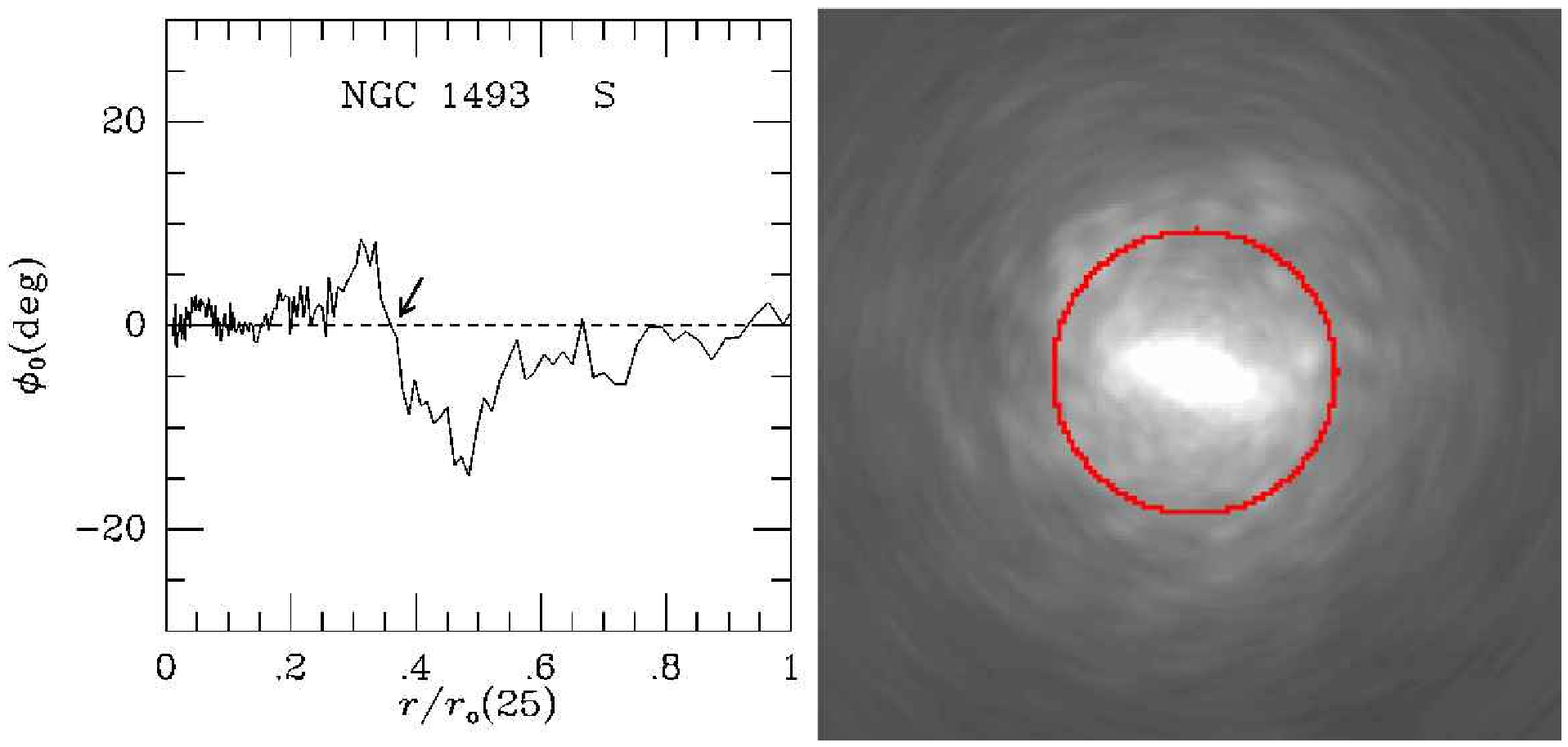}
\vspace{2.0truecm}                                                              
\caption{Same as Figure 2.1 for NGC 1493.}                                        
\label{ngc1493}                                                                 
\end{figure}

NGC 1493 (Figure~\ref{ngc1493}) - There is only a single strong crossing inside $r_o$ for this
galaxy. This crossing circles well beyond the apparent ends of the bar with
faint spiral arms emanating from it. 
A weak crossing near $r$=10\arcs\ (0.10) could signify decoupling within the main
bar itself.

\begin{figure}                                                                  
\figurenum{2.28}
\plotone{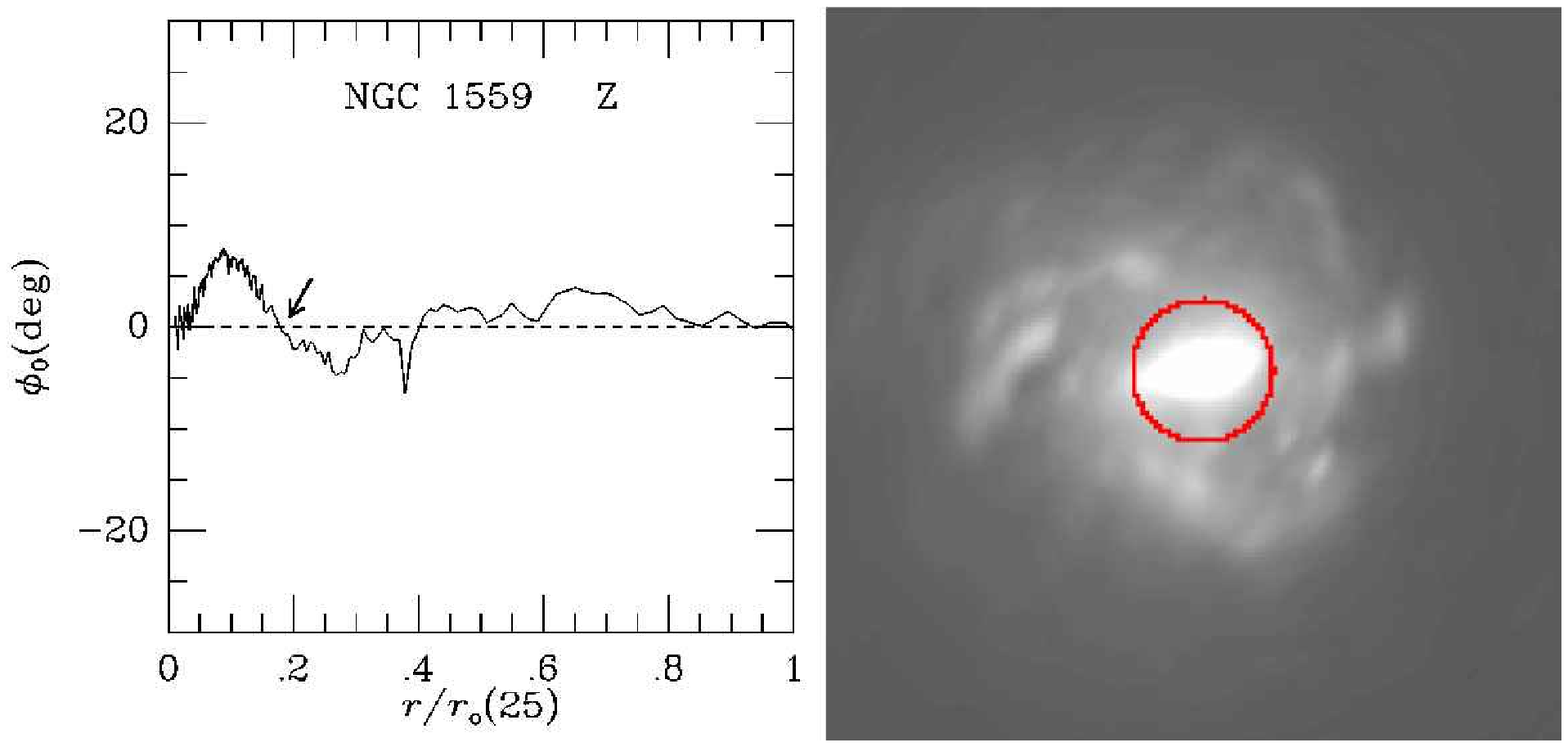}
\vspace{2.0truecm}                                                              
\caption{Same as Figure 2.1 for NGC 1559.}                                        
\label{ngc1559}                                                                 
\end{figure}

NGC 1559 (Figure~\ref{ngc1559}) - There is only a single major crossing, and this circles the
ends of the galaxy's bar. The outer pattern is complicated and may be
unsteady.

\begin{figure}                                                                  
\figurenum{2.29}
\plotone{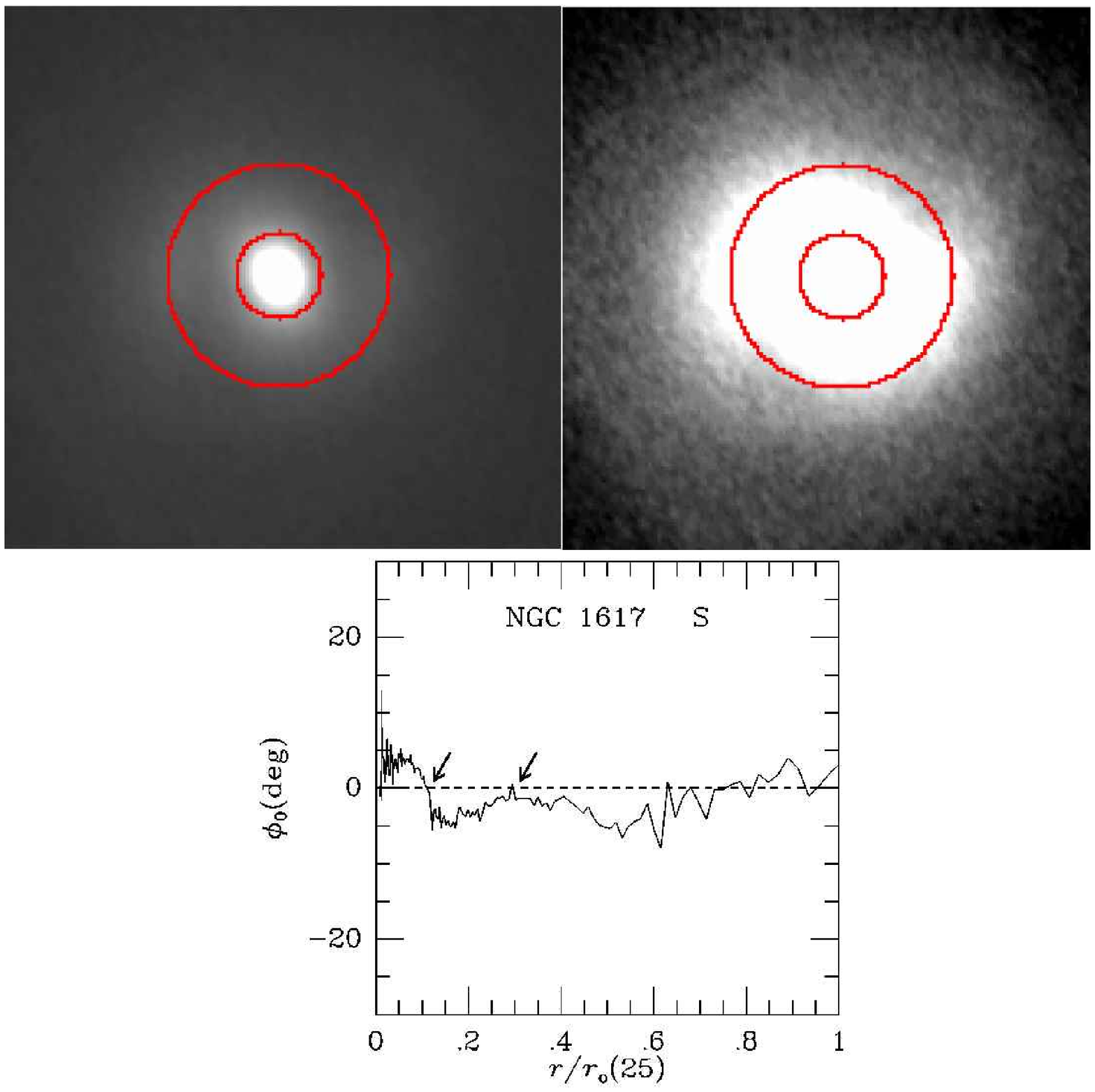}
\vspace{2.0truecm}                                                              
\caption{Same as Figure 2.1 for NGC 1617.}                                        
\label{ngc1617}                                                                 
\end{figure}

NGC 1617 (Figure~\ref{ngc1617}) - Based on an average bulge image
(section 3), we find
two likely significant crossings at $r$=14\arcs\ (0.11) and 38\arcs\
(0.30), although the
latter crossing is weak. The deprojected image shows two misaligned
ovals (Figure~\ref{ngc1617}a,b), although the effects of deprojection are still uncertain as this
galaxy is highly inclined. Each circle lies within the ends of an oval.

\begin{figure}                                                                  
\figurenum{2.30}
\plotone{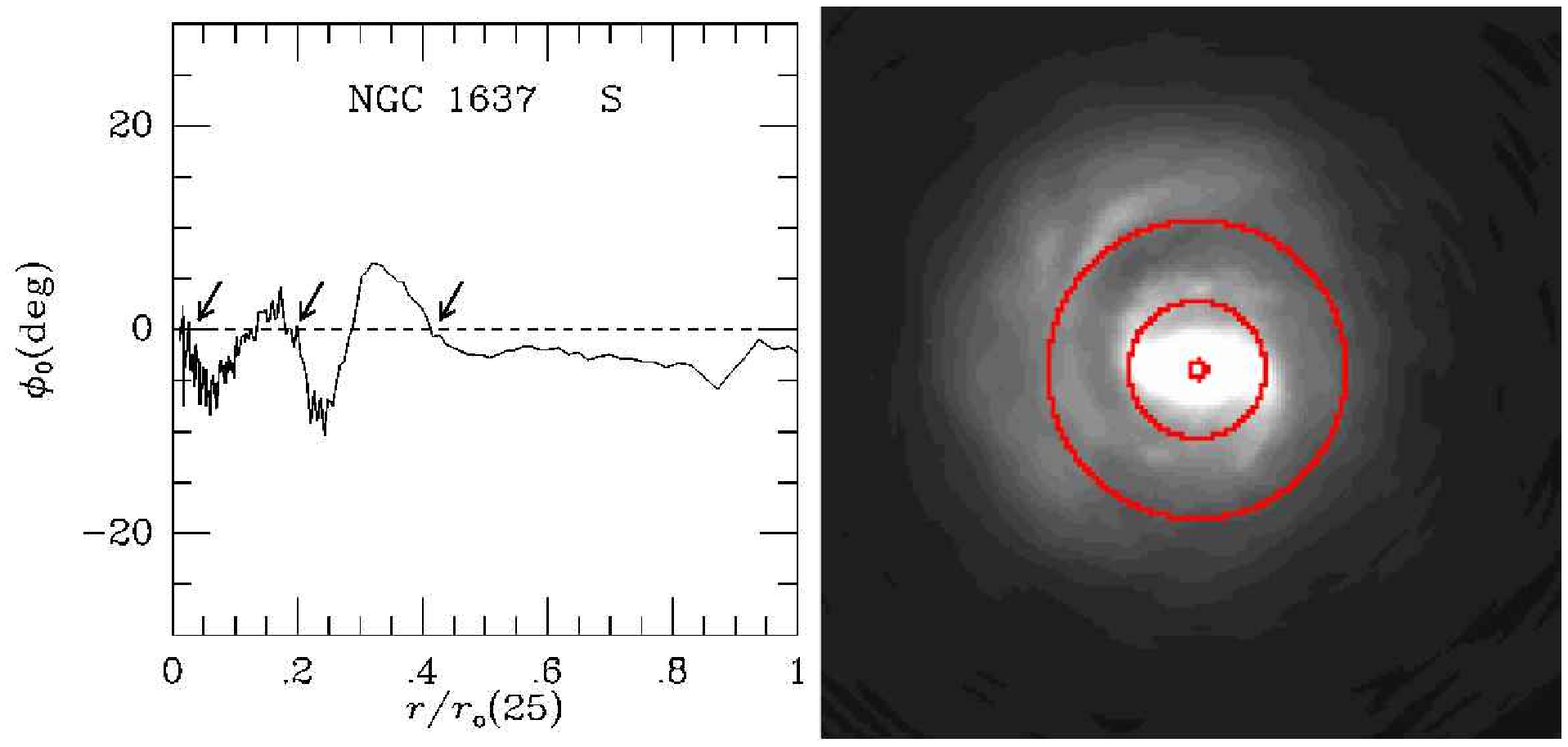}
\vspace{2.0truecm}                                                              
\caption{Same as Figure 2.1 for NGC 1637.}                                        
\label{ngc1637}                                                                 
\end{figure}

NGC 1637 (Figure~\ref{ngc1637}) - The crossing at 
$r$=23\arcs\ (0.19) circles around the ends of the galaxy's bar with two
spiral arms emanating from it. The crossing at 
$r$=51\arcs\ (0.42) also looks significant and is associated
with the main spiral pattern, which is dominated by a single bright arm.
There may also be a small inner crossing.

\begin{figure}                                                                  
\figurenum{2.31}
\plotone{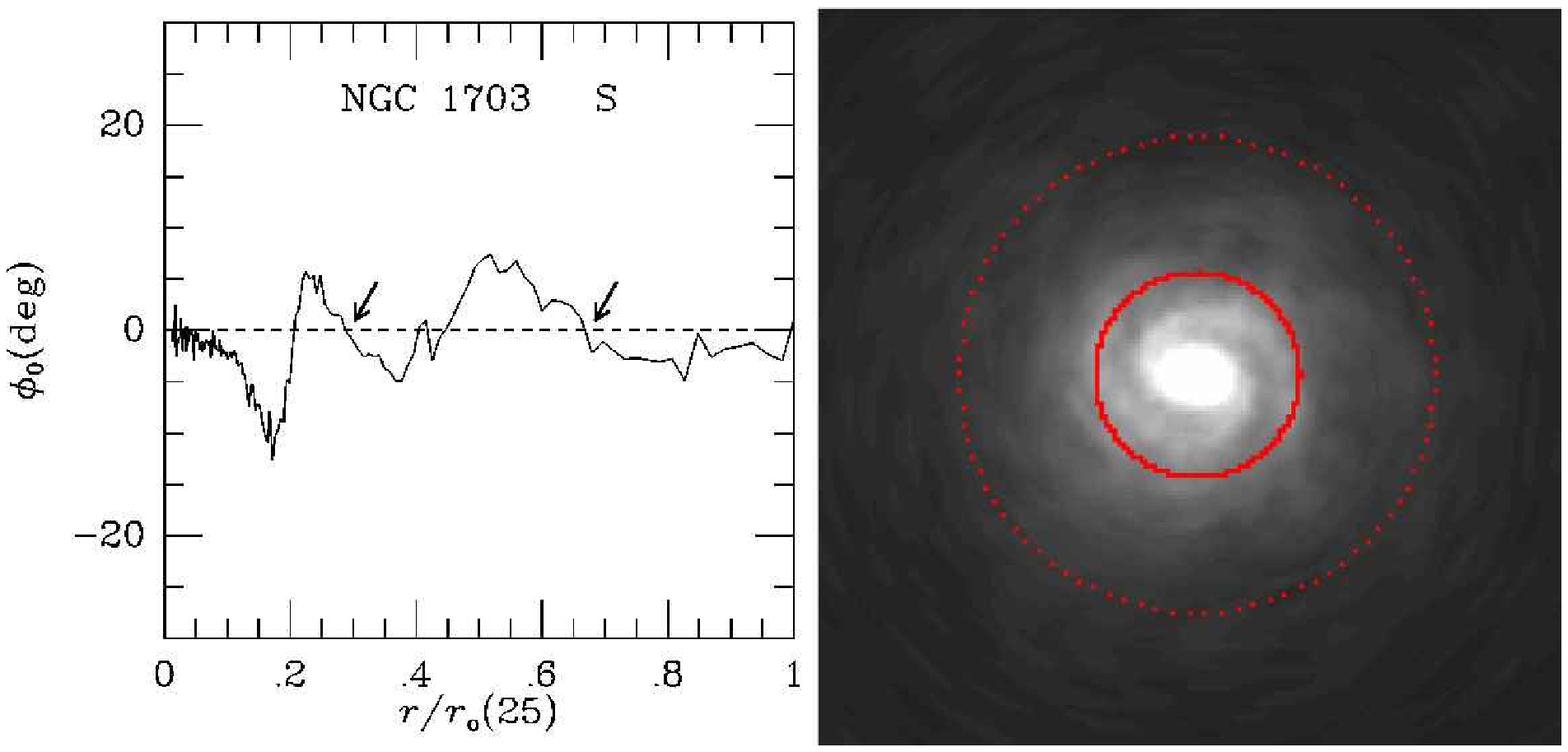}
\vspace{2.0truecm}                                                              
\caption{Same as Figure 2.1 for NGC 1703.}                                        
\label{ngc1703}                                                                 
\end{figure}

NGC 1703 (Figure~\ref{ngc1703}) - The crossing at $r$=25\arcs\ (0.28)
appears associated with the bright
inner spiral pattern. The outer crossing may be associated with the outer,
poorly organized spiral pattern.  There may also be a pattern, an inner oval,
at $r$$\approx$10\arcs\ (0.11)
that is not fully decoupled from the main spiral.

\begin{figure}                                                                  
\figurenum{2.32}
\plotone{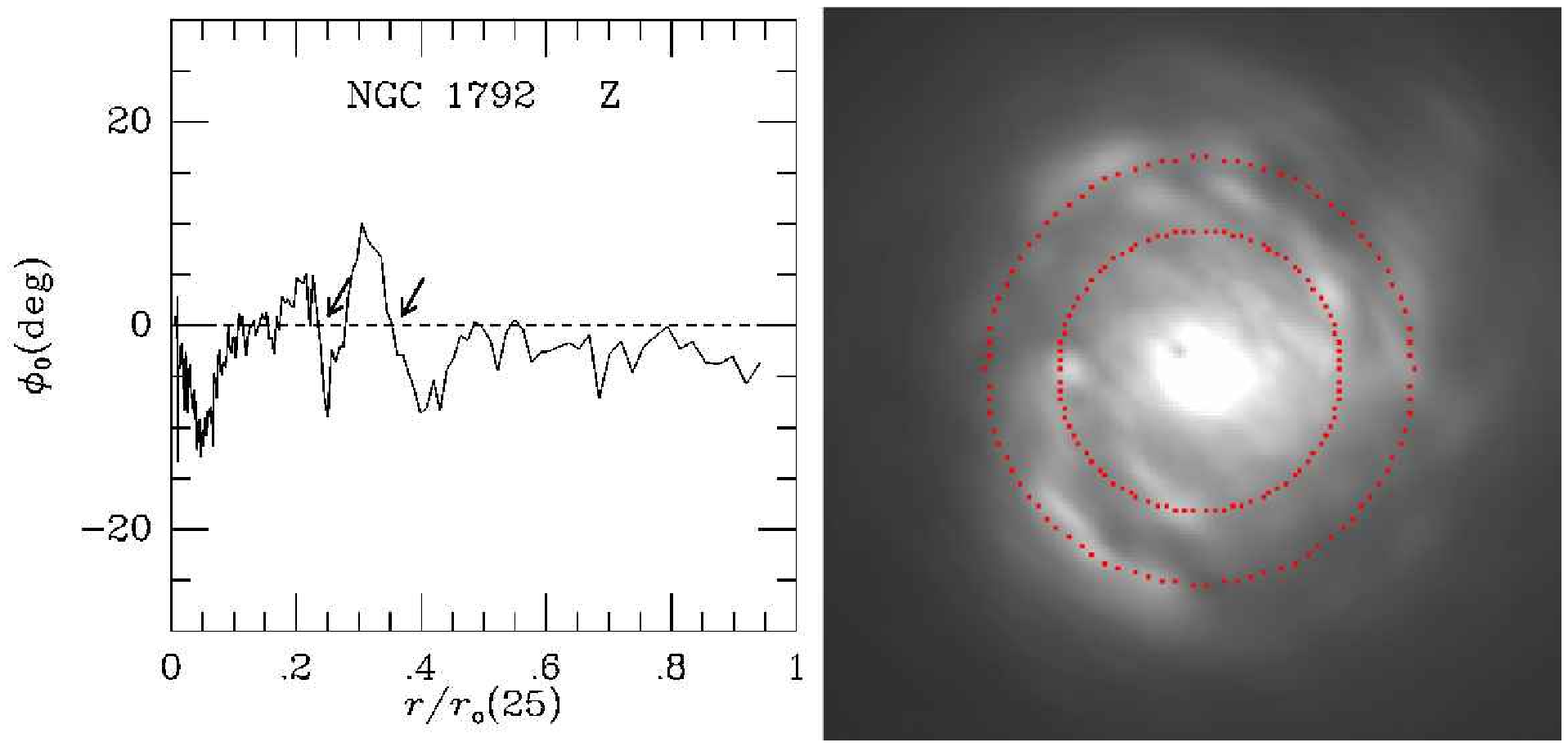}
\vspace{2.0truecm}                                                              
\caption{Same as Figure 2.1 for NGC 1792.}                                        
\label{ngc1792}                                                                 
\end{figure}

NGC 1792 (Figure~\ref{ngc1792}) - The morphology of this galaxy suggests an unsteady pattern 
defined by scattered star-forming regions. Nevertheless, the phase-shift
distribution indicates at least 2 relatively well-defined P/N crossings
within $r_o$, due probably to the skewed distribution of
the density. The analysis is based on an average bulge image.  A nuclear crossing
is suggested but is poorly defined.

\begin{figure}                                                                 
\figurenum{2.33}
\plotone{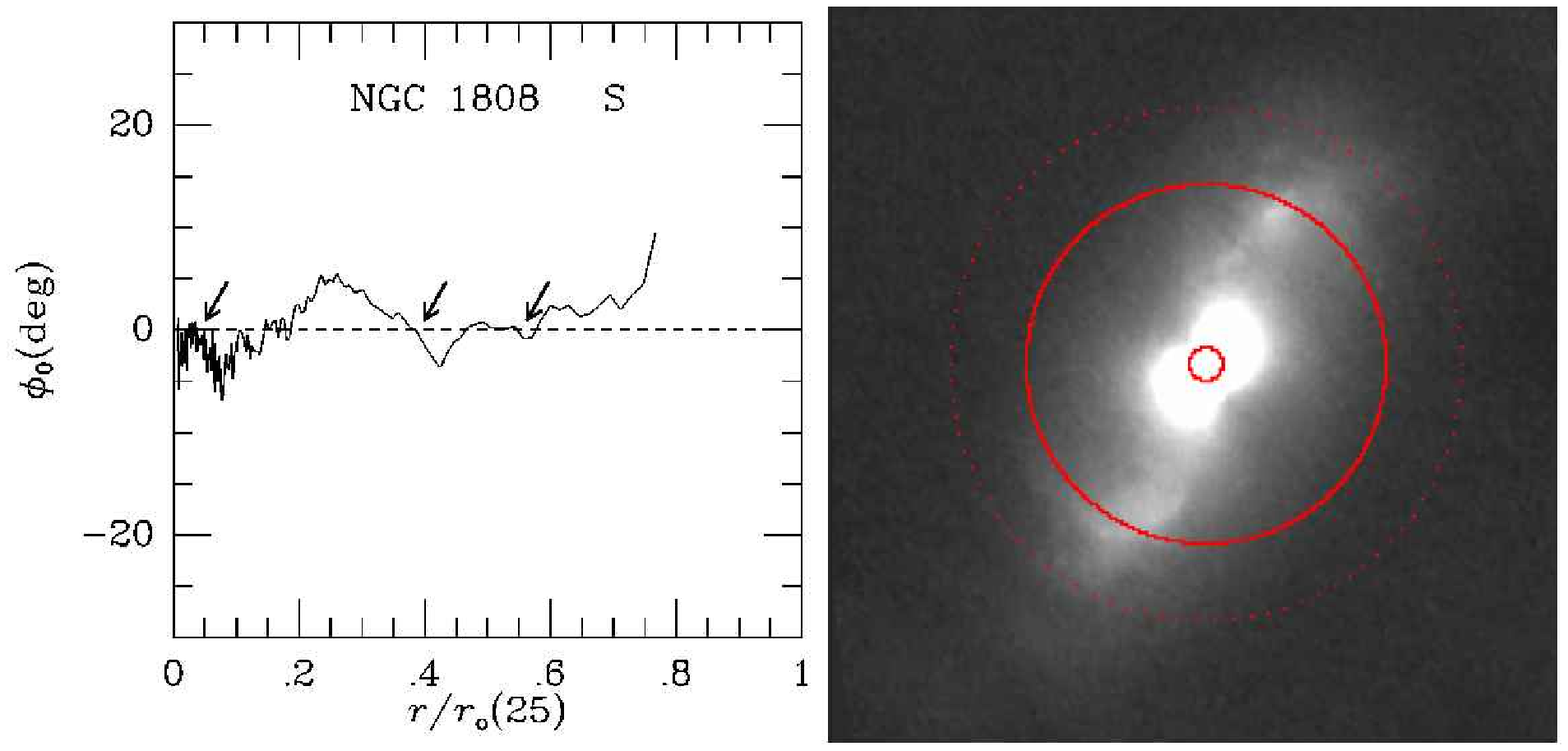}
 \vspace{2.0truecm}                                                             
\caption{Same as Figure 2.1 for NGC 1808}                                         
\label{ngc1808}                                                                 
 \end{figure}

NGC 1808 (Figure~\ref{ngc1808}) - This galaxy overflows the field of view
of the $H$-band image. The main crossing at $r$=74\arcs\ (0.38) lies just
inside the ends of the apparent bar. The small inner pattern at $r$=7\arcs\
could be associated with the galaxy's well-known nuclear structure.  The
outermost crossing is of uncertain significance.

\begin{figure}                                                                  
\figurenum{2.34}
\plotone{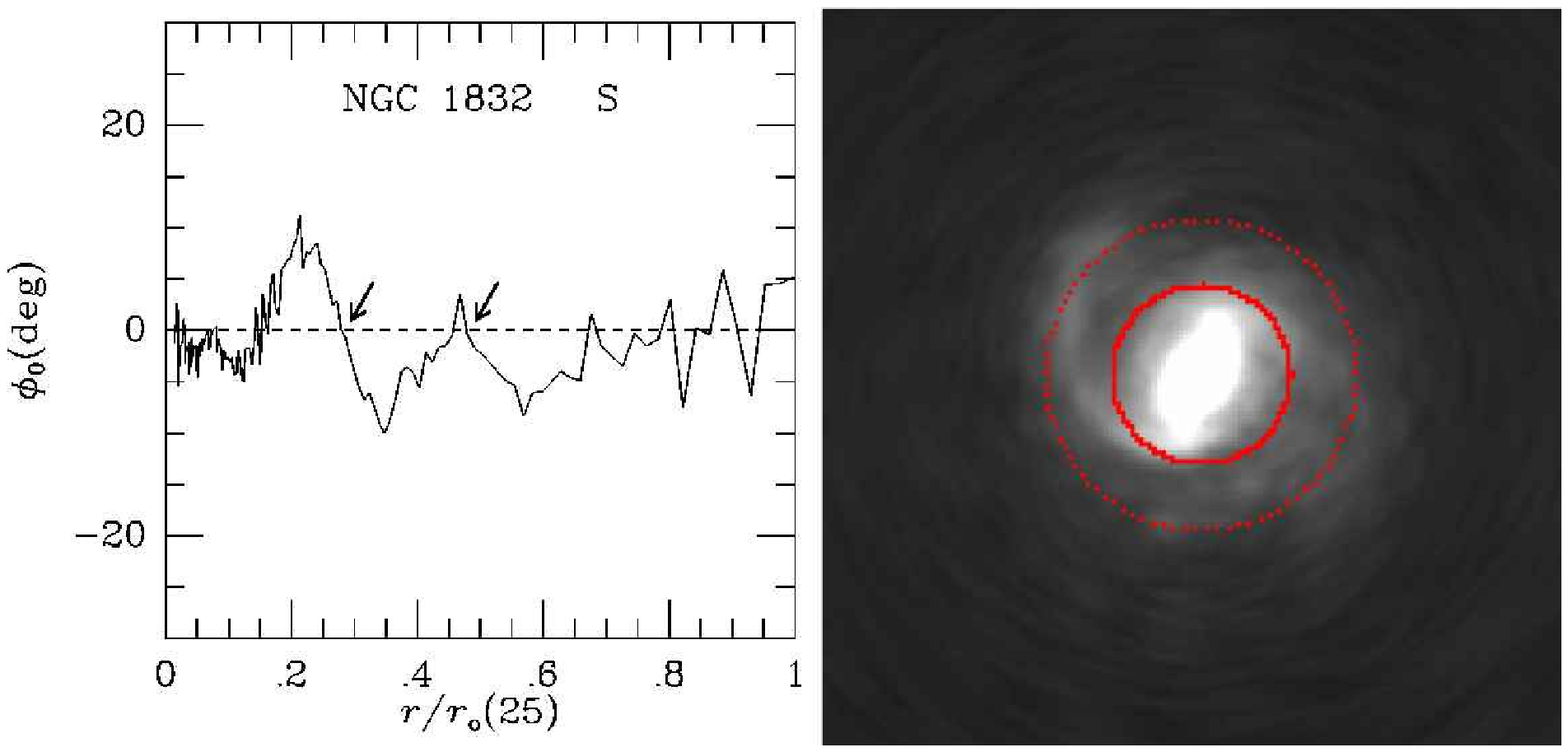}
\vspace{2.0truecm}                                                              
\caption{Same as Figure 2.1 for NGC 1832.}                                        
\label{ngc1832}                                                                 
\end{figure}

NGC 1832 (Figure~\ref{ngc1832}) - The main P/N crossing at $r$=23\arcs\ (0.28)
appears to be associated with the galaxy's bar with emanating spirals. 
The next crossing at $r$=39\arcs\ (0.48) could be associated
with the outer spiral. Other crossings inside and outside these radii
are uncertain, but a possibly decoupling pattern at $r$$\approx$6\arcs\
(0.07) is evident.

RSL08 found a reasonably good simulation of this galaxy that places CR
at 32\arcss 8 (0.41), which is very close to our second CR and is likely 
to be the pattern speed of the spiral, not the bar. Our first crossing
is close to the bar radius of 18\arcss 9 (0.23) estimated by RSL08.
Thus, while RSL08 placed NGC 1832 in the slow bar domain, we place
it in the fast bar domain.

\begin{figure}                                                                  
\figurenum{2.35}
\plotone{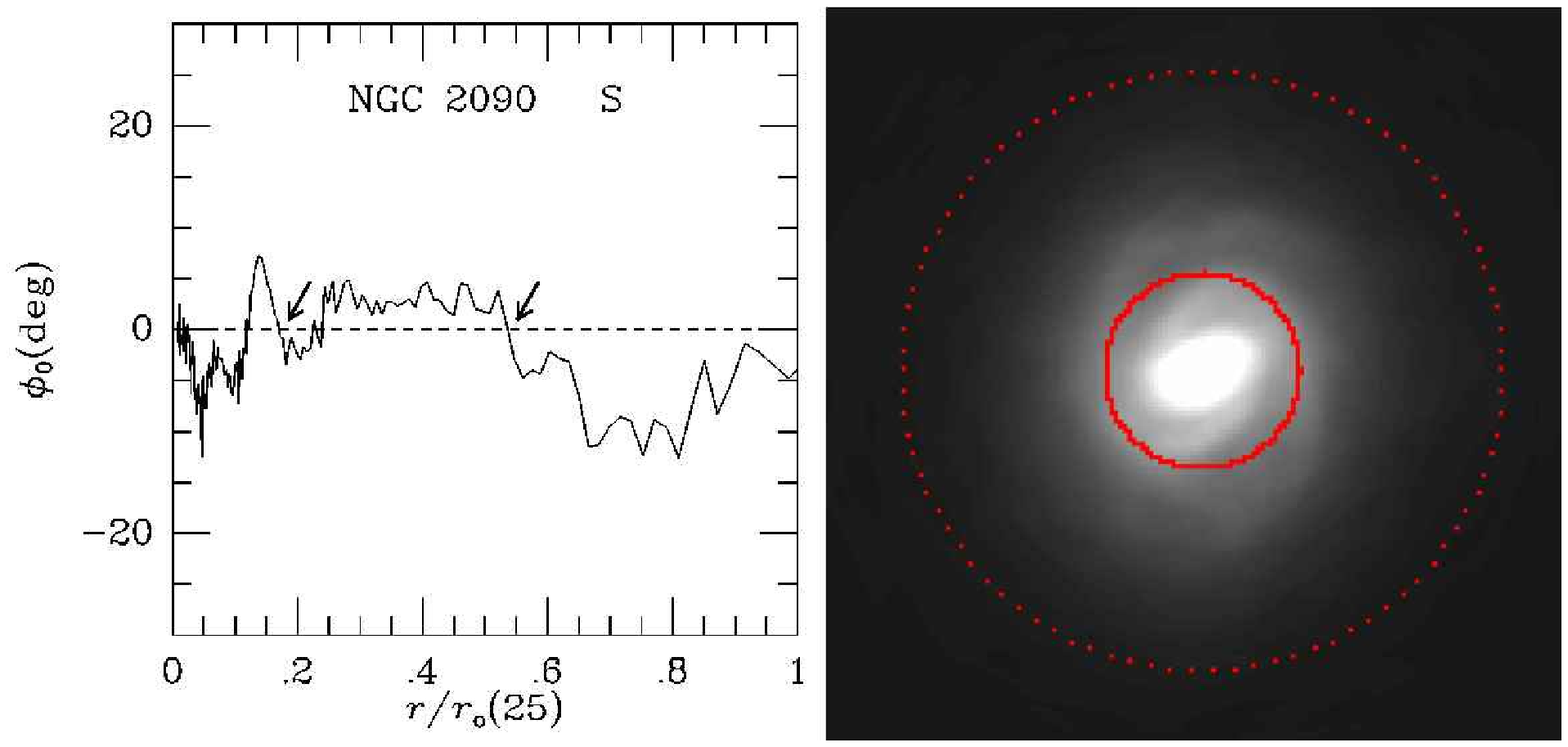}
\vspace{2.0truecm}                                                              
\caption{Same as Figure 2.1 for NGC 2090.}                                        
\label{ngc2090}                                                                 
\end{figure}

NGC 2090 (Figure~\ref{ngc2090}) - The galaxy shows two main P/N
crossings inside $r_o$. The crossing at $r$=25\arcs\ (0.17) is at about
twice the radius of the ends of an inner oval and includes a bright
inner spiral. There is also an indication of a small inner pattern
decoupling at $r$$\approx$11\arcs\ (0.08), as well as a nuclear
pattern.

\begin{figure}                                                                  
\figurenum{2.36}
\plotone{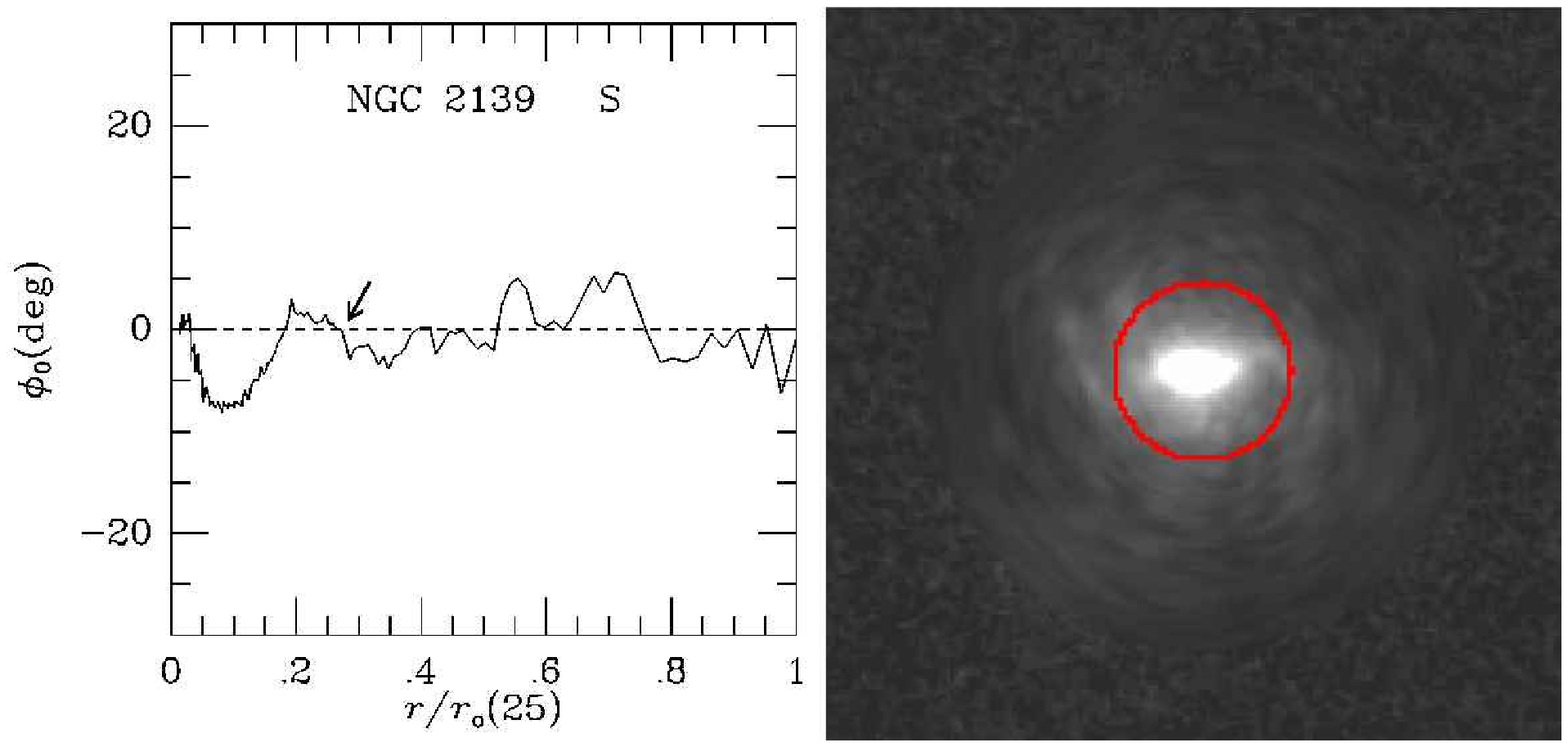}
\vspace{2.0truecm}                                                              
\caption{Same as Figure 2.1 for NGC 2139.}                                        
\label{ngc2139}                                                                 
\end{figure}

NGC 2139 (Figure~\ref{ngc2139}) - 
The spiral pattern in this galaxy is largely chaotic, but there
is a well-defined bar. The main P/N crossing at $r$=22\arcs\ (0.27)
encircles at more than twice the apparent bar radius.

\begin{figure}                                                                  
\figurenum{2.37}
\plotone{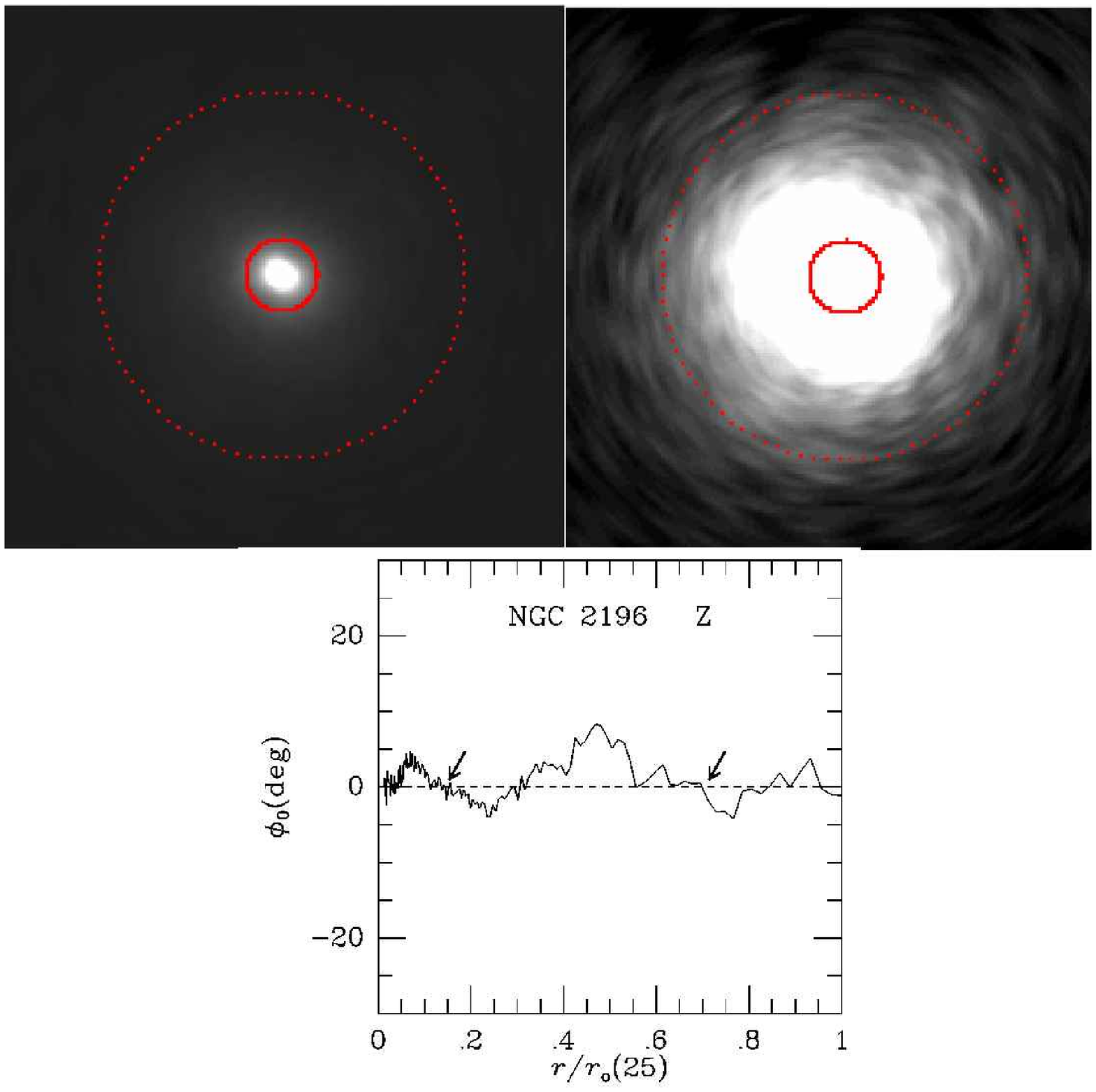}
\vspace{2.0truecm}                                                              
\caption{Same as Figure 2.1 for NGC 2196.}                                        
\label{ngc2196}                                                                 
\end{figure}

NGC 2196 (Figure~\ref{ngc2196}) - The phase-shift distribution shows two
possibly significant crossings, the main pattern appearing to be a large,
broad oval (Figure~\ref{ngc2196}b).

\begin{figure}                                                                  
\figurenum{2.38}
\plotone{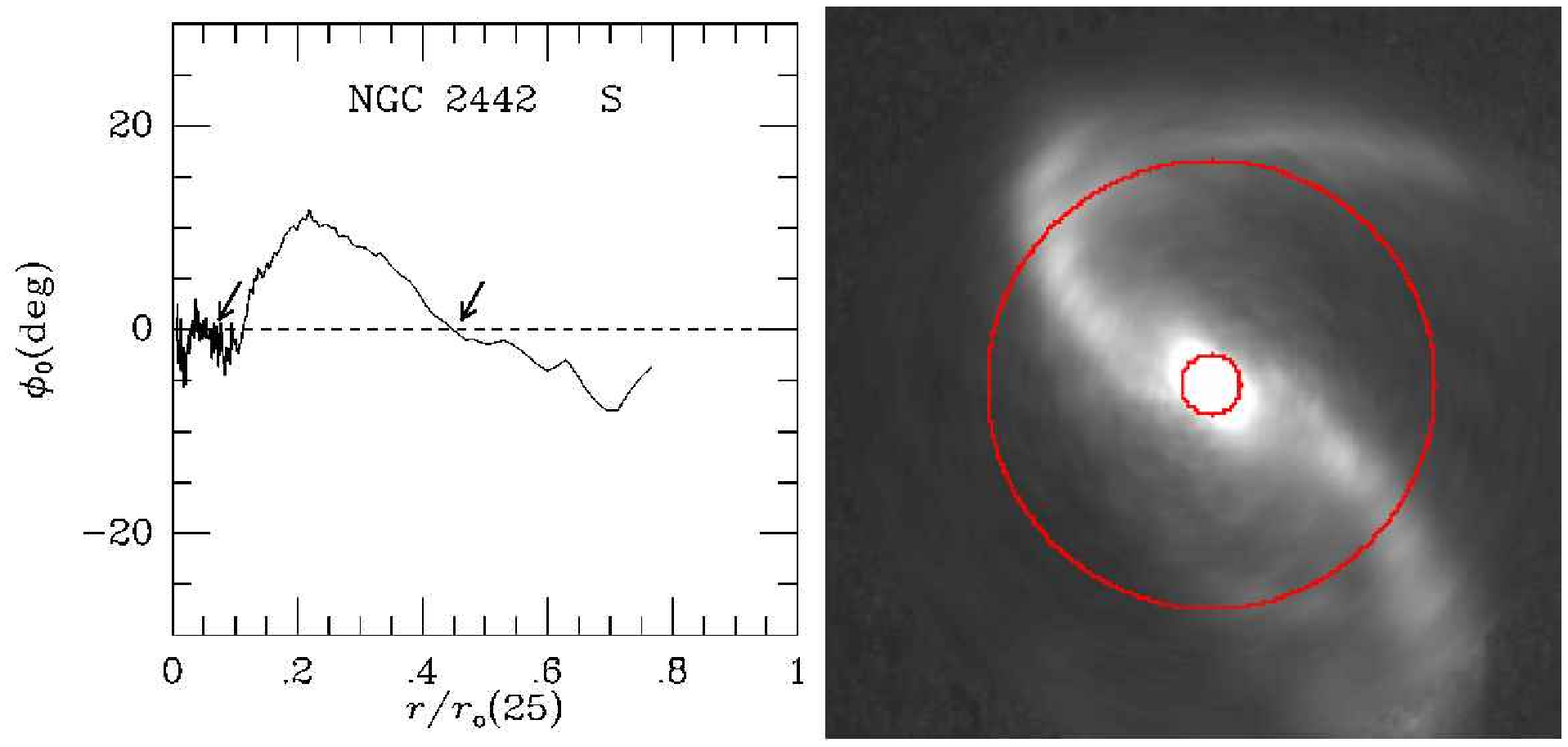}
\vspace{2.0truecm}                                                              
\caption{Same as Figure 2.1 for NGC 2442.}                                        
\label{ngc2442}                                                                 
\end{figure}

NGC 2442 (Figure~\ref{ngc2442}) - 
Although the phase-shift distribution is fairly simple in having
mainly a single major crossing, the galaxy itself is complicated and may
be interacting (Mihos \& Bothun 1997). Thus, we are unsure that all pattern features are in the
same plane. The main P/N crossing encircles just outside the ends of a 
large bar-like feature. There is a small inner pattern that could be
associated with the noisy crossing near $r$=11\arcs\ (0.06).  The presence of
an even smaller nuclear pattern is also indicated but not resolved.

\begin{figure}                                                                 
\figurenum{2.39}
\plotone{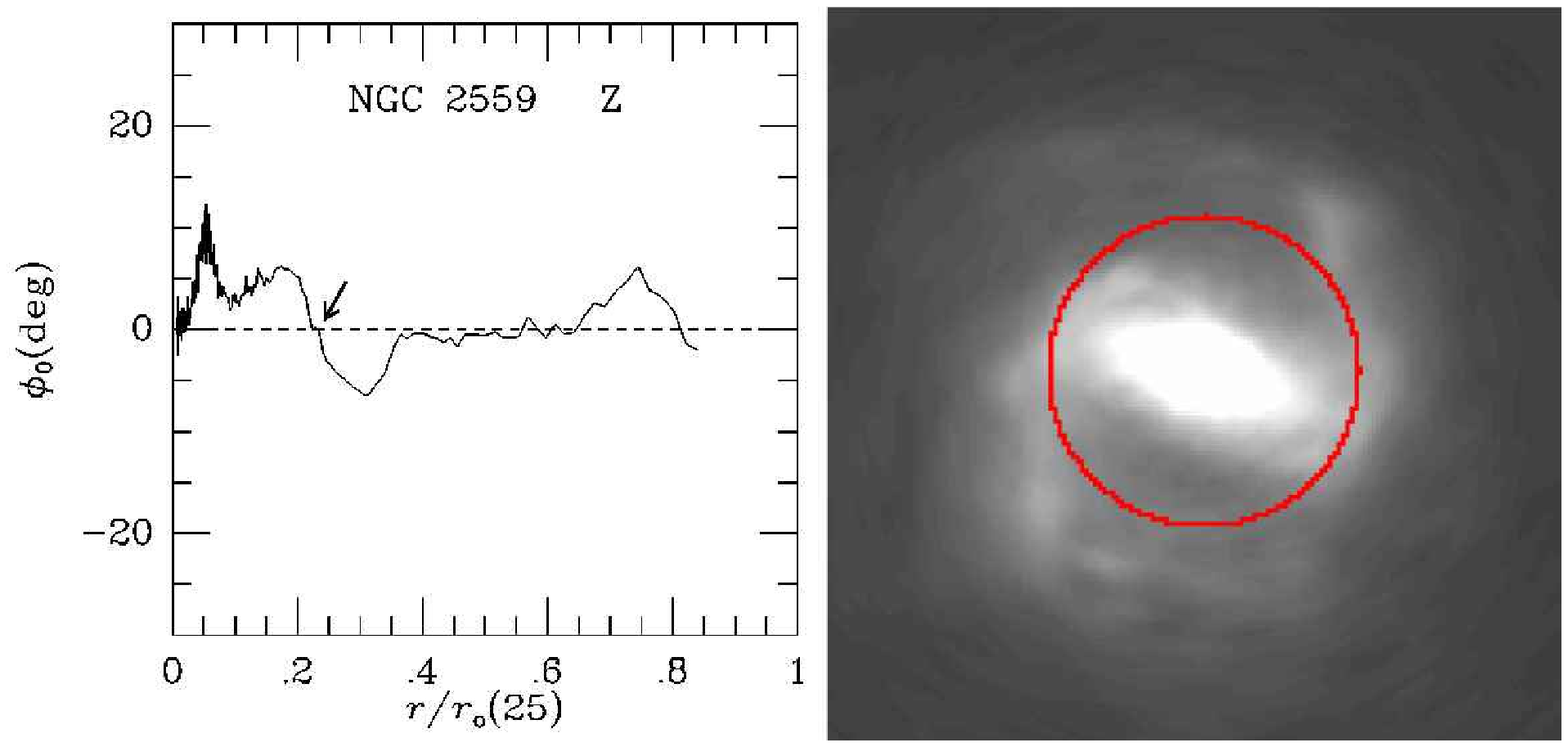}
 \vspace{2.0truecm}                                                             
\caption{Same as Figure 2.1 for NGC 2559}                                         
\label{ngc2559}                                                                 
 \end{figure}

NGC 2559 (Figure~\ref{ngc2559}) - The single strong P/N crossing encircles just outside the ends
of the prominent bar of this galaxy. The spiral appears to be bar-driven,
but the dip in phase-shift near $r$=16\arcs\ (0.09) 
could imply an inner decoupling bar.

\begin{figure}                                                                 
\figurenum{2.40}
\plotone{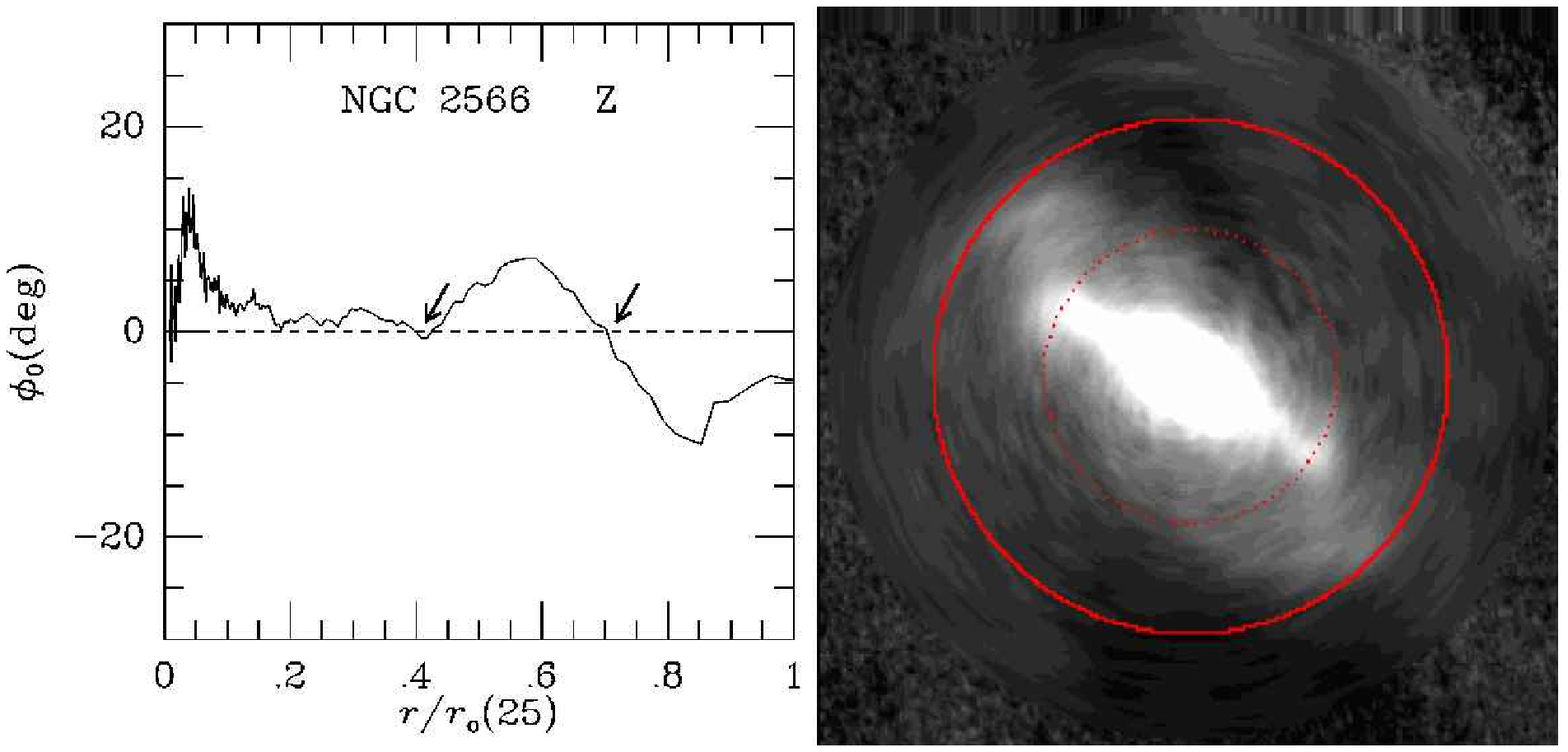}
 \vspace{2.0truecm}                                                             
\caption{Same as Figure 2.1 for NGC 2566}                                         
\label{ngc2566}                                                                 
 \end{figure}

NGC 2566 (Figure~\ref{ngc2566})- An interesting case where the apparent
bar in the near-IR shows two clearly misaligned components. The
phase-shift distribution shows two P/N crossings, but only the outer
one is strong. The crossing at $r$=56\arcs\ (0.40) appears to encircle
the ends of the inner part of the bar, while the crossing at
$r$=99\arcs\ (0.71) circles just outside the ends of the outer part of
the bar. The weakness of the inner crossing could signify a newly
decoupled pattern. This galaxy has very faint outer structure.

\begin{figure}                                                                 
\figurenum{2.41}
\plotone{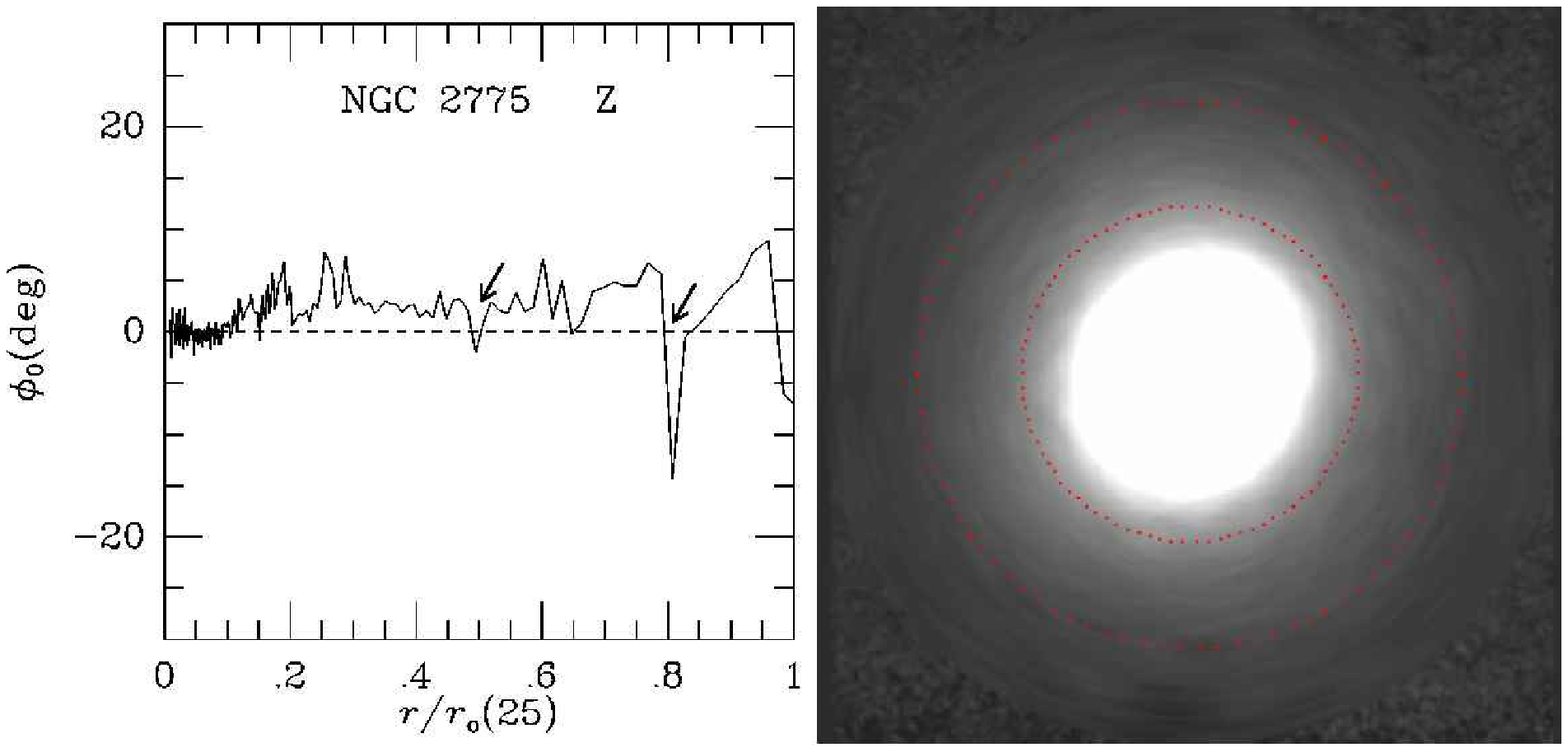}
 \vspace{2.0truecm}                                                             
\caption{Same as Figure 2.1 for NGC 2775}                                         
\label{ngc2775}                                                                 
 \end{figure}

NGC 2775 (Figure~\ref{ngc2775}) - The weak crossing at $r$=64\arcs\ (0.49)
appears associated with a large, broad oval. The crossing at $r$=104\arcs\
(0.79) is of uncertain significance. 

\begin{figure}                                                                 
\figurenum{2.42}
\plotone{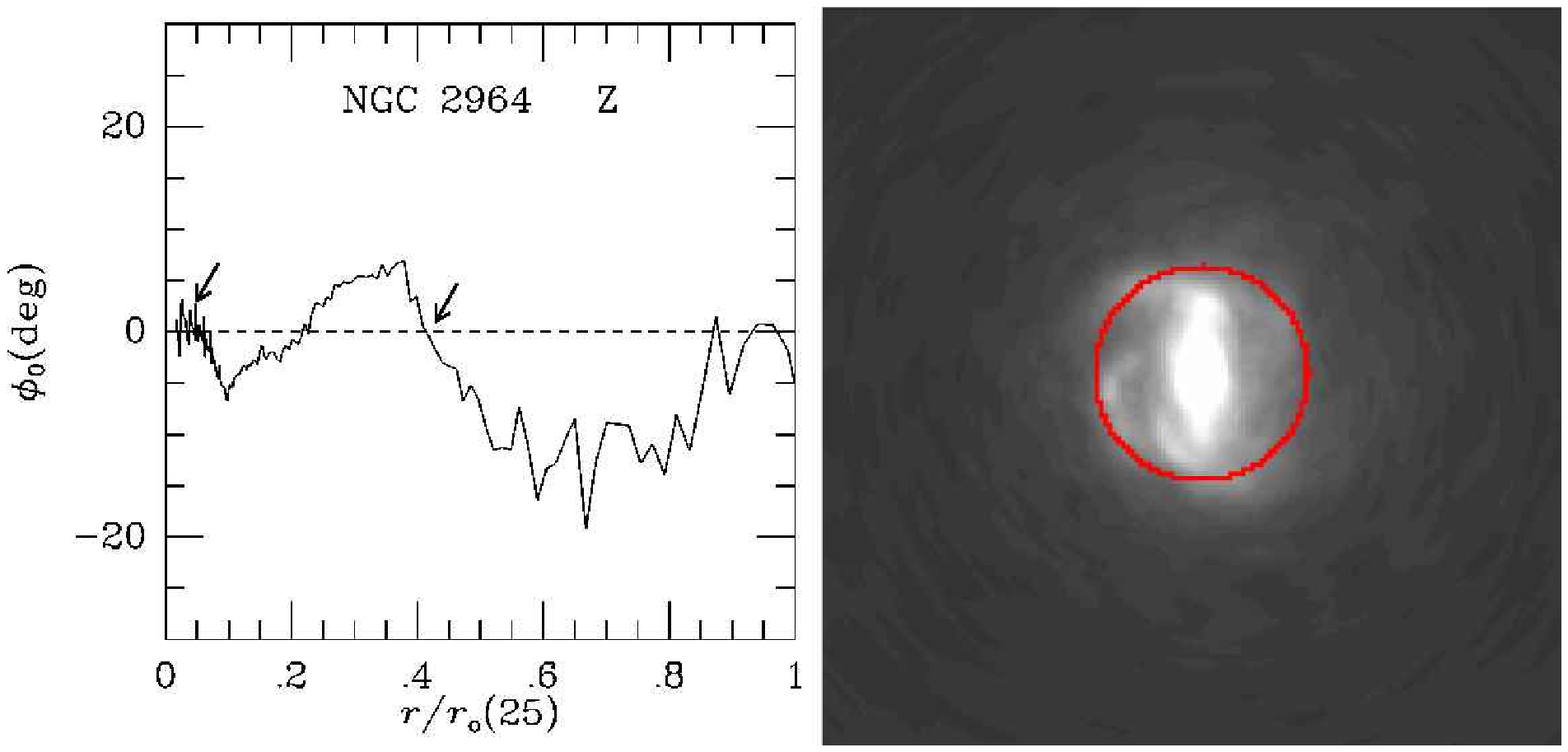}
 \vspace{2.0truecm}                                                             
\caption{Same as Figure 2.1 for NGC 2964. Only CR$_2$ from                        
Table 1 is shown overlaid on the image.}                                        
\label{ngc2964}                                                                 
 \end{figure}

NGC 2964 (Figure~\ref{ngc2964}) - The P/N crossing at
$r$=37\arcs\ (0.42) appears to circle just outside the ends of the
galaxy's main bar. At least two separate spiral patterns break from the
bar region, but neither shows its own crossing, suggesting the system
is defined mostly by a single pattern speed. The phase-shift
distribution also suggests a small inner pattern near
$r$=3\arcs\ (0.03; not shown on the overlay image). The N/P crossing
associated with this inner pattern actually comes close to but just
inside the ends of the bar, and we cannot rule out this galaxy as a
possible bar extending well beyond its CR.

\begin{figure}                                                                 
\figurenum{2.43}
\plotone{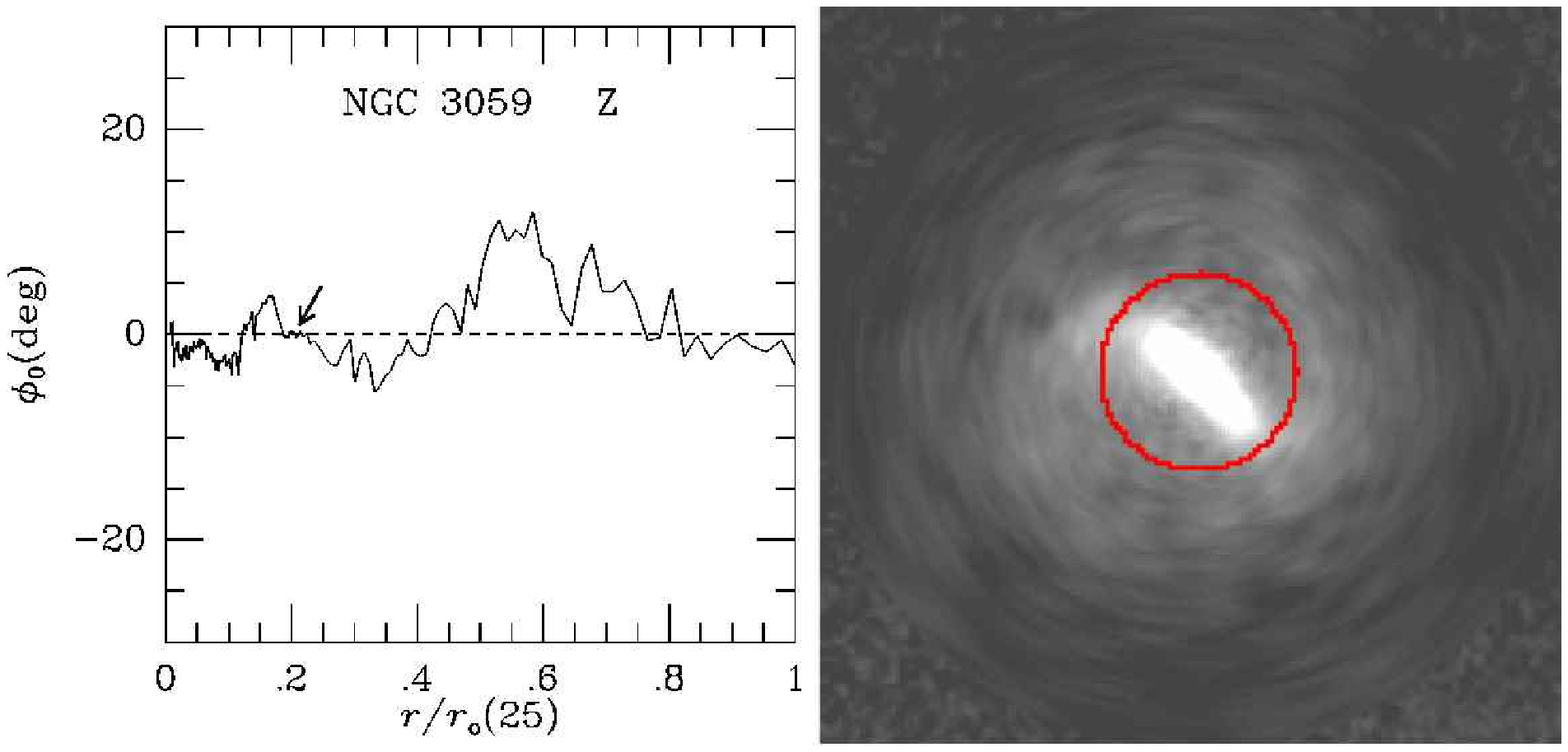}
 \vspace{2.0truecm}                                                             
\caption{Same as Figure 2.1 for NGC 3059}                                         
\label{ngc3059}                                                                 
 \end{figure}

NGC 3059 (Figure~\ref{ngc3059}) - The single main P/N crossing at $r$=26\arcs\ 
(0.20) encircles around
the ends of the strong apparent bar. A second crossing near $r$=102\arcs\
(0.78) is close to $r_o$ and of uncertain significance.

\begin{figure}                                                                 
\figurenum{2.44}
\plotone{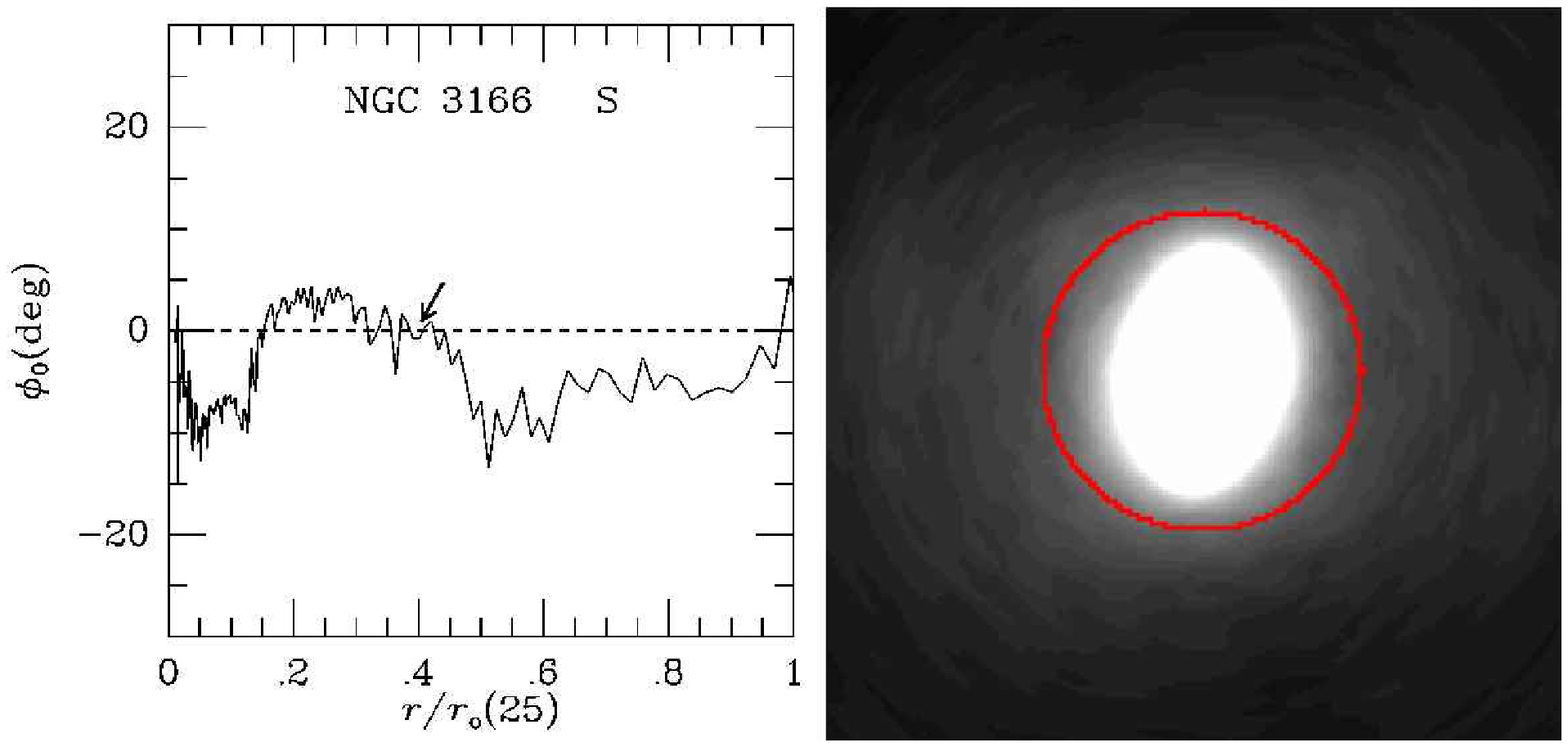}
 \vspace{2.0truecm}                                                             
\caption{Same as Figure 2.1 for NGC 3166}                                         
\label{ngc3166}                                                                 
 \end{figure}

NGC 3166 (Figure~\ref{ngc3166}) - 
The main crossing at $r$$\approx$55\arcs\ (0.39) is broad and somewhat
ill-defined, but appears to circle the ends of the strong inner oval.
There is also the indication of an unresolved nuclear pattern.

\begin{figure}                                                                  
\figurenum{2.45}
\plotone{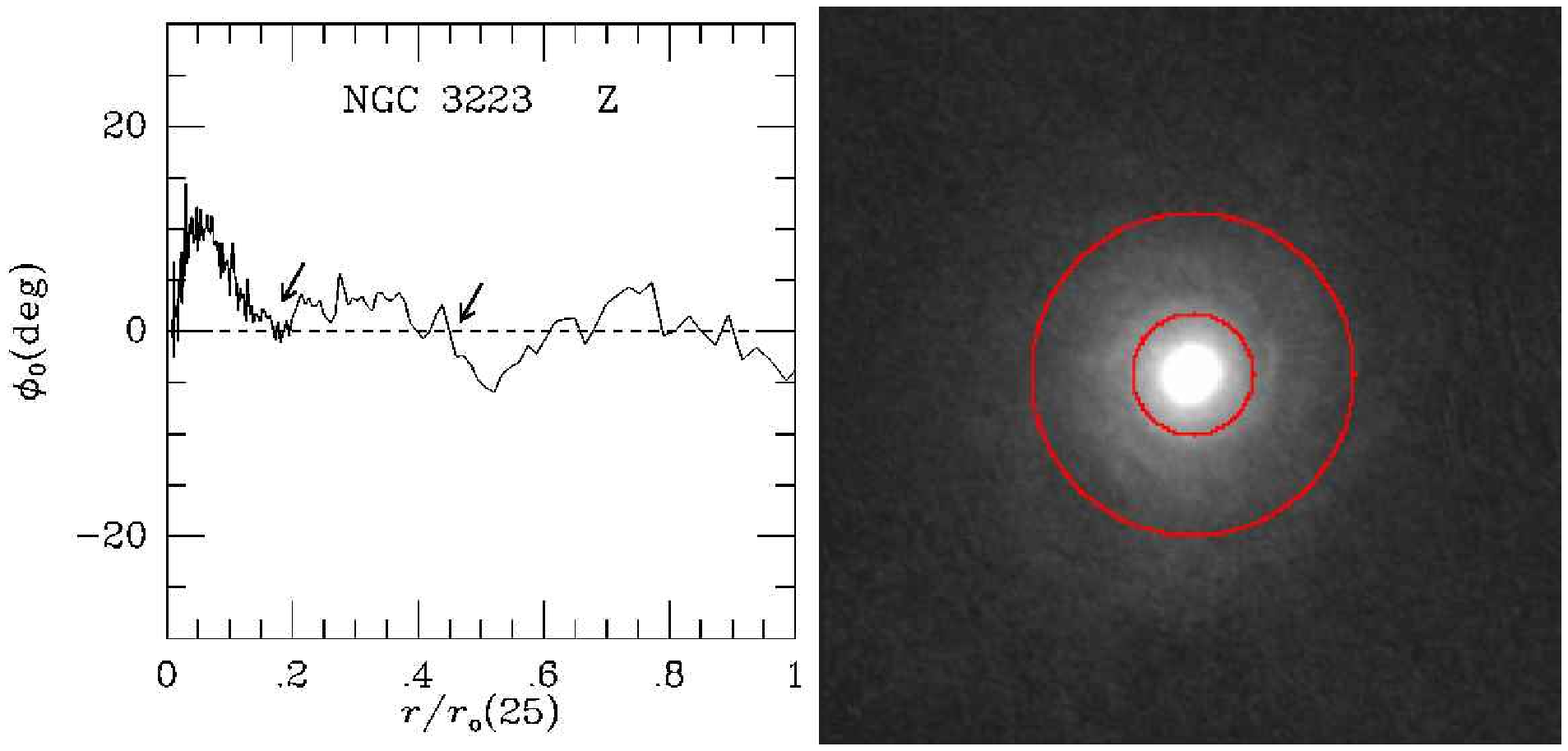}
\vspace{2.0truecm}                                                              
\caption{Same as Figure 2.1 for NGC 3223}                                         
\label{ngc3223}                                                                 
\end{figure}

NGC 3223 (Figure~\ref{ngc3223}) - The crossing at $r$=23\arcs\ (0.17) could be
associated with a small inner oval. The crossing at $r$=62\arcs\ (0.45)
is likely to be associated with the spiral arms. 

\begin{figure}                                                                  
\figurenum{2.46}
\plotone{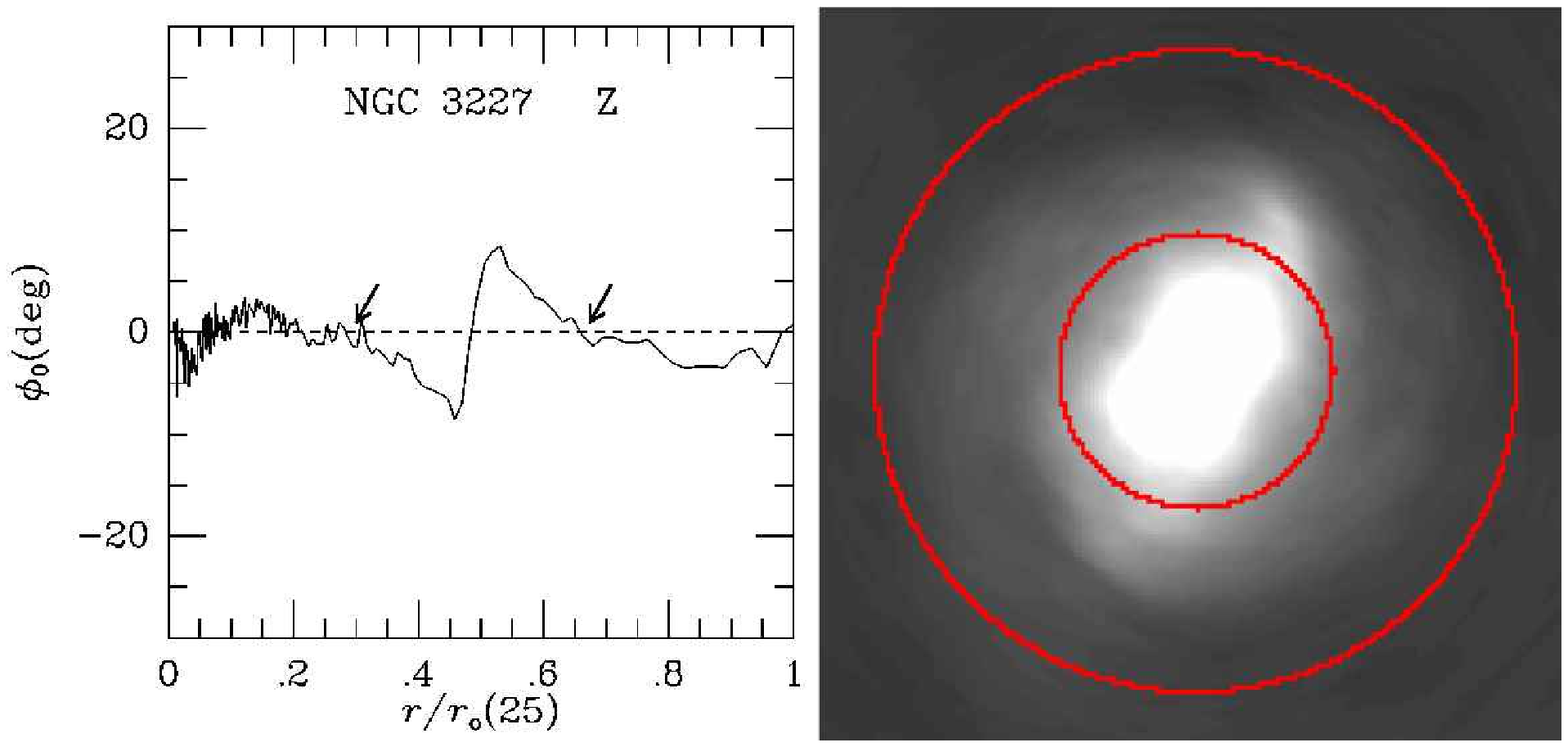}
\vspace{2.0truecm}                                                              
\caption{Same as Figure 2.1 for NGC 3227}                                         
\label{ngc3227}                                                                 
\end{figure}

NGC 3227 (Figure~\ref{ngc3227}) - This galaxy is the spiral component
of an E-S pair with NGC 3226. The phase-shift distribution suggests two
significant crossings. The one at $r$=47\arcs\ (0.29) is noisy but circles just
outside the ends of an inner boxy bar zone. The crossing at 108\arcs\
(0.66) lies outside the main spiral arms but still well inside $r_o$.
There is the suggestion of a nuclear crossing as well.

\begin{figure}                                                                 
\figurenum{2.47}
\plotone{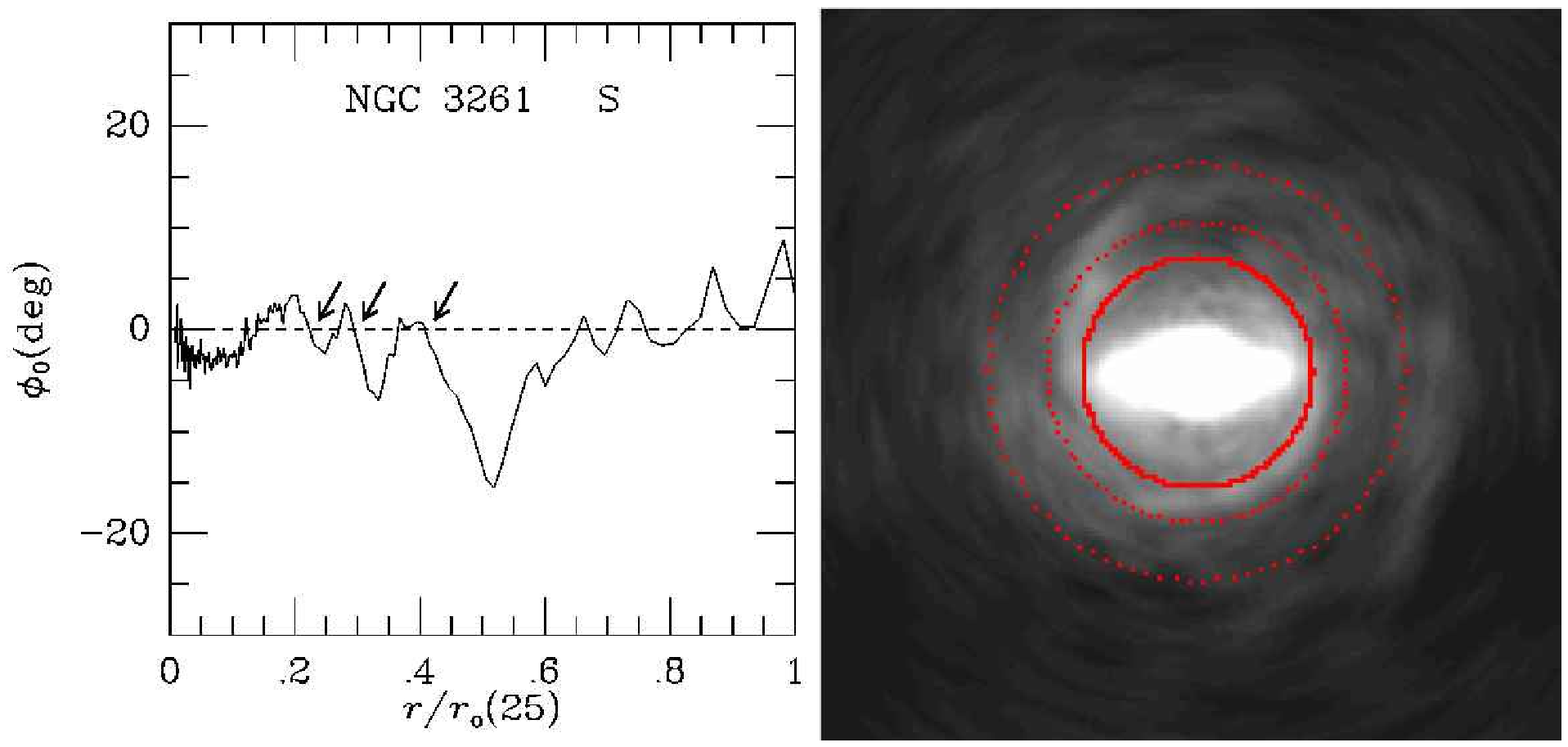}
 \vspace{2.0truecm}                                                             
\caption{Same as Figure 2.1 for NGC 3261}                                         
\label{ngc3261}                                                                 
 \end{figure}

NGC 3261 (Figure~\ref{ngc3261}) - A multi-armed barred spiral showing three possibly significant
P/N crossings within $r_o$. The crossing at $r$=29\arcs\ (0.23) 
circles around the
ends of the prominent bar. The crossings at $r$=38\arcs\ (0.30)
and 52\arcs\ (0.41) could
indicate non-steady spirals, since they are so close together, but we do see
different spiral patterns in this region and these two CRs could be real.

RSL08 obtained a good simulation representation of NGC 3261 with $r(CR)$ =
44\arcss 1 (0.34), intermediate between our second and third CRs and
likely associated with the spiral and not the bar. Our
first CR is comparable to their estimate of the bar radius, 28\arcss 3 (0.22).

\begin{figure}                                                                 
\figurenum{2.48}
\plotone{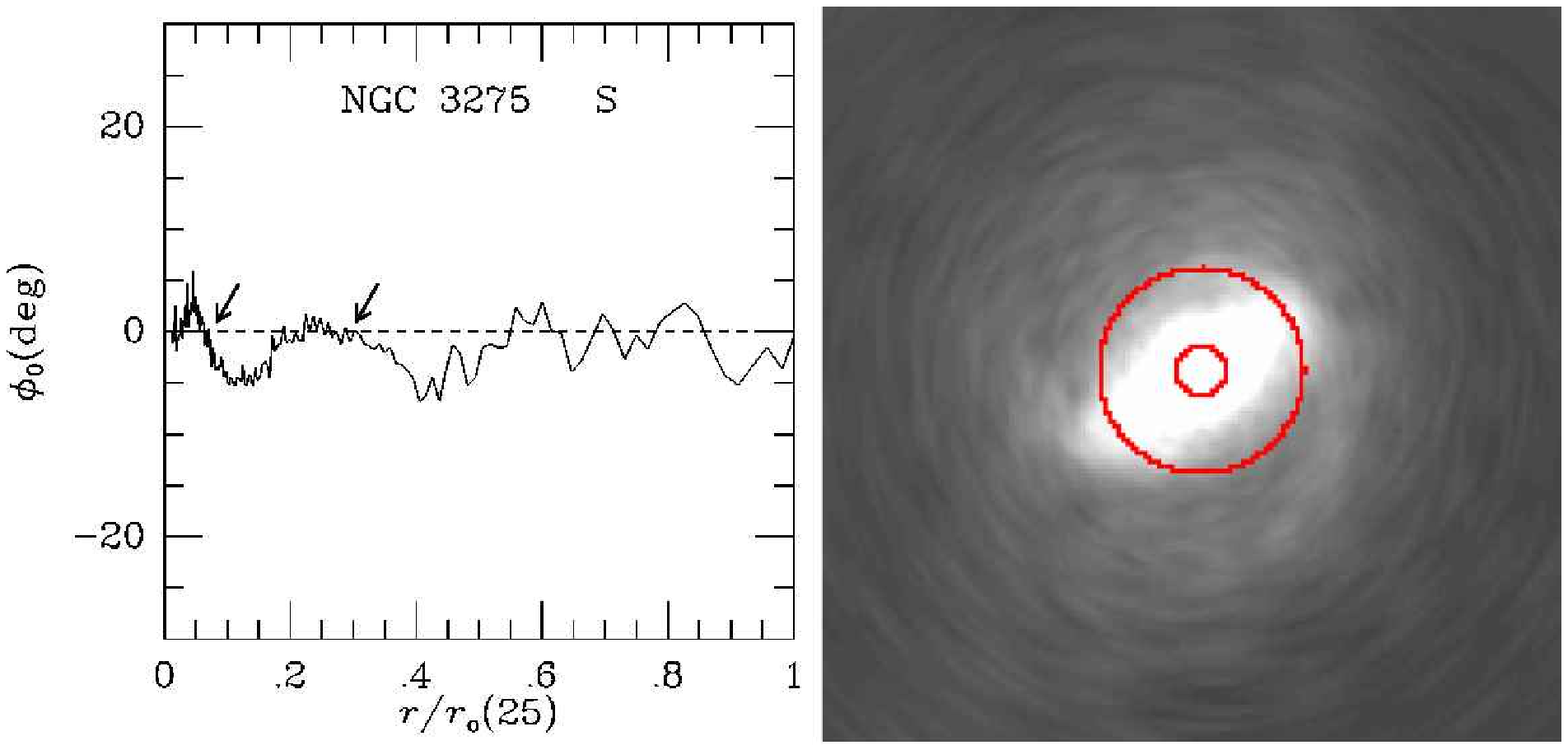}
 \vspace{2.0truecm}                                                             
\caption{Same as Figure 2.1 for NGC 3275}                                         
\label{ngc3275}                                                                 
 \end{figure}

NGC 3275 (Figure~\ref{ngc3275}) - 
Two clear crossings can be identified in the phase-shift plot.
The larger radius, at $r$=26\arcs\ (0.29), circles near the ends of the
bar while the smaller radius is well inside the bar ends. In this case,
the suggestion is that the bar may extend to near its OLR radius, or else
the outer portion is in the process of evolving into a bar-driven
spiral. None of the crossings at larger radii looks significant. 

The RSL08 simulation of this galaxy gives $r(CR)$ = 44\arcss 2 (0.50),
much larger than our second CR which is close to the bar radius of
28\arcss 9 (0.33) estimated by those authors.

\begin{figure}                                                                 
\figurenum{2.49}
\plotone{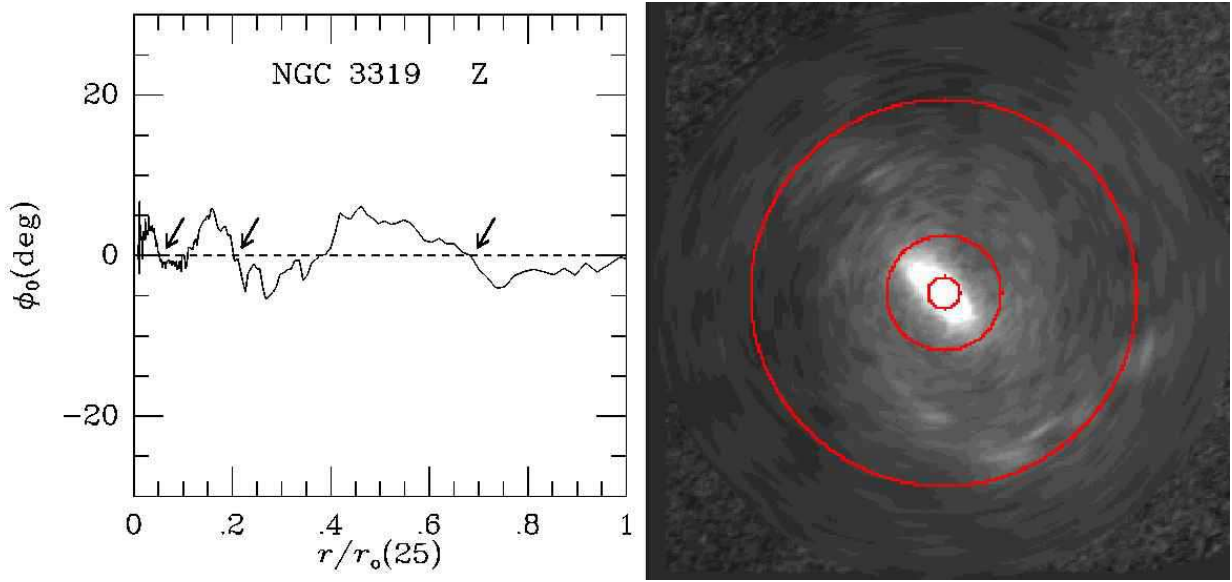}
 \vspace{2.0truecm}                                                             
\caption{Same as Figure 2.1 for NGC 3319}                                         
\label{ngc3319}                                                                 
 \end{figure}

NGC 3319 (Figure~\ref{ngc3319}) - The phase-shift distribution for this
late-type galaxy shows three likely significant P/N crossings. The
crossing at $r$=38\arcs\ (0.21) circles the ends of the prominent bar,
but the much larger crossing at $r$=126\arcs\ (0.68) lies near the
outer extent of the main spiral pattern. The crossing at $r$=9\arcss\ 5
(0.05) lies well inside the bar ends as does its associated N/P crossing. 

\begin{figure}                                                                 
\figurenum{2.50}
\plotone{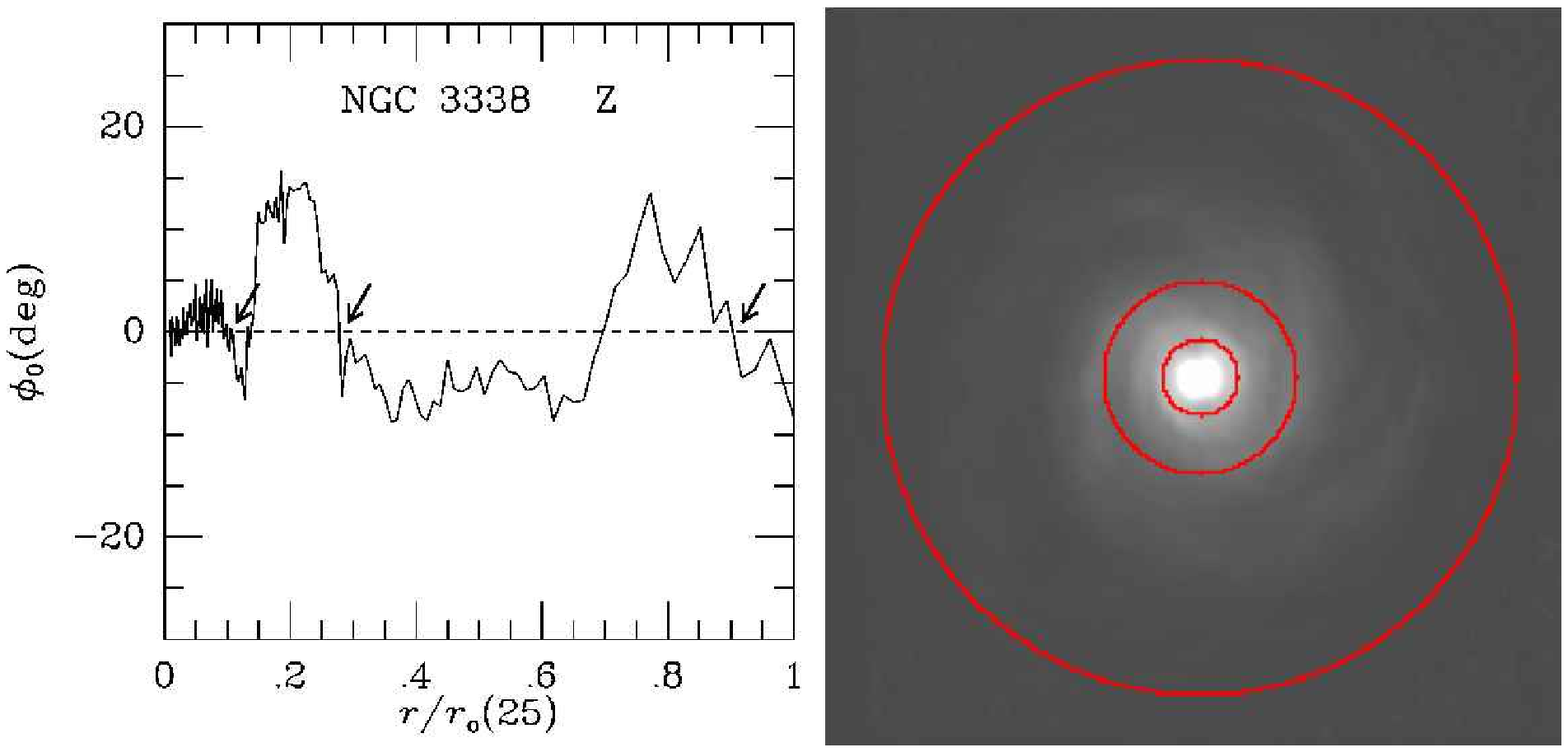}
 \vspace{2.0truecm}                                                             
\caption{Same as Figure 2.1 for NGC 3338}                                         
\label{ngc3338}                                                                 
 \end{figure}

NGC 3338 (Figure~\ref{ngc3338}) - This galaxy has clear patterns on
different scales. The three indicated crossings are all likely to be
significant.

\begin{figure}                                                                 
\figurenum{2.51}
\plotone{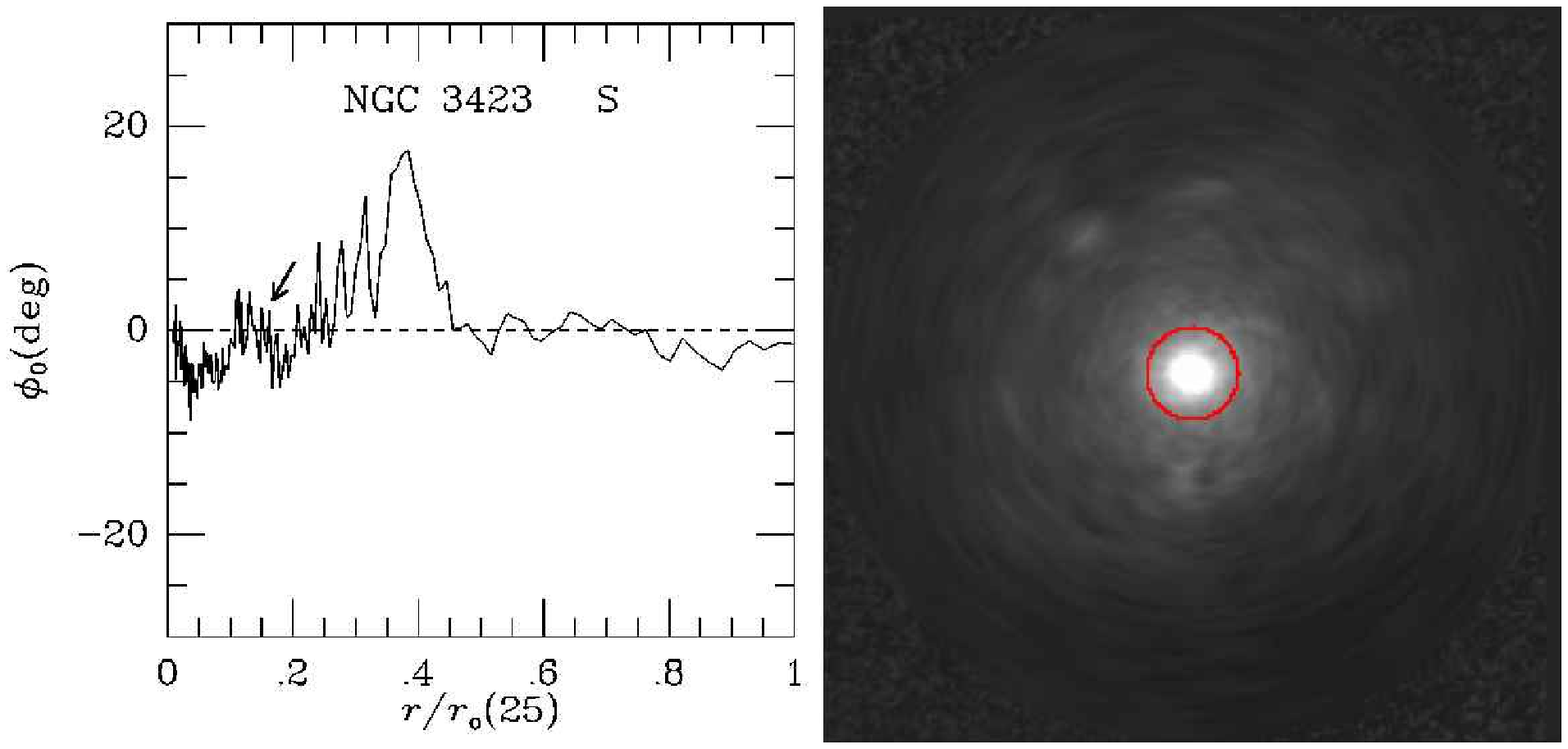}
 \vspace{2.0truecm}                                                             
\caption{Same as Figure 2.1 for NGC 3423}                                         
\label{ngc3423}                                                                 
 \end{figure}

NGC 3423 (Figure~\ref{ngc3423}) - A late-type galaxy with weak features. 
There is a possibly
significant but noisy P/N crossing near $r$=18\arcs\ (0.15), and there may also
be one at smaller radii, but there is no clearly-defined pattern in the outer disk
and the outer crossings may not be significant.

\begin{figure}                                                                 
\figurenum{2.52}
\plotone{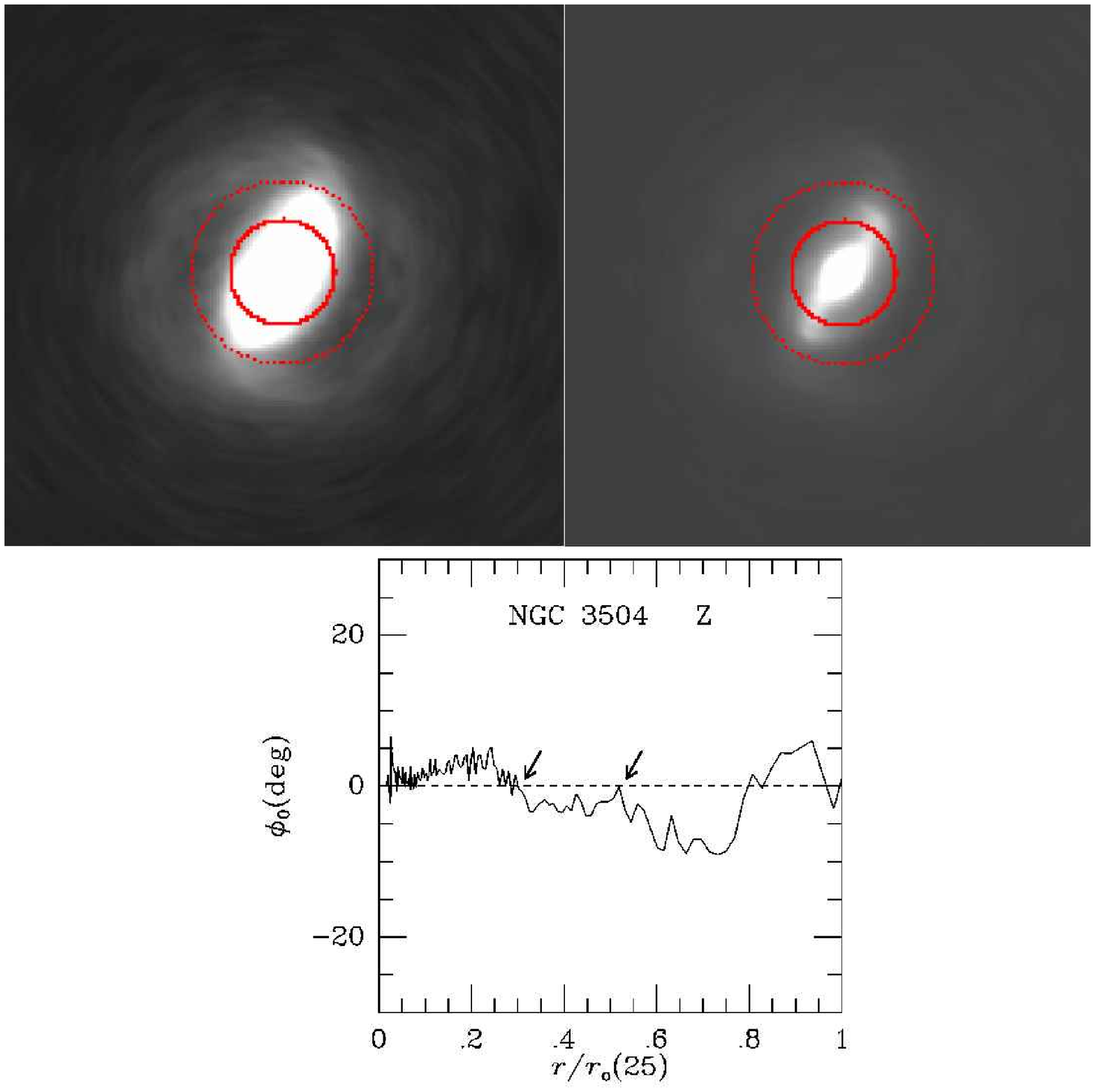}
 \vspace{2.0truecm}                                                             
\caption{Same as Figure 2.1 for NGC 3504}                                         
\label{ngc3504}                                                                 
 \end{figure}

NGC 3504 (Figure~\ref{ngc3504}) - The single 
P/N crossing at $r$=24\arcs\ (0.30) circles well inside the
apparent bar ends of this relatively normal-looking barred spiral. The
primary bar is noticeably twisted (Figure~\ref{ngc3504}a) and the 24\arcs\
crossing appears to circle an inner oval. However,
the near-crossing at $r$$\approx$42\arcs\ (0.52) coincides with the ends of the
main bar/oval (Figure~\ref{ngc3504}b) and could indicate a nearly decoupled 
pattern. The 42\arcs\ near-crossing also passes through the gap regions of the
prominent R$_1^{\prime}$ outer pseudoring and seems to support the idea that
the feature is linked to an OLR. 

RSL08 were able to obtain a good simulation of this galaxy that gave
$r(CR)$ = 44\arcss 5 (0.55), which is very close to our second CR. Their
bar radius of 37\arcss 4 (0.46) is intermediate between our two CRs.

\begin{figure}                                                                 
\figurenum{2.53}
\plotone{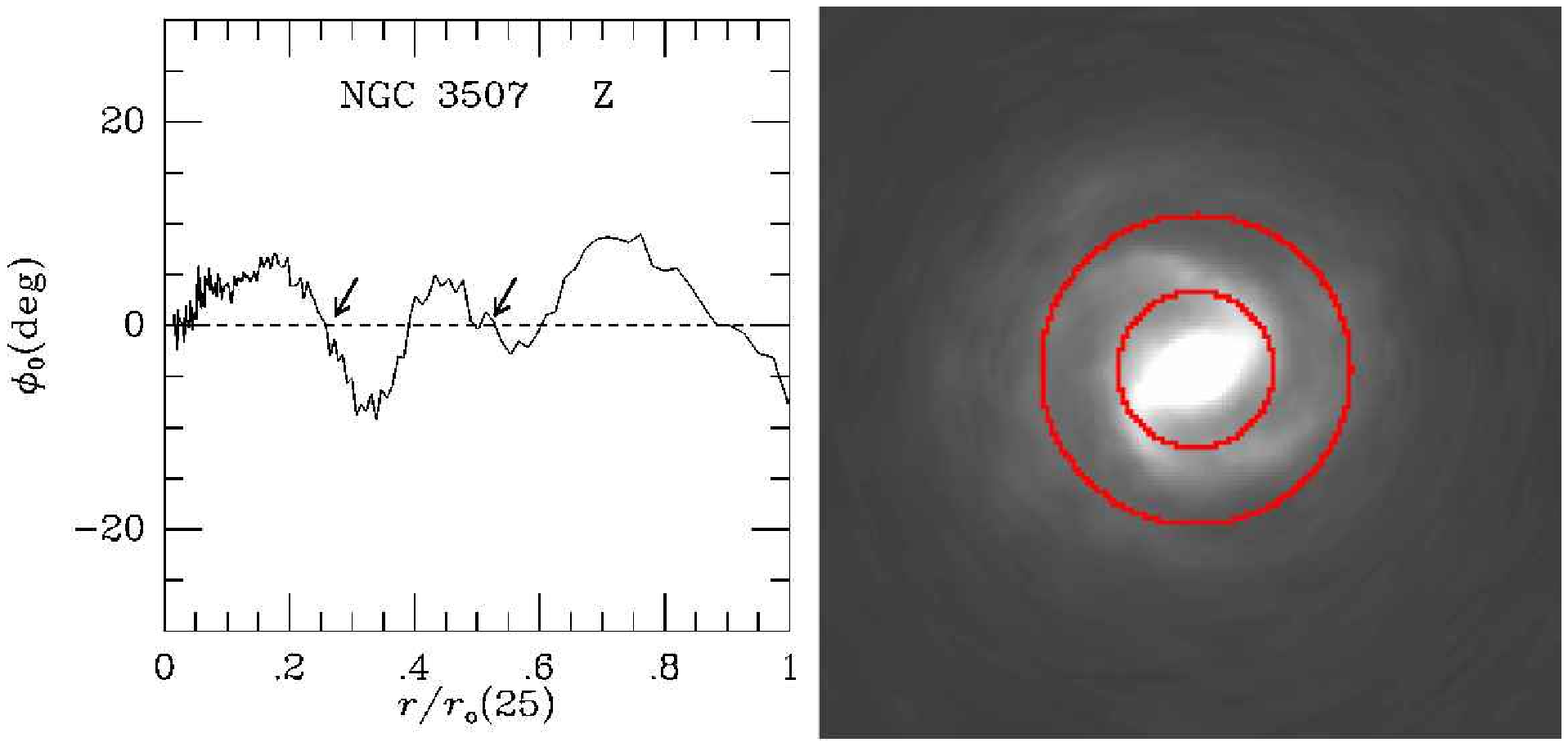}
 \vspace{2.0truecm}                                                             
\caption{Same as Figure 2.1 for NGC 3507}                                         
\label{ngc3507}                                                                 
 \end{figure}

NGC 3507 (Figure~\ref{ngc3507}) - The crossing at $r$=26\arcs\ (0.26)
appears to circle just outside the
bar ends, while the one at $r$=52\arcs\ (0.51) circles just outside the bright
spiral pattern.

An RSL08 simulation model that best matches the $B$-band spiral morphology
gave $r(CR)$ = 34\arcss 2 (0.34), which is well inside our second CR 
but close to our first CR. Nevertheless, our first CR matches well
the bar radius of 28\arcss 1 (0.28) estimated by RSL08. The RSL08
model makes the outer spiral more prominent than the inner spiral,
while the near-IR image indicates a stronger inner spiral.

\begin{figure}                                                                 
\figurenum{2.54}
\plotone{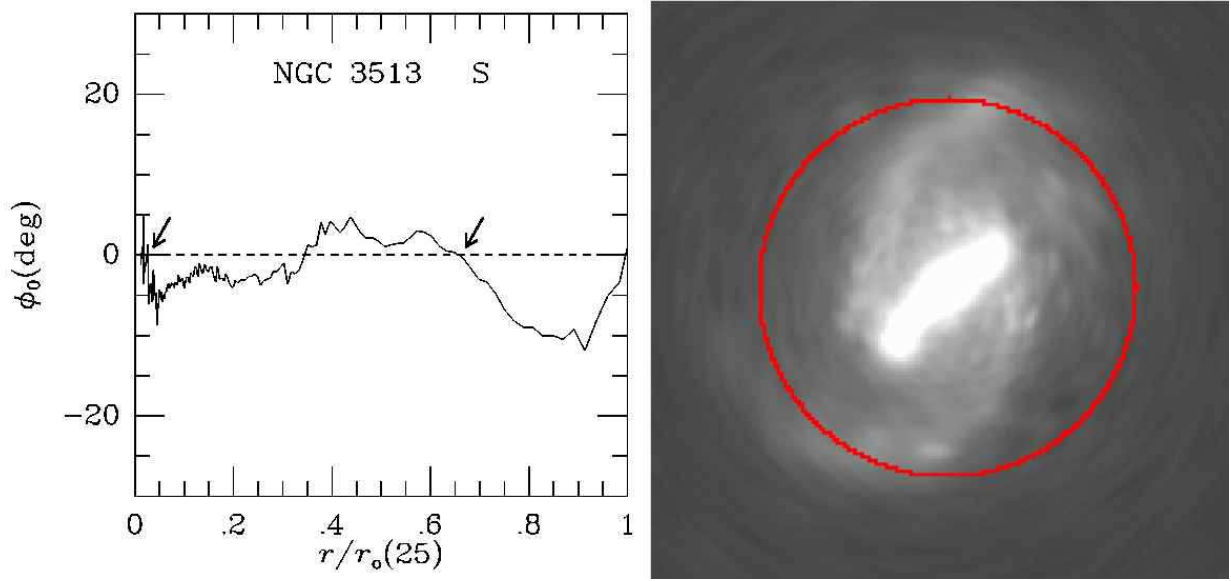}
 \vspace{2.0truecm}                                                             
\caption{Same as Figure 2.1 for NGC 3513. Only CR$_2$ from                        
Table 1 is shown overlaid on the image.}                                        
\label{ngc3513}                                                                 
 \end{figure}

NGC 3513 (Figure~\ref{ngc3513}) - The single major P/N crossing circles
far outside the ends of the bar, close to the outer edge of a strong
spiral. The ends of the bar are delineated in this case by the N/P
crossing at $r$$\approx$31\arcs\ (0.34). Although we have selected the
main P/N crossing as the bar CR in Table 1, the galaxy appears to be a
transition type where an inner pattern is decoupling from the outer
pattern as indicated by the hump in the negative-valued inner portion
of the phase-shift curve. The bar may then be ending at close to its
OLR after decoupling, not its CR.

An excellent RSL08 model gives $r(CR)$ = 43\arcss 5 (0.48), which is
well inside our single main CR of 60\arcs\ (0.66). 

\begin{figure}                                                                 
\figurenum{2.55}
\plotone{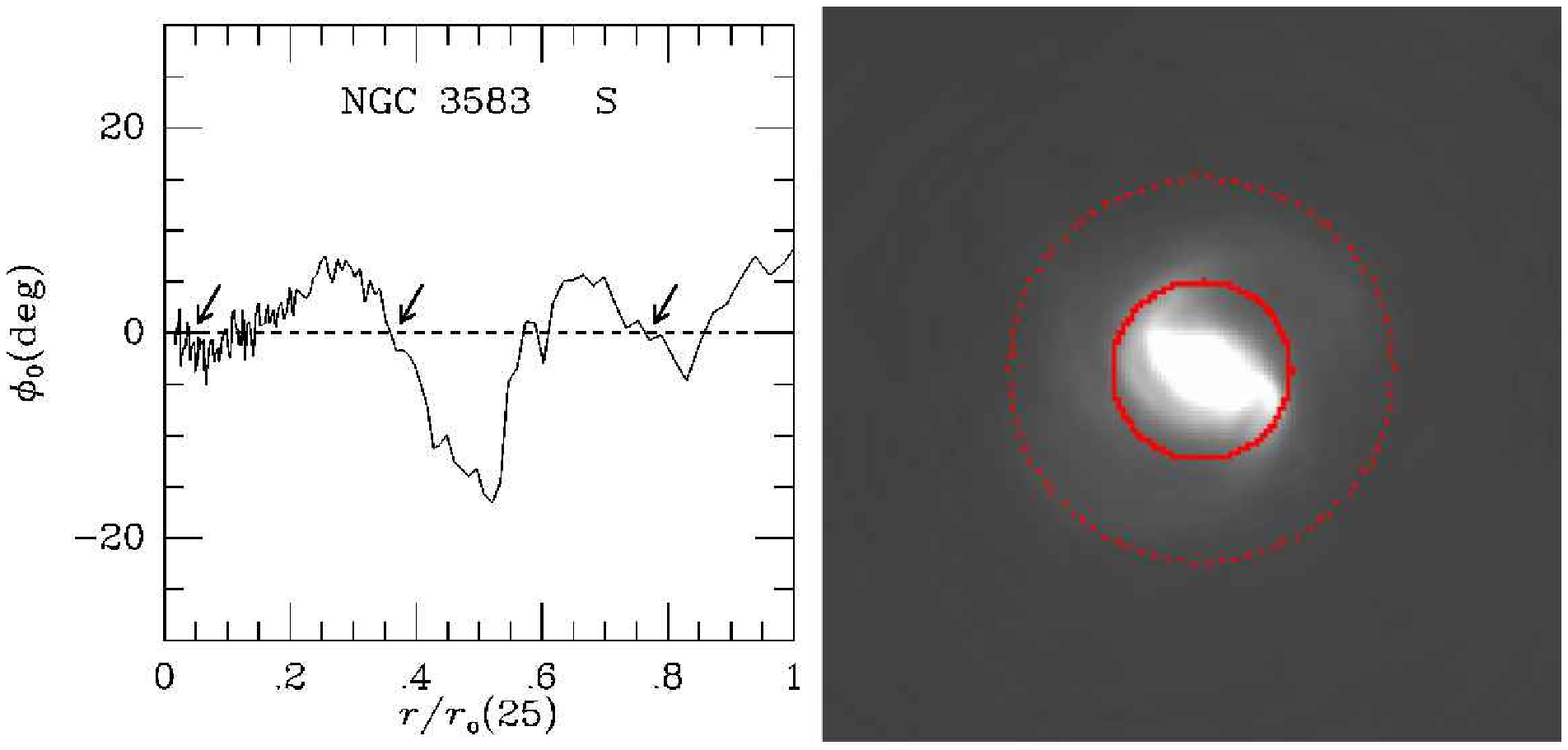}
 \vspace{2.0truecm}                                                             
\caption{Same as Figure 2.1 for NGC 3583. Only CR$_2$ and                         
CR$_3$ 	from                                                                    
Table 1 are shown overlaid on the image.}                                       
\label{ngc3583}                                                                 
 \end{figure}

NGC 3583 (Figure~\ref{ngc3583}) - The main P/N crossing at $r$=30\arcs\ (0.35)
circles around the ends of the bar in this galaxy. The outer crossing appears to be
associated with the outer spiral arms.

The RSL08 model of this galaxy reproduces the shape of the inner pseudoring 
well for $r(CR)$ = 32\arcss 1 (0.38), which is very close to our main
CR crossing. 

\begin{figure}                                                                  
\figurenum{2.56}
\plotone{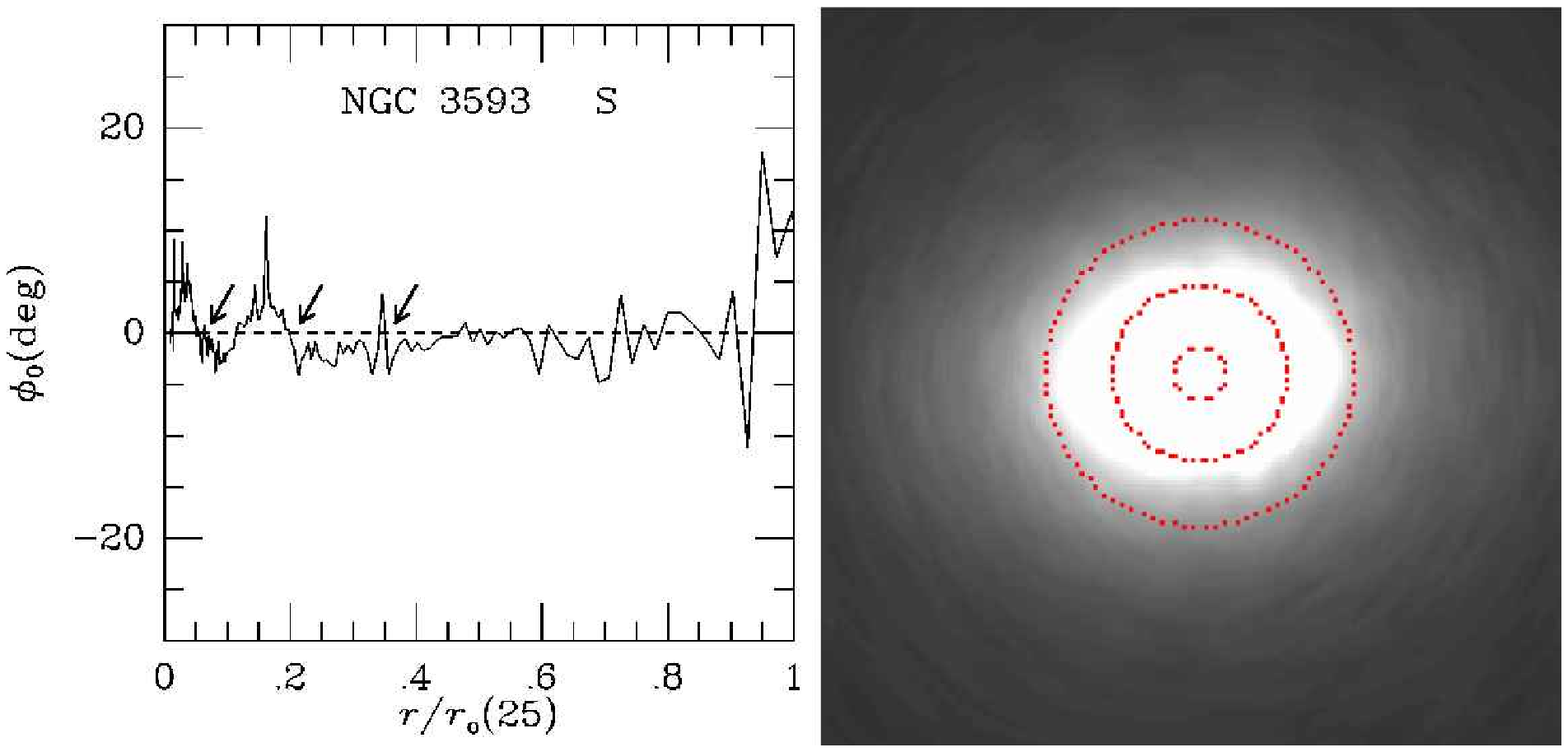}
\vspace{2.0truecm}                                                              
\caption{Same as Figure 2.1 for NGC 3593}                                         
\label{ngc3593}                                                                 
\end{figure}

NGC 3593 (Figure~\ref{ngc3593}) - The phase-shift distribution is based
on an average bulge image (section 3)
since the inclination of this galaxy is high.
The plot shows three crossings of uncertain significance. Deprojection
based on outer isophote shapes still leaves an extended oval distribution
in the galaxy's inner regions. 

\begin{figure}                                                                 
\figurenum{2.57}
\plotone{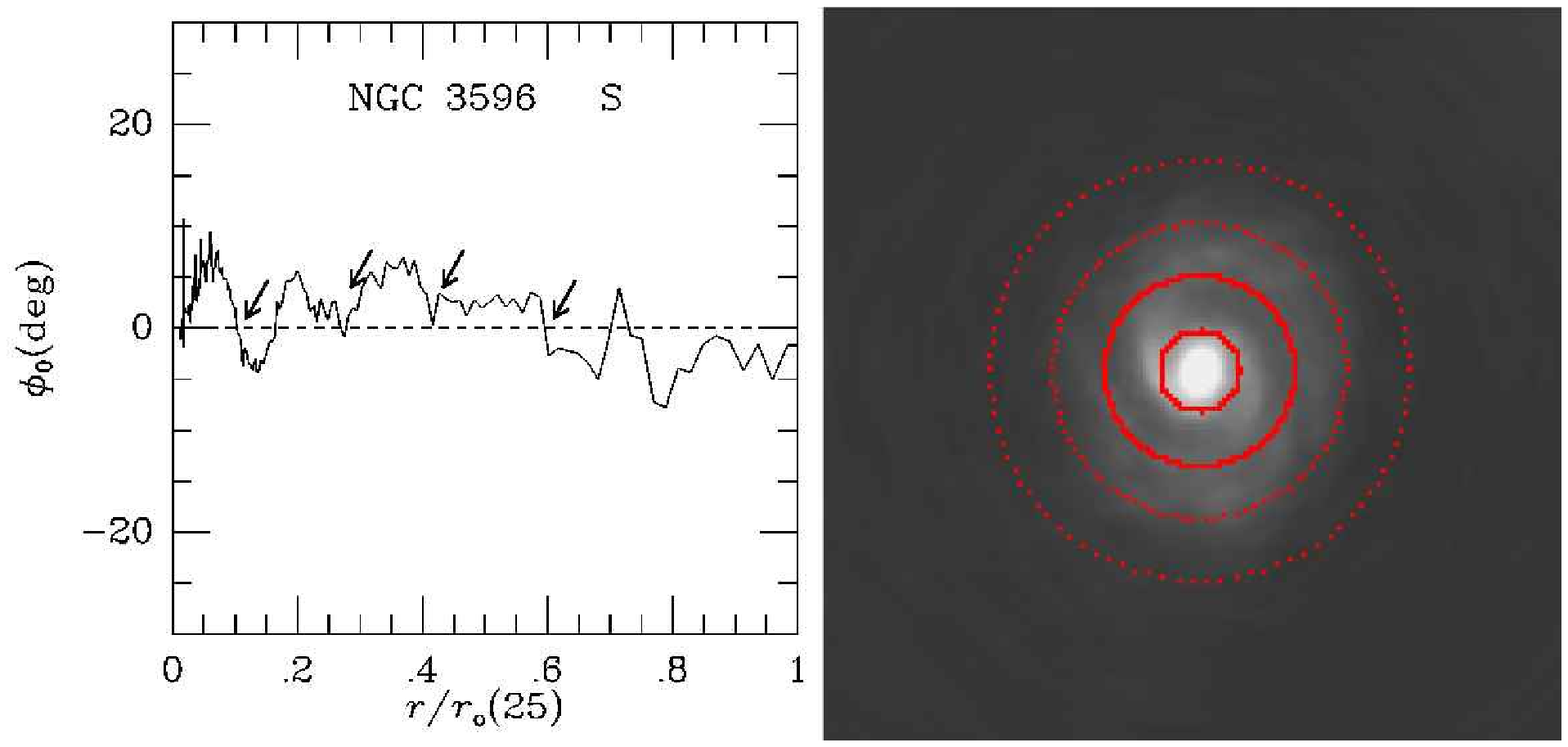}
 \vspace{2.0truecm}                                                             
\caption{Same as Figure 2.1 for NGC 3596}                                         
\label{ngc3596}                                                                 
 \end{figure}

NGC 3596 (Figure~\ref{ngc3596}) - The innermost crossing at
$r$=12\arcs\  (0.10) appears related to a small inner oval. The
crossing at $r$=32\arcs\ (0.27) lies in the middle of the main spiral
pattern and its weakness could indicate an ongoing decoupling. The
remaining two CRs lie near the edge of the near-IR disk, but an optical
color image of NGC 3596 from the Sloan Digital Sky Survey (SDSS) shows
considerable structure at larger radii. From comparisons of the
optical image, the NIR image, and the phase-shift plot we 
consider all four individual crossings to be significant; they
appear to refer to different density wave patterns.

\begin{figure}                                                                 
\figurenum{2.58}
\plotone{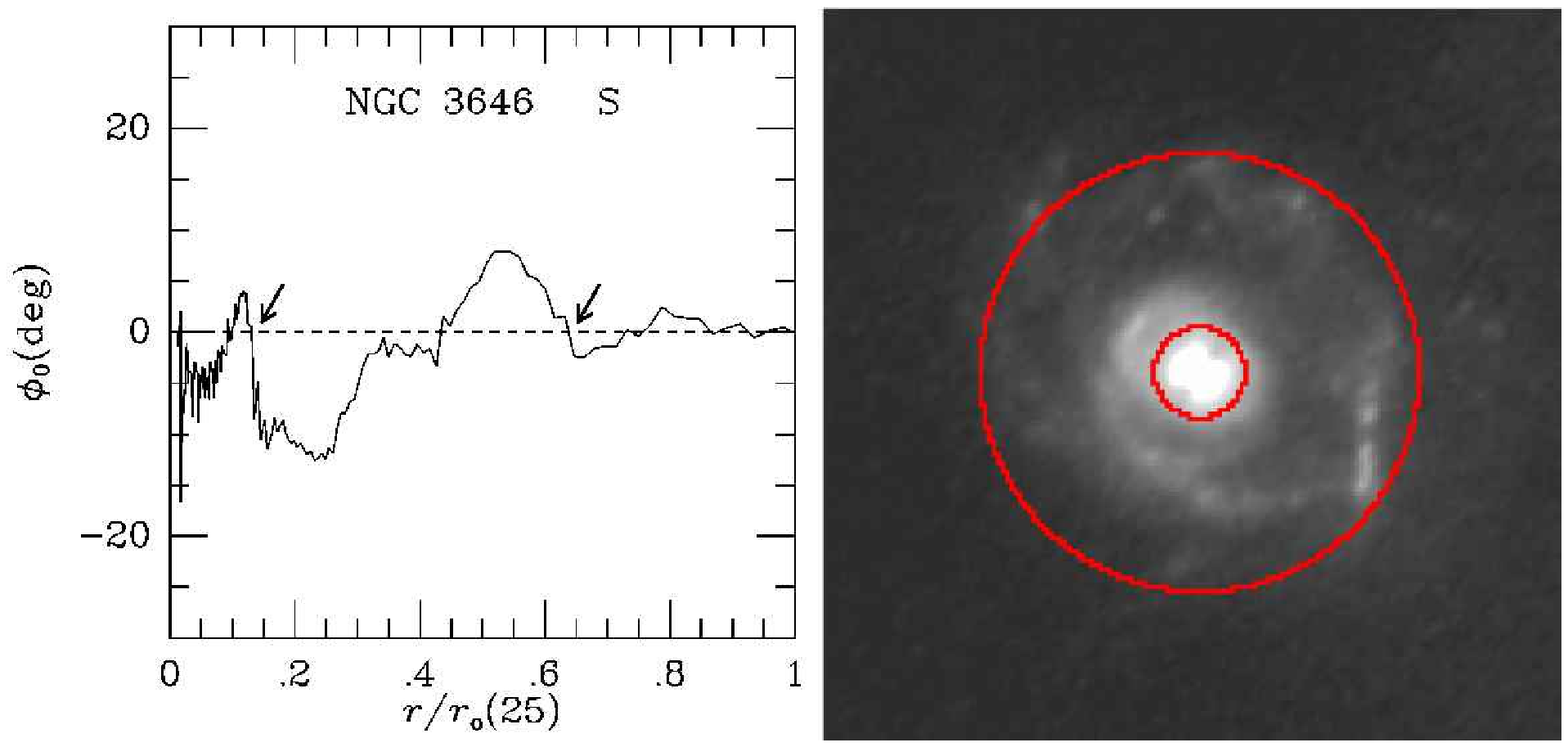}
 \vspace{2.0truecm}                                                             
\caption{Same as Figure 2.1 for NGC 3646}                                         
\label{ngc3646}                                                                 
 \end{figure}

NGC 3646 (Figure~\ref{ngc3646}) - The inner CR at $r$=15\arcs\ (0.13)
appears to encircle an area strongly affected by decomposition
pinch. The crossing at $r$=74\arcs\ (0.63) lies in the outer spiral.
There are also indications from both the phase-shift plot and the image
of a decoupling bar in the intermediate-radius region.

\begin{figure}                                                                 
\figurenum{2.59}
\plotone{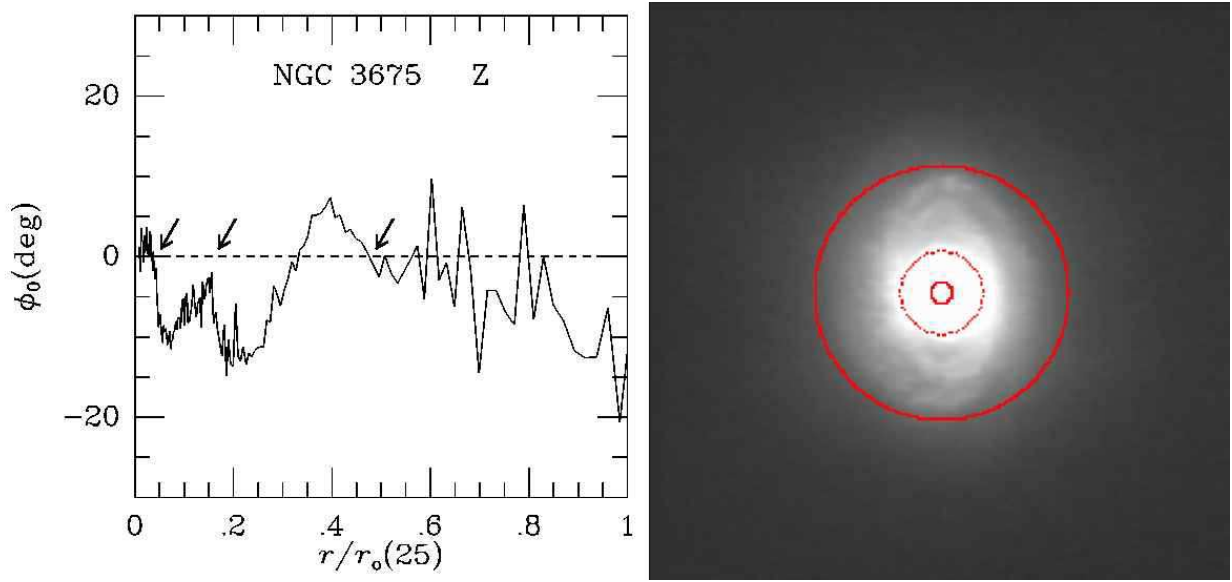}
 \vspace{2.0truecm}                                                             
\caption{Same as Figure 2.1 for NGC 3675. The dotted circle,                      
CR$_2$, is a ``near" crossing.}                                                 
\label{ngc3675}                                                                 
 \end{figure}

NGC 3675 (Figure~\ref{ngc3675}) - The third crossing at $r$=82\arcs\ 
(0.46) encircles a broad oval with tightly wrapped spiral structure emerging. 
The near-crossing at $r$=27\arcs\ (0.15) could be associated with a
decoupling inner oval. The central crossing is not resolved in the NIR
image.

\begin{figure}                                                                 
\figurenum{2.60}
\plotone{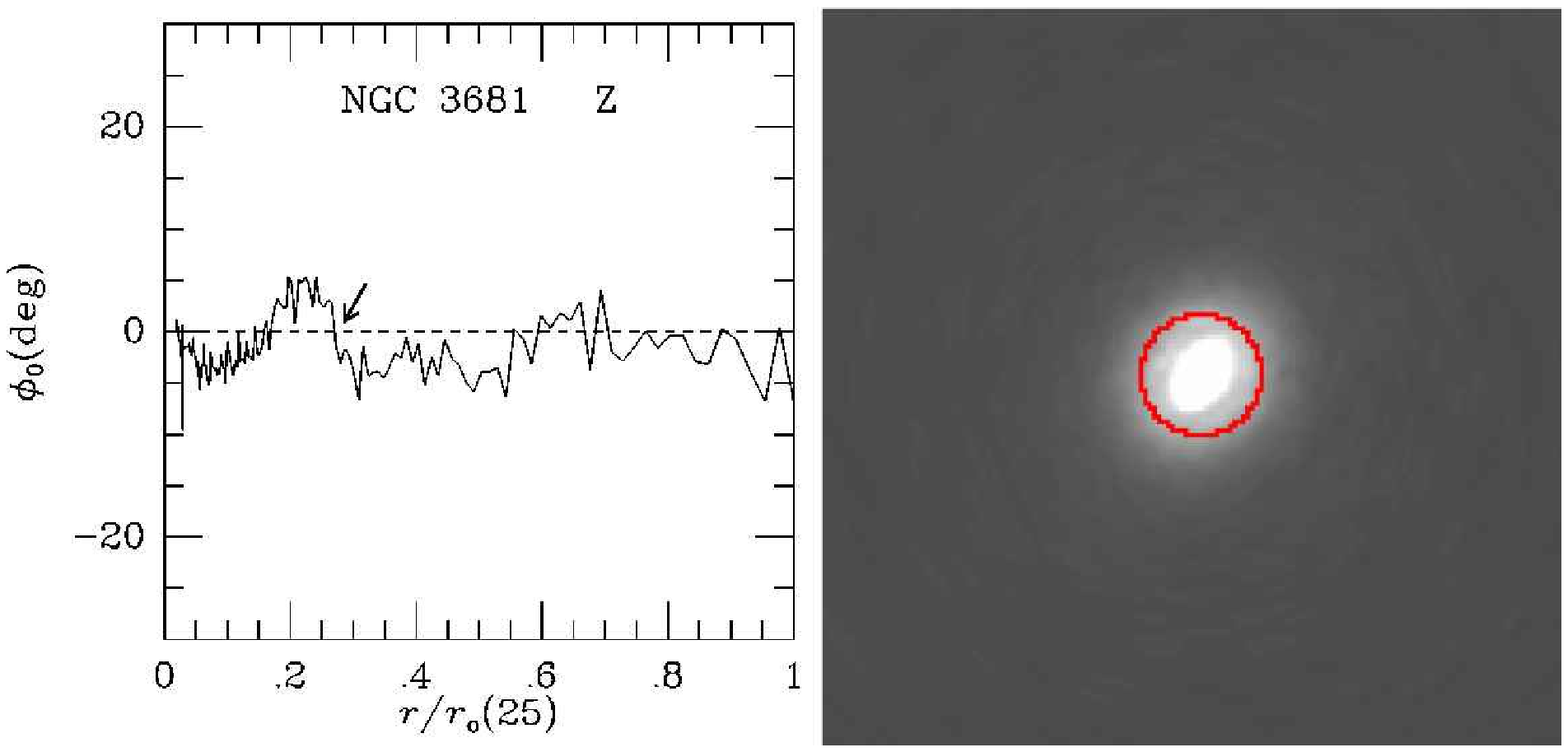}
 \vspace{2.0truecm}                                                             
\caption{Same as Figure 2.1 for NGC 3681}                                         
\label{ngc3681}                                                                 
 \end{figure}

NGC 3681 (Figure~\ref{ngc3681}) - The main crossing lies just beyond the
ends of a well-defined inner oval.  An unresolved nuclear crossing is also
indicated.

\begin{figure}                                                                 
\figurenum{2.61}
\plotone{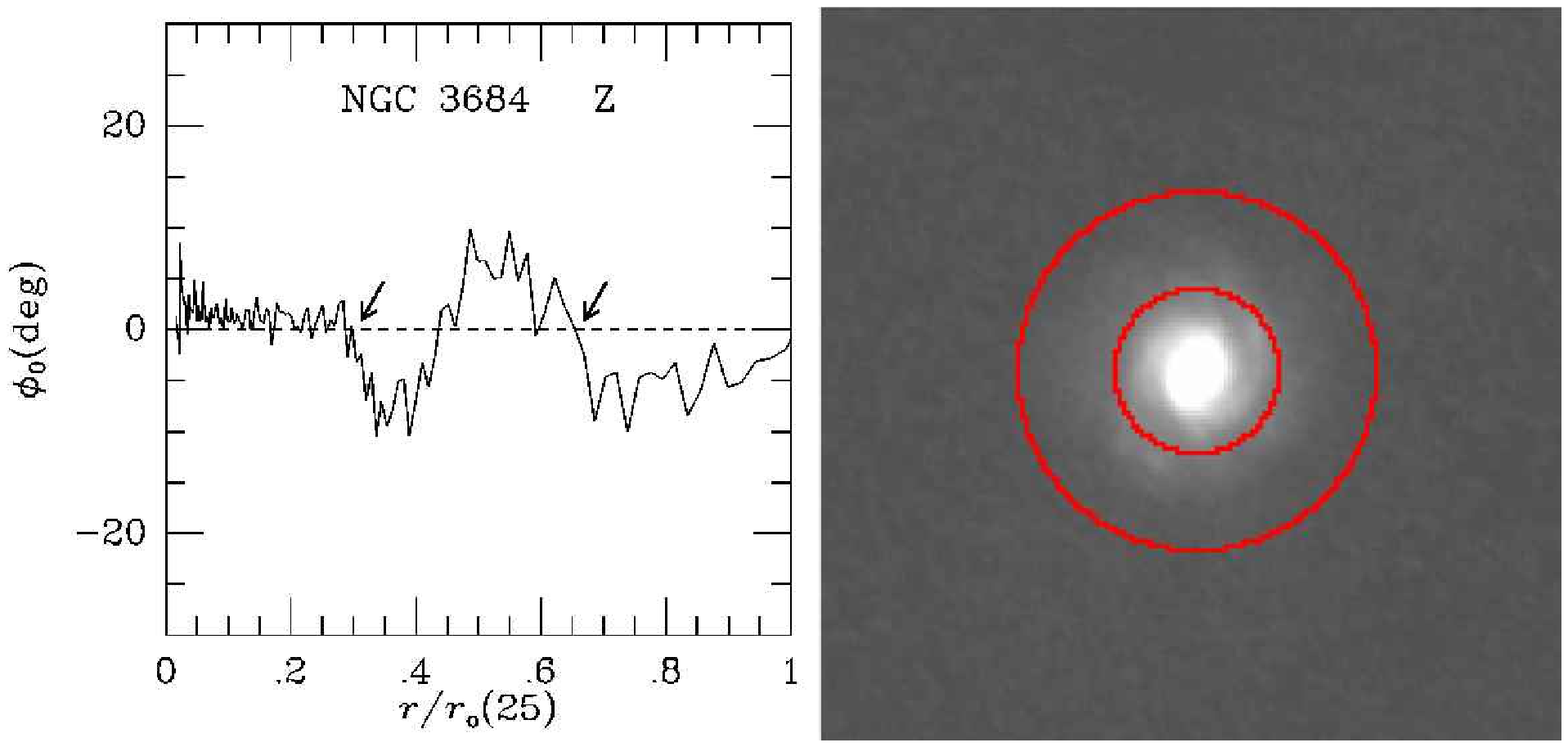}
 \vspace{2.0truecm}                                                             
\caption{Same as Figure 2.1 for NGC 3684}                                         
\label{ngc3684}                                                                 
 \end{figure}

NGC 3684 (Figure~\ref{ngc3684}) - The crossing at $r$=28\arcs\ (0.30) encircles
an inner oval. The outer crossing is likely related to the faint outer
structure.

\begin{figure}                                                                 
\figurenum{2.62}
\plotone{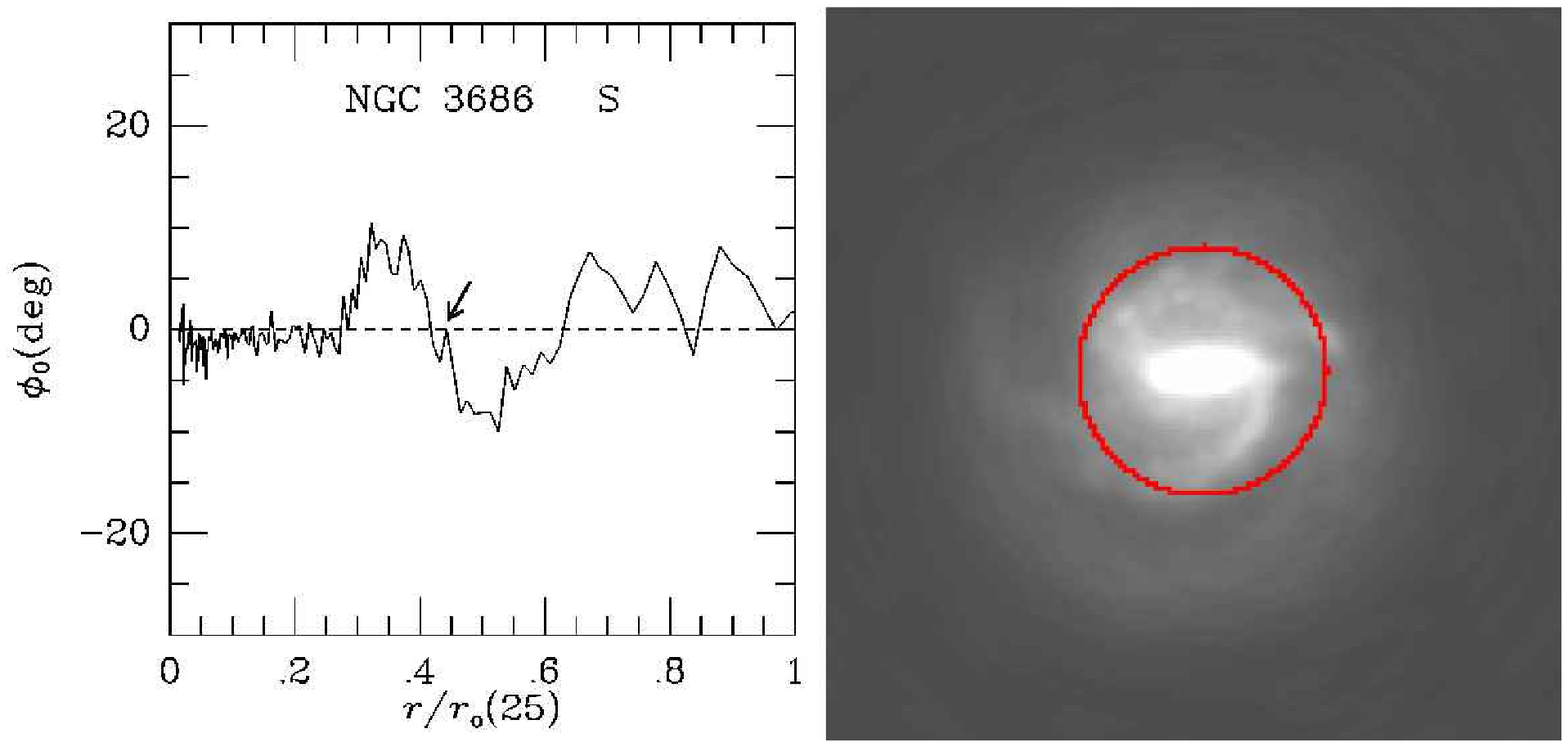}
 \vspace{2.0truecm}                                                             
\caption{Same as Figure 2.1 for NGC 3686}                                         
\label{ngc3686}                                                                 
 \end{figure}

NGC 3686 (Figure~\ref{ngc3686}) - The main crossing at 42\arcs\ (0.43)
lies largely outside
a prominent bar and inner spiral. This could be an example of a genuine
bar-driven spiral, but the prominent N/P crossing near $r$=27\arcs\
(0.28), and the undulations in the phase-shift curve in that region 
suggest that the inner bar pattern is going through a decoupling process.

The best-fitting RSL08 model for this galaxy gives $r(CR)$ = 35\arcss 6
(0.37), which is close to our main CR. Our two studies are in agreement
that this galaxy may be a slow bar case.

\begin{figure}                                                                 
\figurenum{2.63}
\plotone{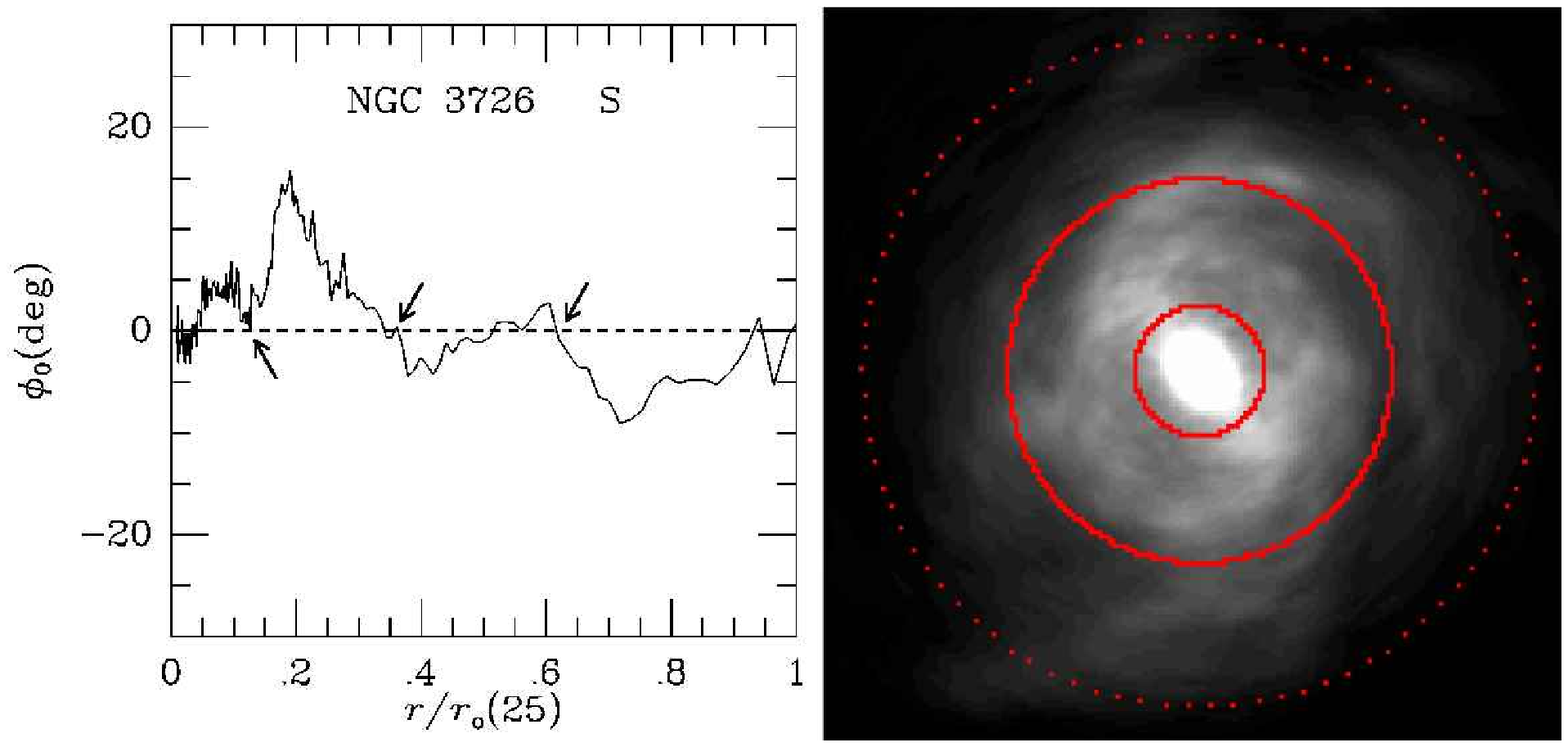}
 \vspace{2.0truecm}                                                             
\caption{Same as Figure 2.1 for NGC 3726}                                         
\label{ngc3726}                                                                 
 \end{figure}

NGC 3726 (Figure~\ref{ngc3726}) - The inner shallow-crossing suggests a
nearly-decoupled inner bar. The main crossing at $r$=65\arcs\ (0.35)
lies well outside the bar ends. An SDSS image shows a fragmentary outer 
spiral.

An RSL08 model of this galaxy has $r(CR)$ = 83\arcss 5 $\pm$ 13\arcss 9
(0.45$\pm$0.075), which is intermediate between our two outer CRs.

\begin{figure}                                                                 
\figurenum{2.64}
\plotone{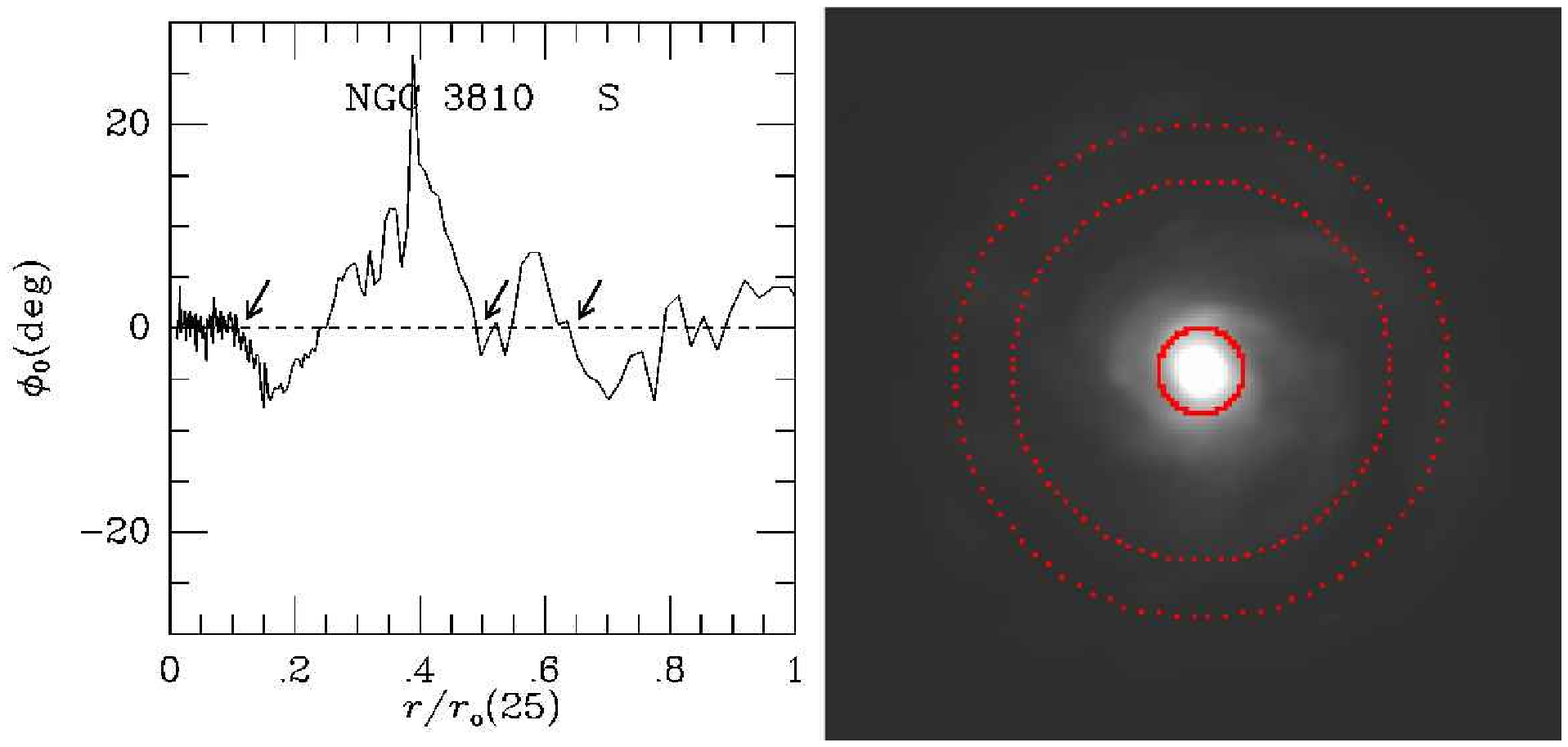}
 \vspace{2.0truecm}                                                             
\caption{Same as Figure 2.1 for NGC 3810}                                         
\label{ngc3810}                                                                 
 \end{figure}

NGC 3810 (Figure~\ref{ngc3810}) - The crossing at $r$=14\arcs\ (0.11) appears
to encircle a small inner oval. The other two crossings are in the main
spiral arms. The spiral is multi-armed and lacks bisymmetry but is still
regular-looking. 

\begin{figure}                                                                 
\figurenum{2.65}
\plotone{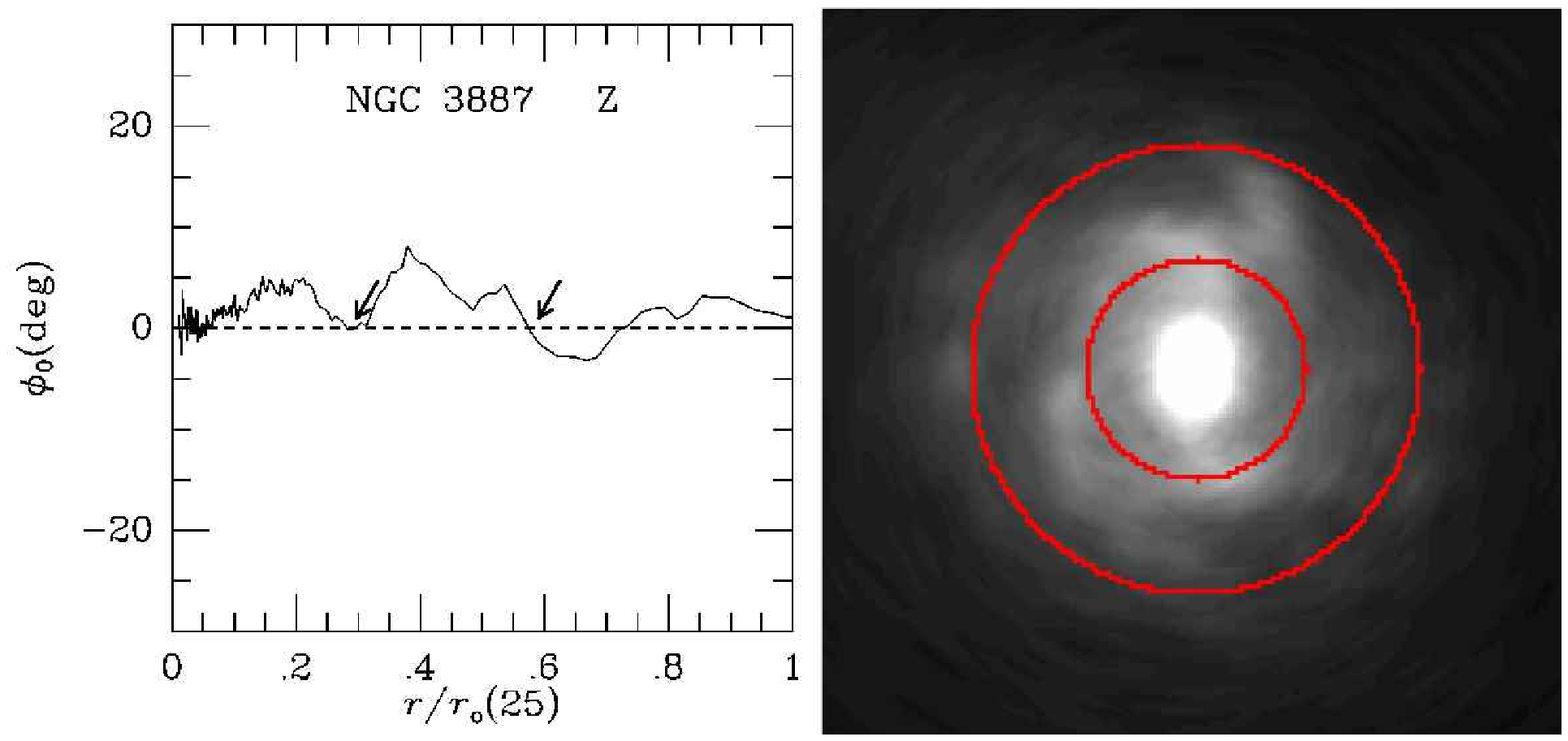}
 \vspace{2.0truecm}                                                             
\caption{Same as Figure 2.1 for NGC 3887}                                         
\label{ngc3887}                                                                 
 \end{figure}

NGC 3887 (Figure~\ref{ngc3887}) - The crossing at $r$=28\arcs\ (0.28)
circles outside a large, inner oval. The crossing at
$r$=57\arcs\ (0.57) encircles the main spiral pattern. On an optical
SDSS image, there is a well-defined two-armed spiral beyond 60\arcs\
(0.60).

\begin{figure}                                                                 
\figurenum{2.66}
\plotone{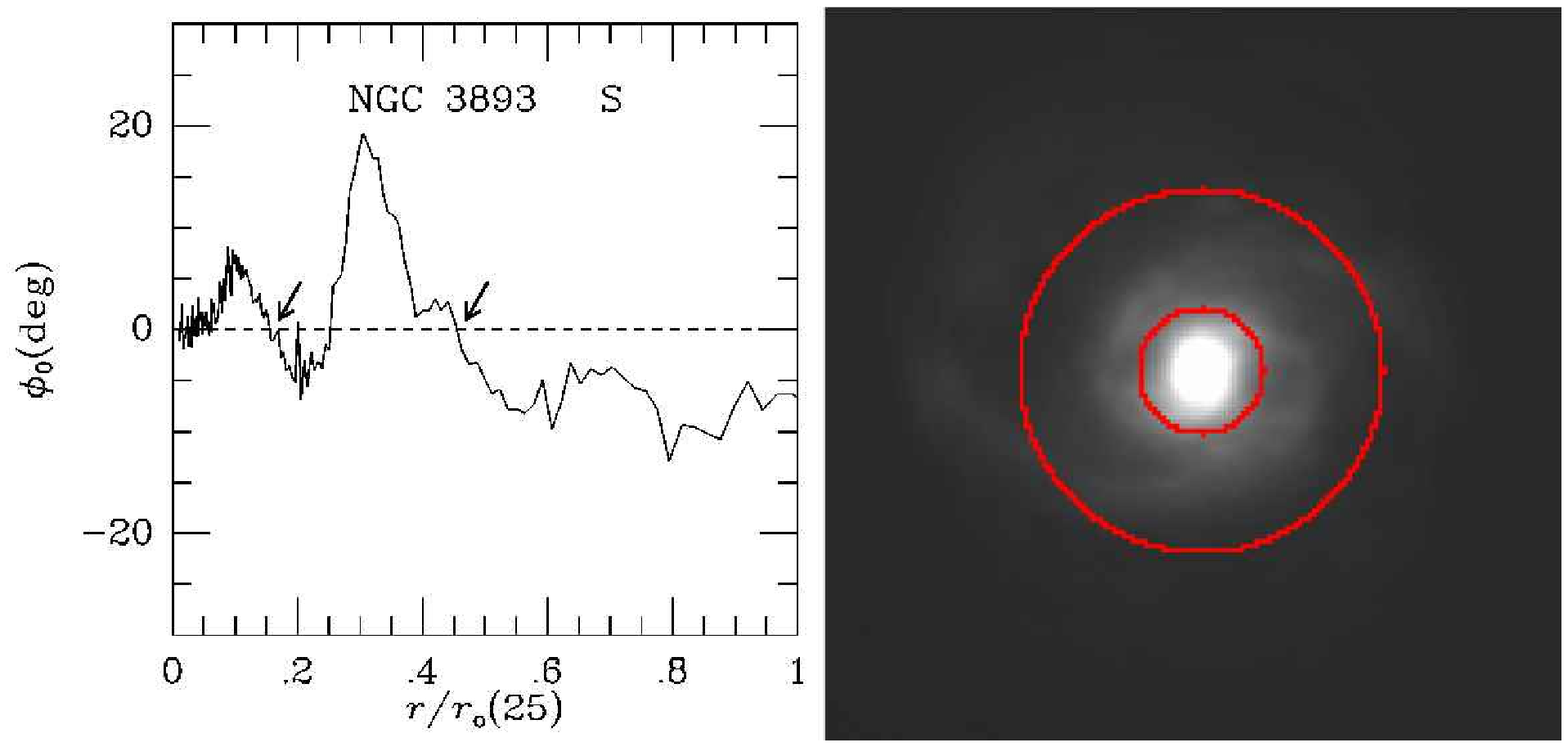}
 \vspace{2.0truecm}                                                             
\caption{Same as Figure 2.1 for NGC 3893}                                         
\label{ngc3893}                                                                 
 \end{figure}

NGC 3893 (Figure~\ref{ngc3893}) - The phaseshift plot shows two fairly
well-defined crossings. The crossing at $r$=21\arcs\ (0.16) encircles a small
inner oval, while the crossing at $r$=61\arcs\ (0.46) lies in the middle of
the main outer spiral. 

\begin{figure}                                                                 
\figurenum{2.67}
\plotone{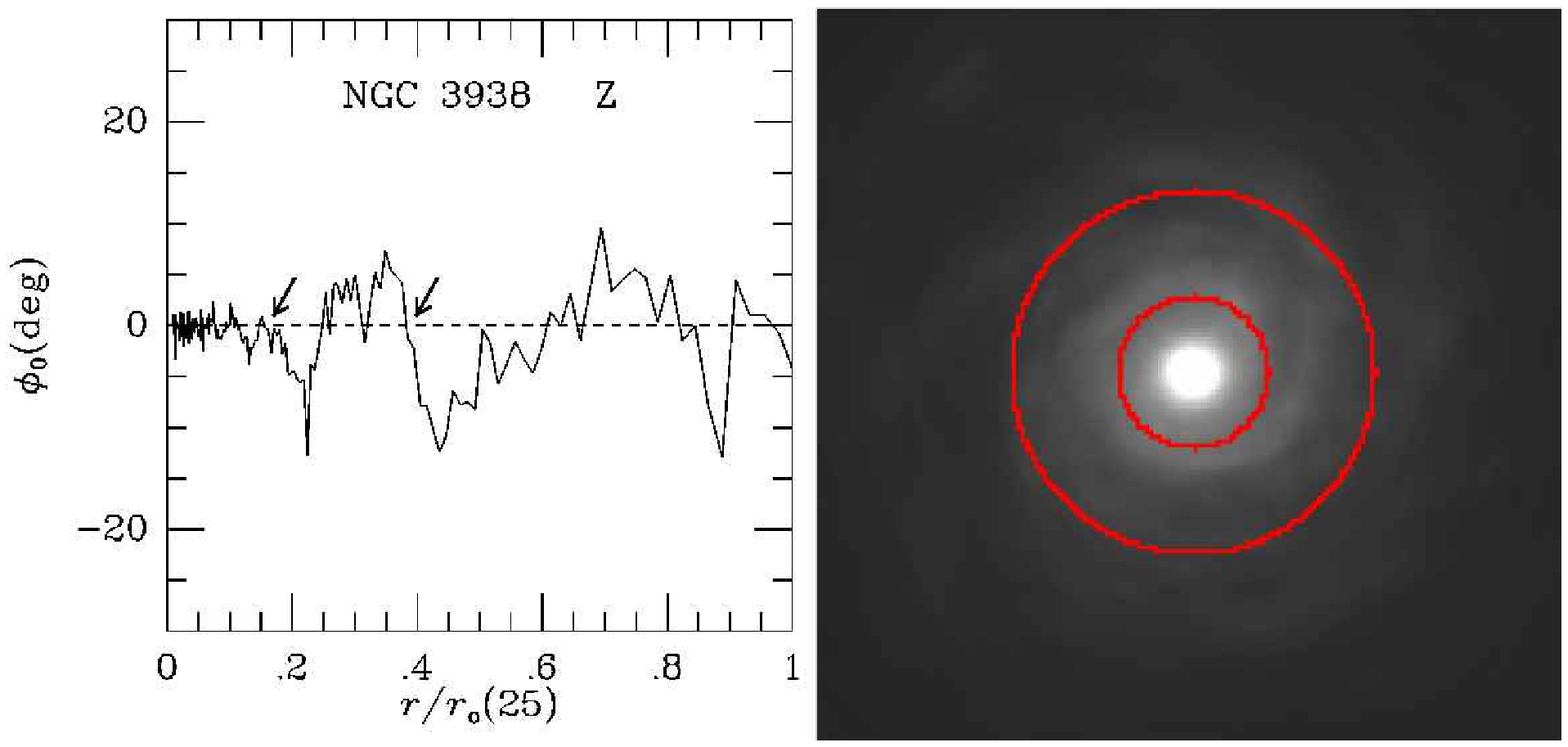}
 \vspace{2.0truecm}                                                             
\caption{Same as Figure 2.1 for NGC 3938}                                         
\label{ngc3938}                                                                 
 \end{figure}

NGC 3938 (Figure~\ref{ngc3938}) - The inner crossing at $r$=25\arcs\ (0.16)
appears
to encompass a broad oval, outside of which there is a bright inner spiral
emanating from this CR circle.
The crossing at $r$=62\arcs\ (0.38)
lies in the middle of the main spiral pattern.

\begin{figure}                                                                 
\figurenum{2.68}
\plotone{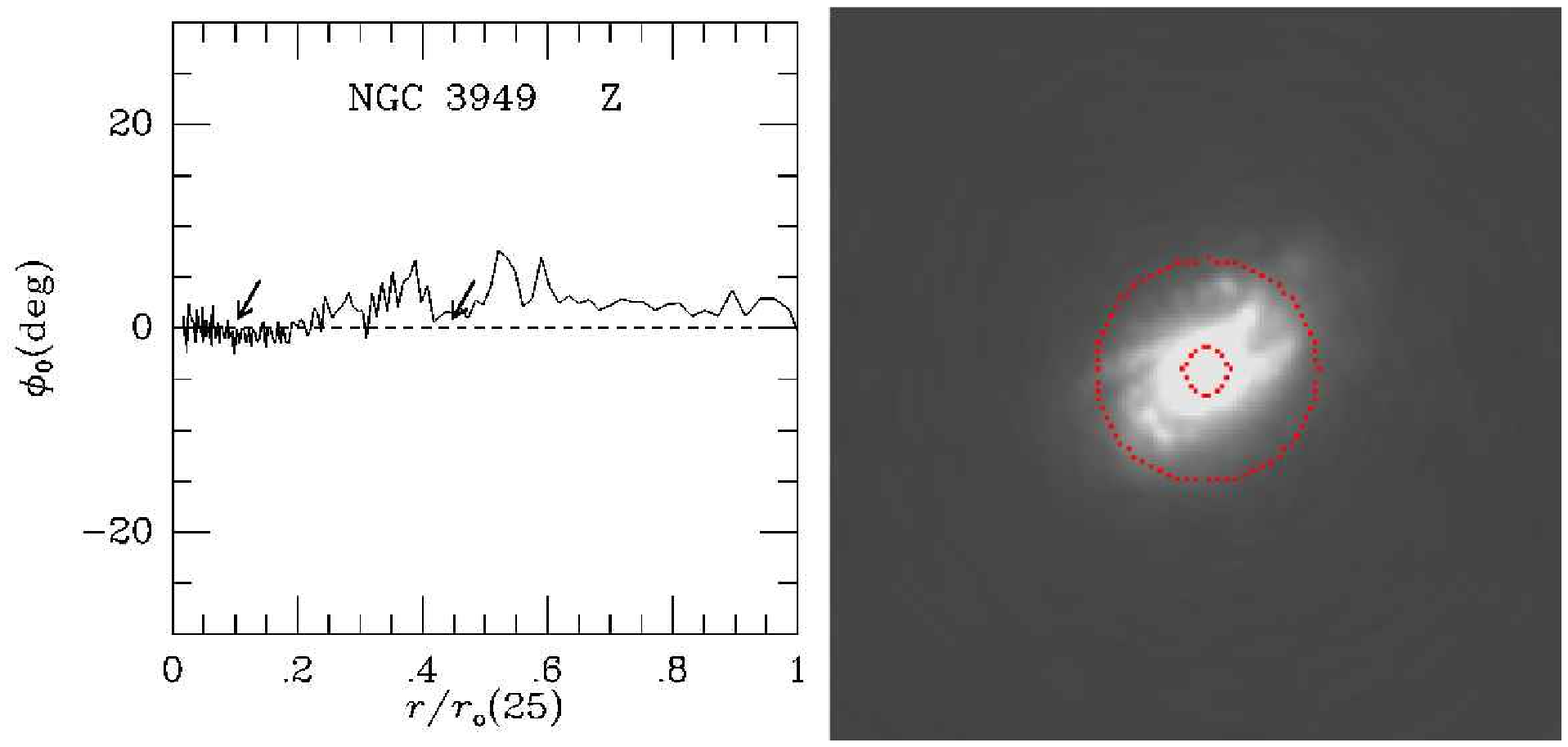}
\vspace{2.0truecm}                                                              
\caption{Same as Figure 2.1 for NGC 3949}                                         
\label{ngc3949}                                                                 
\end{figure}

NGC 3949 (Figure~\ref{ngc3949}) - The crossing at 8\arcs\ (0.09)
circles just inside the ends of a small oval. The near-crossing at
38\arcs\ (0.44) is just around the chaotic spiral.

\begin{figure}                                                                 
\figurenum{2.69}
\plotone{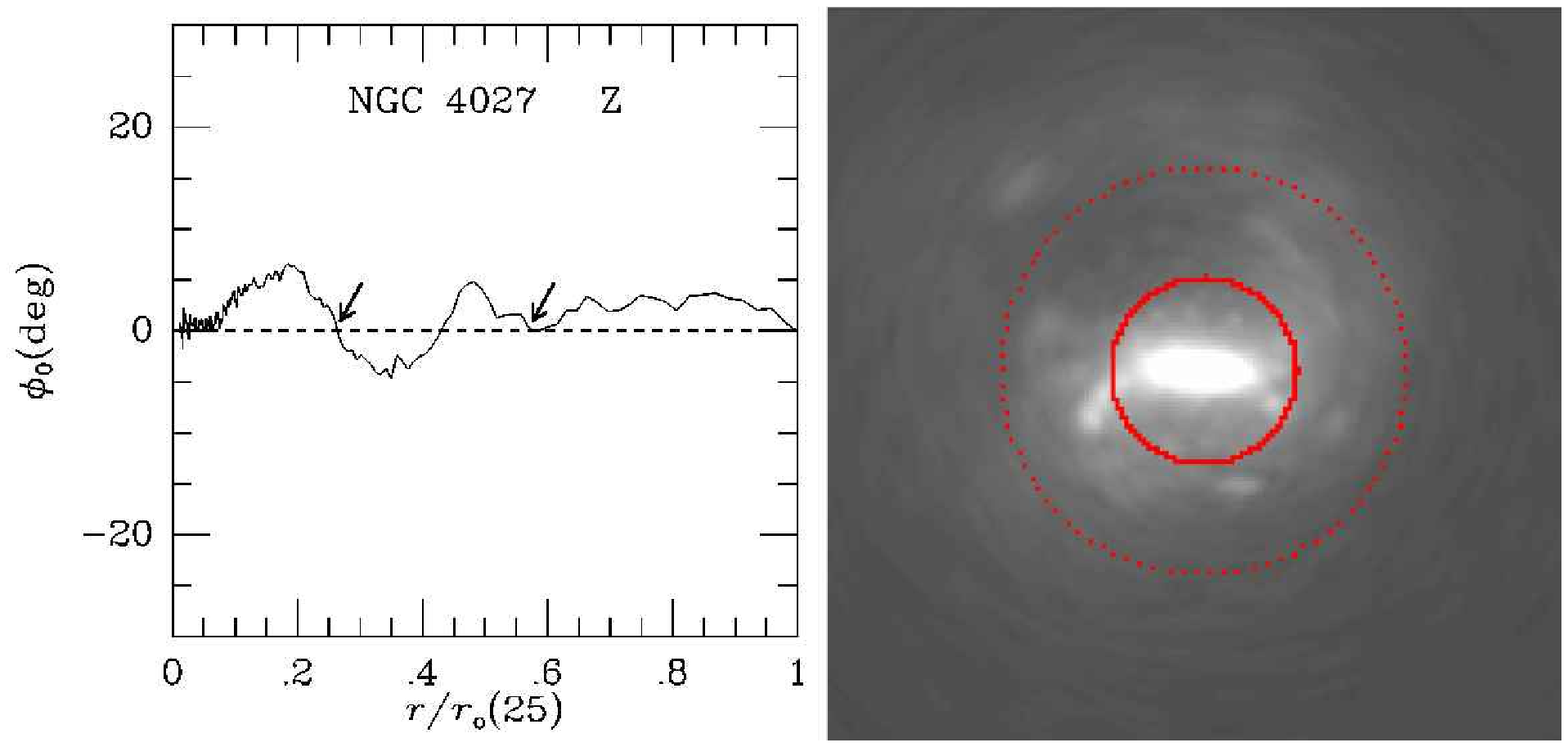}
 \vspace{2.0truecm}                                                             
\caption{Same as Figure 2.1 for NGC 4027}                                         
\label{ngc4027}                                                                 
 \end{figure}

NGC 4027 (Figure~\ref{ngc4027}) - This small late-type spiral considered
of Magellanic type by de Vaucouleurs shows a well-defined single crossing
at $r$=24\arcs\ (0.25) that encircles just outside the ends of the prominent
bar. The second crossing is barely evident but could signify a decoupling
of the outer, asymmetric spiral pattern.

\begin{figure}                                                                 
\figurenum{2.70}
\plotone{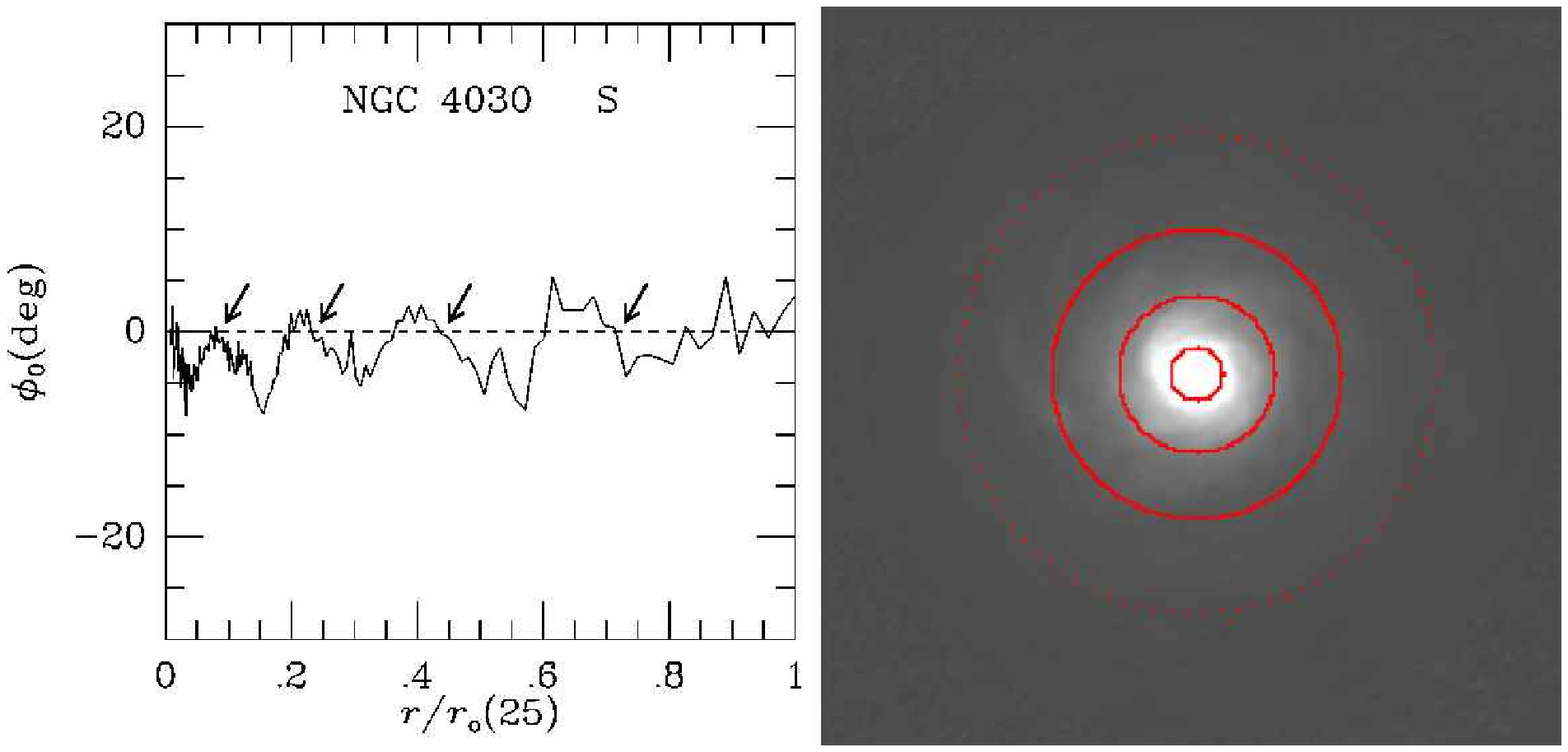}
 \vspace{2.0truecm}                                                             
\caption{Same as Figure 2.1 for NGC 4030}                                         
\label{ngc4030}                                                                 
 \end{figure}

NGC 4030 (Figure~\ref{ngc4030}) - This galaxy shows several significant-looking
P/N crossings. The one at $r$=10\arcs\ (0.08) 
appears to be associated with an inner oval. The other outer crossings
are connected with the prominent, complex spiral pattern
which has up to 6 arms. There may also be a significant pattern at very
small radii. The outer crossings are not affected by the fact that the
multiplicity is greater than 2. 

\begin{figure}                                                                 
\figurenum{2.71}
\plotone{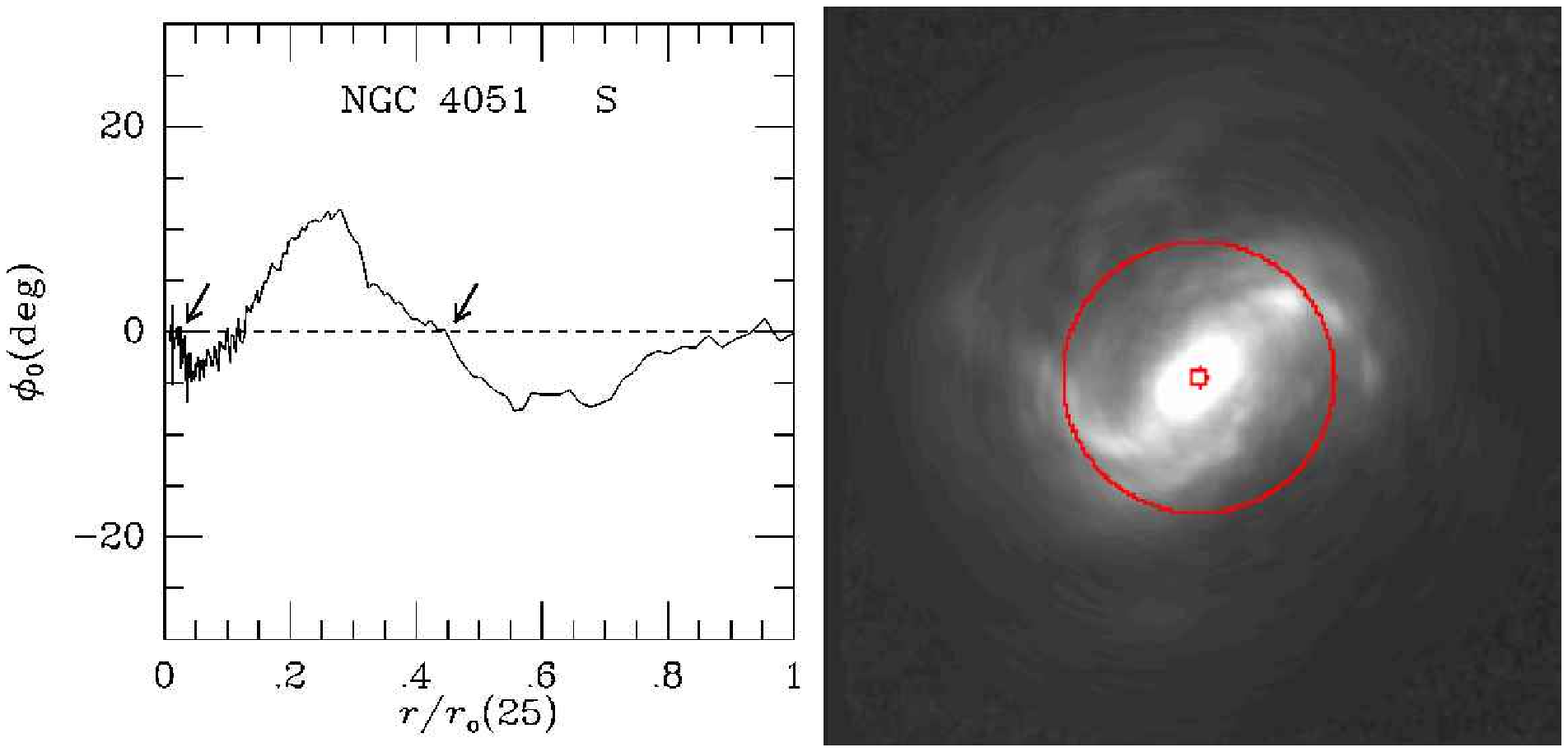}
 \vspace{2.0truecm}                                                             
\caption{Same as Figure 2.1 for NGC 4051}                                         
\label{ngc4051}                                                                 
 \end{figure}

NGC 4051 (Figure~\ref{ngc4051}) - This galaxy has a single main crossing
at $r$=70\arcs\ (0.44) and another possibly significant crossing at $r$=3\arcs\
(0.02).
The main crossing circles beyond the ends of the apparent bar in the middle
of the spiral pattern. In this case, the spiral and the bar should have the
same pattern speed.

An RSL08 model gives $r(CR)$ = 98\arcs $\pm$ 14\arcs\ 
(0.62$\pm$0.09), which is larger than our single main CR crossing. For the
RSL08 value of the bar radius of 54\arcss 1 (0.34), our analysis makes
NGC 4051 a fast bar case, while the RSL08 model makes it a possible slow
bar case.

\begin{figure}                                                                 
\figurenum{2.72}
\plotone{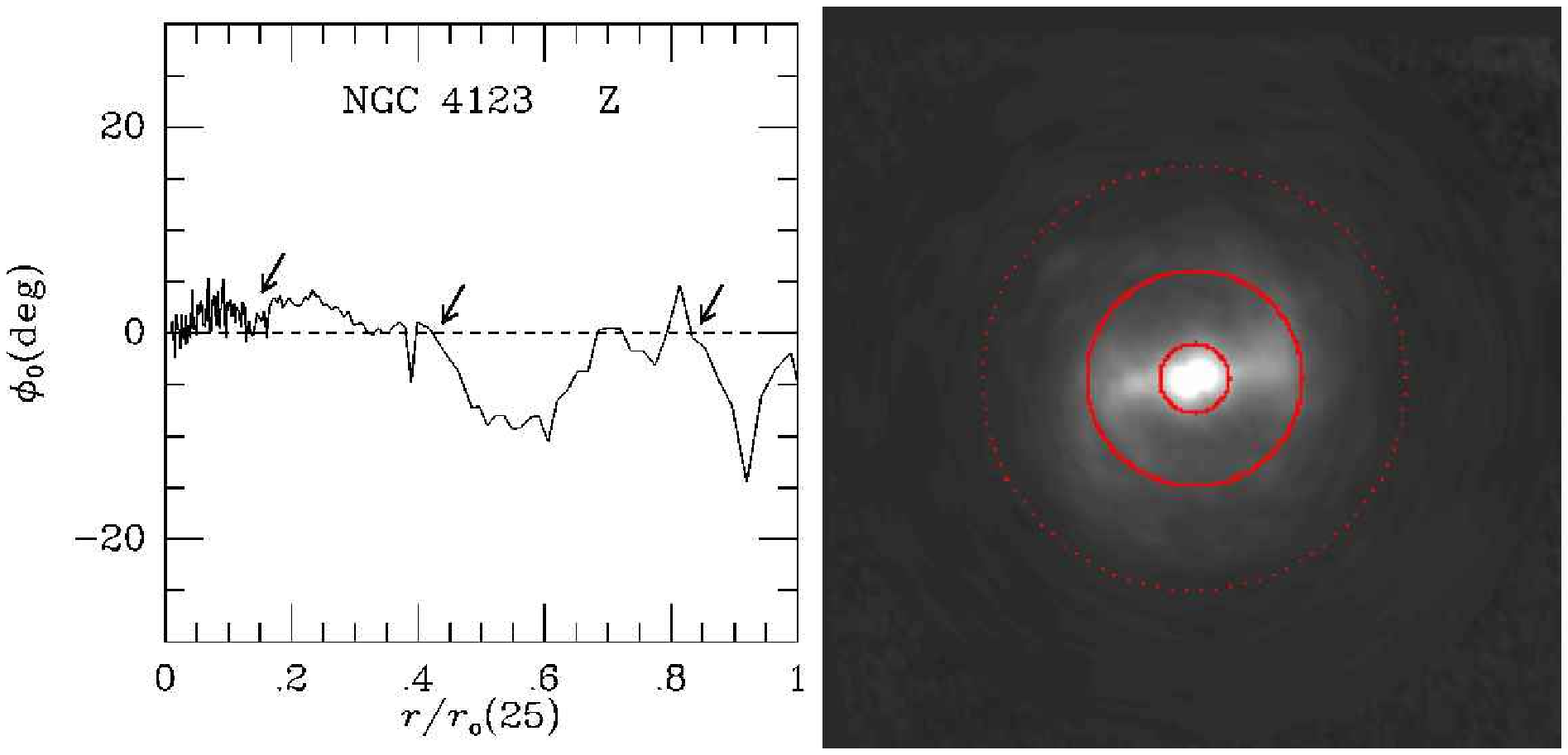}
 \vspace{2.0truecm}                                                             
\caption{Same as Figure 2.1 for NGC 4123}                                         
\label{ngc4123}                                                                 
 \end{figure}

NGC 4123 (Figure~\ref{ngc4123}) - The main crossing at $r$=55\arcs\ (0.42)
circles
around the ends of the bar. The near crossing at $r$=18\arcs\ (0.14) could be 
associated with an inner misaligned pattern (affected slightly by 
decomposition pinch). The outer crossing lies mostly outside the
spiral structure.

RSL08 show that a simulation model having $r(CR)$ = 69\arcss 1 (0.53)
matches the spiral structure well. Our main CR at 55\arcs\ is closer
to their estimate of the bar radius, 59\arcss 2 (0.45).

\begin{figure}                                                                 
\figurenum{2.73}
\plotone{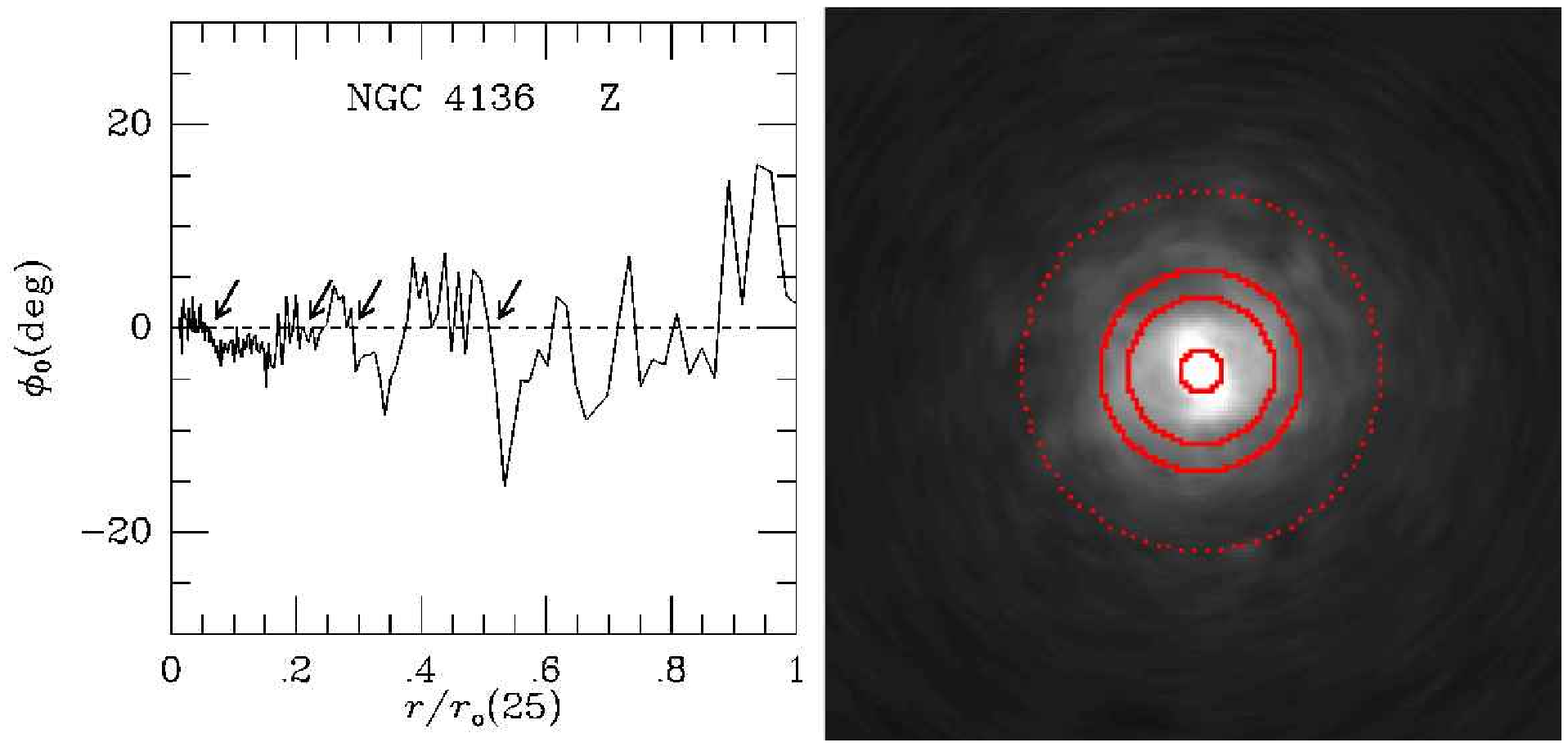}
 \vspace{2.0truecm}                                                             
\caption{Same as Figure 2.1 for NGC 4136}                                         
\label{ngc4136}                                                                 
 \end{figure}

NGC 4136 (Figure~\ref{ngc4136}) - This is possibly an evolving, unsettled
system. The inner crossing lies well inside the ends of the inner bar.
The next crossing encircles around the galaxy's inner ring and is well
outside the bar ends. The remaining crossings are in the spiral arms.



\begin{figure}                                                                 
\figurenum{2.74}
\plotone{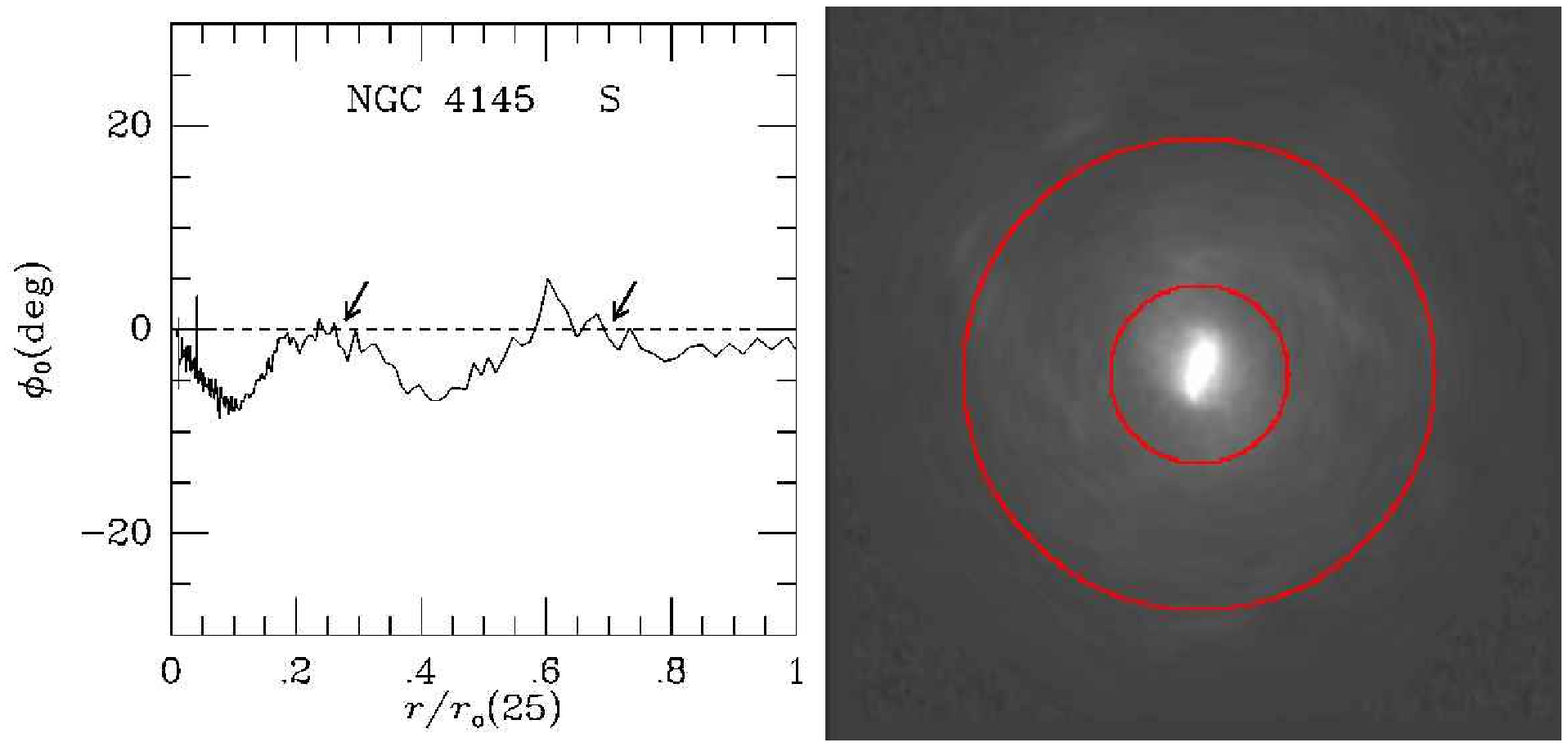}
 \vspace{2.0truecm}                                                             
\caption{Same as Figure 2.1 for NGC 4145}                                         
\label{ngc4145}                                                                 
 \end{figure}

NGC 4145 (Figure~\ref{ngc4145}) - This very late-type spiral has a 
conspicuous bar and two apparent P/N crossings. The one at $r$=47\arcs\
(0.27) circles well outside the ends of the bar, while the crossing at $r$
= 122\arcs\ (0.69) lies in the outer arms.

\begin{figure}                                                                 
\figurenum{2.75}
\plotone{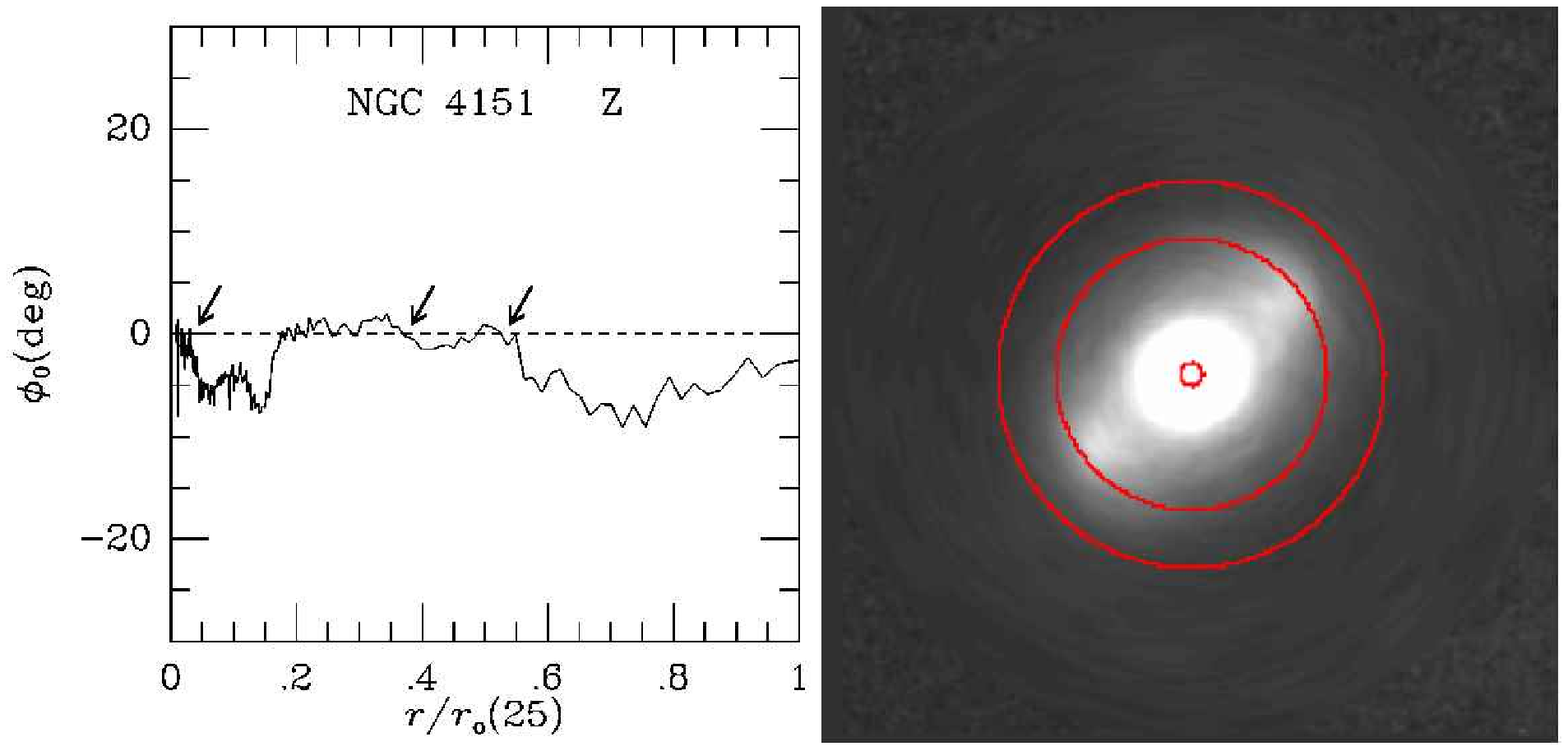}
 \vspace{2.0truecm}                                                             
\caption{Same as Figure 2.1 for NGC 4151}                                         
\label{ngc4151}                                                                 
 \end{figure}

NGC 4151 (Figure~\ref{ngc4151}) - The main P/N crossing at $r$=99\arcs\
(0.52) appears to circle just outside the ends of the bright bar/oval. The
crossing at $r$=70\arcs\ (0.37) could signify some decoupling of the
inner part of the bar. However, there is no misalignment between the
inner parts of the bar and the extent of the outer oval. The phase-shift
curve also indicates an unresolved nuclear pattern, and the possible
decoupling of the inner oval in progress.

\begin{figure}                                                                 
\figurenum{2.76}
\plotone{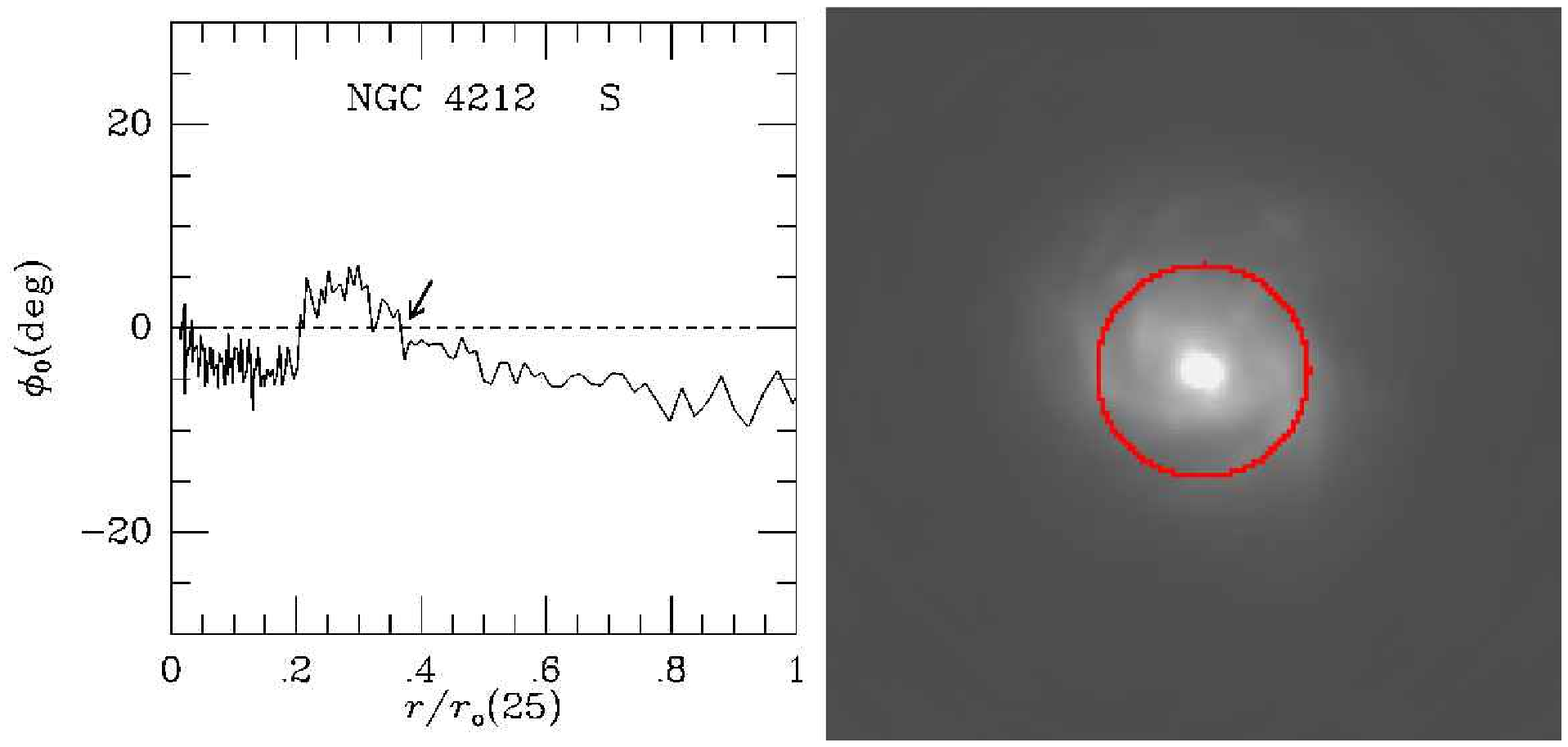}
 \vspace{2.0truecm}                                                             
\caption{Same as Figure 2.1 for NGC 4212}                                         
\label{ngc4212}                                                                 
 \end{figure}

NGC 4212 (Figure~\ref{ngc4212}) - The main CR circle is connected with
the spiral arms, but no crossing is seen for the small inner oval. This
pattern may not yet be decoupled from the spiral, and a hump on the negative
portion of the phase-shift plot near the central region may indicate potential
future decoupling. 

\begin{figure}                                                                 
\figurenum{2.77}
\plotone{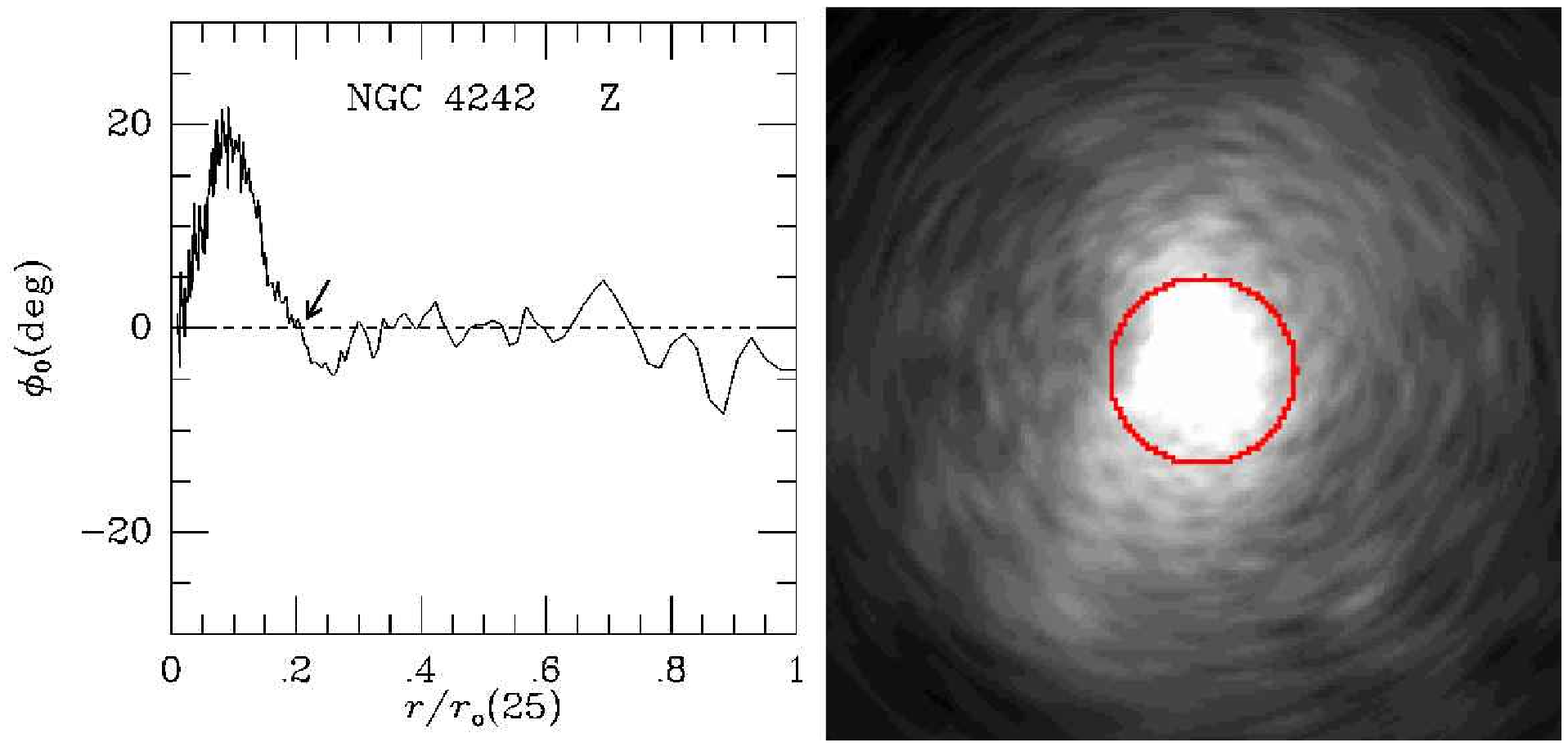}
 \vspace{2.0truecm}                                                             
\caption{Same as Figure 2.1 for NGC 4242}                                         
\label{ngc4242}                                                                 
 \end{figure}

NGC 4242 (Figure~\ref{ngc4242}) - The main inner crossing encompasses a
weak, broad oval, but other than this the galaxy has no organized
pattern in the near-IR.  

\begin{figure}                                                                 
\figurenum{2.78}
\plotone{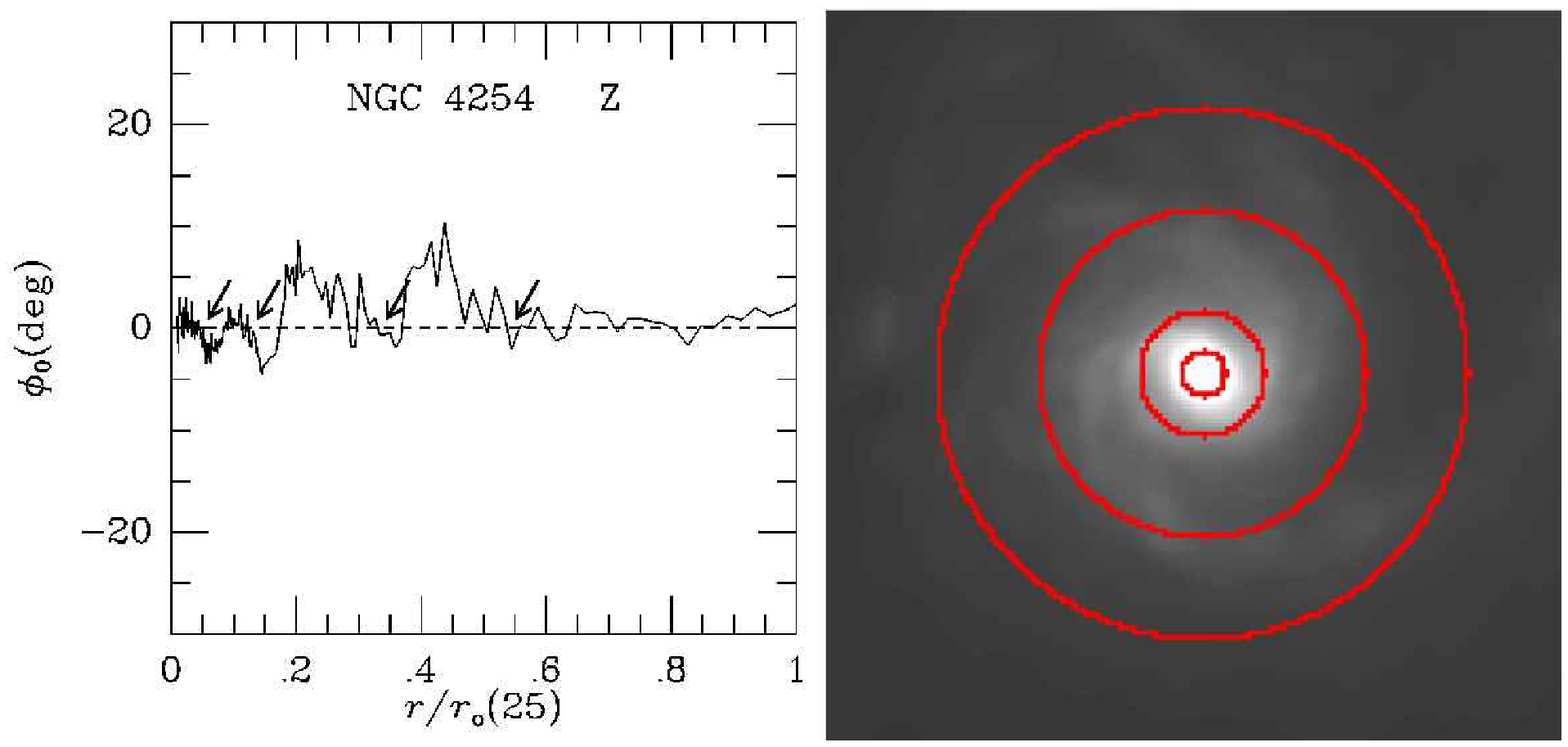}
 \vspace{2.0truecm}                                                             
\caption{Same as Figure 2.1 for NGC 4254}                                         
\label{ngc4254}                                                                 
 \end{figure}

NGC 4254 (Figure~\ref{ngc4254}) - The phase-shift plot shows mostly weak
P/N crossings. The one at $r$=21\arcs\ (0.12) is related to a small inner oval.
The crossing at $r$=56\arcs\ (0.33) may be connected with the inner spiral arms,
while that at $r$=91\arcs\ (0.54) could be connected with the outer arms. 
There is also a nuclear crossing.

\begin{figure}                                                                 
\figurenum{2.79}
\plotone{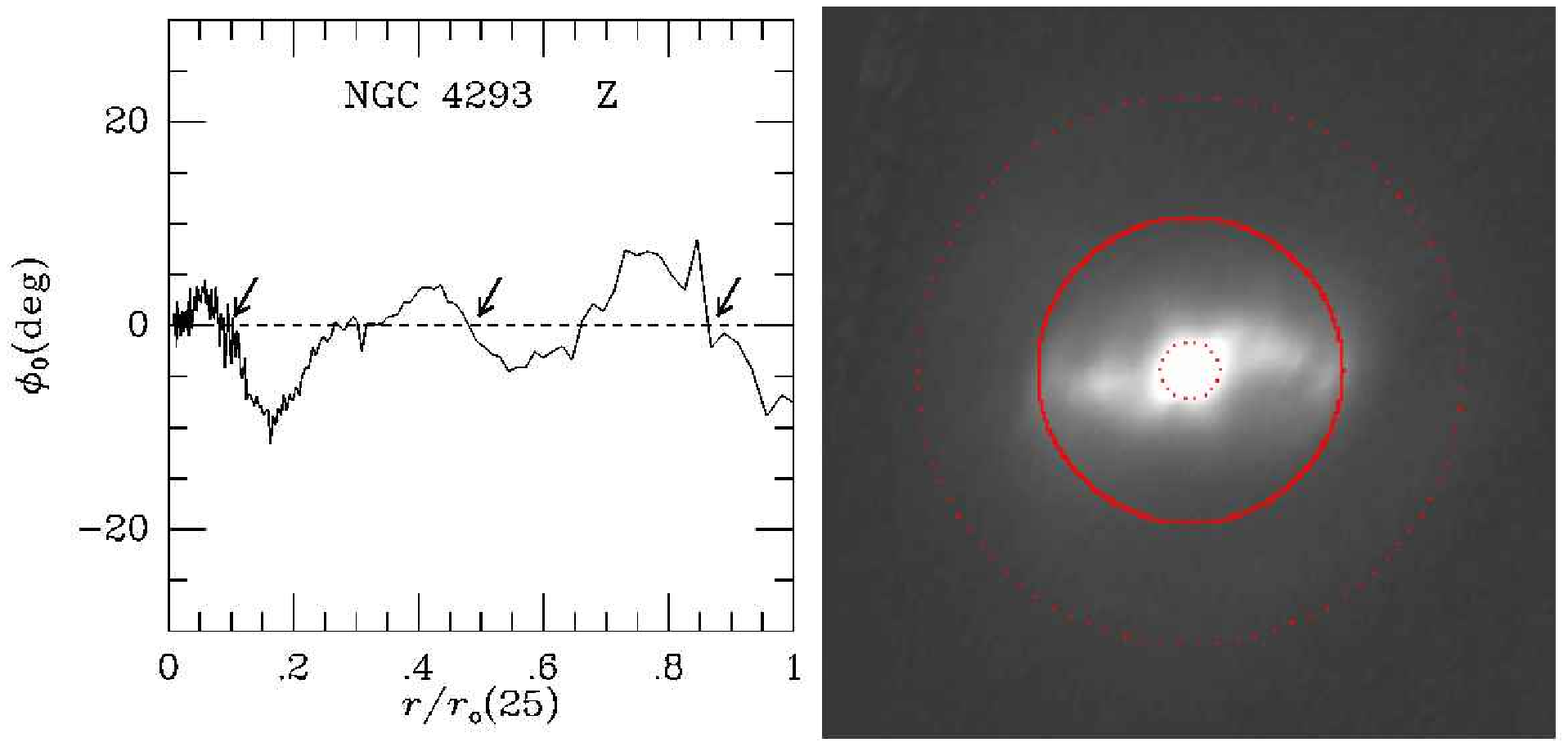}
 \vspace{2.0truecm}                                                             
\caption{Same as Figure 2.1 for NGC 4293}                                         
\label{ngc4293}                                                                 
 \end{figure}

NGC 4293 (Figure~\ref{ngc4293}) - The crossing at $r$=79\arcs\ (0.48) circles
near the ends of the bar. The innermost crossing is in an area affected
by strong decomposition pinch. We used an average bulge image for the
analysis, but the crossing is still of uncertain accuracy. The 
outermost crossing is well-defined but close to $r_o$.

\begin{figure}                                                                 
\figurenum{2.80}
\plotone{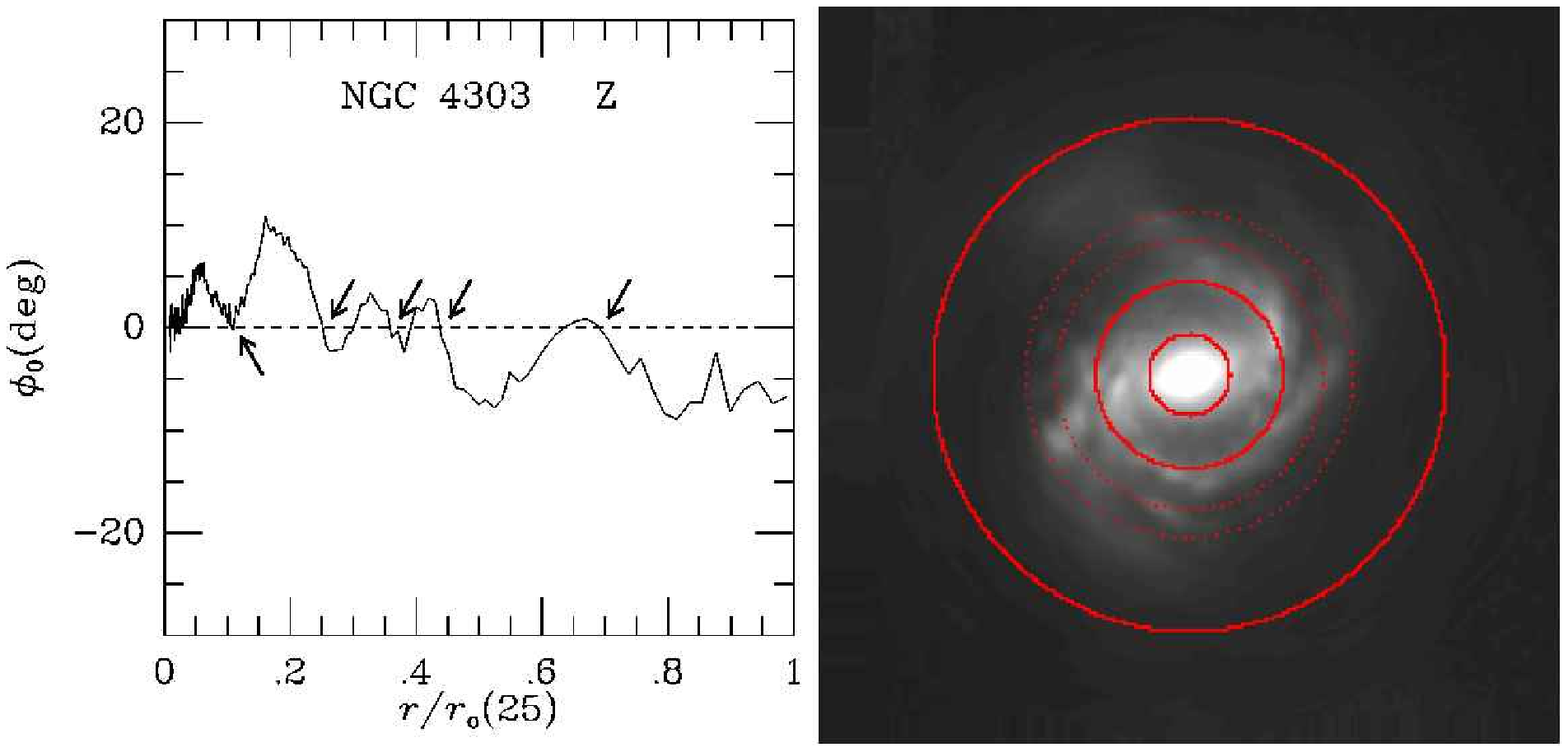}
 \vspace{2.0truecm}                                                             
\caption{Same as Figure 2.1 for NGC 4303}                                         
\label{ngc4303}                                                                 
 \end{figure}

NGC 4303 (Figure~\ref{ngc4303}) - The phase-shift method yields 5 
significant-looking crossings for this excellent face-on spiral.
The smallest crossing encircles the ends
of an inner oval, the inner part of the bar of this galaxy. 
The second crossing encircles the ends of the whole bar and part of
the inner spiral.  The next two
crossings appear to be related to the inner pseudoring spiral arms, while
the largest crossing is related to the outer arms. 

RSL08 obtained an excellent fit to the $B$-band spiral structure of this
galaxy with a
single pattern speed having $r(CR)$=89\arcs\ (0.46), which is very close
to our 4th CR at 85\arcs\ (0.44). Their estimate of the bar radius, 52\arcss 5
(0.27),
is closer to our second CR radius at 49\arcs\ (0.25). Although their model
gives a good match to the outer arms, there is a third arm beyond the
inner pseudoring that is not reproduced at all. Thus, our finding of a
CR at 134\arcs\ (0.69) could indicate that there is indeed an independent
pattern in this region.

\begin{figure}                                                                 
\figurenum{2.81}
\plotone{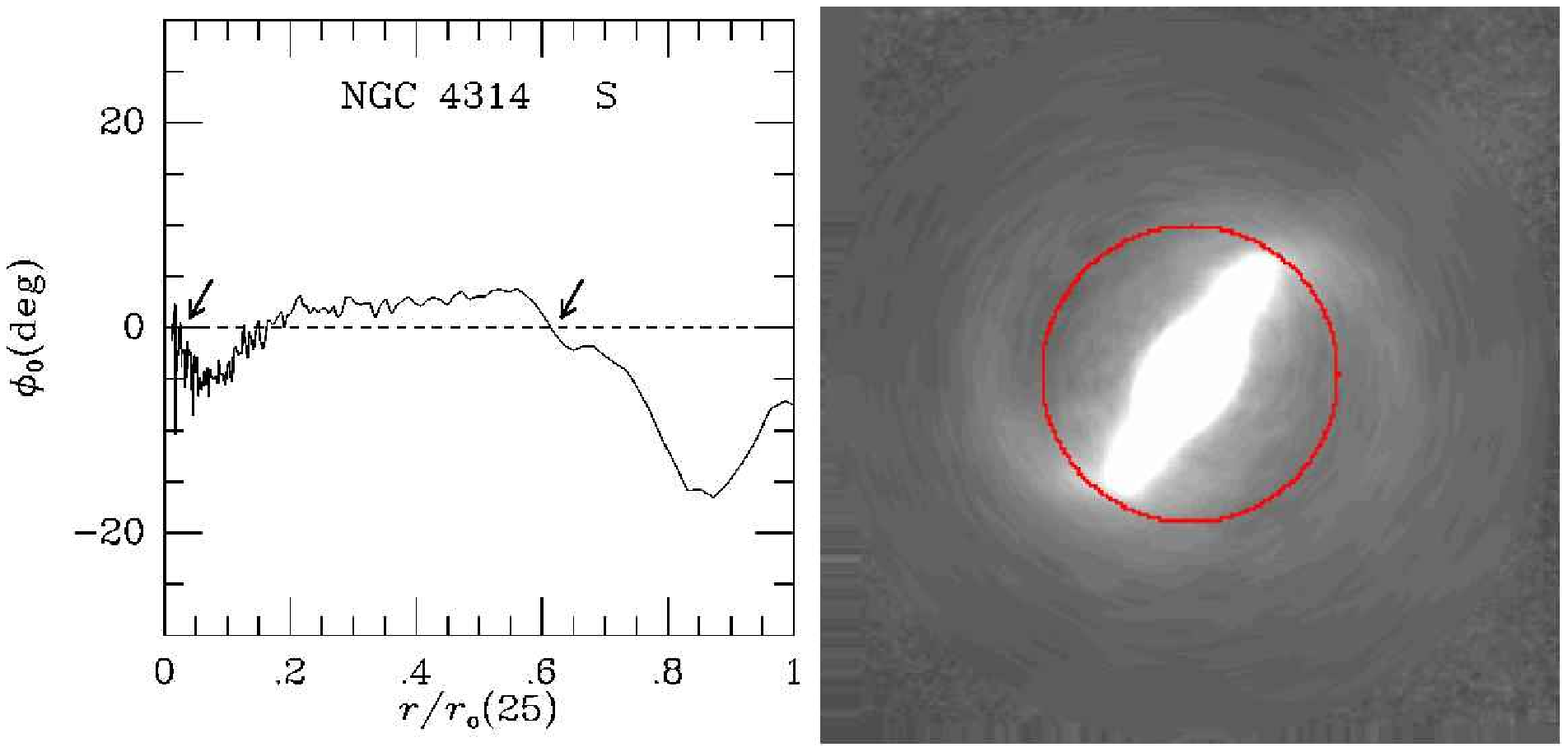}
 \vspace{2.0truecm}                                                             
\caption{Same as Figure 2.1 for NGC 4314. Only CR$_2$                             
is shown overlaid on the image.}                                                
\label{ngc4314}                                                                 
 \end{figure}

NGC 4314 (Figure~\ref{ngc4314}) - As shown in paper 1, this galaxy has
a well-defined P/N crossing at 77\arcs\ (0.61) that circles almost exactly around
the ends of the bar. In paper 1, we also detected an inner crossing,
but our revised analysis only weakly detects it. The N/P crossing 
at $r$=20\arcs\ (0.16) indicates the presence of an inner pattern in any case.

An excellent RSL08 model that reproduces this galaxy very well gives 
$r(CR)$ = 81\arcss 7 (0.65), in good agreement not only with our main
CR estimate but also with their estimate of the
 radius of the bar at 82\arcss 9 (0.66).

\begin{figure}                                                                 
\figurenum{2.82}
\plotone{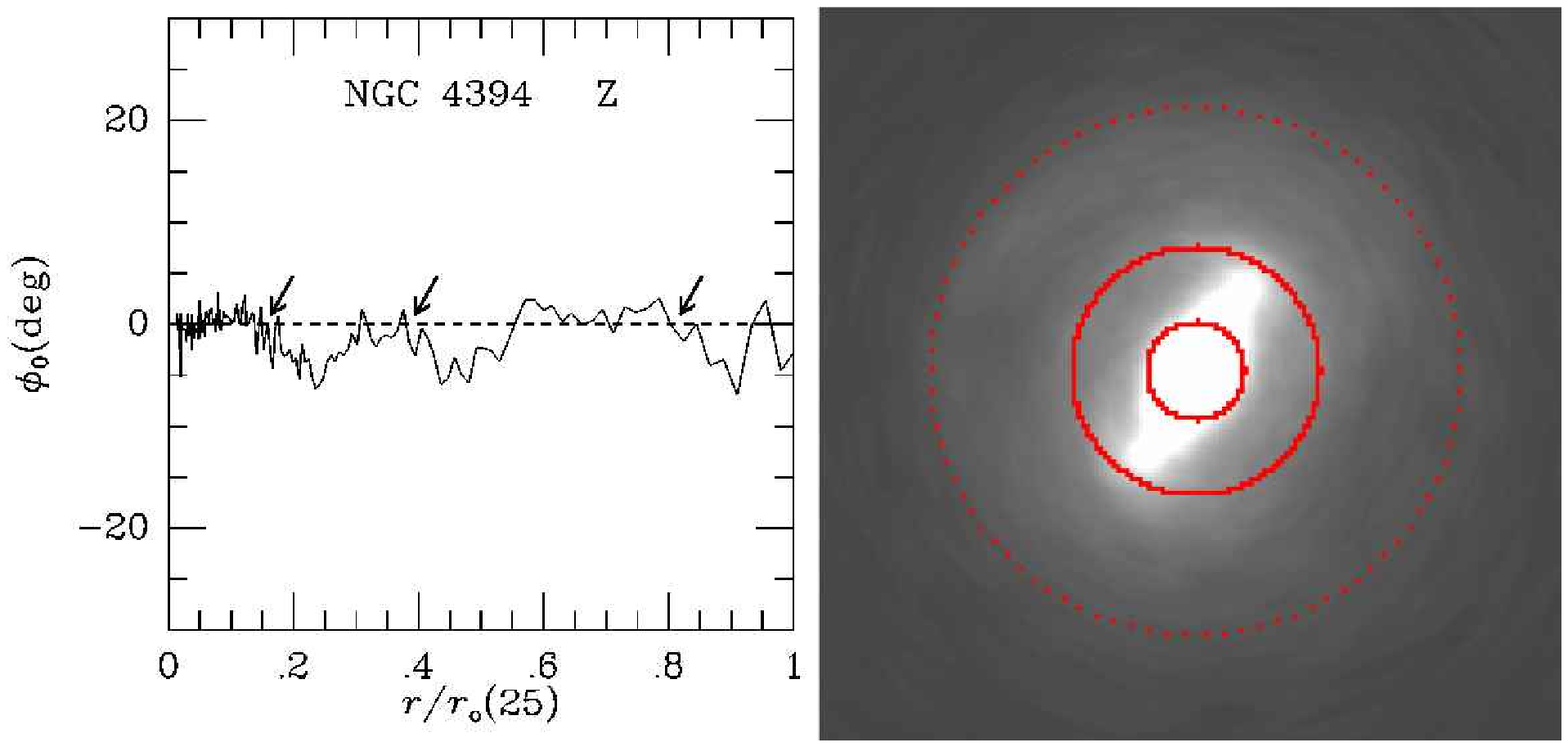}
 \vspace{2.0truecm}                                                             
\caption{Same as Figure 2.1 for NGC 4394}                                         
\label{ngc4394}                                                                 
 \end{figure}

NGC 4394 (Figure~\ref{ngc4394}) - There are at least three possibly
significant crossings for this galaxy. The one at $r$=42\arcs\ (0.38)
is noisy but appears to circle around the ends of the bar. The crossing
at $r$=17\arcs\ (0.15) encompasses a broad inner part of the bar, while
the crossing at $r$=90\arcs\ (0.81) is close to $r_o$, just outside a
faint outer ring, and may not be significant. The weak N/P crossing at
$r$=33\arcs (0.30) circles somewhat inside the bar ends, suggesting
that much of the bar may actually belong to the inner mode. This galaxy
is interesting in that its prominent bar is skewed slightly in a
leading sense, opposite the winding of its main outer arms.

An RSL08 model gives $r(CR)$ = 76\arcss 7 $\pm$ 10\arcss 0 (0.69$\pm$0.09), 
which agrees within uncertainties with our third CR value.

\begin{figure}                                                                 
\figurenum{2.83}
\plotone{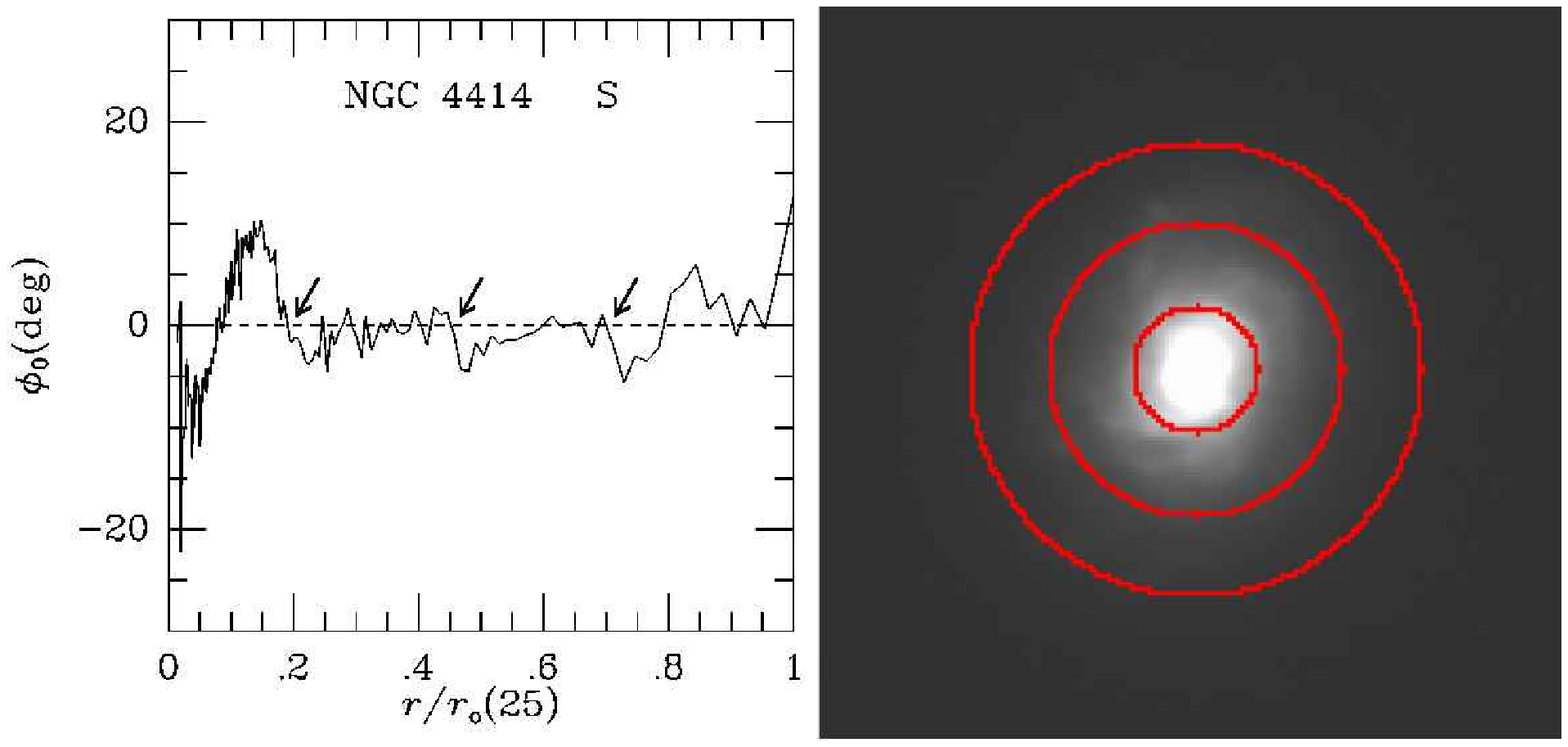}
 \vspace{2.0truecm}                                                             
\caption{Same as Figure 2.1 for NGC 4414}                                         
\label{ngc4414}                                                                 
 \end{figure}

NGC 4414 (Figure~\ref{ngc4414}) - The innermost crossing encircles an
inner oval. The pattern is weak in the outer regions, and the outer
crossings are uncertain.  There is also the indication of an unresolved
nuclear pattern.

\begin{figure}                                                                 
\figurenum{2.84}
\plotone{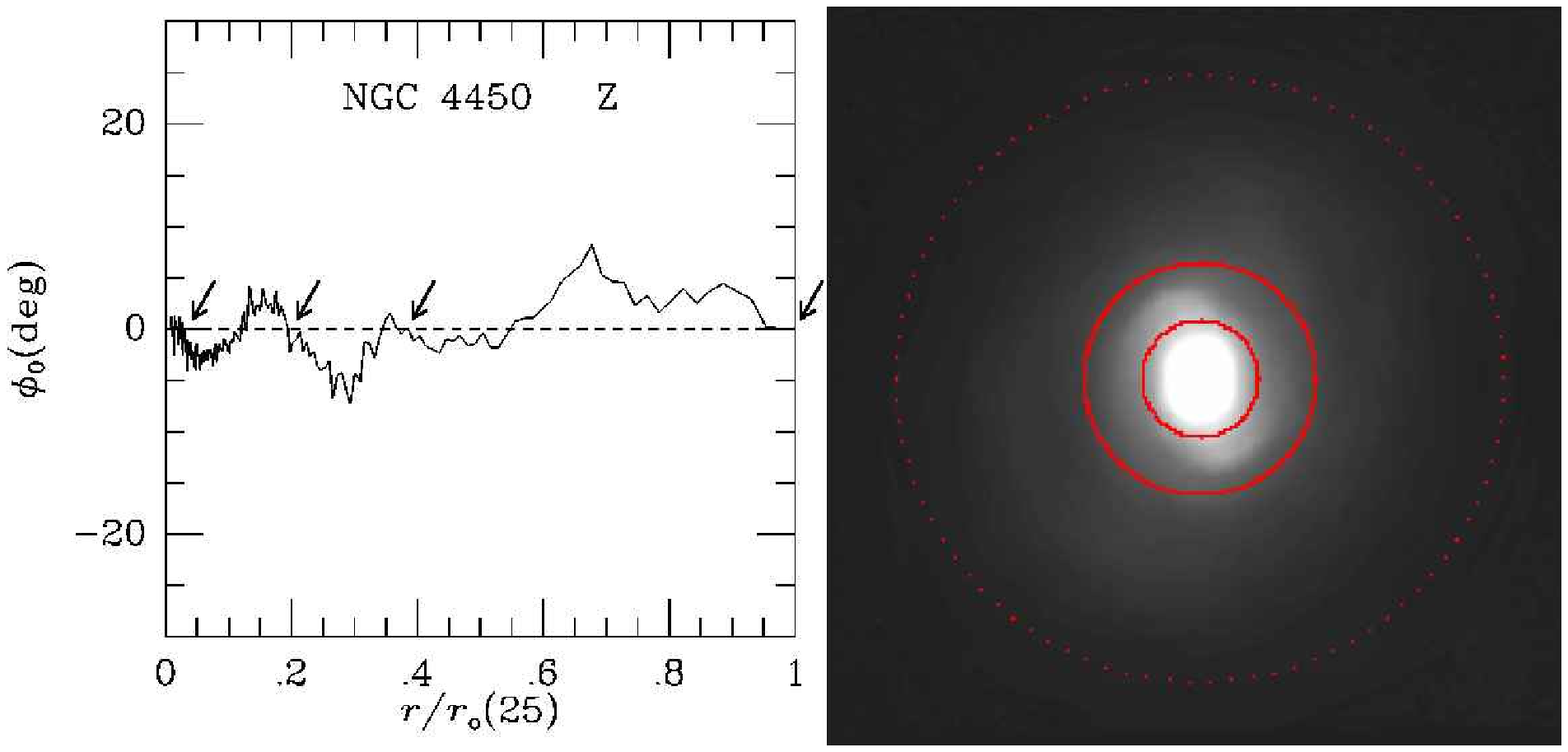}
 \vspace{2.0truecm}                                                             
\caption{Same as Figure 2.1 for NGC 4450. Only CR$_2$,                            
CR$_3$, and CR$_4$ in Table 1 are overlaid on the image.}                       
\label{ngc4450}                                                                 
 \end{figure}

NGC 4450 (Figure~\ref{ngc4450}) - The crossing at $r$=31\arcs\ (0.20) appears
related to a broad inner oval, while the one at $r$=59\arcs\ (0.37) could be
associated with the spiral emanating from the ends of this oval. There
may also be a small inner crossing. The outermost crossing is near $r_o$
and is of uncertain significance although it is well-defined.

An RSL08 model places CR in NGC 4450 at 52\arcss 3 (0.33), close to our
third CR. RSL08 give a bar radius of 49\arcss 7 (0.32), considerably larger
than the inner oval and probably affected by the inner arms. 

\begin{figure}                                                                 
\figurenum{2.85}
\plotone{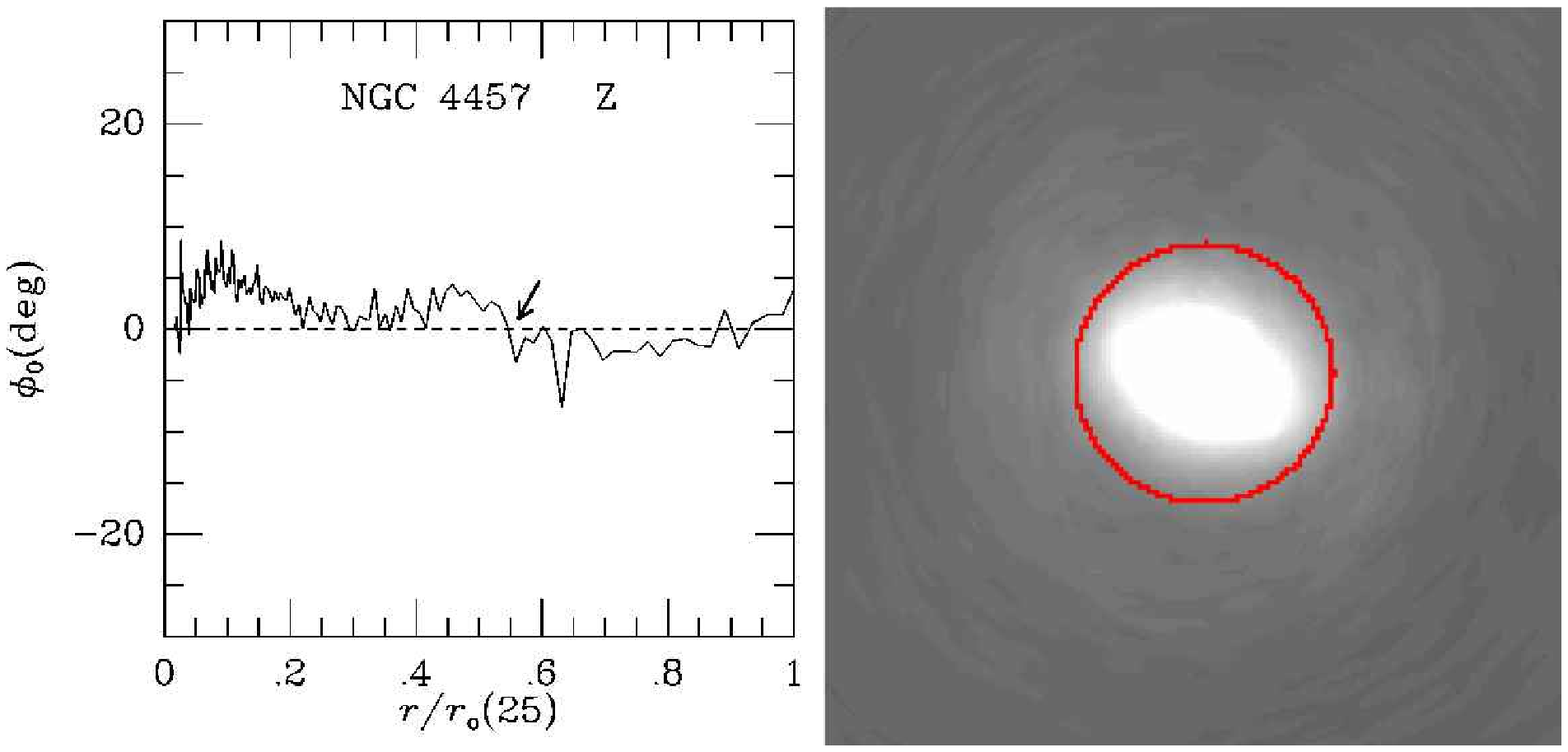}
 \vspace{2.0truecm}                                                             
\caption{Same as Figure 2.1 for NGC 4457}                                         
\label{ngc4457}                                                                 
 \end{figure}

NGC 4457 (Figure~\ref{ngc4457}) - The main P/N crossing at $r$=44\arcs\ (0.55) 
is noisy but appears to circle almost exactly around the ends of the galaxy's
prominent inner oval. The phase-shift plot also suggests pattern decoupling
near $r$=18\arcs\ (0.22).

The best-fitting simulation of RSL08 places CR at 39\arcss 9 (0.49), close to
our single crossing. 

\begin{figure}                                                                 
\figurenum{2.86}
\plotone{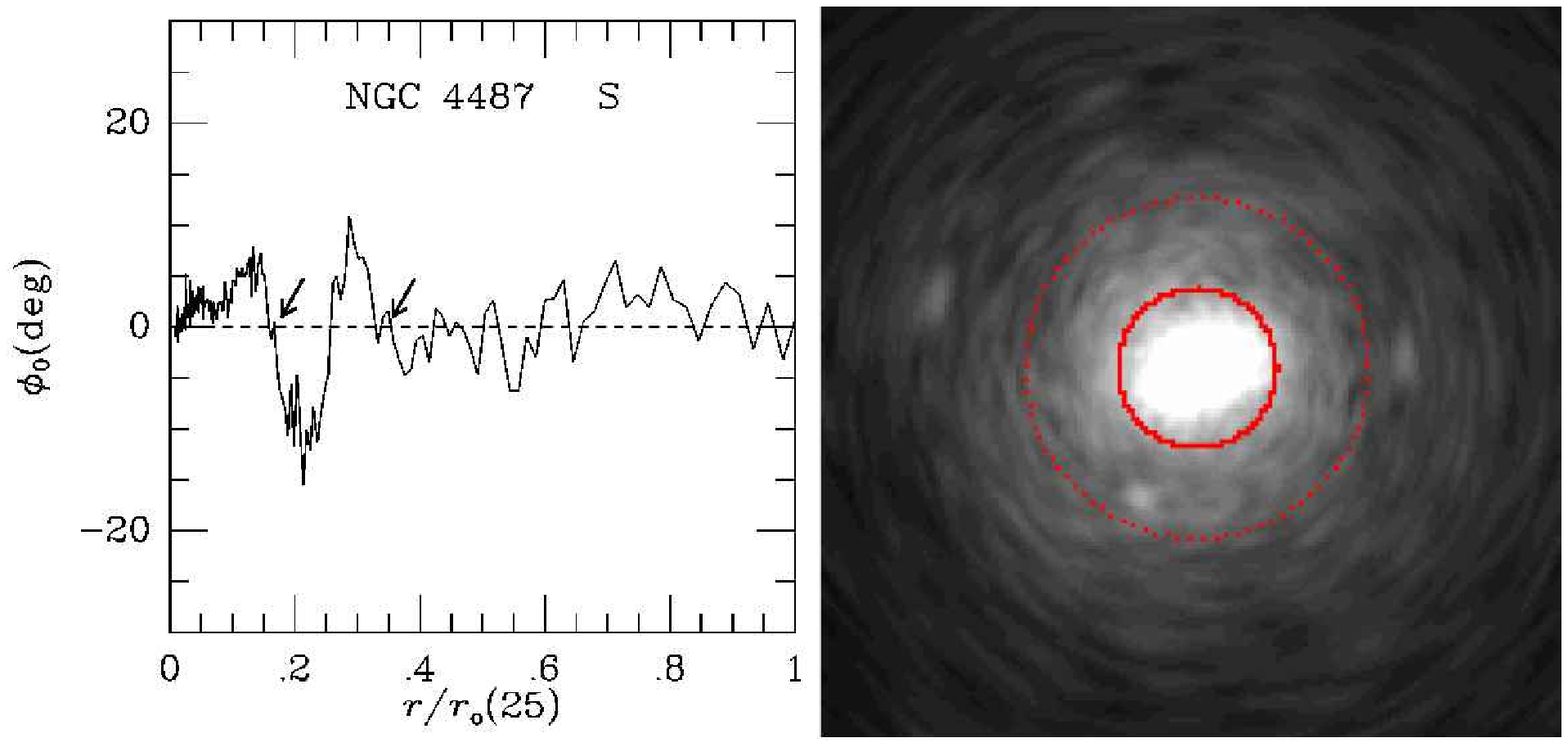}
 \vspace{2.0truecm}                                                             
\caption{Same as Figure 2.1 for NGC 4487}                                         
\label{ngc4487}                                                                 
 \end{figure}

NGC 4487 (Figure~\ref{ngc4487}) - The main crossing at $r$=20\arcs\ (0.16)
is associated with a bright inner oval. The outer crossing is in an area
where there is little coherent pattern and may not be significant. 

\begin{figure}                                                                 
\figurenum{2.87}
\plotone{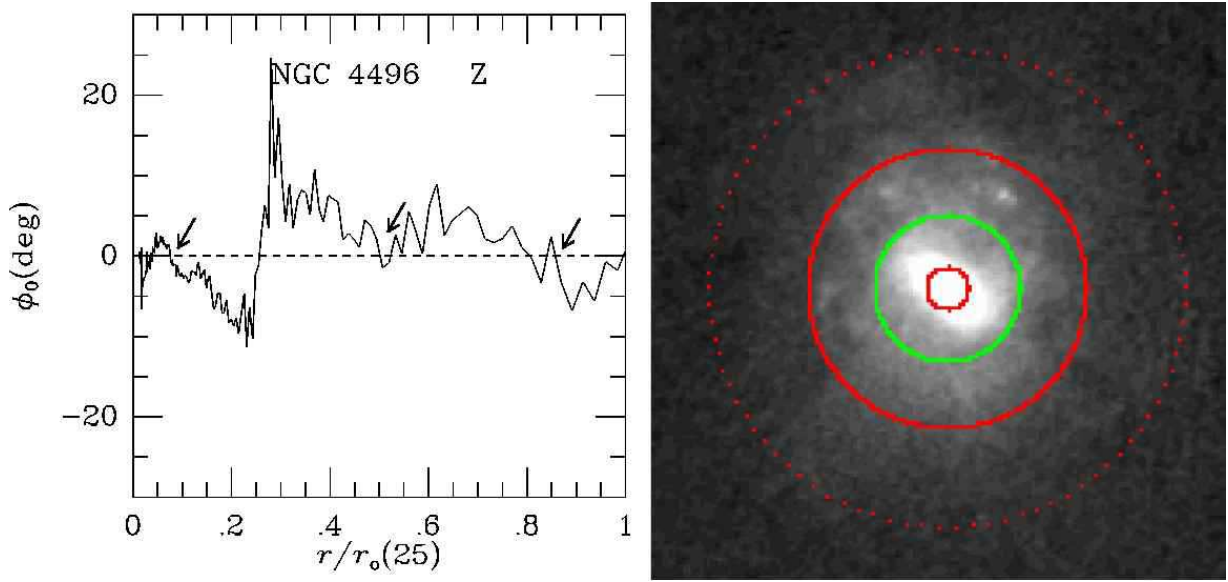}
 \vspace{2.0truecm}                                                             
\caption{Same as Figure 2.1 for NGC 4496. The green circle                        
indicates the major N/P crossing at                                             
$r/r_o(25)$$\approx$0.25.}                                                      
\label{ngc4496}                                                                 
 \end{figure}

NGC 4496 (Figure~\ref{ngc4496}) - Surprisingly, the most significant
crossing is the N/P at $r$=31\arcs\ (0.26), shown as a green circle in 
Figure~\ref{ngc4496}. This crossing encircles just outside the bar ends,
and could be close to the location of the OLR of the small inner pattern. The
two outer crossings are of uncertain significance.

\begin{figure}                                                                 
\figurenum{2.88}
\plotone{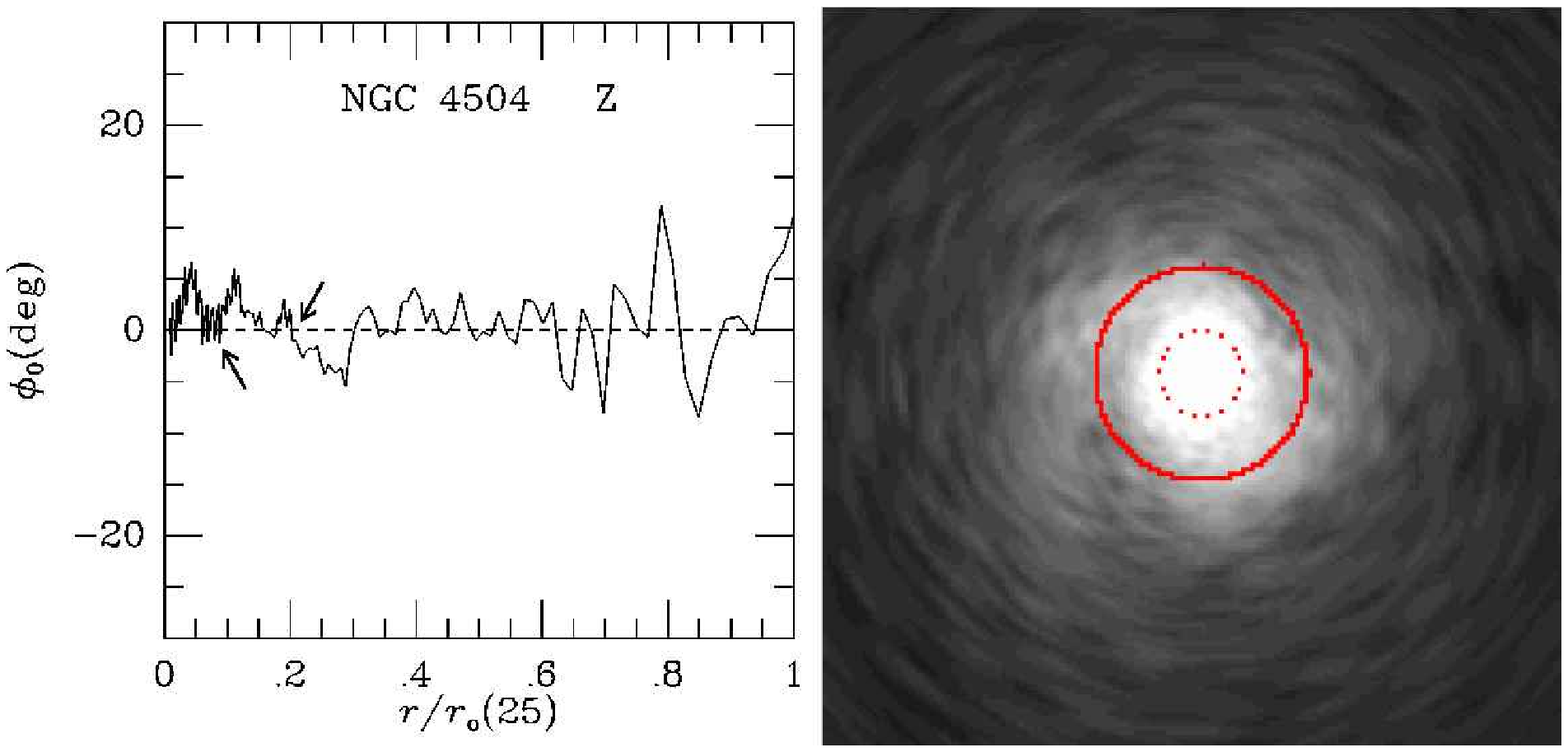}
 \vspace{2.0truecm}                                                             
\caption{Same as Figure 2.1 for NGC 4504}                                         
\label{ngc4504}                                                                 
 \end{figure}

NGC 4504 (Figure~\ref{ngc4504}) - The main crossing at $r$=27\arcs\ (0.21)
lies just beyond the ends of a broad inner oval. There is an indication of a
decoupling pattern near $r$=10\arcs.

\begin{figure}                                                                 
\figurenum{2.89}
\plotone{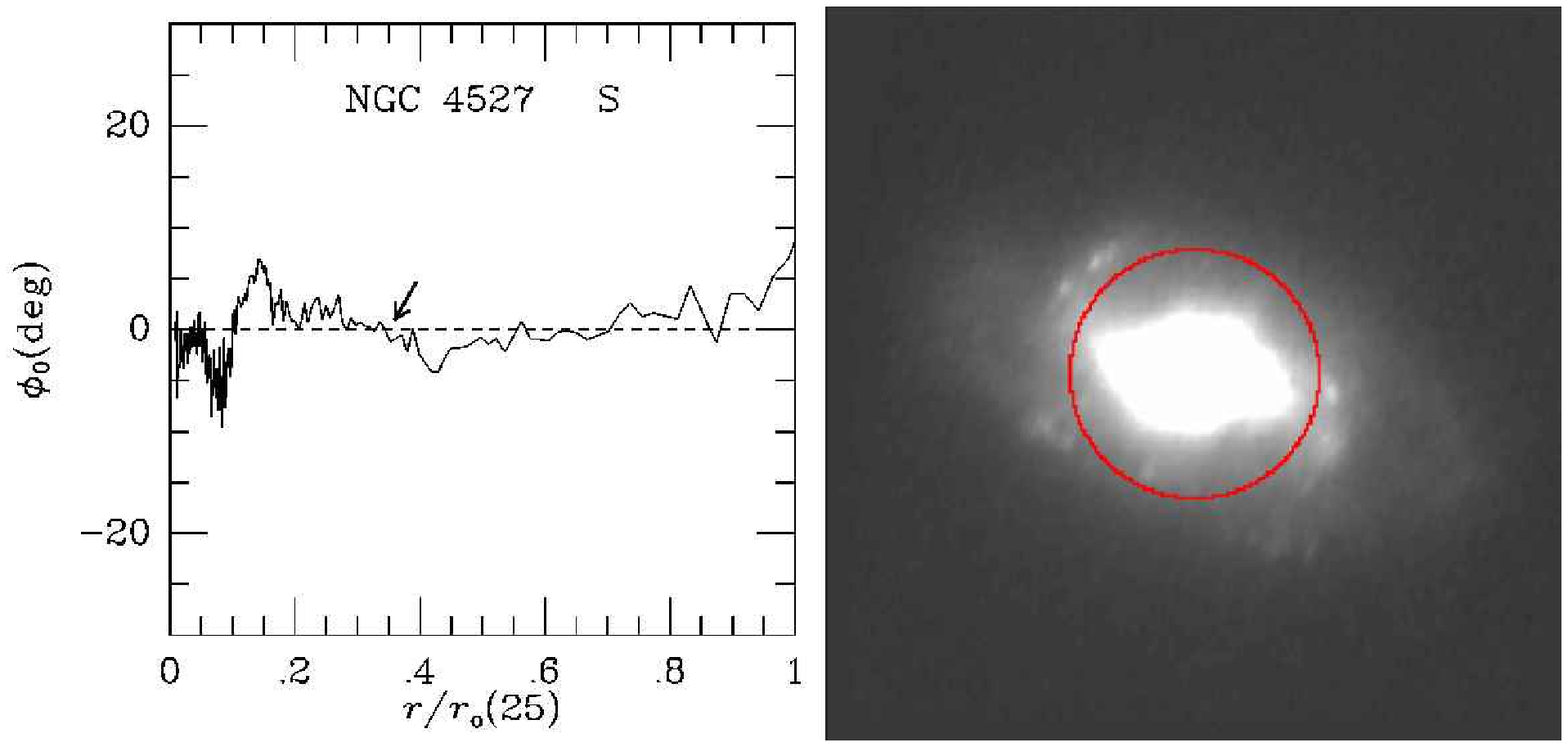}
 \vspace{2.0truecm}                                                             
\caption{Same as Figure 2.1 for NGC 4527}                                         
\label{ngc4527}                                                                 
 \end{figure}

NGC 4527 (Figure~\ref{ngc4527}) - The crossing at $r$=65\arcs\ (0.34)
encircles a broad bar-like feature. The central area is affected
by strong decomposition pinch, and our analysis is based on an 
average bulge image. The apparent crossing near $r$=10\arcs\ (0.05) is 
uncertain for this reason. Also, deprojection leaves an extensive
oval zone beyond 65\arcs\ that may not be significant.

\begin{figure}                                                                 
\figurenum{2.90}
\plotone{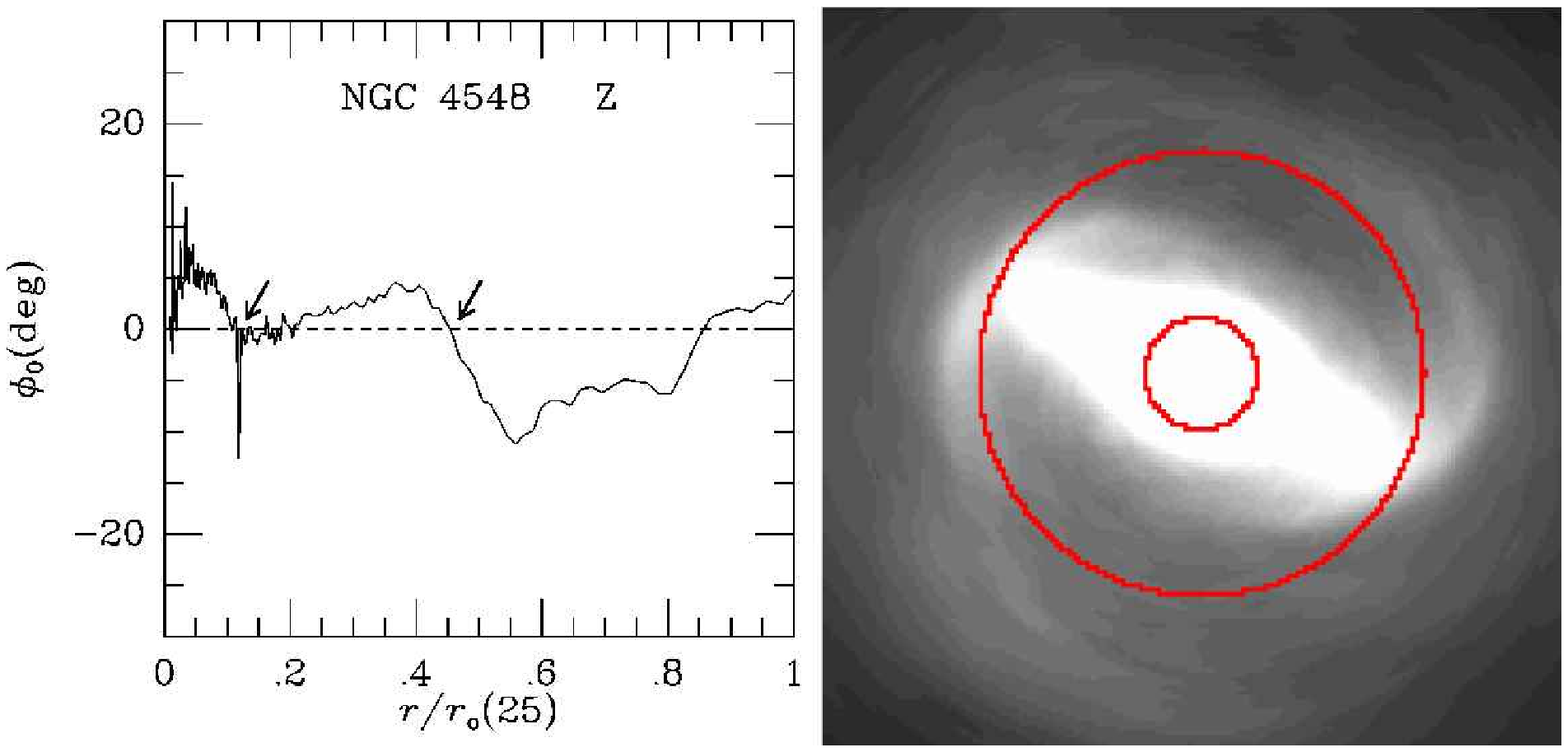}
 \vspace{2.0truecm}                                                             
\caption{Same as Figure 2.1 for NGC 4548}                                         
\label{ngc4548}                                                                 
 \end{figure}

NGC 4548 (Figure~\ref{ngc4548}) - The inner crossing at
$r$=19\arcs\ (0.12) coincides with a small inner oval, seen as a nuclear ring
in the optical, and the outer crossing coincides with the ends of the
primary bar. In this case, the bar and the main spiral would have the
same pattern speed, and the galaxy can be thought of as having a 
``bar-driven" spiral.

A very interesting RSL08 model reproduces the structure of NGC 4548 for a
corotation radius of 95\arcss 2 (0.58). The main phase-shift CR is at
75\arcs\ (0.45), which is closer to the RSL08 bar radius of 75\arcss 8
(0.46).

\begin{figure}                                                                 
\figurenum{2.91}
\plotone{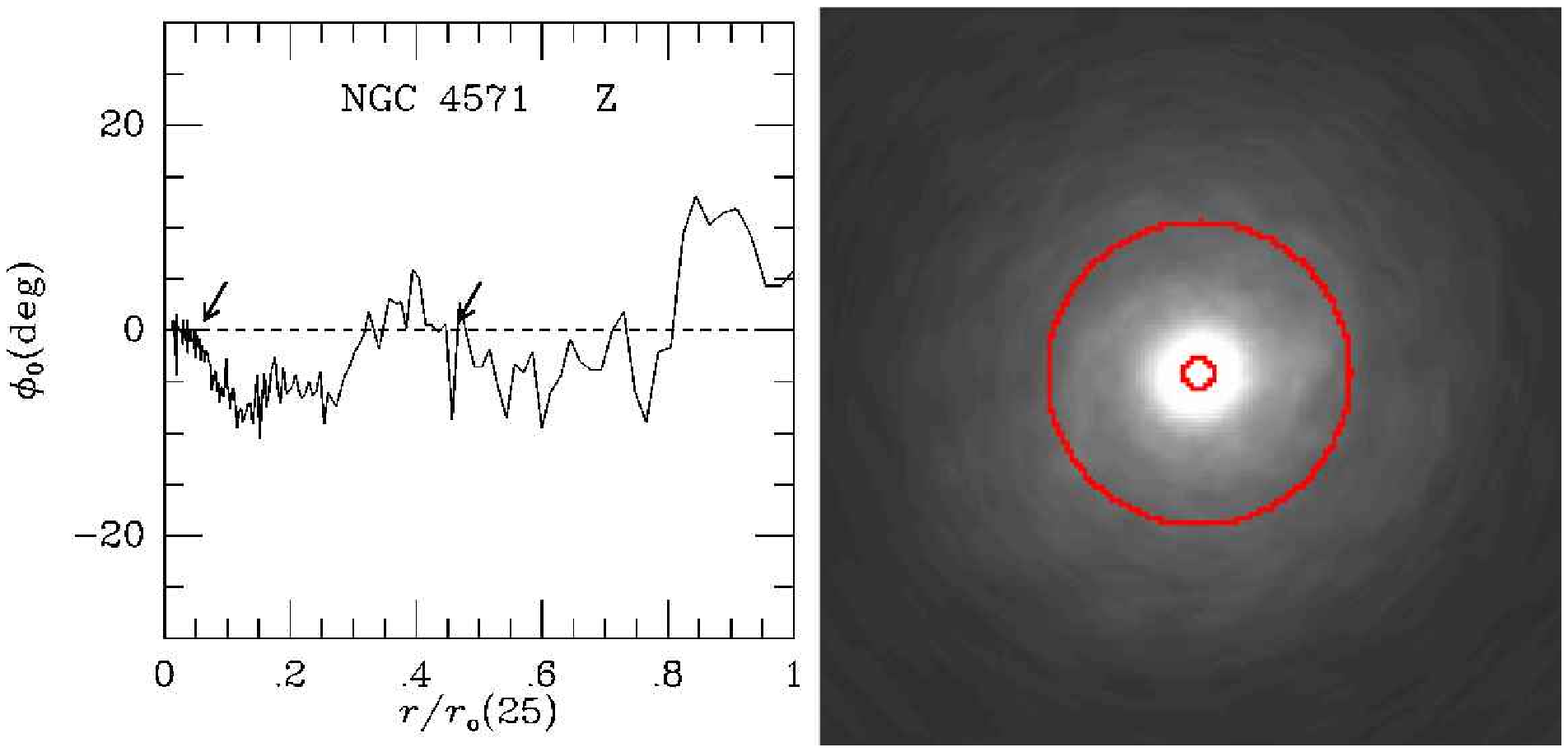}
 \vspace{2.0truecm}                                                             
\caption{Same as Figure 2.1 for NGC 4571}                                         
\label{ngc4571}                                                                 
 \end{figure}

NGC 4571 (Figure~\ref{ngc4571}) - The main crossing at $r$=51\arcs\ (0.46)
lies within a faint spiral pattern. The inner-most crossing is of uncertain
significance.  The inner negative portion of the phase-shift plot indicates
a possible future decoupling pattern, which may correspond to the inner
oval.

\begin{figure}                                                                 
\figurenum{2.92}
\plotone{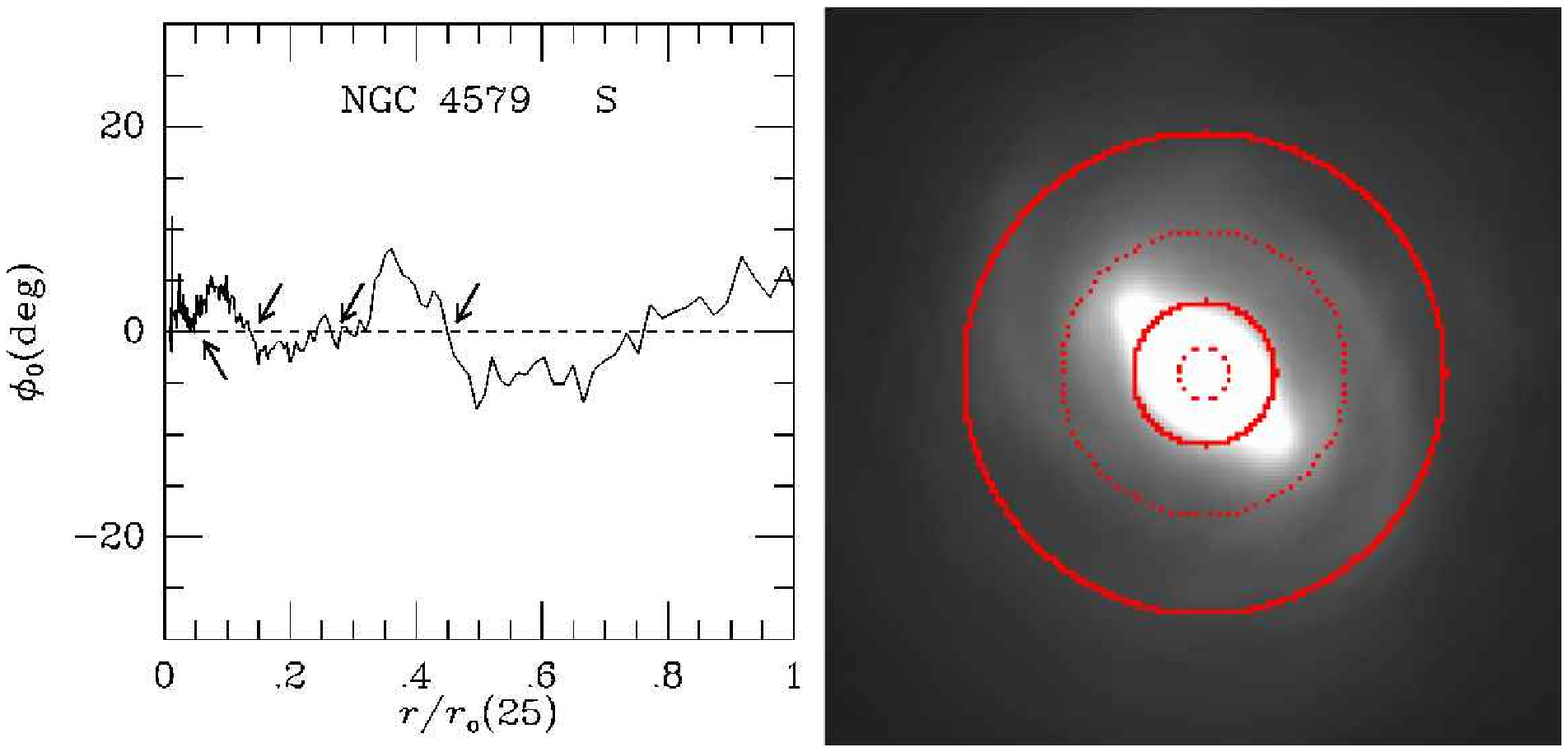}
 \vspace{2.0truecm}                                                             
\caption{Same as Figure 2.1 for NGC 4579}                                         
\label{ngc4579}                                                                 
 \end{figure}

NGC 4579 (Figure~\ref{ngc4579}) - Only two of the four P/N crossings
are well-defined. The one at $r$=25\arcs\ (0.14) encompasses a broad
oval inside the bar, while the crossing at $r$=81\arcs\ (0.44) passes
through the bright outer two-armed pattern. The weak crossing at
$r$=48\arcs\ (0.27) coincides with the approximate ends of the bar, but
if this is just a noise blip, then the dotted circle should be green
and NGC 4579 would be a case where the bar extends to close to the OLR of its
inner pattern. The innermost crossing at $r$=9\arcs\ (0.05) coincides
with a small inner oval.

Garcia-Burillo et al. (2008) have made a detailed study of the
gravitational potential in NGC 4579, using both gas observations and a
$K$-band image. Their goal was to examine the possible mechanism of
fueling of its LINER/Seyfert 1.9 nucleus. These authors also suggested
the possibility of multiple pattern speeds in NGC 4579 because of the
prominent bar and inner oval. They placed the CR of the primary bar at
6$\pm$1kpc by assuming the ends of the bar extend exactly to CR. They
also assumed that a ring in the vicinity of the inner oval lies at the
oval's inner 4:1 resonance, and placed the CR of the oval at 1kpc. In
comparison to our findings, Garcia-Burillo et al.'s adopted bar CR is
intermediate between our 3rd and 4th CRs in Table 1 (at 4.7 and 7.9
kpc, for an assumed distance of 20 Mpc).  Our innermost CR lies close
to their adopted oval CR. Based on azimuthally-averaged torques,
Garcia-Burillo et al. also suggested that the corotation barrier of the
bar has been overcome in NGC 4579 due to secular evolutionary processes,
and that the spiral may be decoupled from the bar. This was concluded
because the measured torques were found to still be negative just
outside their adopted bar CR, allowing an efficient population of the
bar's inner 4:1 resonance. Although there is some general agreement
between our two studies, the phase-shift method makes no {\em a priori}
assumptions about the correspondence of bar CR radii and resonance
features on the image, and as we have noted, NGC 4579 could be a case
where the bar extends to close to the OLR of its inner pattern.


An RSL08 model that reproduces the arm structure of NGC 4579 gives $r(CR)$ =
71\arcss 1 $\pm$ 8\arcss 4 (0.39$\pm$0.05). This is good agreement with 
our largest CR radius of 81\arcs\ (0.45), suggesting that their CR refers to 
the spiral and not the bar.

\begin{figure}                                                                 
\figurenum{2.93}
\plotone{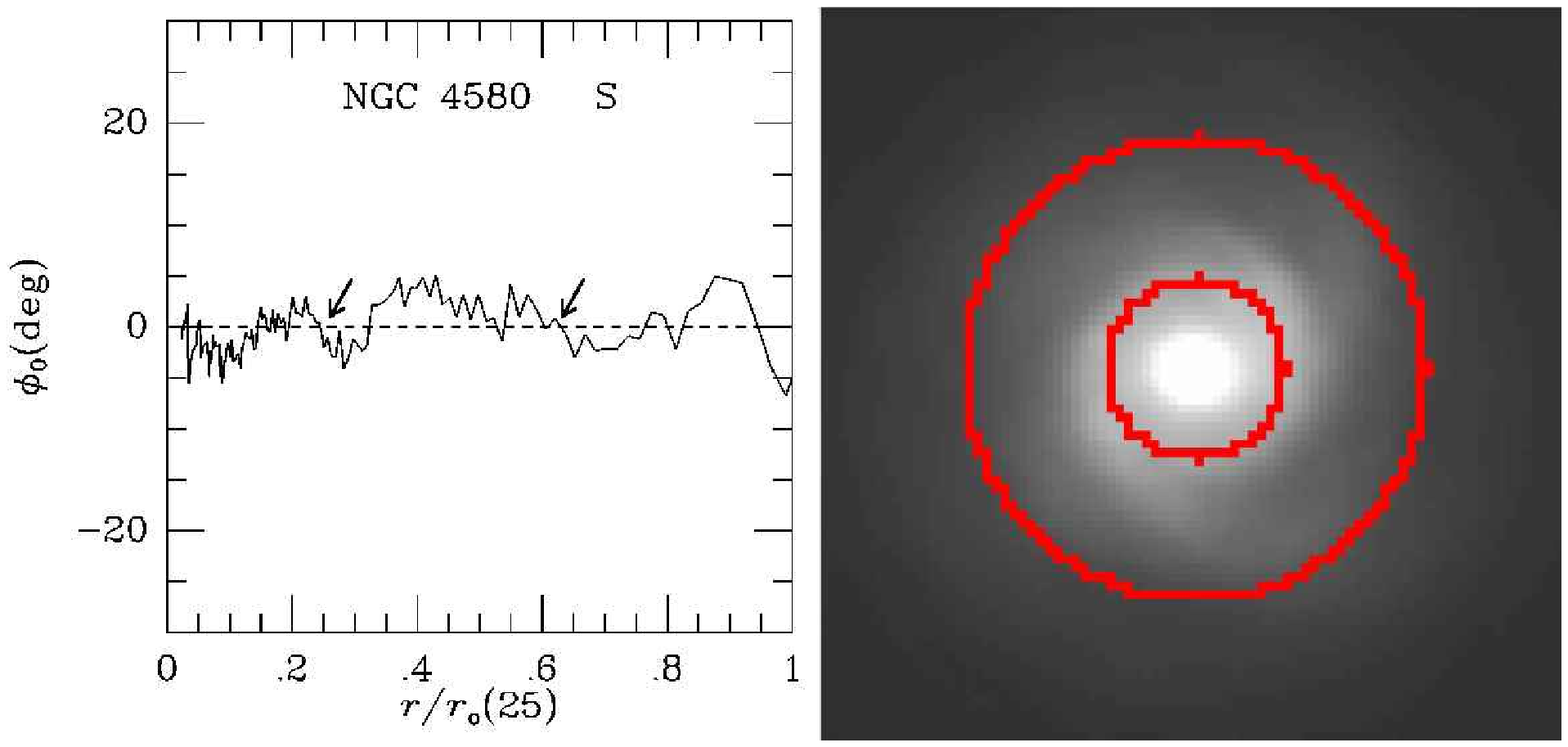}
 \vspace{2.0truecm}                                                             
\caption{Same as Figure 2.1 for NGC 4580}                                         
\label{ngc4580}                                                                 
 \end{figure}

NGC 4580 (Figure~\ref{ngc4580}) - The crossing at $r$=15\arcs\ (0.24) could
be associated with a small inner oval. The crossing at $r$=39\arcs\ (0.62)
could be associated with the main spiral arms.

\begin{figure}                                                                 
\figurenum{2.94}
\plotone{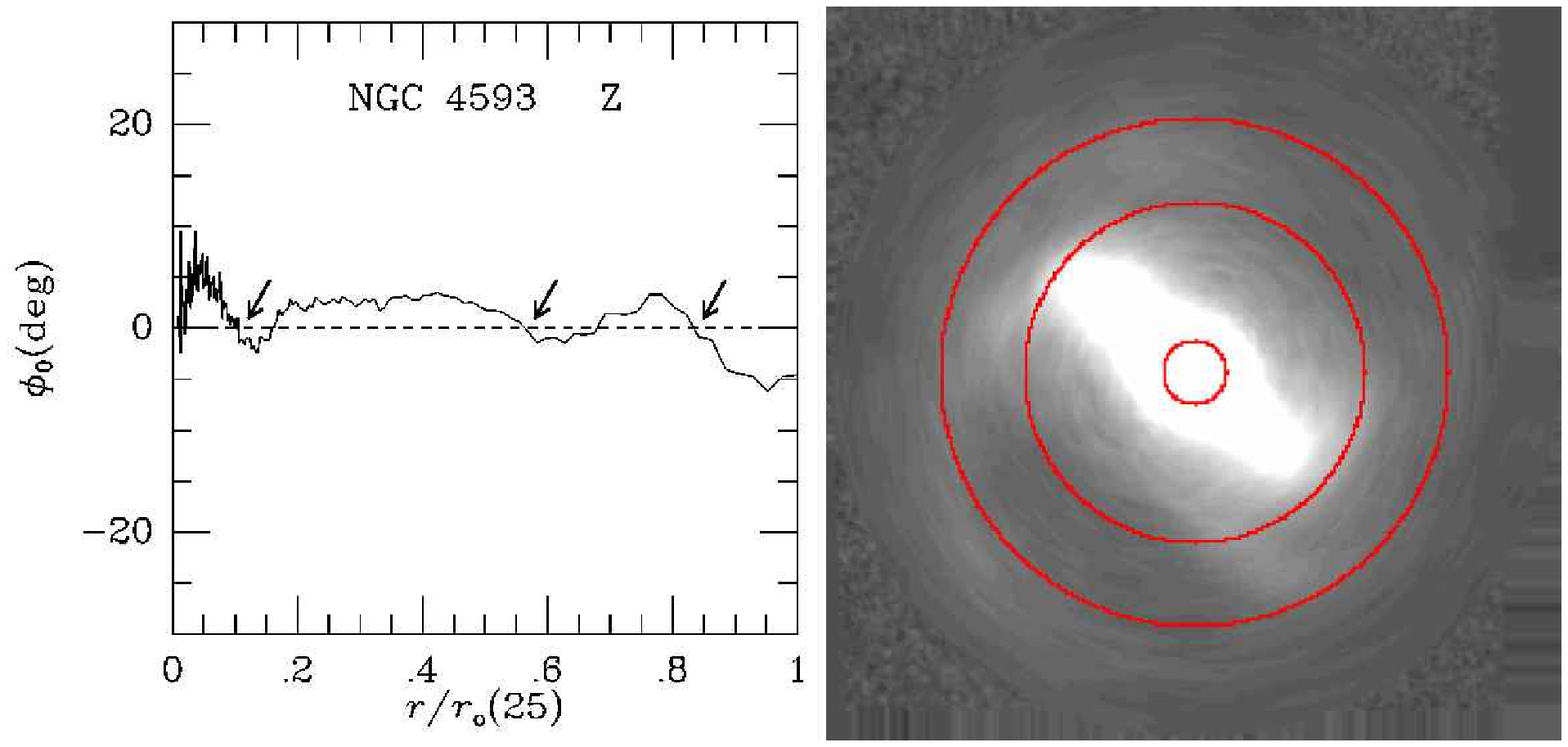}
 \vspace{2.0truecm}                                                             
\caption{Same as Figure 2.1 for NGC 4593}                                         
\label{ngc4593}                                                                 
 \end{figure}

NGC 4593 (Figure~\ref{ngc4593}) - The three P/N crossings are all
well-defined.  The one at $r$=12\arcs\ (0.10) coincides with a small
inner oval. The crossing at $r$=66\arcs\ (0.57) circles just around the
ends of the bar, while the outermost crossing is within the outer arms.
This could be a case where the spiral pattern is fully decoupled from
the bar pattern (see Figure~\ref{schematics}).

\begin{figure}                                                                 
\figurenum{2.95}
\plotone{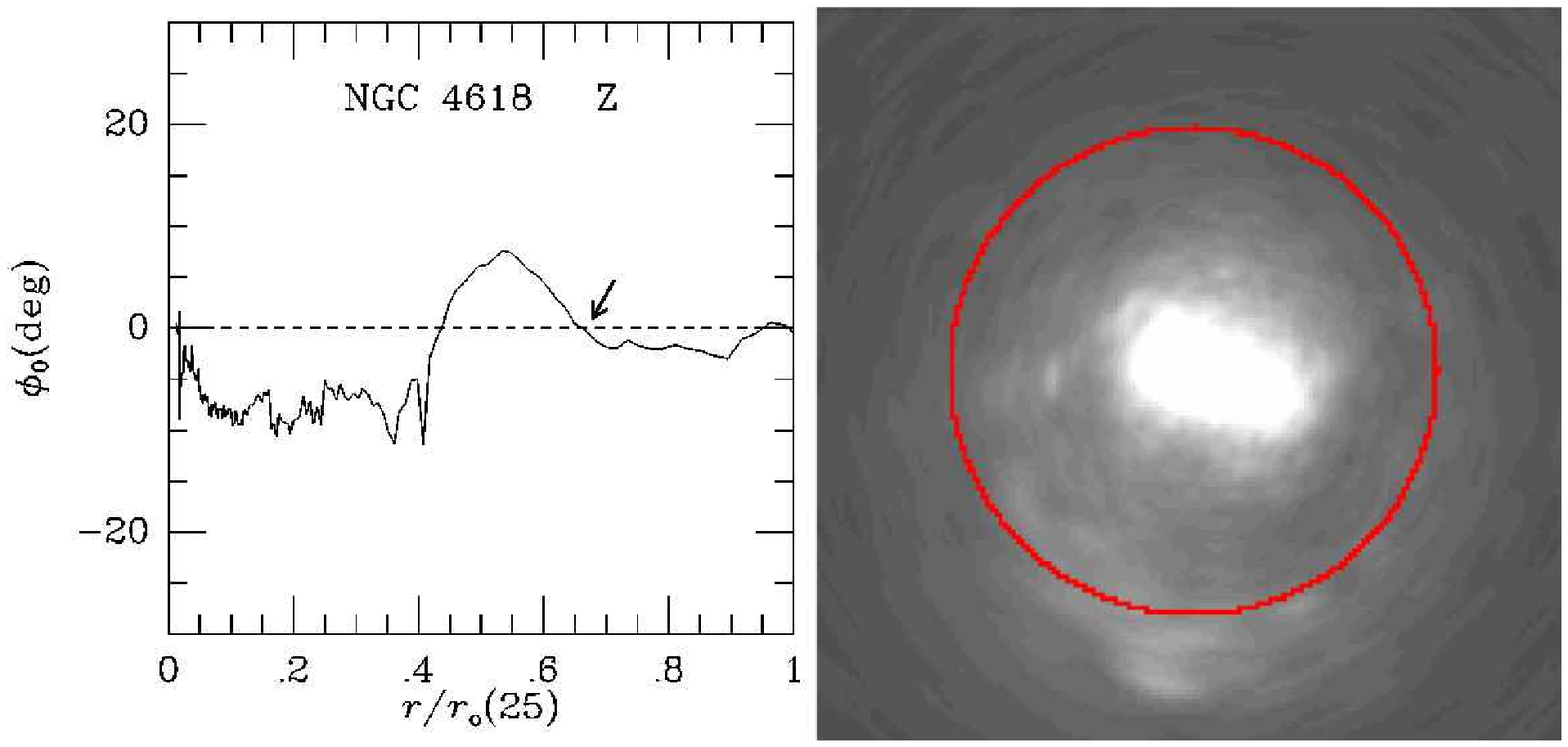}
 \vspace{2.0truecm}                                                             
\caption{Same as Figure 2.1 for NGC 4618}                                         
\label{ngc4618}                                                                 
 \end{figure}

NGC 4618 (Figure~\ref{ngc4618}) - The phase-shift plot is unusual in
having a broad negative region. In the $H$-band, the apparent nucleus
is not centered within the bar. The plot shows only one major crossing
that lies well beyond the bar ends. An inner fast bar might be
emerging under the background of an outer, slower-rotating
chaotic spiral.

\begin{figure}                                                                 
\figurenum{2.96}
\plotone{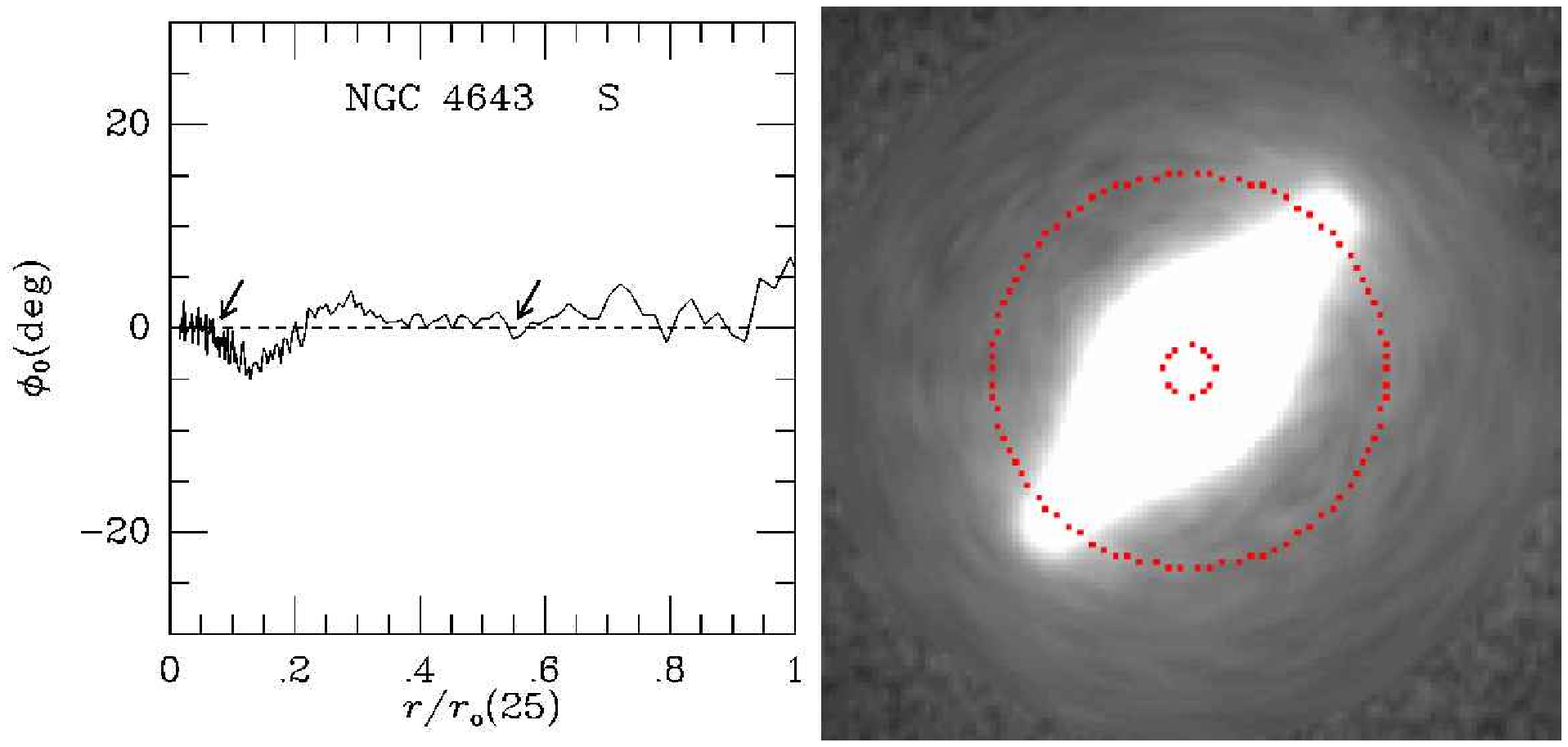}
 \vspace{2.0truecm}                                                             
\caption{Same as Figure 2.1 for NGC 4643}                                         
\label{ngc4643}                                                                 
 \end{figure}

NGC 4643 (Figure~\ref{ngc4643}) - None of the crossings for this galaxy
appears well-defined. Although the bar is strong, the pattern is so 
straight that the phase-shift method has difficulty locating the corotation
radii. Our CR at 50\arcs\ (0.54) is considerably less than that provided
by a model of RSL08: 69\arcss 1 (0.75). The overlay also shows that the
bar extends outside our CR estimate.

\begin{figure}                                                                 
\figurenum{2.97}
\plotone{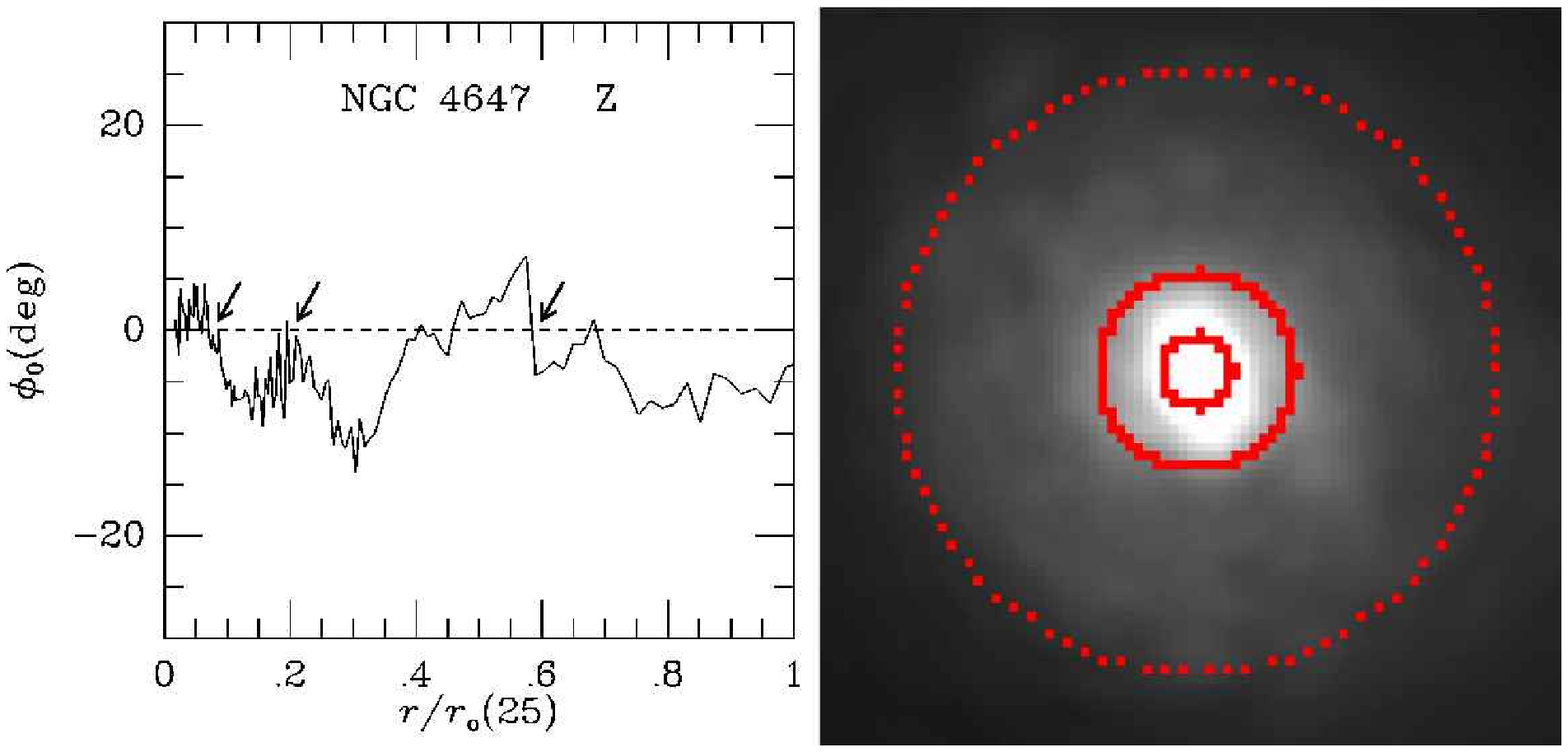}
 \vspace{2.0truecm}                                                             
\caption{Same as Figure 2.1 for NGC 4647}                                         
\label{ngc4647}                                                                 
 \end{figure}

NGC 4647 (Figure~\ref{ngc4647}) - The crossing at $r$=17\arcs\ (0.20)
circles directly around the ends of the main bar. There is no coherent
outer pattern in the $H$-band image, and the outer crossing is
uncertain. Although we have interpreted the $r$=17\arcs\ crossing as
the bar CR in Table 1, the negative phase-shifts inside this crossing
suggest that all or most of the bar may belong to the innermost mode.

\begin{figure}                                                                 
\figurenum{2.98}
\plotone{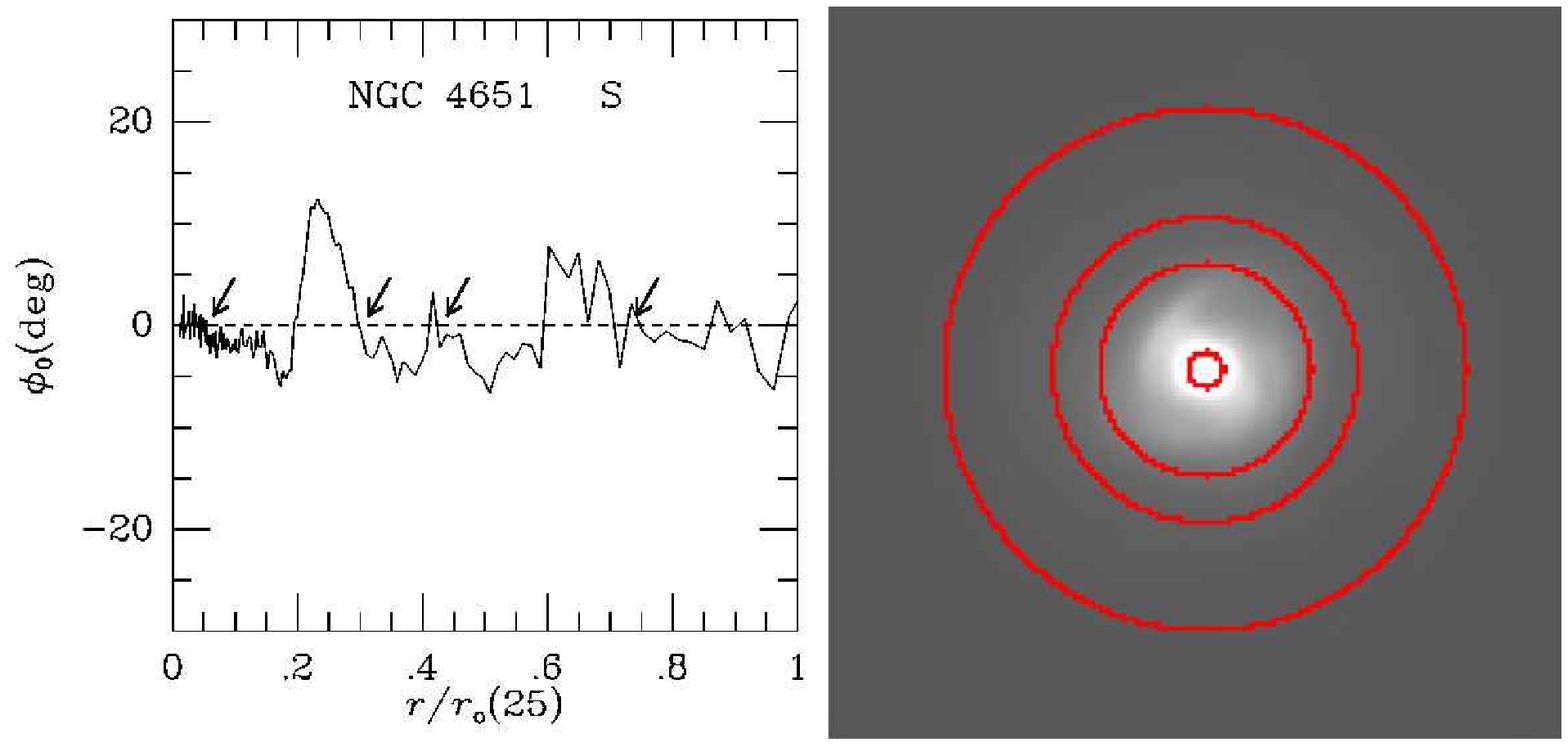}
 \vspace{2.0truecm}                                                             
\caption{Same as Figure 2.1 for NGC 4651}                                         
\label{ngc4651}                                                                 
 \end{figure}

NGC 4651 (Figure~\ref{ngc4651}) - The strong crossing at $r$=37\arcs\
(0.30) appears associated with the well-defined inner spiral pseudoring of
the galaxy. A deep optical SDSS image shows other patterns outside
this pseudoring that could account for the other outer crossings.
The exact location of the innermost crossing is uncertain although the
trend of the phase-shift curve indicates that a nuclear pattern is
likely to be present.   There is also the possibility of another decoupling
pattern just outside the nuclear pattern both from the image and from
the phase-shift plot.

\begin{figure}                                                                 
\figurenum{2.99}
\plotone{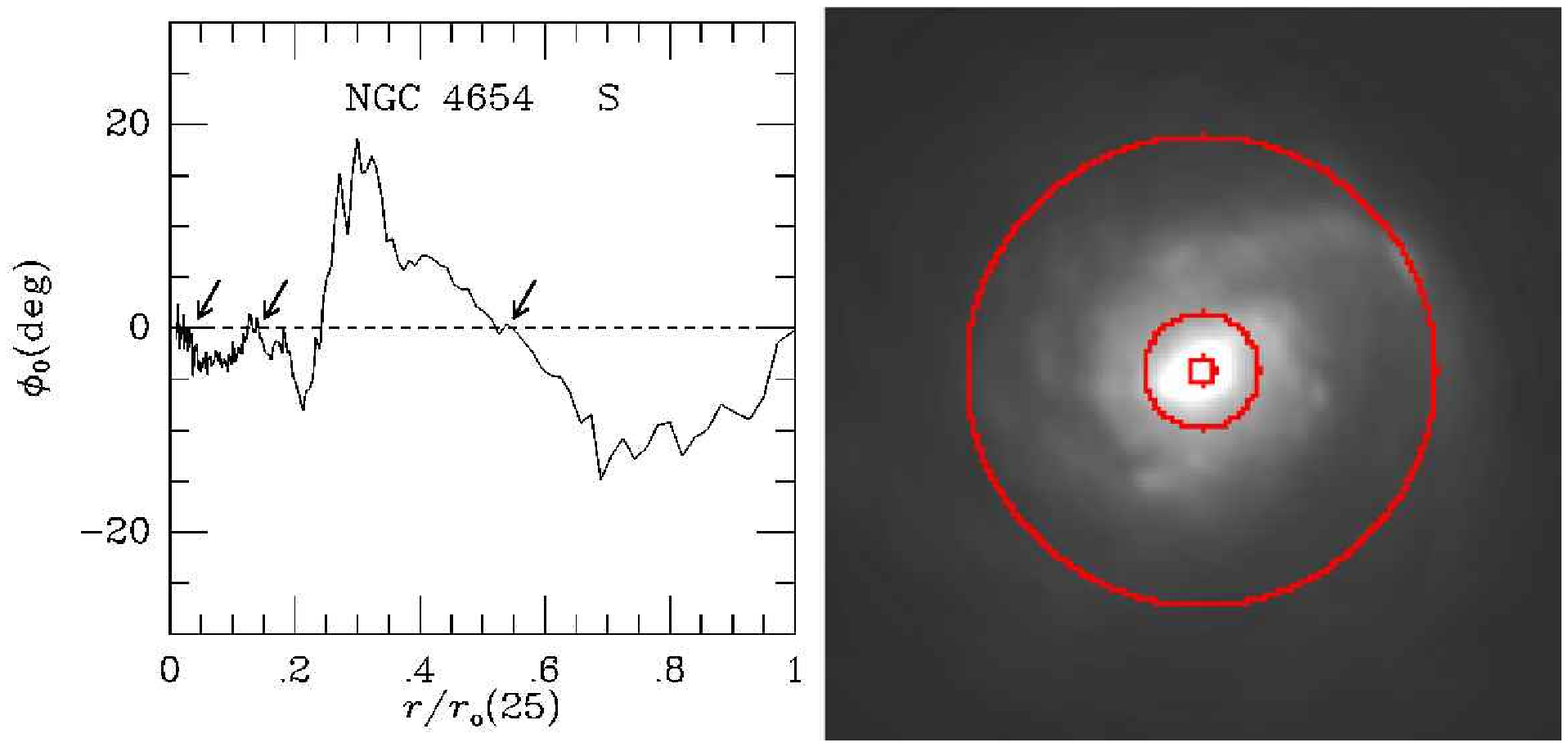}
 \vspace{2.0truecm}                                                             
\caption{Same as Figure 2.1 for NGC 4654}                                         
\label{ngc4654}                                                                 
 \end{figure}

NGC 4654 (Figure~\ref{ngc4654}) - The crossing at $r$=20\arcs\ (0.14) encircles
the ends of an inner oval. The outer crossing may be associated with the
outer spiral arms. There is also the indication of a nuclear pattern.

\begin{figure}                                                                 
\figurenum{2.100}
\plotone{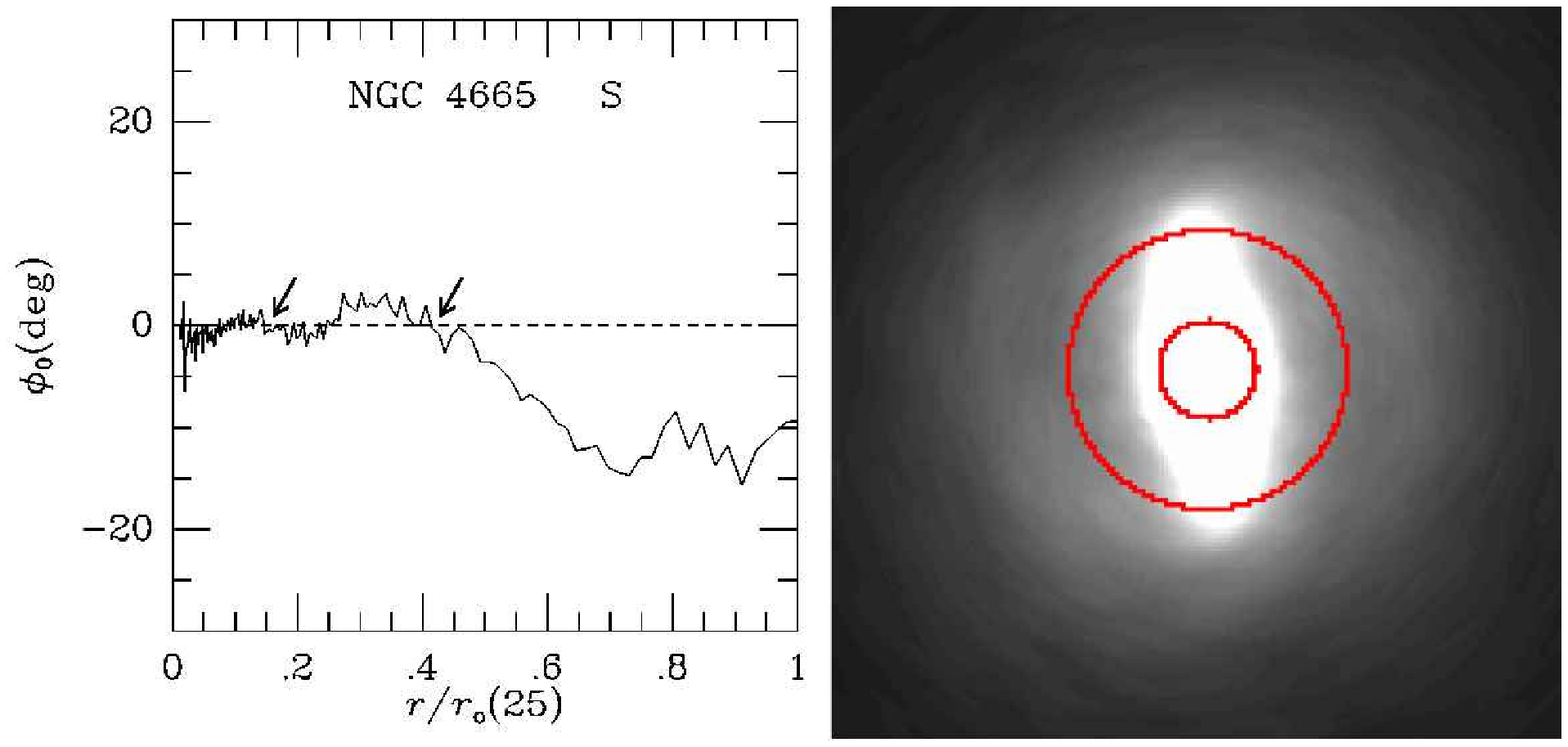}
 \vspace{2.0truecm}                                                             
\caption{Same as Figure 2.1 for NGC 4665}                                         
\label{ngc4665}                                                                 
 \end{figure}

NGC 4665 (Figure~\ref{ngc4665}) - As shown in paper 1, the main crossing
for this galaxy occurs just inside the ends of the very strong bar. In
paper 1, we argued that this could be a genuine case where the bar 
extends to beyond its CR radius.
A second crossing in the inner regions may be associated
with a small inner oval.

An RSL08 model of this galaxy places CR at 55\arcs 2 $\pm$ 11\arcss 8 
(0.48$\pm$0.10), comparable
to our main CR radius of 47\arcss 2 (0.41). In both cases, the CR radius
is less than the RSL08 bar radius of 62\arcss 5 (0.55). 

\begin{figure}                                                                 
\figurenum{2.101}
\plotone{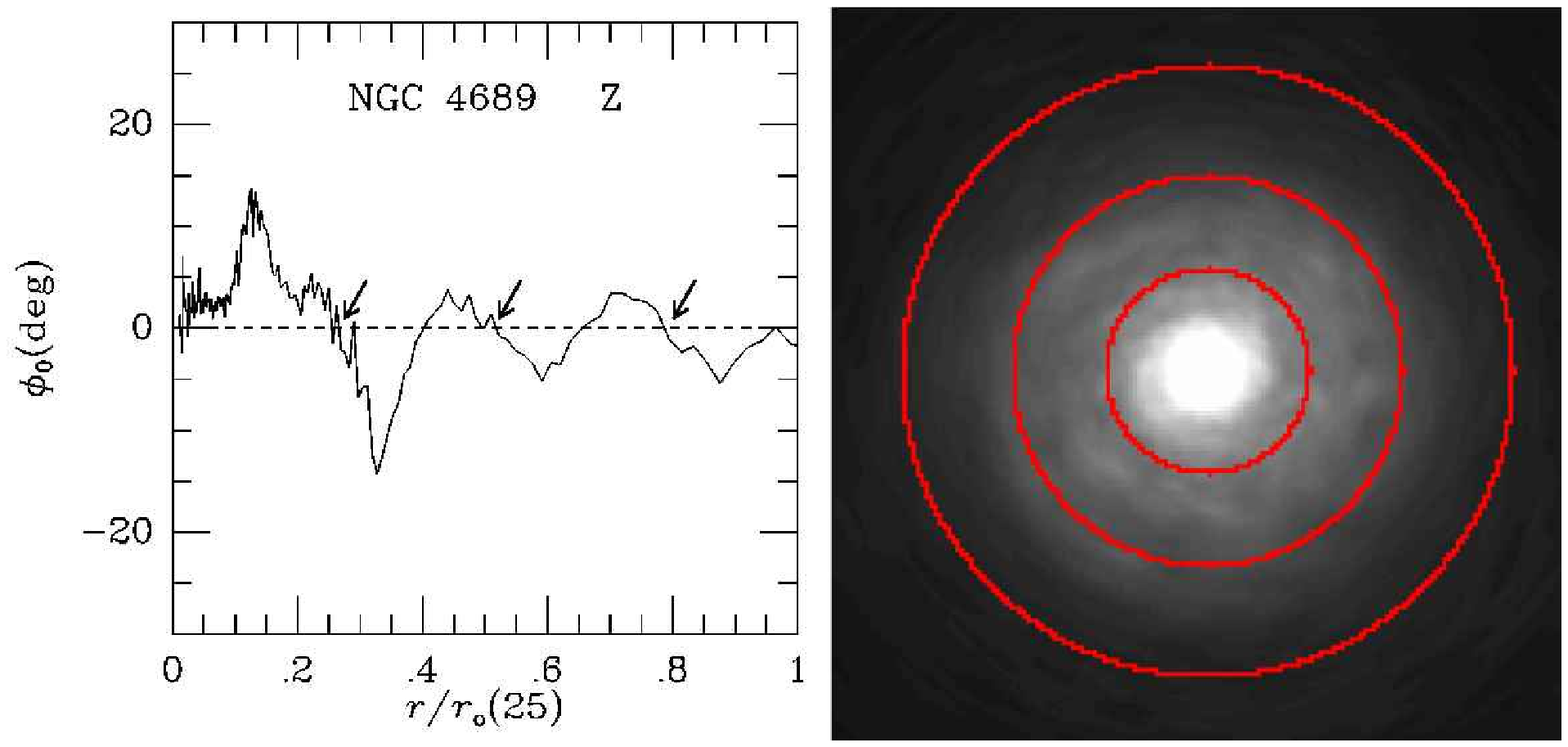}
 \vspace{2.0truecm}                                                             
\caption{Same as Figure 2.1 for NGC 4689}                                         
\label{ngc4689}                                                                 
 \end{figure}

NGC 4689 (Figure~\ref{ngc4689}) - In an optical SDSS image, this galaxy
has an inner pseudoring pattern, and intermediate pseudoring pattern,
and an outer ring or pseudoring pattern. The three main crossings in the
phase-shift plot could be related to these features.  Further decouplings
of patterns in the inner region are also possibly going on.

\begin{figure}                                                                 
\figurenum{2.102}
\plotone{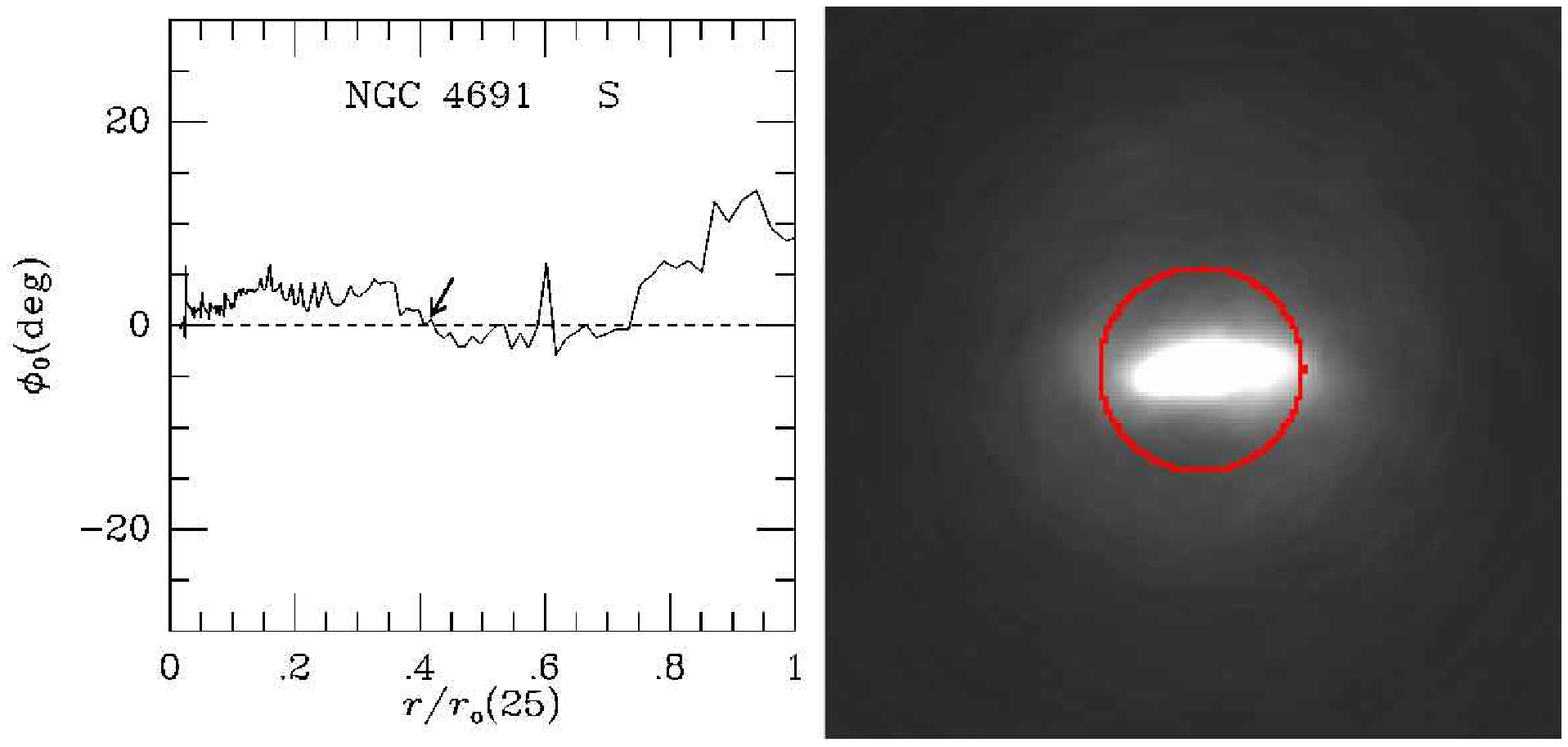}
 \vspace{2.0truecm}                                                             
\caption{Same as Figure 2.1 for NGC 4691}                                         
\label{ngc4691}                                                                 
 \end{figure}

NGC 4691 (Figure~\ref{ngc4691}) - This galaxy has one main crossing inside
$r_o$, but the overlay of this circle on the images shows that the bar
is asymmetric. The crossing circles near the ends of the bar in any case,
but at low light levels, the bar gets less elliptical and extends well 
outside the circle.

\begin{figure}                                                                 
\figurenum{2.103}
\plotone{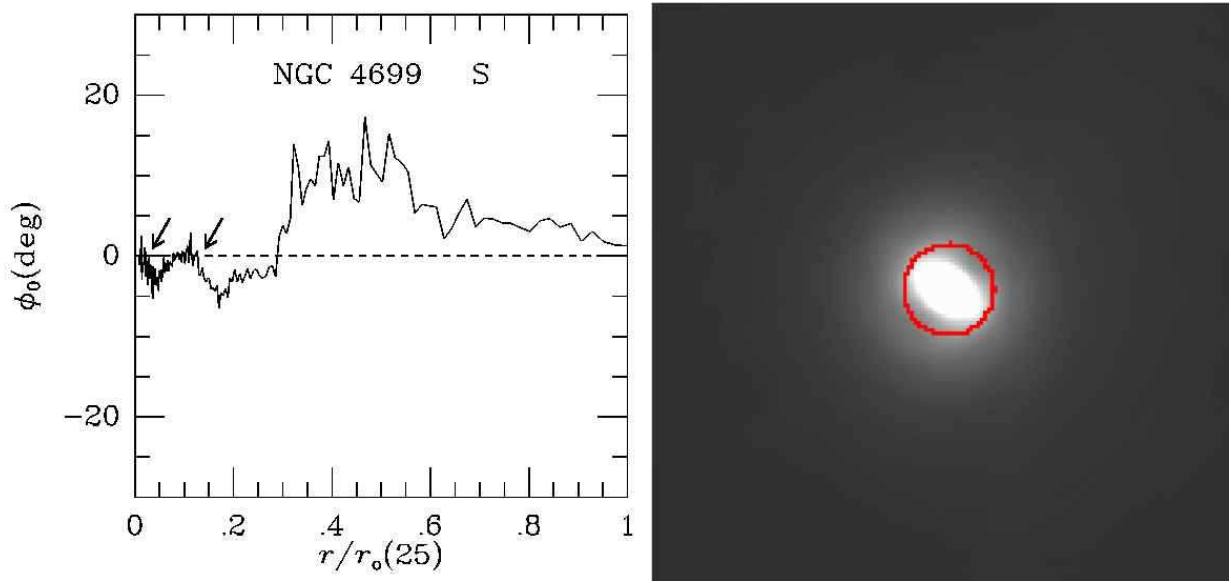}
 \vspace{2.0truecm}                                                             
\caption{Same as Figure 2.1 for NGC 4699. Only CR$_2$ from                        
Table 1 is overlaid on the image.}                                              
\label{ngc4699}                                                                 
 \end{figure}

NGC 4699 (Figure~\ref{ngc4699}) - The main crossing for this galaxy circles
around the ends of the prominent inner bar. A smaller crossing indicates
an unresolved nuclear pattern.

\begin{figure}                                                                 
\figurenum{2.104}
\plotone{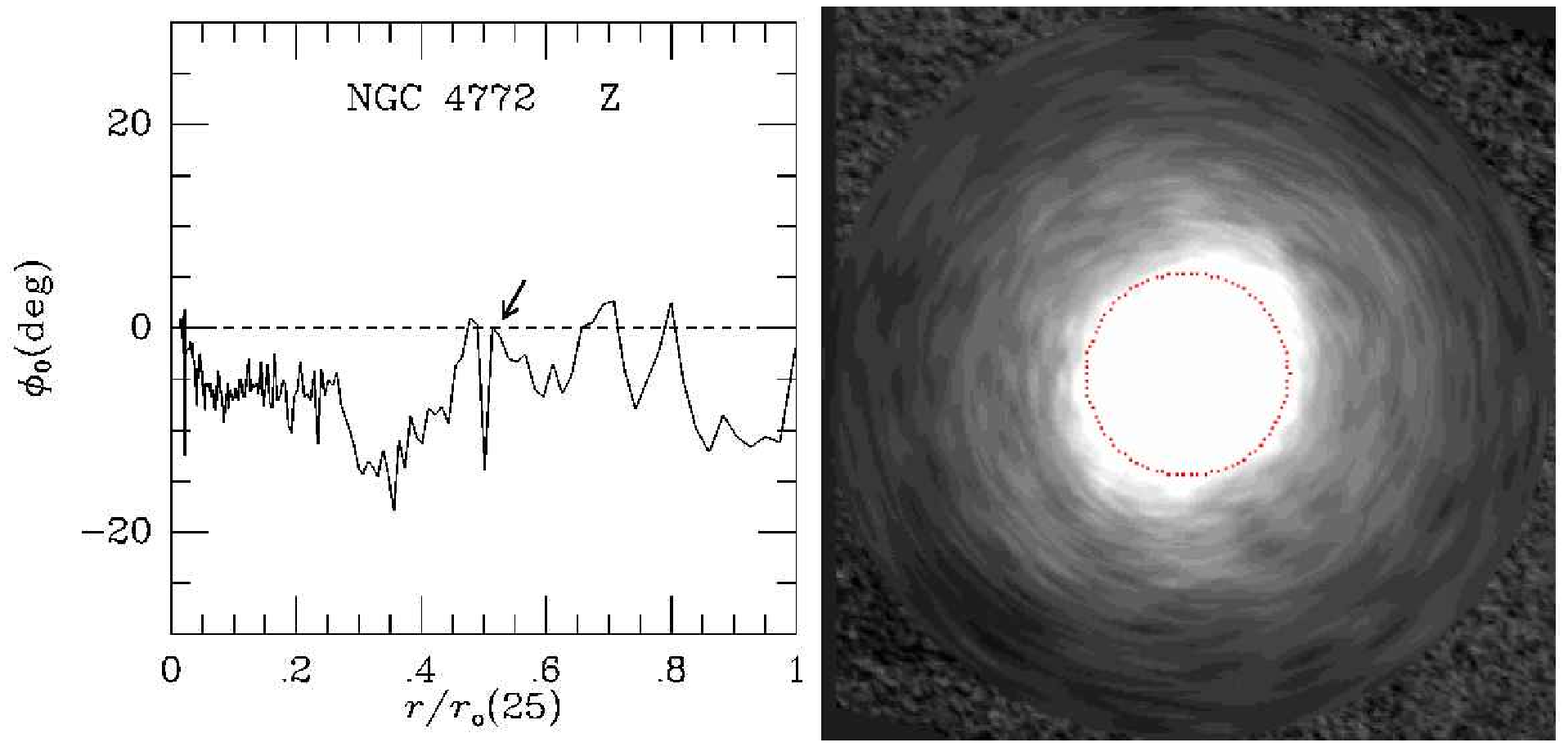}
 \vspace{2.0truecm}                                                             
\caption{Same as Figure 2.1 for NGC 4772}                                         
\label{ngc4772}                                                                 
 \end{figure}

NGC 4772 (Figure~\ref{ngc4772}) - This galaxy could be 
related to polar ring systems. Most of the phase differences are 
negative, which is very unusual.  The patterns in this galaxy might
not have reached quasi-steady state.
The weak crossing at $r$=53\arcs\ (0.52) is shown as the red circle in 
Figure~\ref{ngc4772}.

\begin{figure}                                                                 
\figurenum{2.105}
\plotone{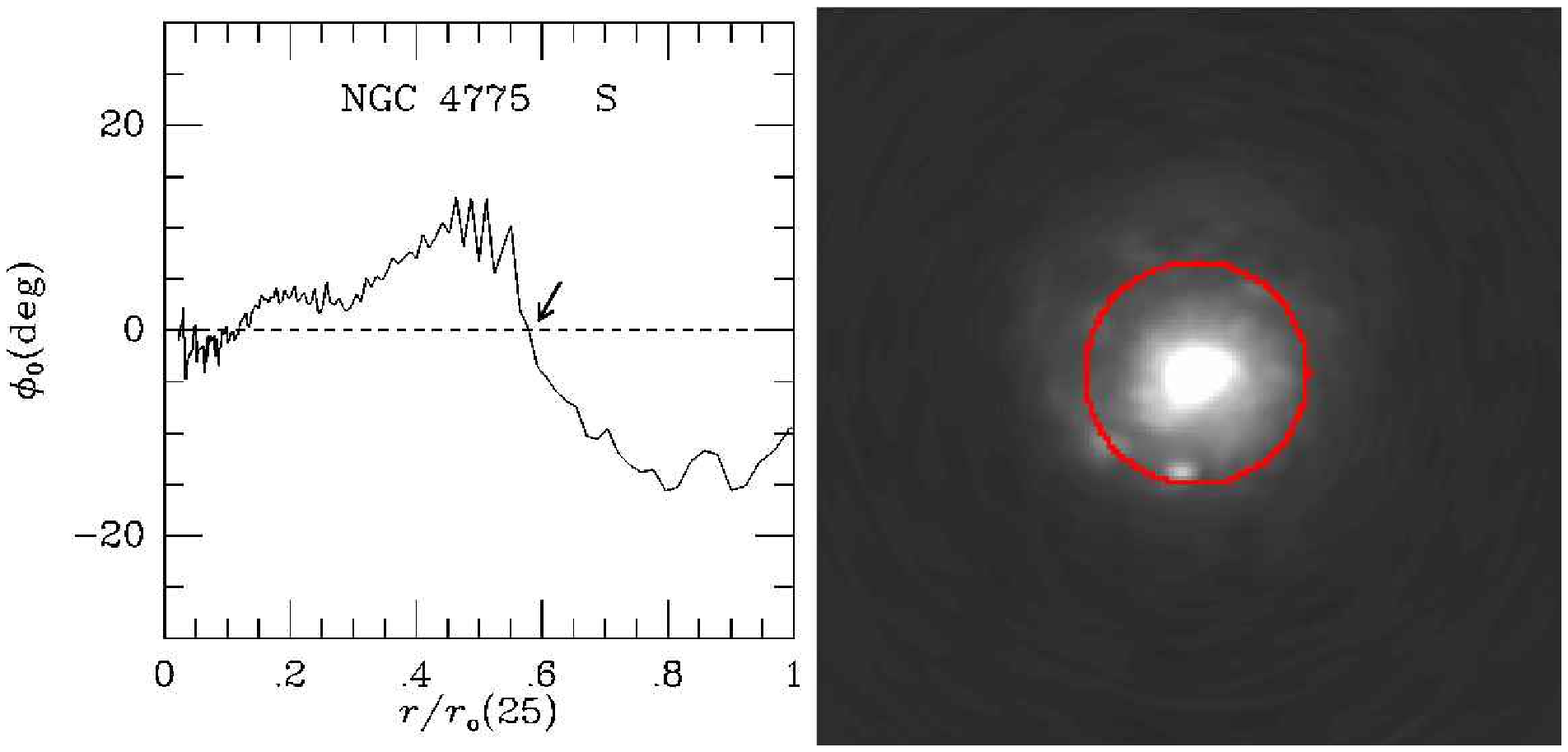}
 \vspace{2.0truecm}                                                             
\caption{Same as Figure 2.1 for NGC 4775}                                         
\label{ngc4775}                                                                 
 \end{figure}

NGC 4775 (Figure~\ref{ngc4775}) - This galaxy lacks a coherent spiral
pattern in visible light, yet the phase-shift plot shows one very
well-defined crossing. This crossing could be related to a small inner
oval, although the circle lies well beyond this oval. 

\begin{figure}                                                                 
\figurenum{2.106}
\plotone{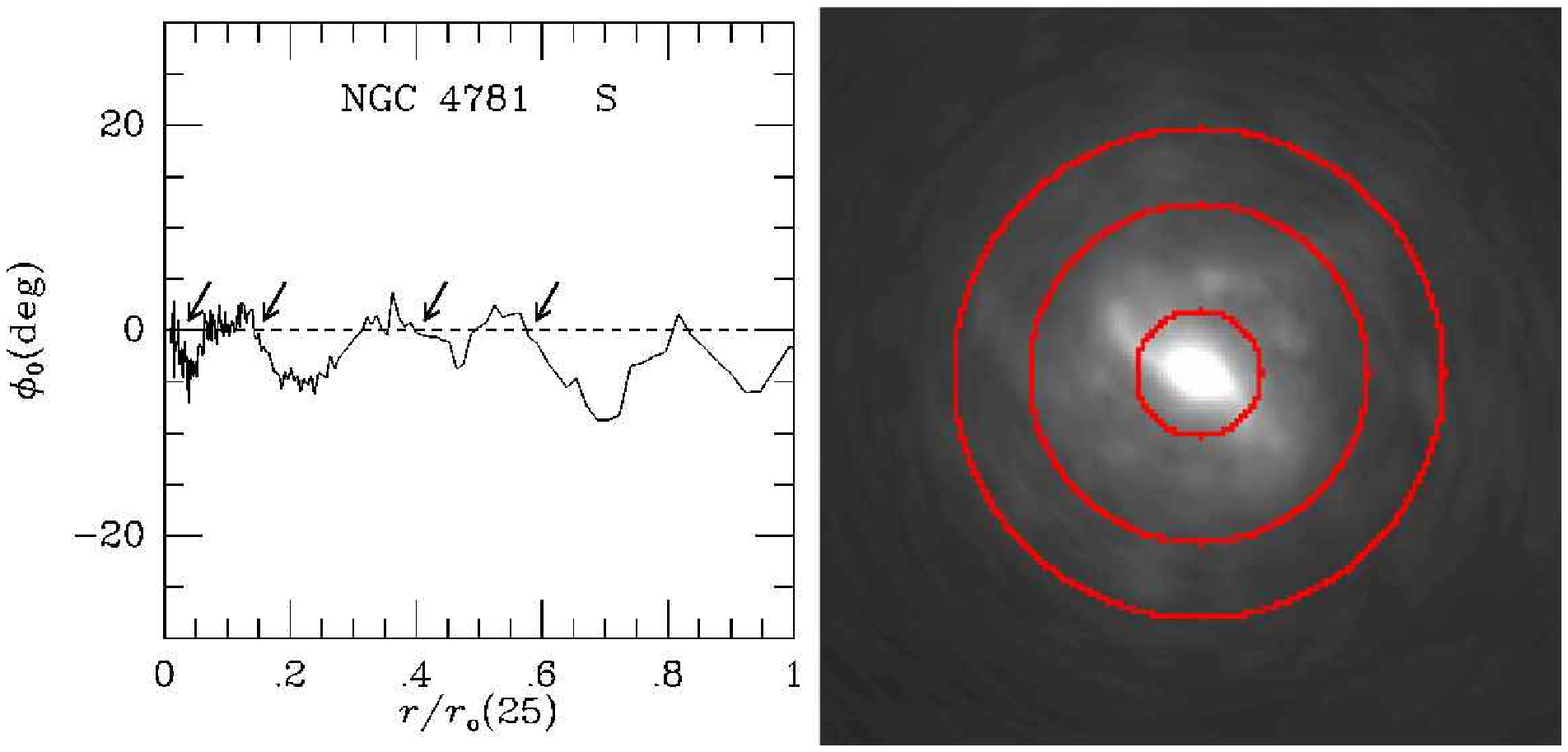}
 \vspace{2.0truecm}                                                             
\caption{Same as Figure 2.1 for NGC 4781.                                         
Only CR$_2$,                                                                    
CR$_3$, and CR$_4$ in Table 1 are overlaid on the image.}                       
\label{ngc4781}                                                                 
 \end{figure}

NGC 4781 (Figure~\ref{ngc4781}) - The crossing at $r$=15\arcs\ (0.14)
appears associated with a small inner oval. The next crossing lies
within the spiral pattern, and well beyond the ends of the rest of the
bar. The outer crossing lies in a region of chaotic pattern. There is
also the indication of a nuclear pattern. Although we have interpreted
the $r$=17\arcs\ crossing as the bar CR in Table 1, the negative
phase-shifts inside this crossing suggest that all or most of the bar
may belong to this innermost mode.

\begin{figure}                                                                 
\figurenum{2.107}
\plotone{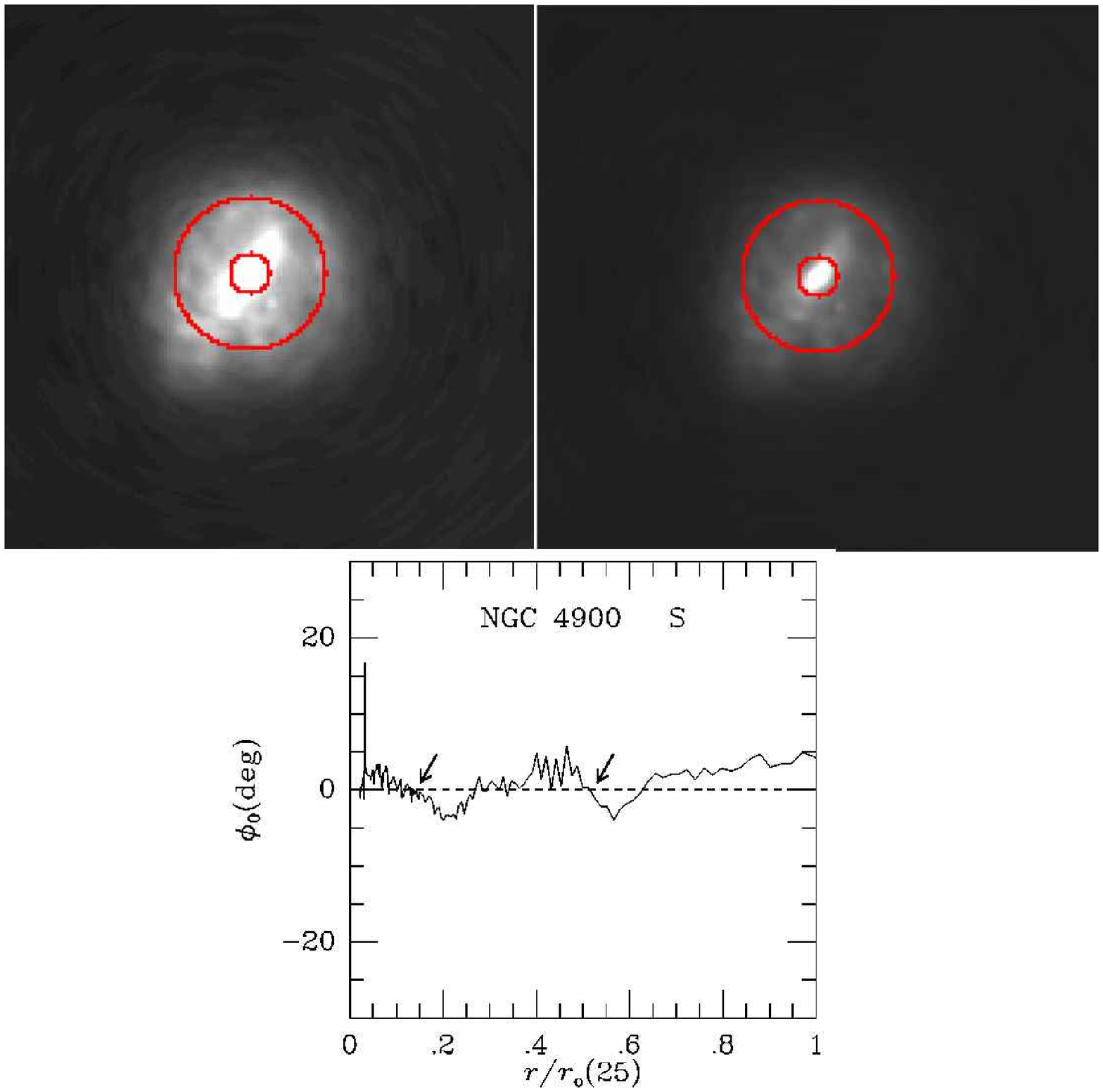}
 \vspace{2.0truecm}                                                             
\caption{Same as Figure 2.1 for NGC 4900}                                         
\label{ngc4900}                                                                 
 \end{figure}

NGC 4900 (Figure~\ref{ngc4900}) - The crossing at $r$=9\arcs\ (0.13)
lies inside the bar and encompasses an inner oval. Although we have
interpreted the $r$=35\arcs\ (0.52) crossing as the bar CR in Table 1, the
negative phase-shifts inside this crossing suggest that all or most of
the bar may belong to this innermost mode.

\begin{figure}                                                                 
\figurenum{2.108}
\plotone{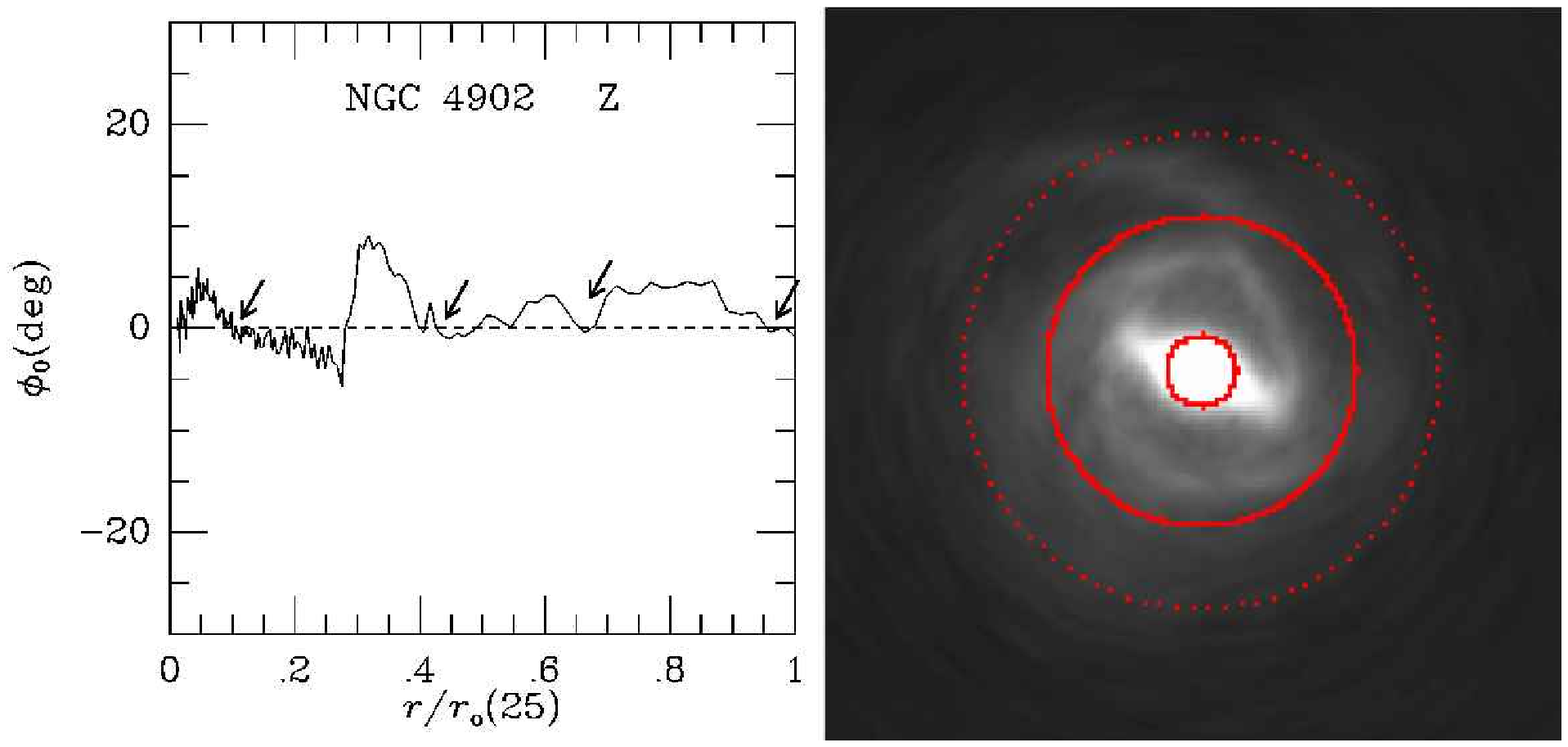}
 \vspace{2.0truecm}                                                             
\caption{Same as Figure 2.1 for NGC 4902. CR$_4$ from Table 1                     
is not overlaid on the image.}                                                  
\label{ngc4902}                                                                 
 \end{figure}

NGC 4902 (Figure~\ref{ngc4902}) - The crossing at $r$=9\arcs\ (0.10)
encircles the ends of a small inner oval that is slightly misaligned
with the main bar. The crossing at $r$=39\arcs\ (0.43) lies well beyond
the bar ends and encompasses the main inner spiral. The crossing at
$r$=60\arcs\ (0.66) could be associated with the outer spiral arms,
although this crossing is weak and may indicate an ongoing decoupling.
Although we have interpreted the $r$=39\arcs\ (0.43) crossing as the
bar CR in Table 1, the negative phase-shifts inside this crossing
suggest that the bar may actually belong to this innermost mode.  This
is especially suggested by the fact that the strong N/P crossing
between the two CRs circles almost exactly around the bar ends. We
suspect that in this case, the bar extends to close to the OLR of the
innermost pattern (see sections 5.1.3 and 6.1). A fourth crossing at
$r$ = 87\arcs\ (0.96) is close to $r_o$ and not within the field of the
overlay.

An excellent RSL08 model of this galaxy places CR at 44\arcss 3 (0.49),
close to our second CR.

\begin{figure}                                                                 
\figurenum{2.109}
\plotone{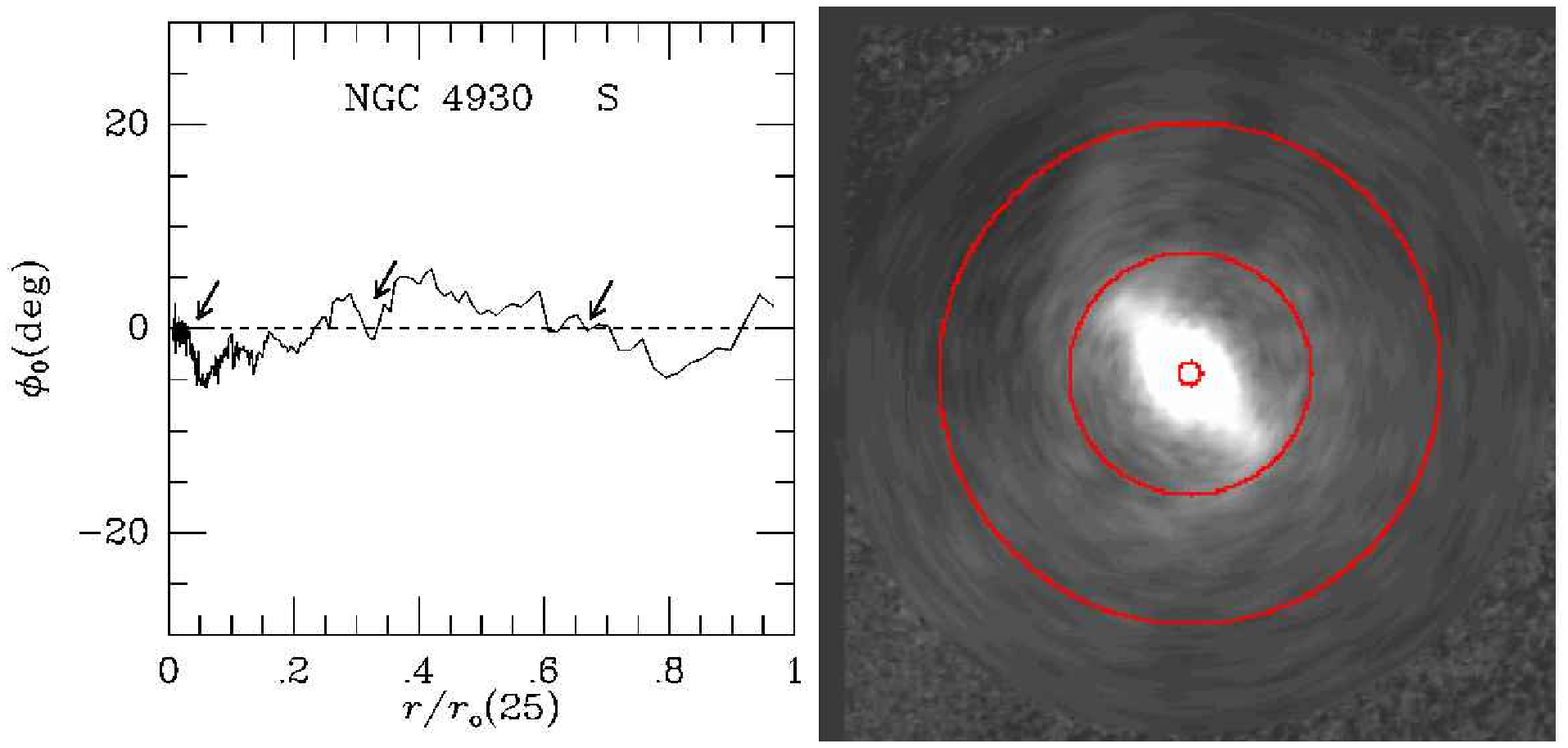}
 \vspace{2.0truecm}                                                             
\caption{Same as Figure 2.1 for NGC 4930}                                         
\label{ngc4930}                                                                 
 \end{figure}

NGC 4930 (Figure~\ref{ngc4930}) - The crossing at $r$=46\arcs\ (0.31) 
encompasses
the ends of the bar. The one at $r$=97\arcs\ (0.66) may be associated with the
outer spiral arms. There is definitely also an inner oval, but the innermost
crossing is mostly inside this oval. The phase-shift plot suggests ongoing
decoupling in the inner parts of the bar that might eventually result in the oval
and the bar having different pattern speeds.

An RSL08 model places CR at 46\arcss 6 (0.32), close to our second crossing
although the latter is weak.

\begin{figure}                                                                 
\figurenum{2.110}
\plotone{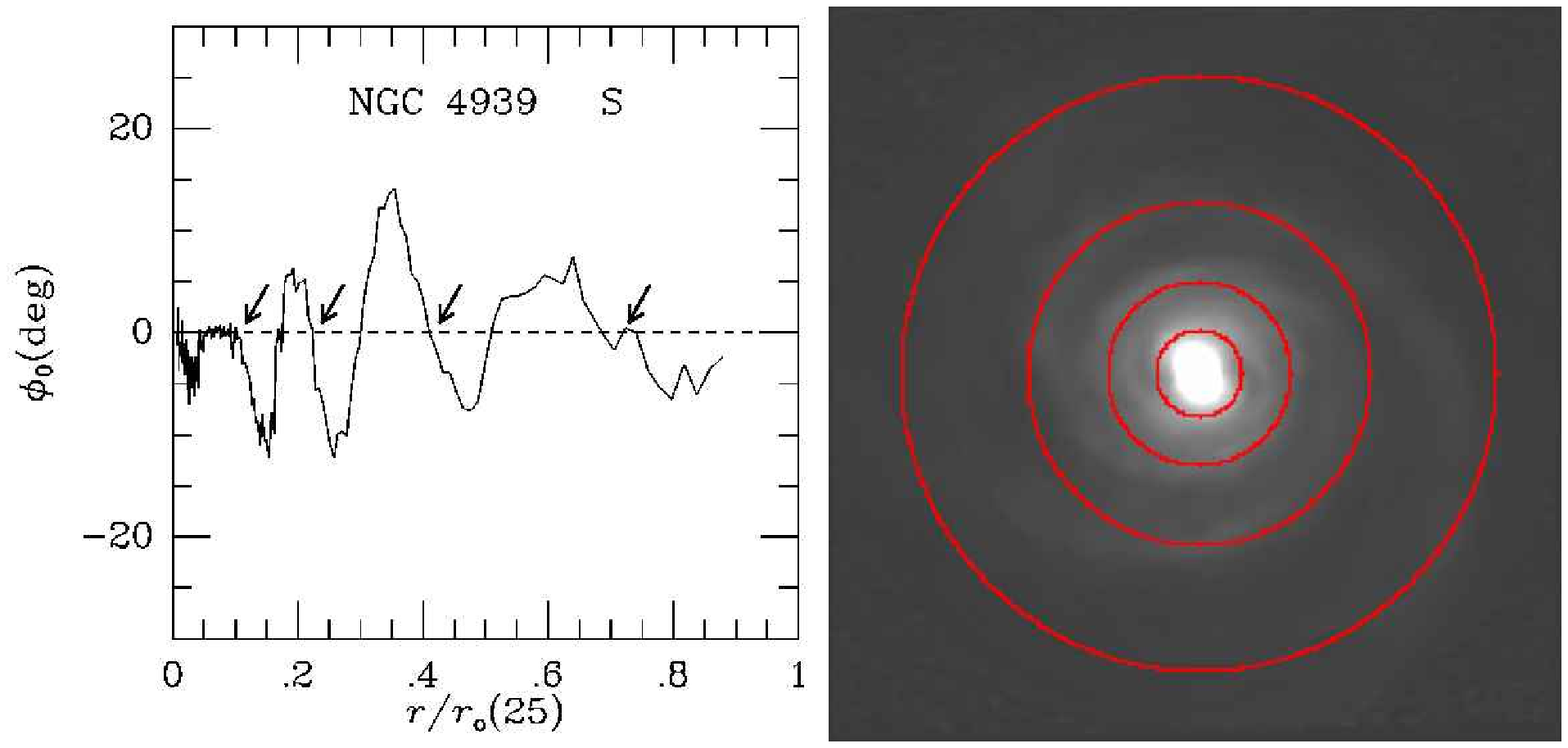}
 \vspace{2.0truecm}                                                             
\caption{Same as Figure 2.1 for NGC 4939}                                         
\label{ngc4939}                                                                 
 \end{figure}

NGC 4939 (Figure~\ref{ngc4939}) - This galaxy has a strong and
complicated spiral pattern that yields several P/N crossings in the
phase-shift plot. The crossing at $r$=17\arcs\ (0.10) appears associated
with a small inner oval or bar. The crossing at $r$=38\arcs\ (0.23)
could be associated with two inner spiral arms, while that at
$r$=69\arcs\ (0.41) could be associated with the intermediate, still
tightly-wrapped pattern. The crossing at $r$=120\arcs\ (0.82) could
refer to the outer, more open arms. An optical image on the NASA/IPAC Extragalactic Database
(NED) website suggests
all the crossings refer to distinct patterns.  There is indication also of a
nuclear pattern that is not well resolved.

\begin{figure}                                                                 
\figurenum{2.111}
\plotone{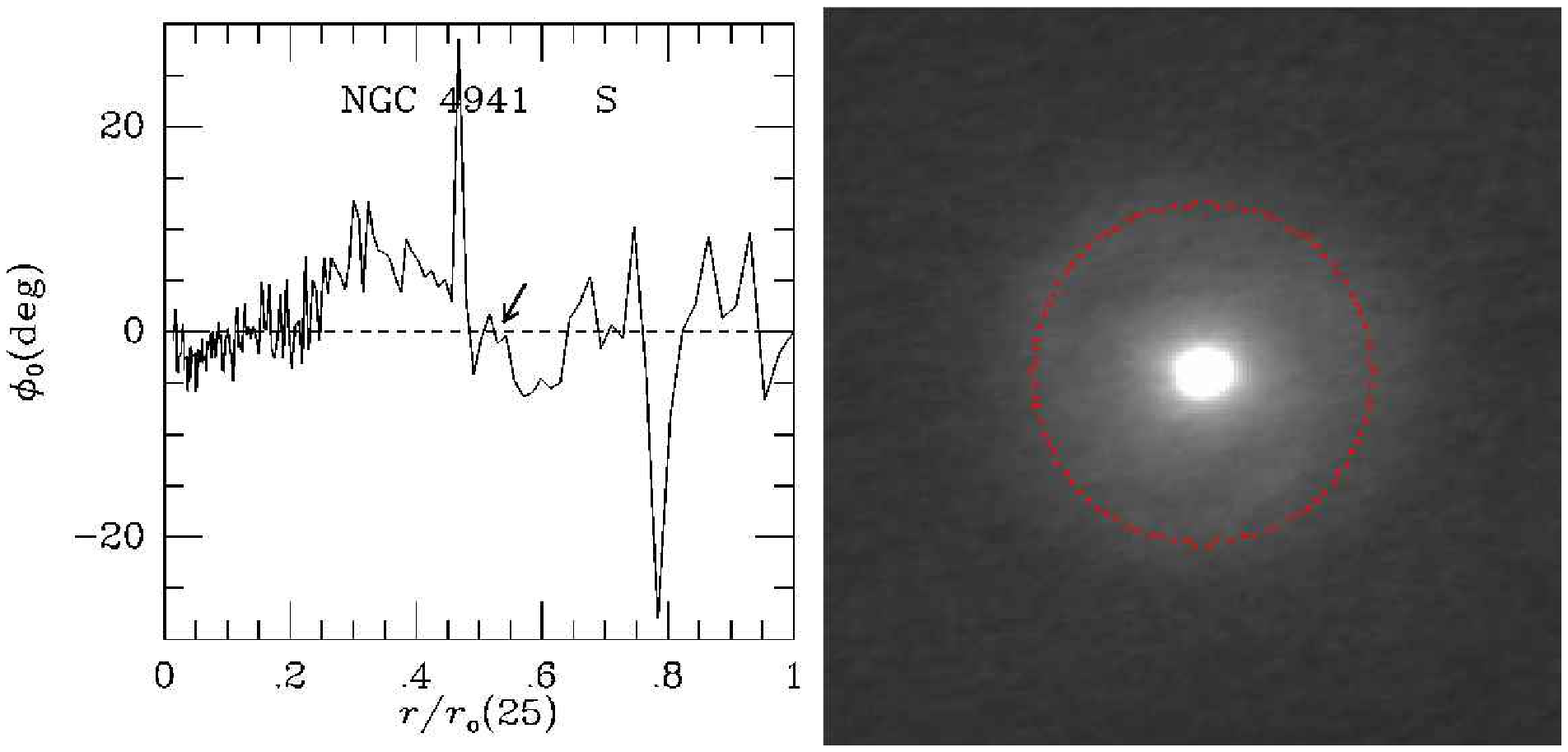}
 \vspace{2.0truecm}                                                             
\caption{Same as Figure 2.1 for NGC 4941}                                         
\label{ngc4941}                                                                 
 \end{figure}

NGC 4941 (Figure~\ref{ngc4941}) - The phaseshift distribution is noisy 
for this galaxy which includes an ovally-distorted spiral zone surrounded
by a well-defined outer ring. The crossing at $r$=57\arcs\ (0.52) could be
associated with the broad oval disk and spiral. The analysis is based
on an average bulge image.

\begin{figure}                                                                 
\figurenum{2.112}
\plotone{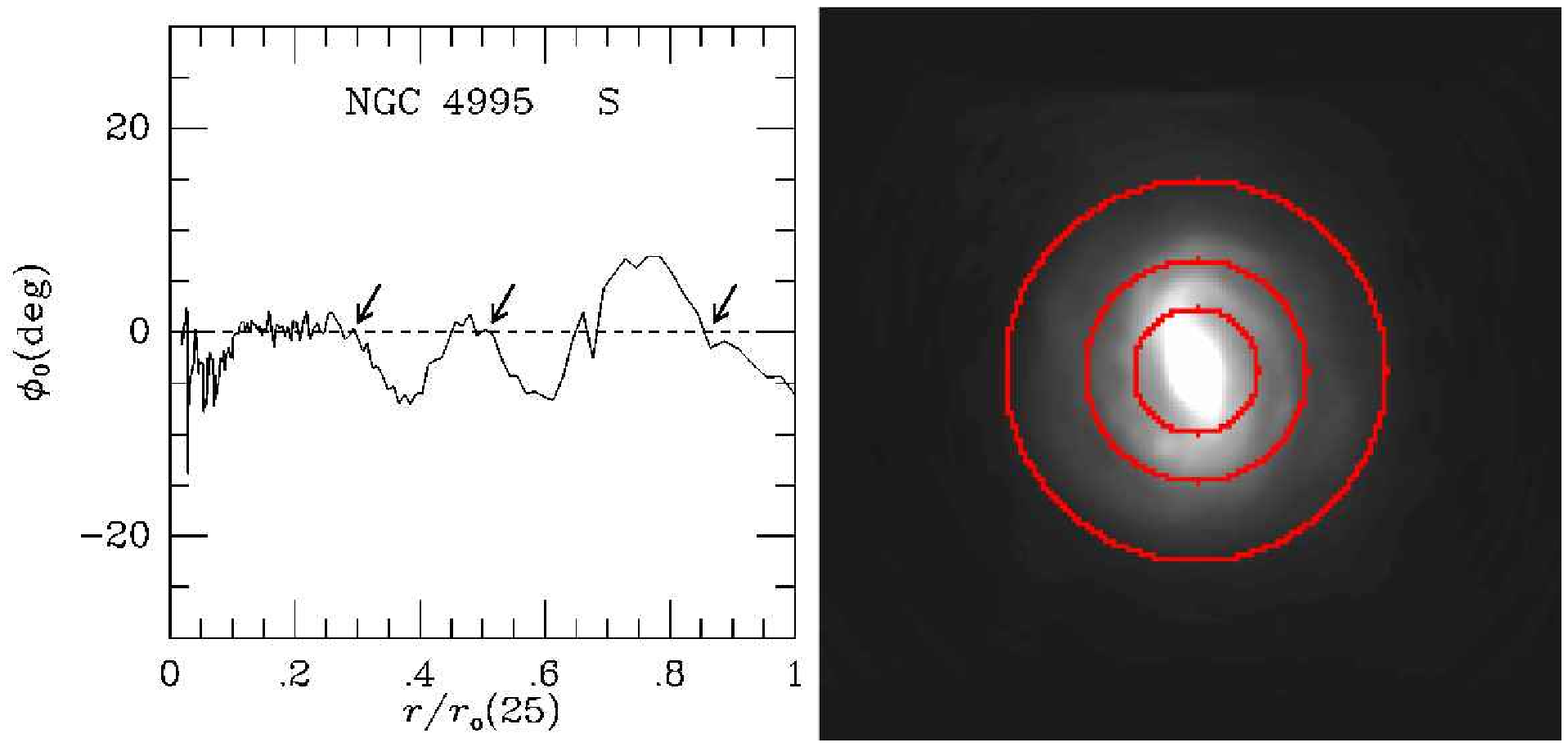}
 \vspace{2.0truecm}                                                             
\caption{Same as Figure 2.1 for NGC 4995}                                         
\label{ngc4995}                                                                 
 \end{figure}

NGC 4995 (Figure~\ref{ngc4995}) - The crossing at $r$=22\arcs\ (0.29) encircles
the ends of the bar of this galaxy. The other crossings may be associated
with different parts of the spiral arms, a bright inner pseudoring and very
faint outer arms. There is indication of a nuclear pattern as well.

An RSL08 model places CR at 64\arcss 2 (0.85), very close to our outermost
CR of 64\arcss 5 (0.86). Since we also find a crossing around the ends of
the bar, it is likely that the RSL08 value refers only to the spiral, and
that the galaxy is a fast bar case. 

\begin{figure}                                                                 
\figurenum{2.113}
\plotone{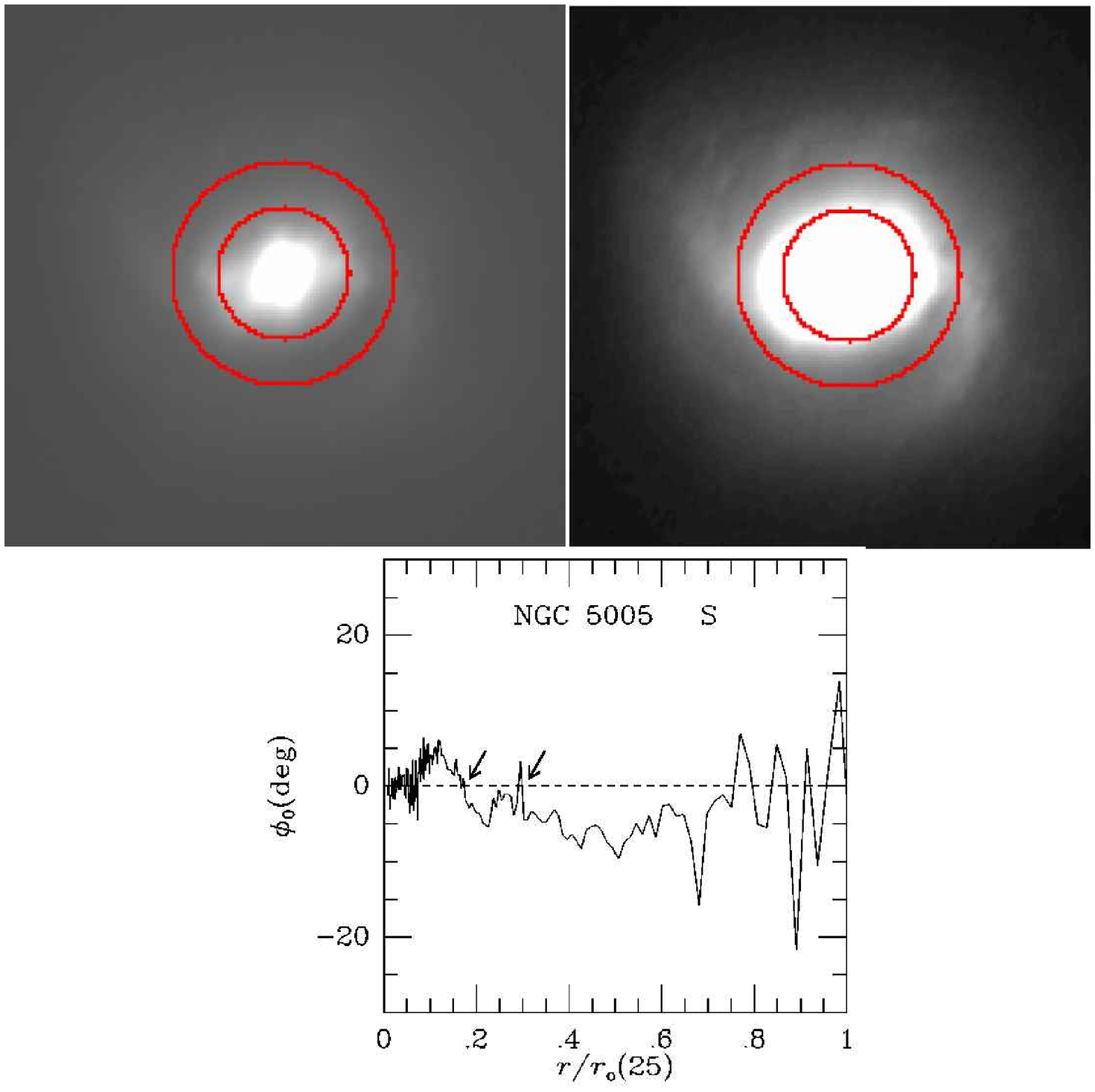}
 \vspace{2.0truecm}                                                             
\caption{Same as Figure 2.1 for NGC 5005}                                         
\label{ngc5005}                                                                 
 \end{figure}

NGC 5005 (Figure~\ref{ngc5005}) - The phase-shift distribution is
based on an average bulge image, owing to a high inclination. The
deprojected image shows an oval with a twisted inner region. The
crossing at $r$=29\arcs\ (0.17) encircles the twisted inner oval,
while that at $r$=52\arcs\ (0.30) lies just outside the ends of the oval/bar,
and from this outer CR two spiral arms are seen emanating.

\begin{figure}                                                                 
\figurenum{2.114}
\plotone{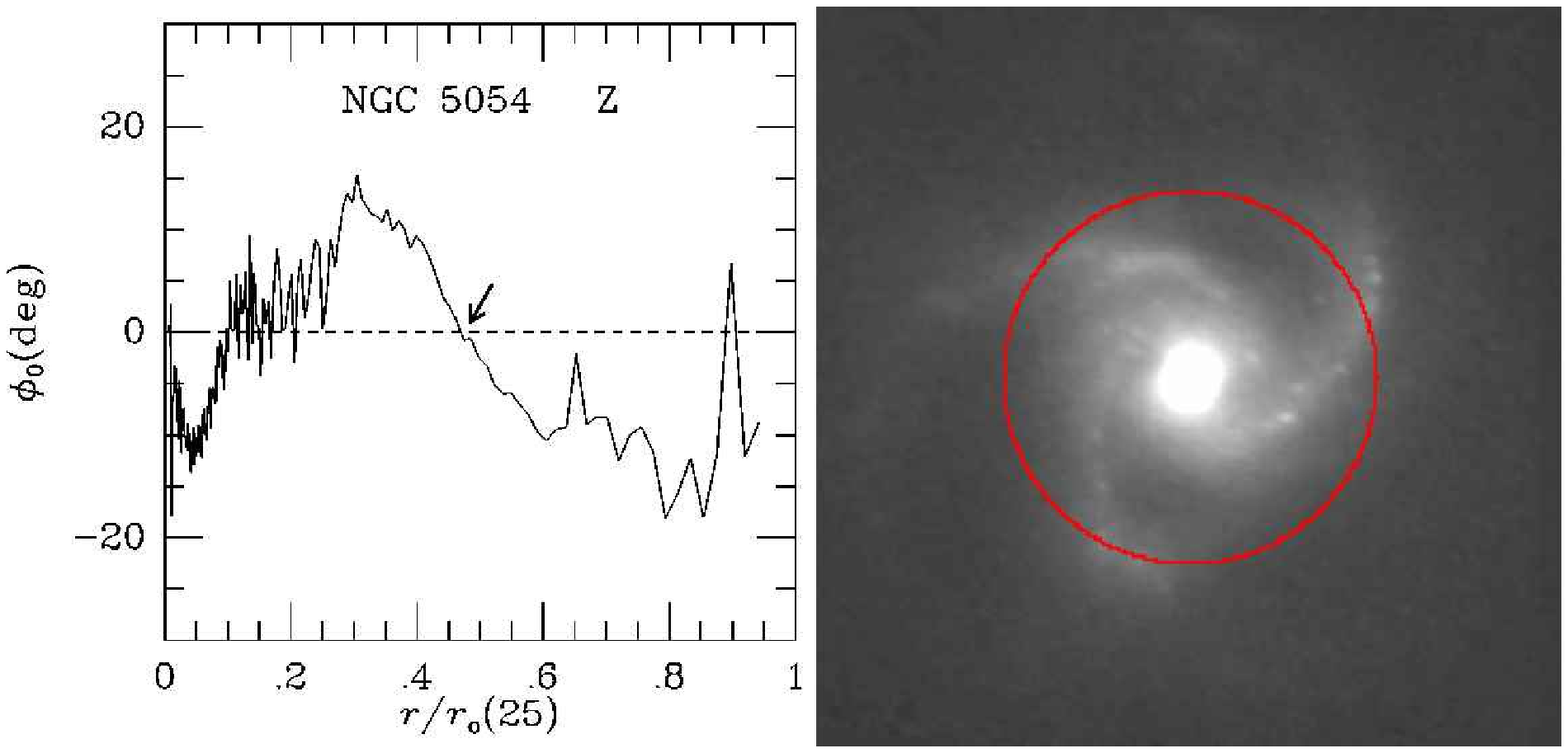}
 \vspace{2.0truecm}                                                             
\caption{Same as Figure 2.1 for NGC 5054}                                         
\label{ngc5054}                                                                 
 \end{figure}

NGC 5054 (Figure~\ref{ngc5054}) - Our analysis is based on an average
bulge image. The apparent oval in the center of
this galaxy could be an artifact of decomposition pinch. There is no
strong P/N crossing in this region, although the phase-shifts curve
of the positive hump indicates
a decoupling process in progress. The main crossing at $r$=74\arcs\ (0.47) lies
in the middle of the main spiral. Figure~\ref{fourier_plots} shows
that this radius lies near the maximum of the $m$=3 relative
Fourier amplitudes of the $H$-band light distribution.  There is
also indication of an unresolved nuclear pattern.

\begin{figure}                                                                 
\figurenum{2.115}
\plotone{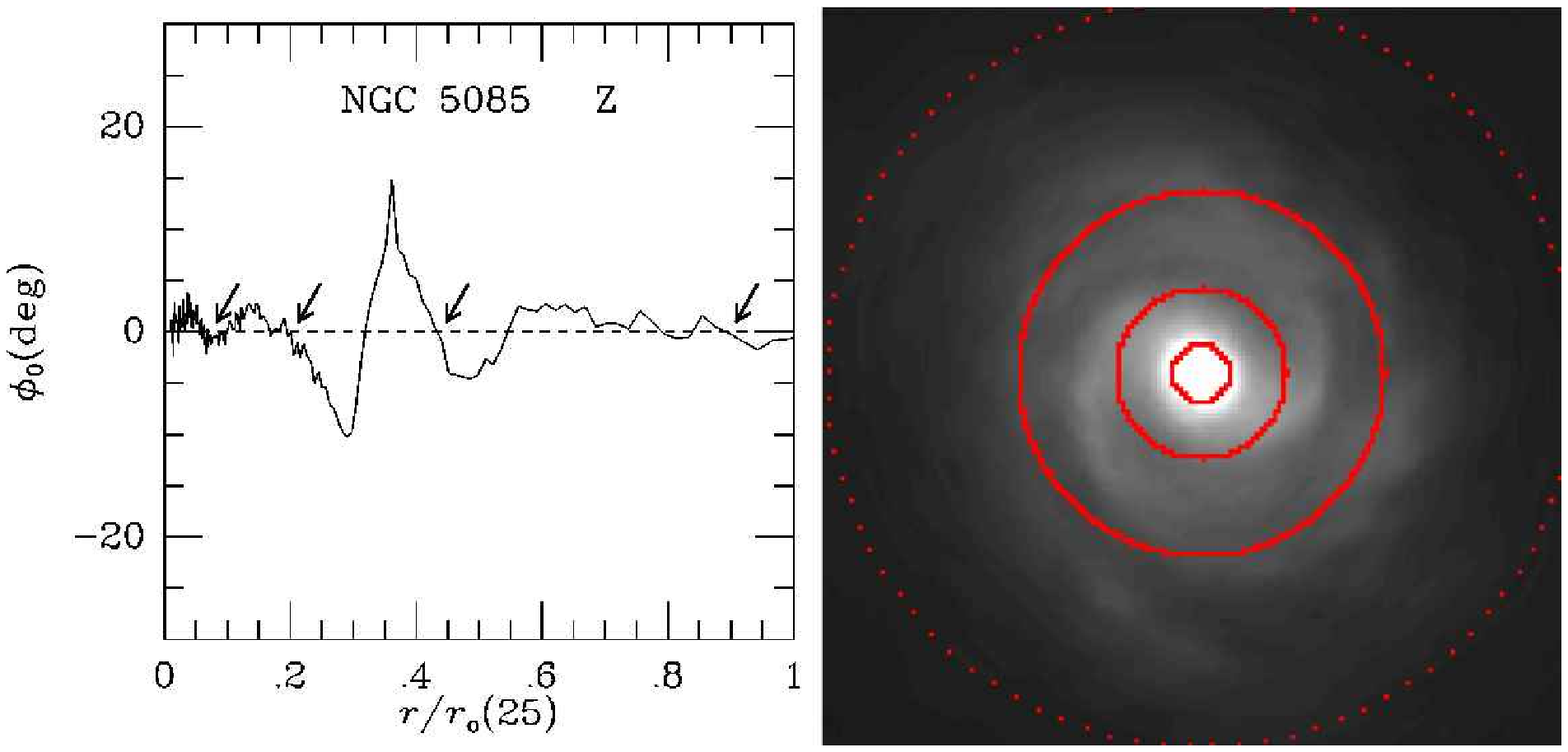}
 \vspace{2.0truecm}                                                             
\caption{Same as Figure 2.1 for NGC 5085}                                         
\label{ngc5085}                                                                 
 \end{figure}

NGC 5085 (Figure~\ref{ngc5085}) - The galaxy shows several
significant-looking crossings. The crossing at $r$=6\arcs\ (0.06) may
be associated with a small inner oval. The crossing at
$r$=22\arcs\ (0.20) seems associated with an oval that joins with the
inner spiral, while the crossing at $r$=47\arcs\ (0.43) may correspond
to a different, outer spiral pattern. The crossing at
$r$=97\arcs\ (0.89) is of uncertain significance.

\begin{figure}                                                                 
\figurenum{2.116}
\plotone{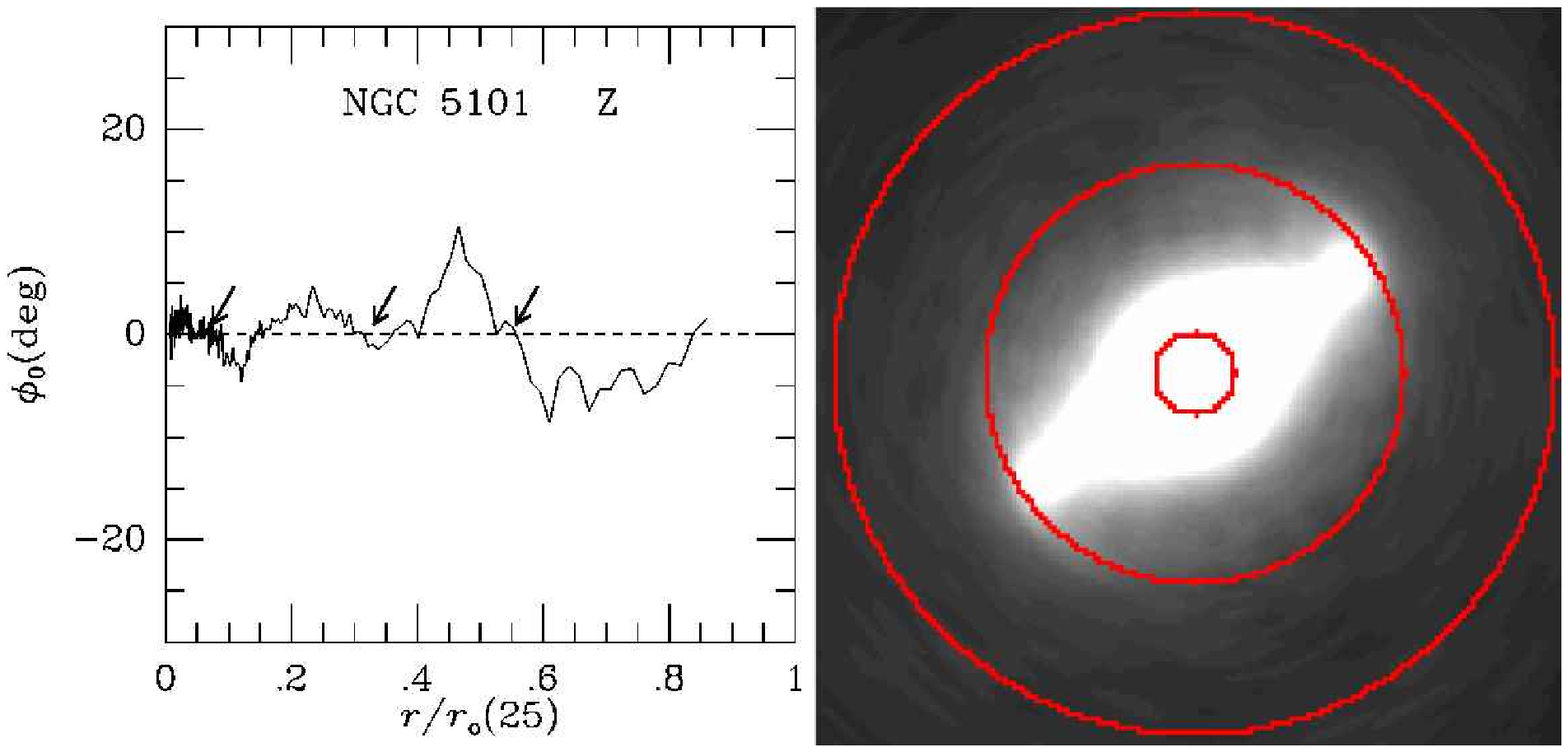}
 \vspace{2.0truecm}                                                             
\caption{Same as Figure 2.1 for NGC 5101}                                         
\label{ngc5101}                                                                 
 \end{figure}

NGC 5101 (Figure~\ref{ngc5101}) - The crossing at $r$=54\arcs\ (0.31)
appears to encompass the ends of the bar. The crossing at
$r$=94\arcs\ (0.54) could be associated with the main outer arms,
while the innermost crossing could refer to a weak inner oval.

\begin{figure}                                                                 
\figurenum{2.117}
\plotone{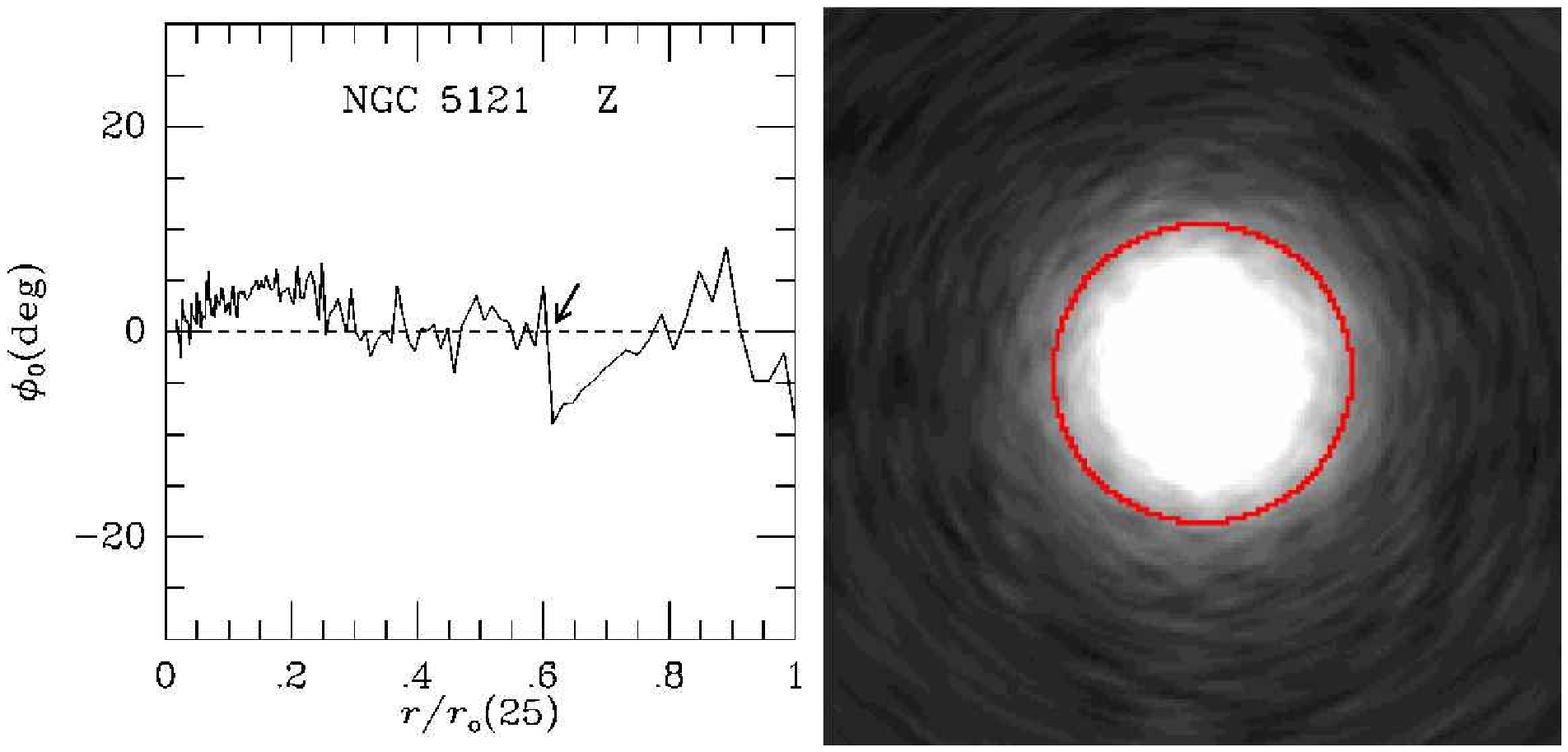}
 \vspace{2.0truecm}                                                             
\caption{Same as Figure 2.1 for NGC 5121}                                         
\label{ngc5121}                                                                 
 \end{figure}

NGC 5121 (Figure~\ref{ngc5121}) - The main crossing at $r$=37\arcs\ (0.60)
could be related to a large, broad oval.

\begin{figure}                                                                 
\figurenum{2.118}
\plotone{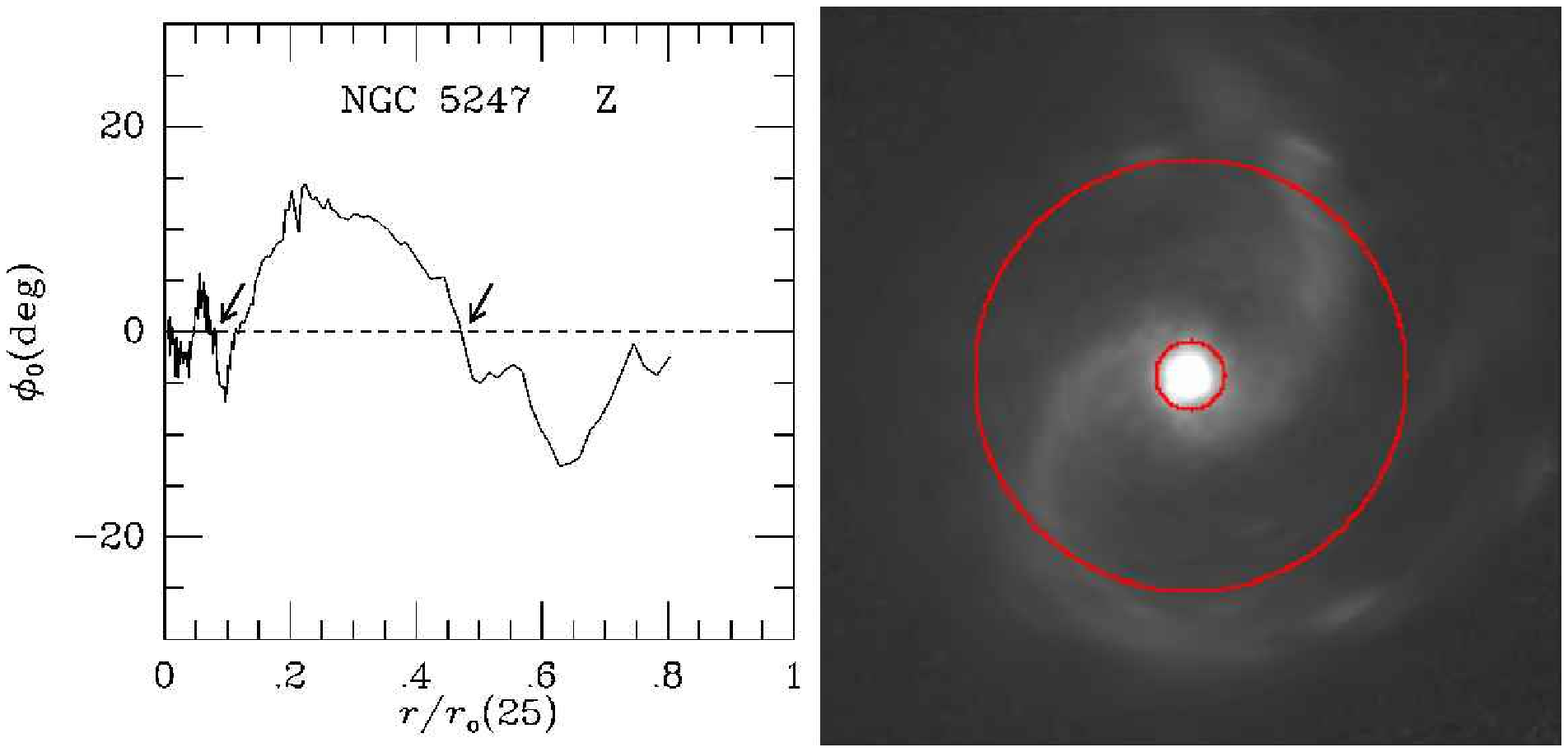}
 \vspace{2.0truecm}                                                             
\caption{Same as Figure 2.1 for NGC 5247}                                         
\label{ngc5247}                                                                 
 \end{figure}

NGC 5247 (Figure~\ref{ngc5247}) - The main crossing at $r$=87\arcs\ (0.47)
lies in the middle of the bright spiral pattern. Figure~\ref{fourier_plots} 
shows that this radius lies near the maximum of the $m$=2 relative
Fourier amplitudes of the $H$-band light distribution. The innermost 
crossing appears associated with a small inner oval. There is also the
indication of an unresolved nuclear pattern. 

\begin{figure}                                                                 
\figurenum{2.119}
\plotone{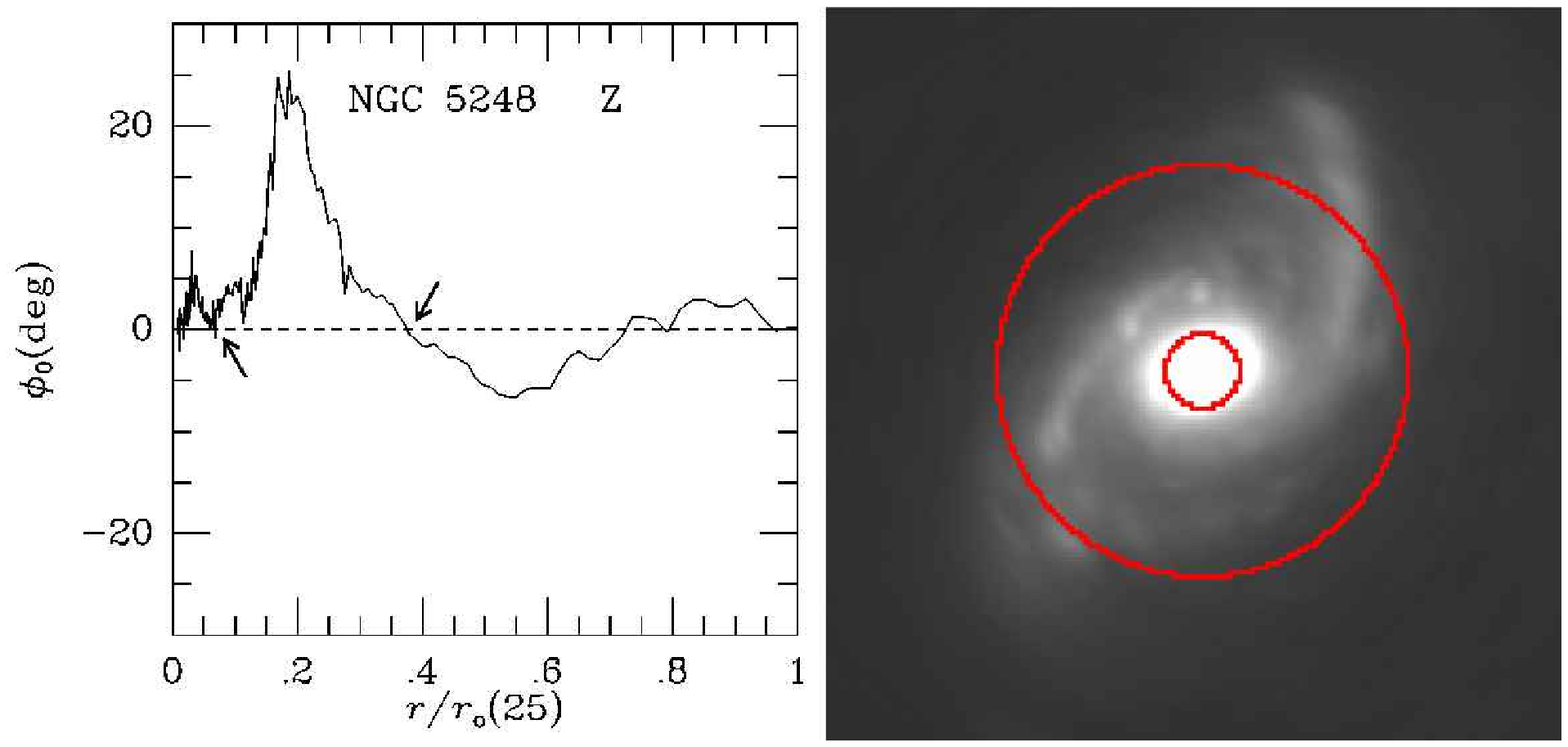}
 \vspace{2.0truecm}                                                             
\caption{Same as Figure 2.1 for NGC 5248}                                         
\label{ngc5248}                                                                 
 \end{figure}

NGC 5248 (Figure~\ref{ngc5248}) - The main crossing at $r$=70\arcs\ (0.38)
lies in the middle of the main inner spiral, while the crossing at $r$=12\arcs\ 
(0.06) encompasses a small inner oval where the well-known nuclear ring of
this galaxy is located. The two weak outer crossings seen in the phase-shift
plot are close to $r_o$ but may be associated with the faint outer pseudoring
of the galaxy. Figure~\ref{fourier_plots} shows
that the main crossing lies near the maximum of the $m$=2 relative
Fourier amplitudes of the $H$-band light distribution.

\begin{figure}                                                                 
\figurenum{2.120}
\plotone{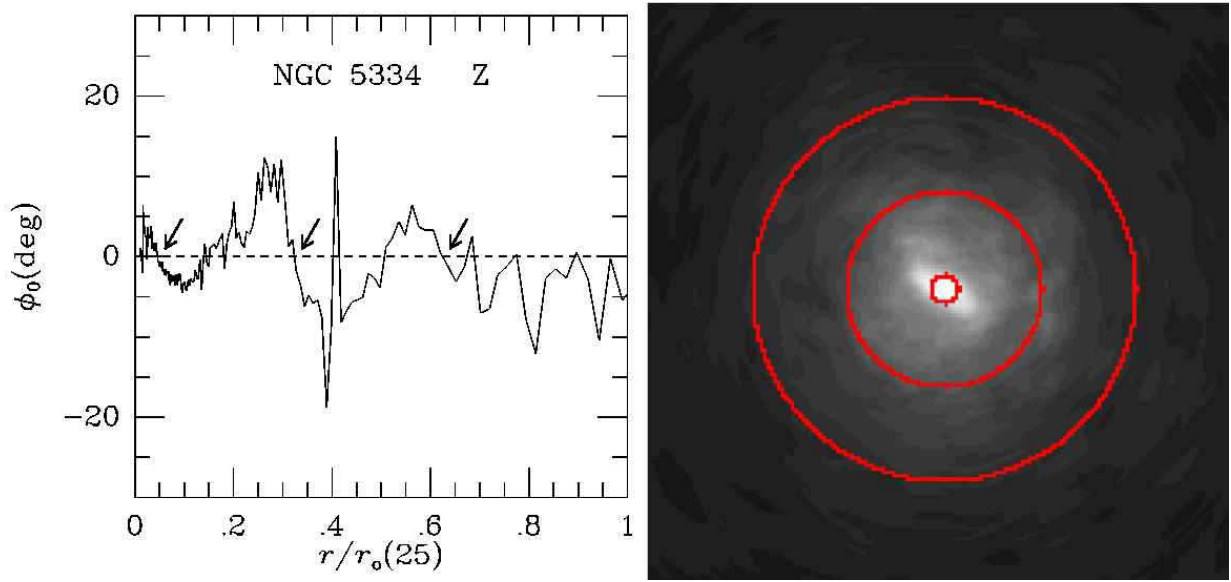}
 \vspace{2.0truecm}                                                             
\caption{Same as Figure 2.1 for NGC 5334}                                         
\label{ngc5334}                                                                 
 \end{figure}

NGC  5334 (Figure~\ref{ngc5334}) - The innermost crossing lies well
inside the bar ends. The crossing at $r$=42\arcs\ (0.32) circles far
outside the bar ends, in the main spiral pattern. The third crossing
may be connected with outer spiral arms. Although we have interpreted
the $r$=42\arcs\ crossing as the bar CR in Table 1, the negative
phase-shifts inside this crossing suggest that the bar may belong to
the innermost mode. There in indication from the phase-shift plot that
the bar is in the process of decoupling. The spike at $r/r_o$=0.4 is of
uncertain significance.

\begin{figure}                                                                 
\figurenum{2.121}
\plotone{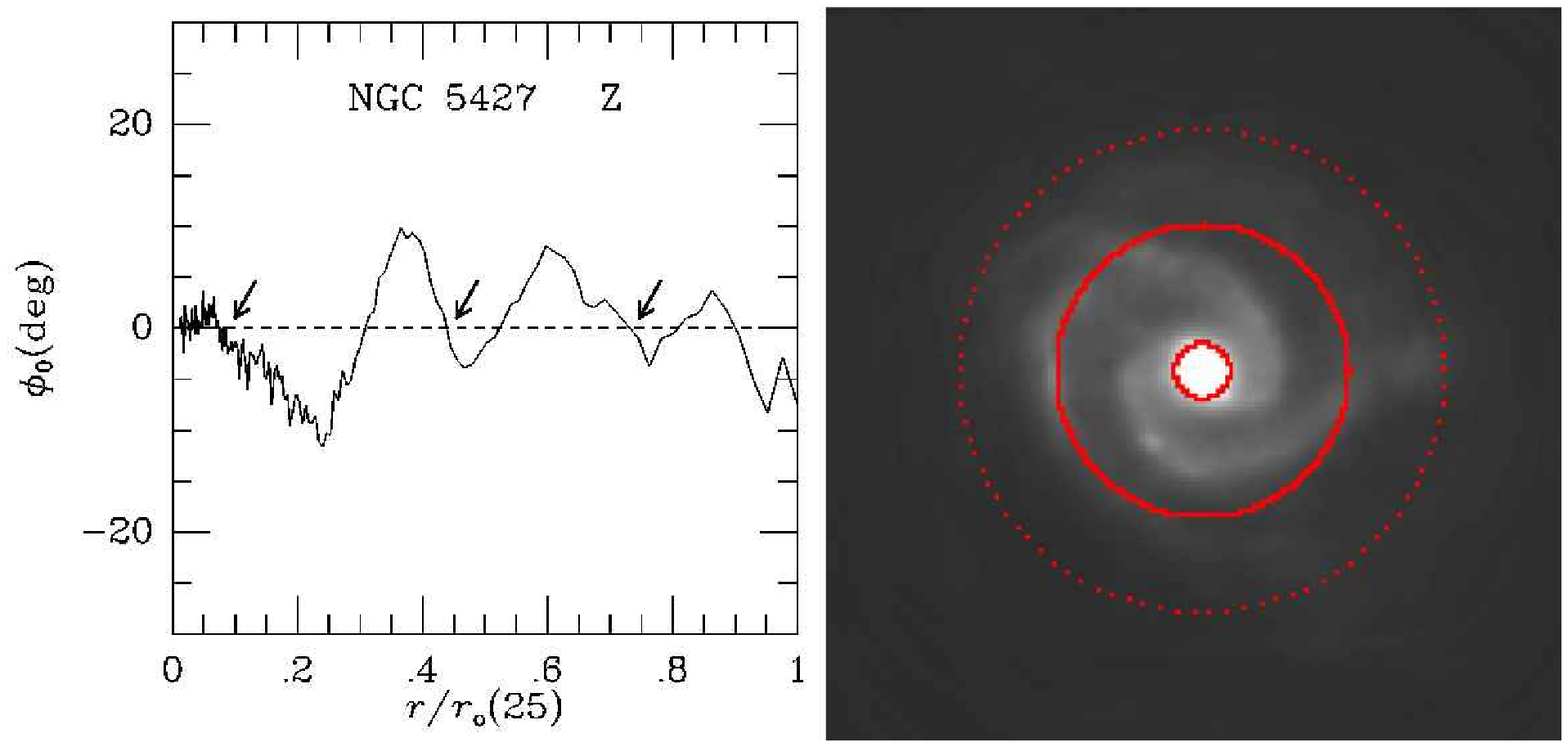}
 \vspace{2.0truecm}                                                             
\caption{Same as Figure 2.1 for NGC 5427}                                         
\label{ngc5427}                                                                 
 \end{figure}

NGC 5427 (Figure~\ref{ngc5427}) - A grand-design spiral with several
P/N crossings. The innermost crossing appears to be connected 
with a small inner oval. The next crossing, at $r$=38\arcs\ (0.44), lies in the
middle of the main spiral pattern. The outermost crossing at $r$=63\arcs\
(0.73) could be associated with faint outer spiral structure.

\begin{figure}                                                                 
\figurenum{2.122}
\plotone{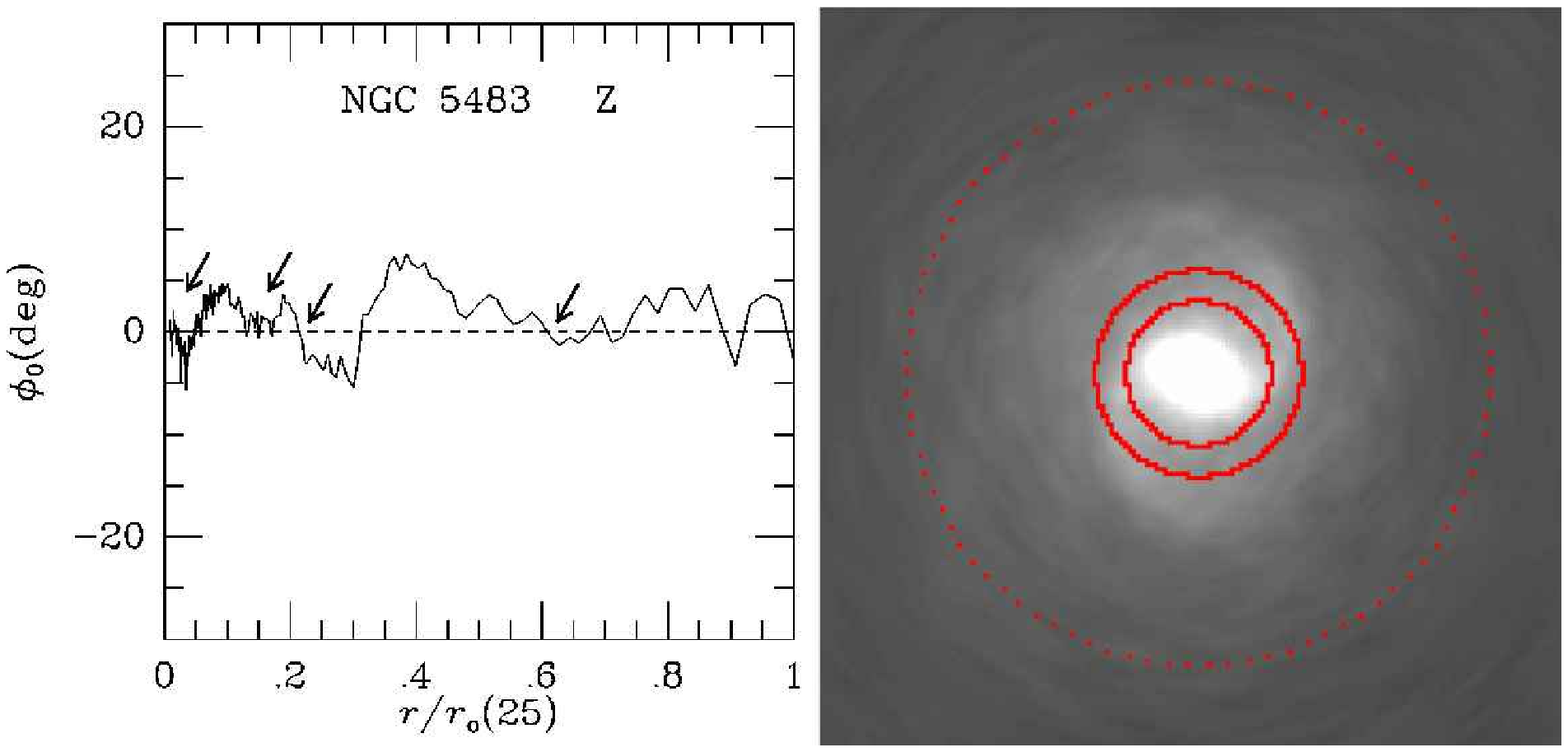}
 \vspace{2.0truecm}                                                             
\caption{Same as Figure 2.1 for NGC 5483.                                         
CR$_1$ from Table 1                                                             
is not overlaid on the image.}                                                  
\label{ngc5483}                                                                 
 \end{figure}

NGC 5483 (Figure~\ref{ngc5483}) - The weak crossing at $r$=18\arcs\ (0.15)
appears
to encircle the ends of an oval bar feature. This is followed closely by
a better-defined crossing at $r$=26\arcs\ (0.22) which circles through the main
inner arms. The phase-shift distribution suggests that the spiral and
the oval are still in the process of decoupling. The weak outer crossing 
at $r$=73\arcs\ (0.61) may indicate a non-steady pattern in this region.
There is also the indication of an unresolved nuclear structure.

\begin{figure}                                                                 
\figurenum{2.123}
\plotone{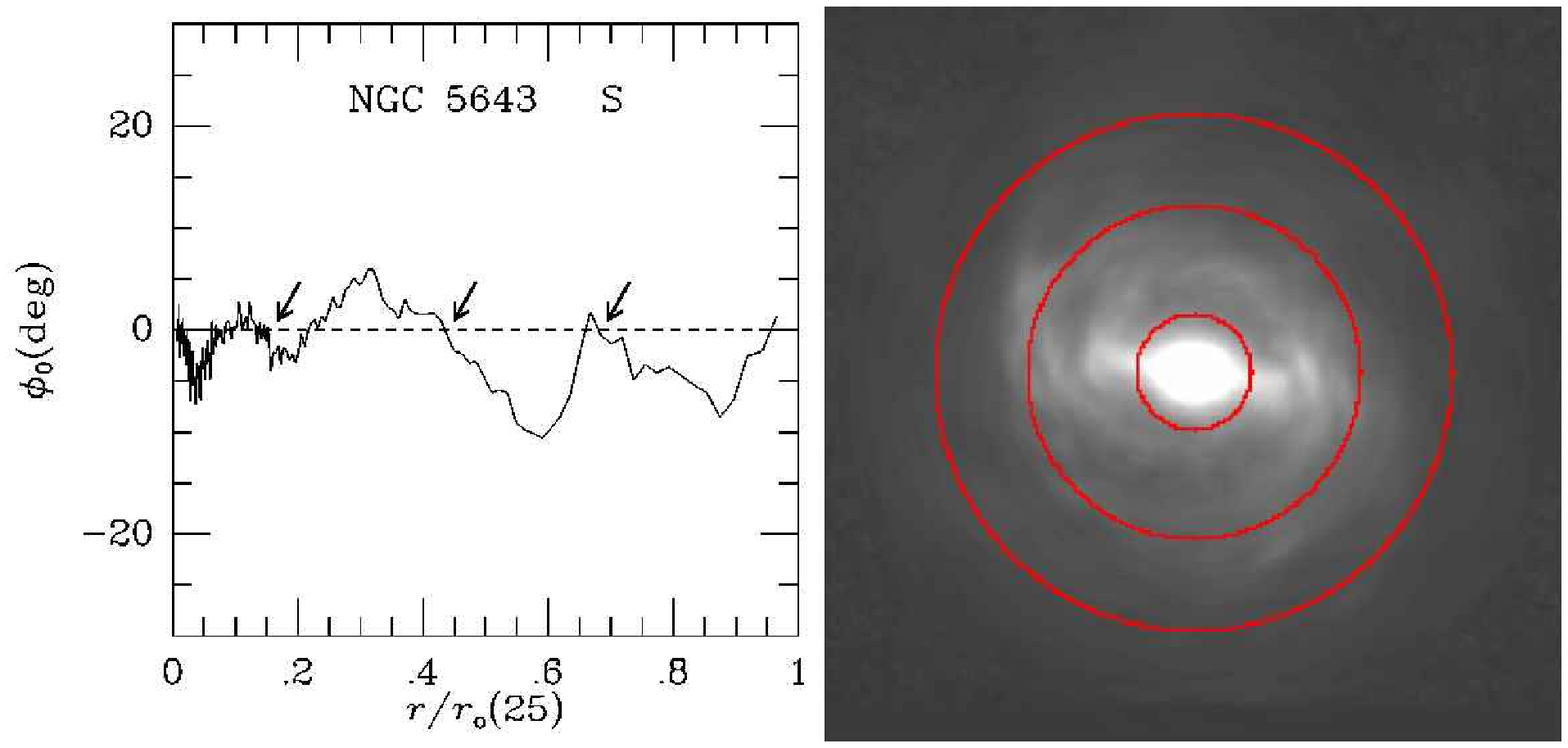}
 \vspace{2.0truecm}                                                             
\caption{Same as Figure 2.1 for NGC 5643}                                         
\label{ngc5643}                                                                 
 \end{figure}

NGC 5643 (Figure~\ref{ngc5643}) - The crossing at $r$=24\arcs\ (0.16)
circles the inner broad oval zone of the prominent bar. The next
crossing, at $r$=67\arcs\ (0.44), is very well-defined, and circles far
beyond the ends of the rest of the bar. This intermediate crossing
could be the CR of the broad inner bar/spiral feature, while the
crossing at $r$=105\arcs\ (0.68) could be associated with the outer
spiral pattern. On the other hand, the inner portion of the bar could
be corotating with the inner oval, since the features are aligned, and
there is a ring-like feature near the N/P crossing indicating possible
shearing between two decoupled patterns at this radius. We consider
this second scenario more likely, making NGC 5643 very much like NGC
4902. There is also the indication of an unresolved nuclear structure.

\begin{figure}                                                                 
\figurenum{2.124}
\plotone{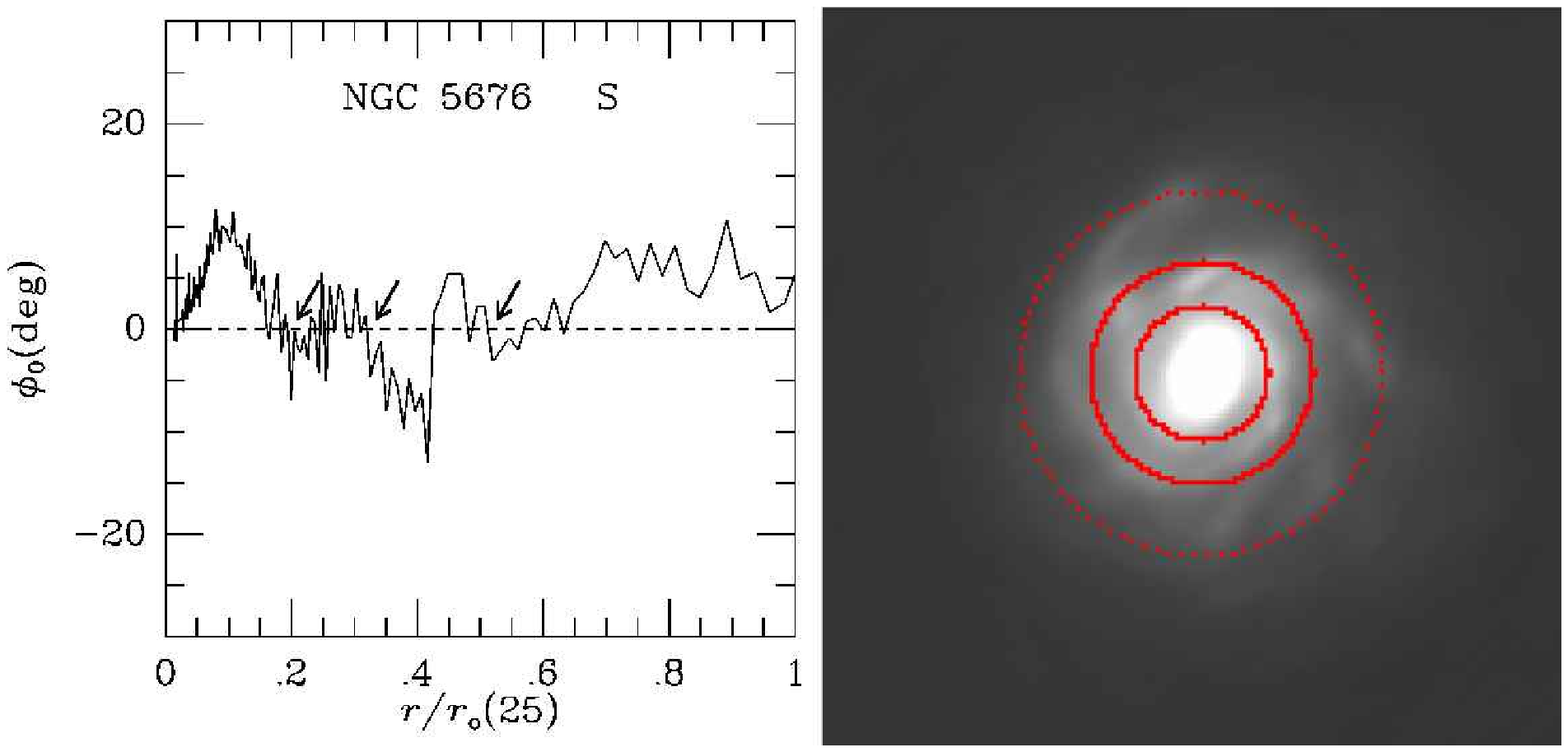}
 \vspace{2.0truecm}                                                             
\caption{Same as Figure 2.1 for NGC 5676}                                         
\label{ngc5676}                                                                 
 \end{figure}

NGC 5676 (Figure~\ref{ngc5676}) - The crossing at $r$=23\arcs\ (0.19) circles
the ends of a prominent oval. The other crossings are connected to the
spiral arms which in optical images extend well beyond the outer circle.

\begin{figure}                                                                 
\figurenum{2.125}
\plotone{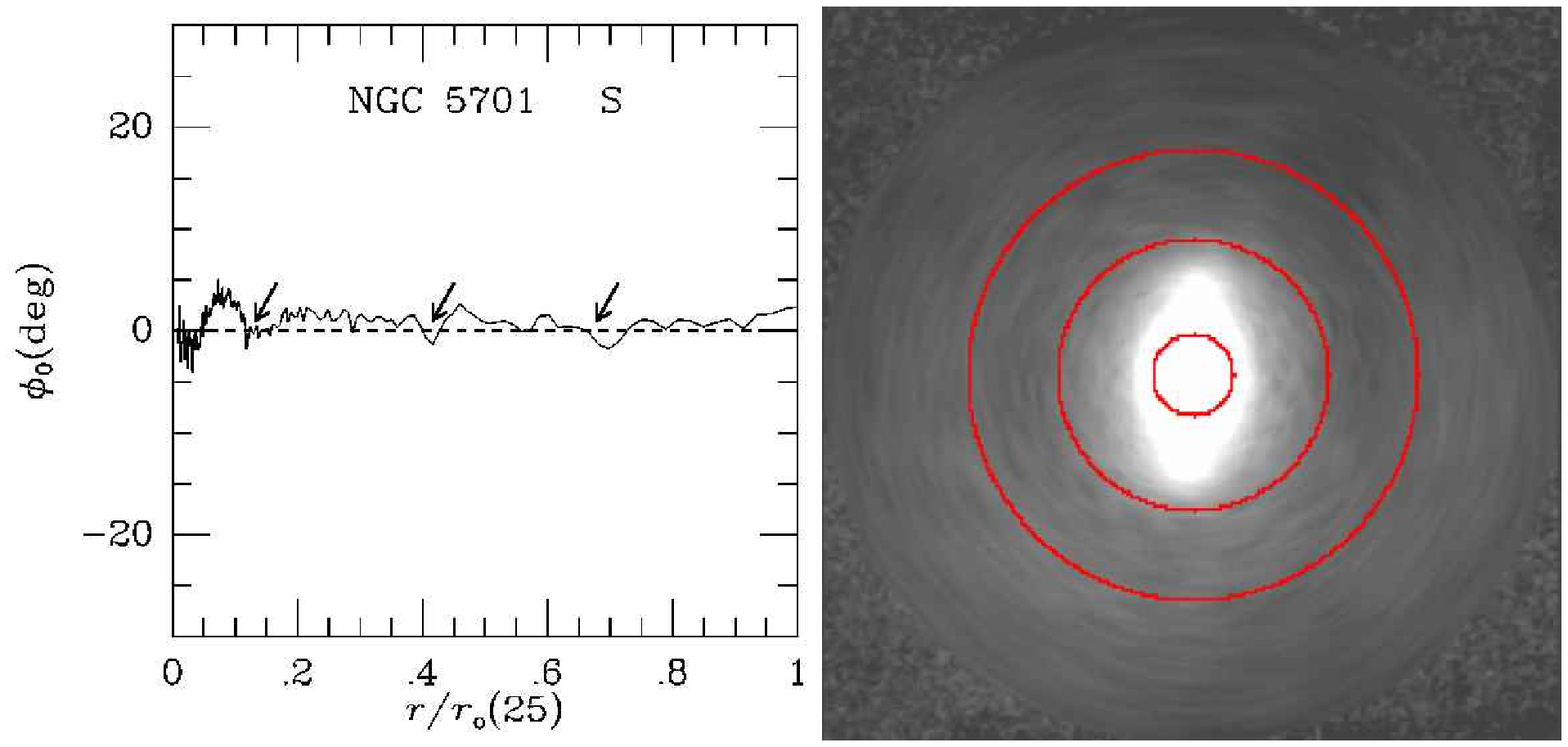}
 \vspace{2.0truecm}                                                             
\caption{Same as Figure 2.1 for NGC 5701}                                         
\label{ngc5701}                                                                 
 \end{figure}

NGC 5701 (Figure~\ref{ngc5701}) - The phase-shift values are small in this
galaxy because the density wave features are only weakly skewed. 
The one at $r$=52\arcs\ (0.40) appears to circle closely around the bar
ends, while that at $r$=87\arcs\ (0.66) could be associated with the outer
spiral arms, which show an R$_1$R$_2^{\prime}$ morphology in optical
images. This structure is barely detectable in the near-IR OSUBGS image. 
There is also an indication of an unresolved nuclear structure.

An RSL08 model of this galaxy places CR at 64\arcss 6 (0.49), intermediate
between our two outer CR values.

\begin{figure}                                                                 
\figurenum{2.126}
\plotone{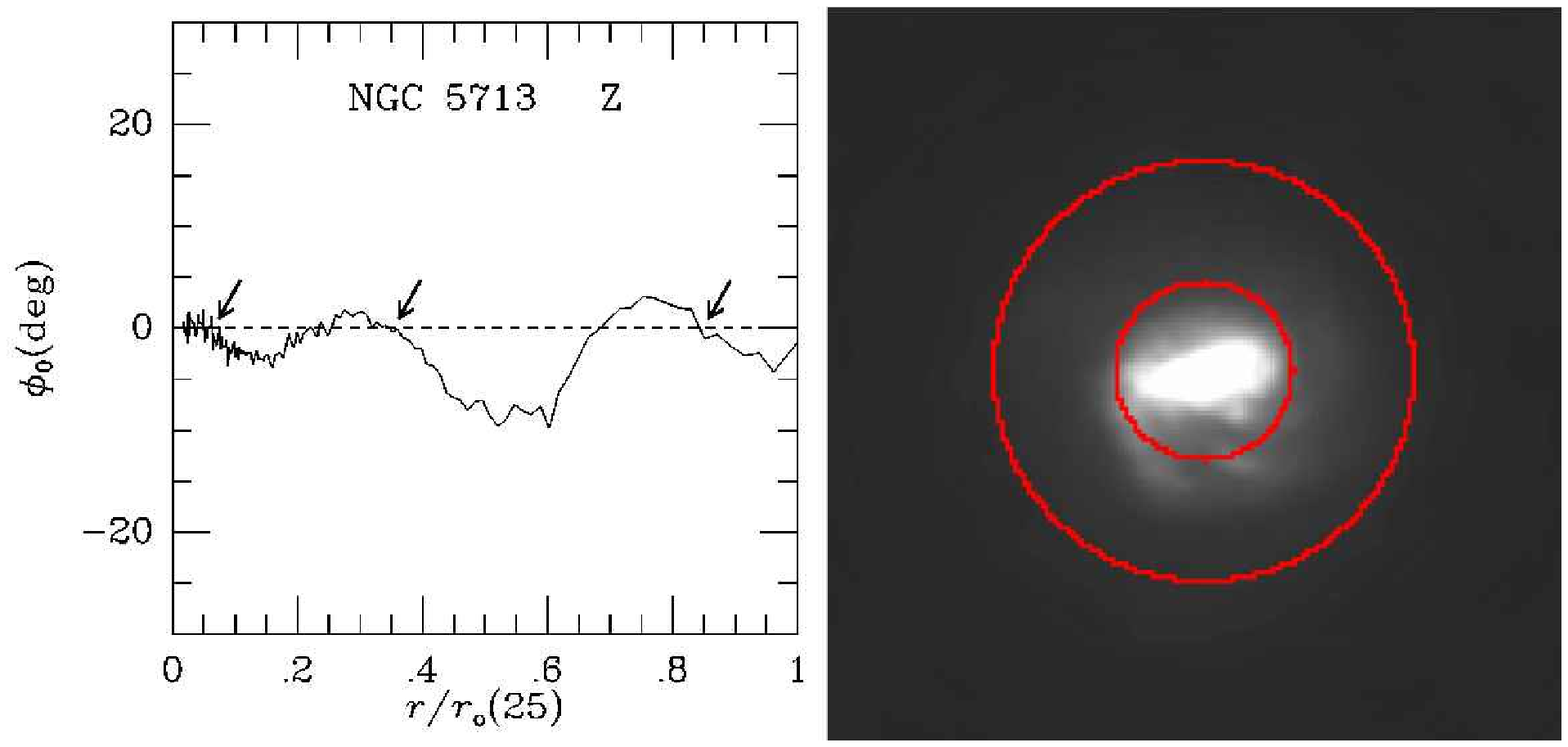}
 \vspace{2.0truecm}                                                             
\caption{Same as Figure 2.1 for NGC 5713}                                         
\label{ngc5713}                                                                 
 \end{figure}

NGC 5713 (Figure~\ref{ngc5713}) - The crossing at $r$=29\arcs\ (0.34)
circles the ends of the bar. The crossing at $r$=71\arcs\ (0.84) may be
associated with the outer arms, which are peculiar in this object.
There is also indication of an unresolved nuclear structure.  Although
we have interpreted the $r$=29\arcs\ crossing as the bar CR in Table 1,
the negative phase-shifts inside this crossing suggest that all or most
of the bar may belong to this innermost mode.

\begin{figure}                                                                 
\figurenum{2.127}
\plotone{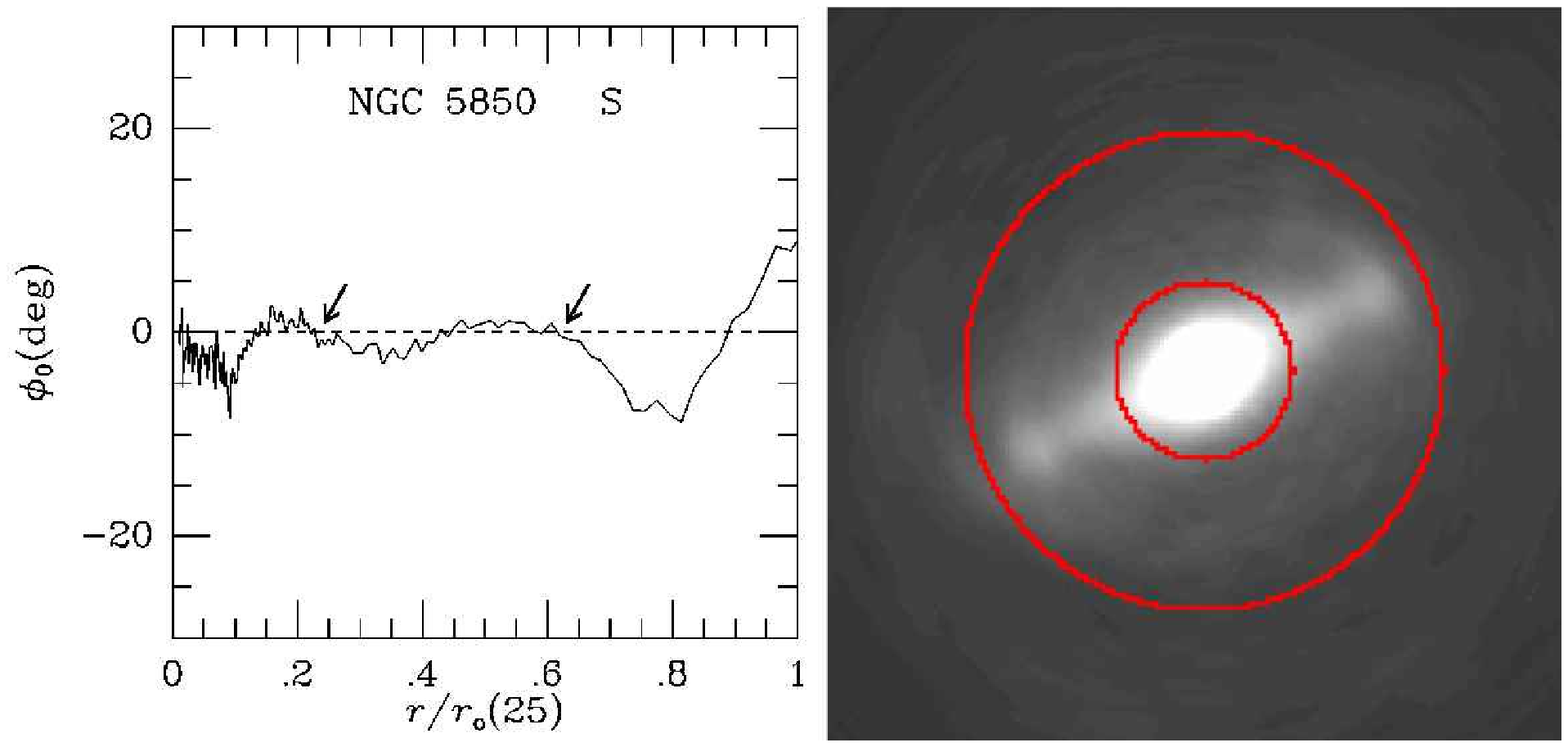}
 \vspace{2.0truecm}                                                             
\caption{Same as Figure 2.1 for NGC 5850}                                         
\label{ngc5850}                                                                 
 \end{figure}

NGC 5850 (Figure~\ref{ngc5850}) - This galaxy is interesting in that it
has a weak N/P crossing that passes almost exactly through a minimum in
the bar profile. The bar has ansae, but these ansae lie outside the N/P
crossing in an area with positive phase-shift. If we take our ``rules"
literally, that modes should come with both positive and negative
phase-differences, with the latter following the former in radius, this
means everything inside the bar minimum belongs to the inner mode,
while the ansae belong to the outer mode. The main crossing at
$r$=81\arcs\ (0.62) circles just beyond the ansae. The inner crossing
at $r$=30\arcs\ (0.23) encircles a broad inner oval and doesn't seem to
be related to the galaxy's well-known nuclear bar which only shows a
hint of decoupling in the phase-shift plot.

An interesting RSL08 model of this galaxy places CR at 105\arcss 1 (0.80).
Our main CR value is close to the RSL08 bar radius of 75\arcss 8 (0.58).

\begin{figure}                                                                 
\figurenum{2.128}
\plotone{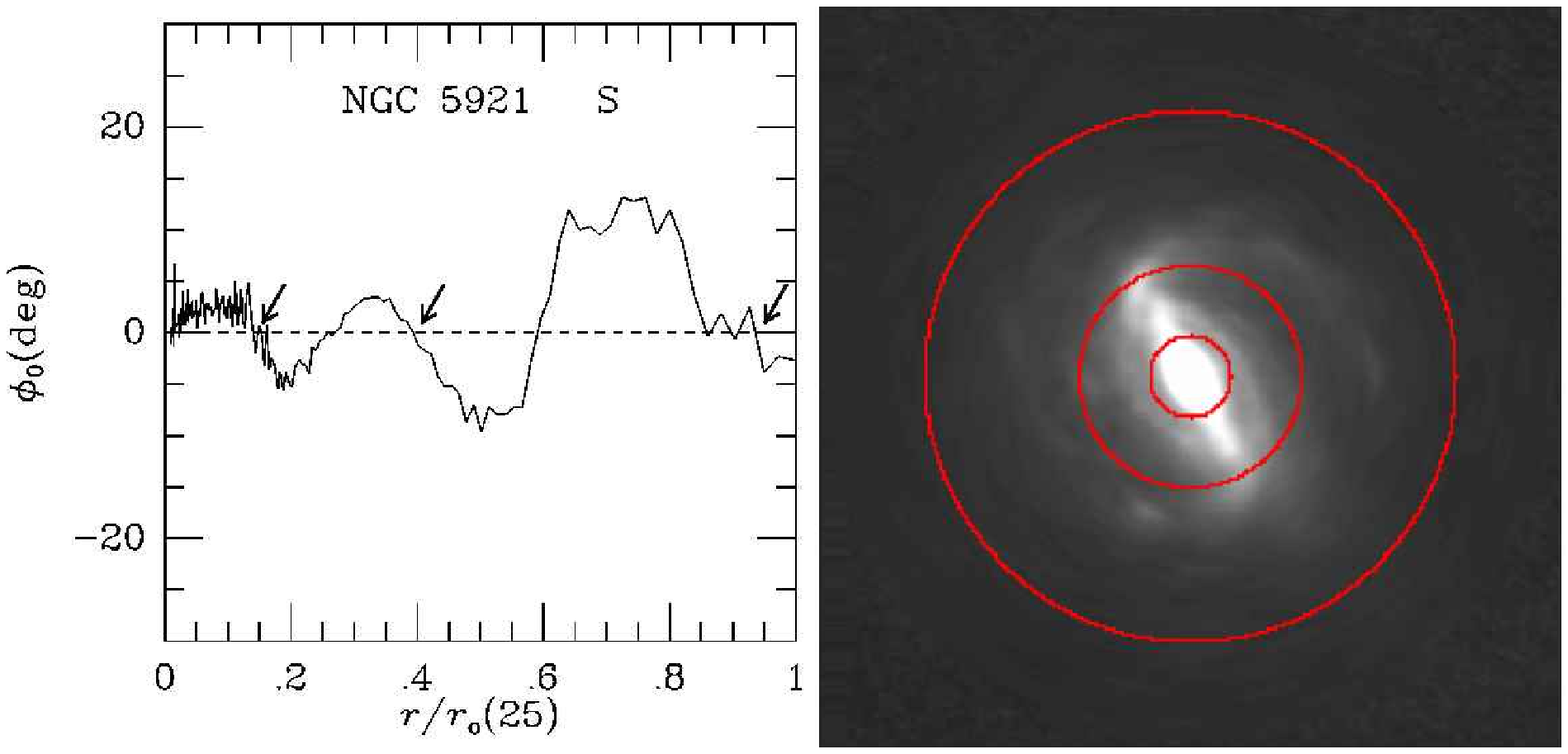}
 \vspace{2.0truecm}                                                             
\caption{Same as Figure 2.1 for NGC 5921}                                         
\label{ngc5921}                                                                 
 \end{figure}

NGC 5921 (Figure~\ref{ngc5921}) - The crossing at $r$=58\arcs\ (0.39)
circles just outside the bar ends, while the one at $r$=21\arcs\ (0.14)
circles an inner oval aligned with the main bar. The crossing at
$r$=137\arcs\ (0.93) may be associated with the outer spiral arms.  The
N/P crossing at $r$=39\arcs\ (0.27) lies just inside the bar ends.
Although we have selected the $r$=58\arcs\ crossing as the bar CR in
Table 1, the substantial negative phase-shifts in the bar region suggests
that much of the bar may belong to the inner mode.

The RSL08 model of this galaxy places CR at 71\arcss 4 (0.49), well
outside our bar CR value. Instead, our value corresponds with their bar
radius of 57\arcss 0 (0.39).

\begin{figure}                                                                 
\figurenum{2.129}
\plotone{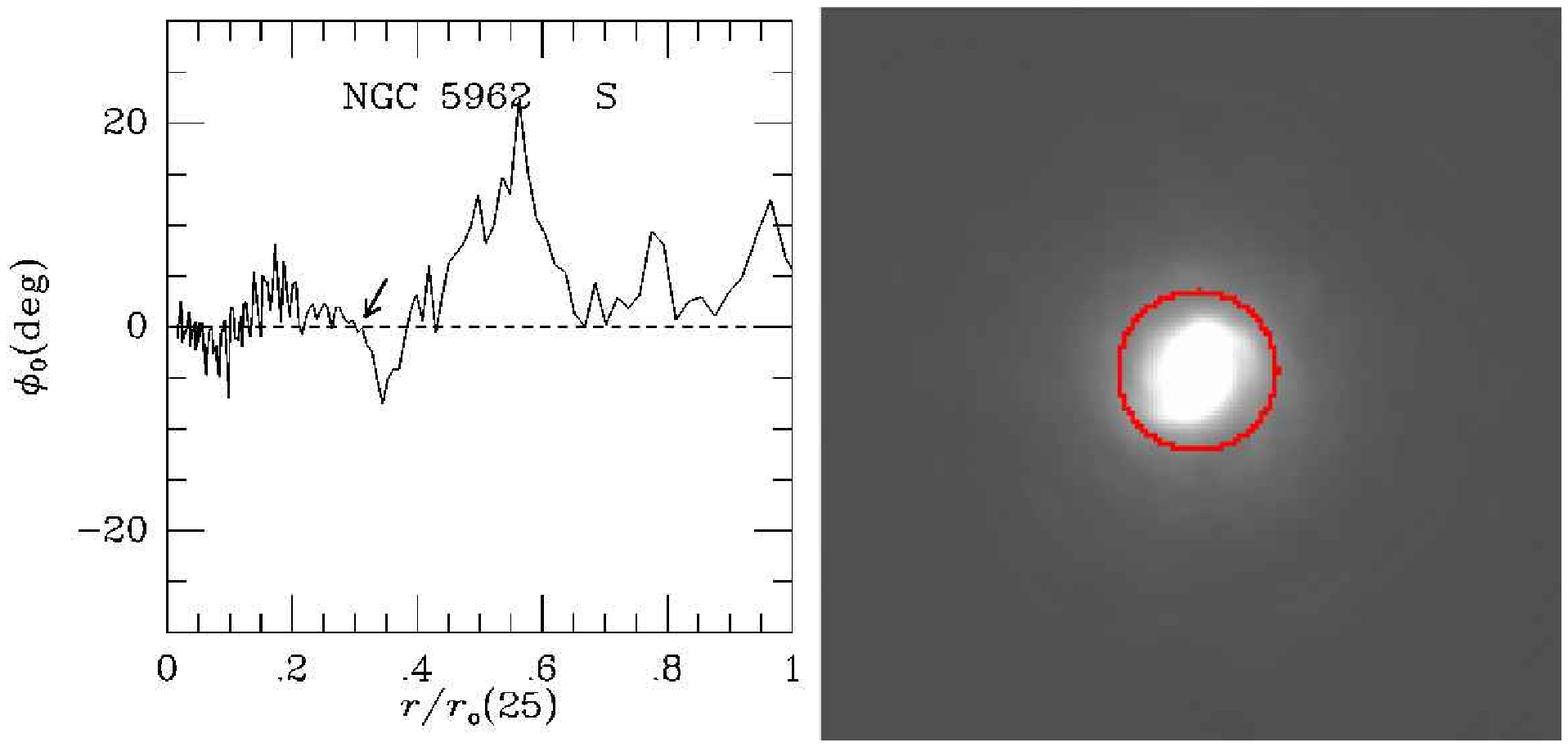}
 \vspace{2.0truecm}                                                             
\caption{Same as Figure 2.1 for NGC 5962}                                         
\label{ngc5962}                                                                 
 \end{figure}

NGC 5962 (Figure~\ref{ngc5962}) - The main crossing at $r$=27\arcs\ (0.30) 
circles just outside the ends of a broad inner oval.  There is indication
of an unresolved inner pattern.

\begin{figure}                                                                 
\figurenum{2.130}
\plotone{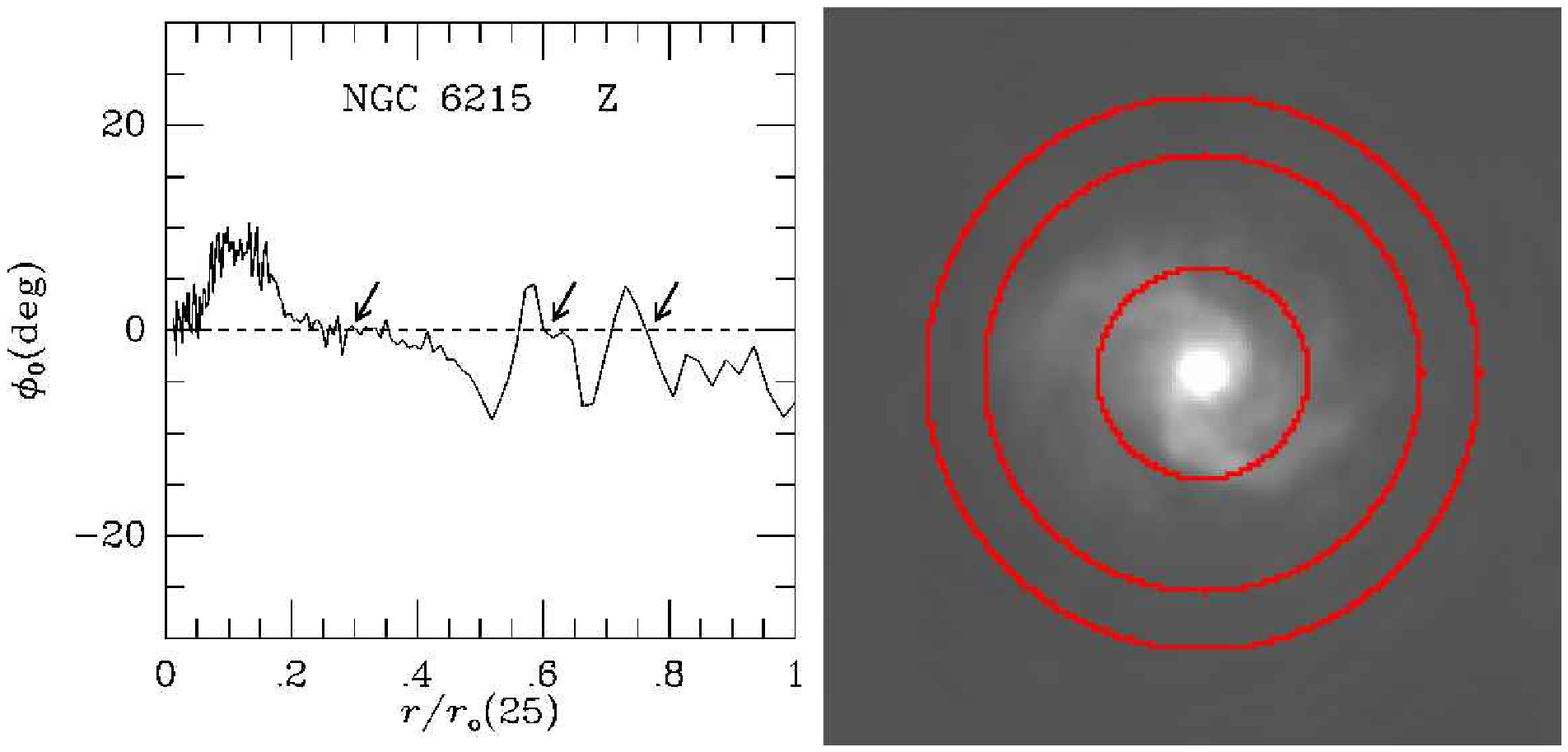}
 \vspace{2.0truecm}                                                             
\caption{Same as Figure 2.1 for NGC 6215}                                         
\label{ngc6215}                                                                 
 \end{figure}

NGC 6215 (Figure~\ref{ngc6215}) - There is a small inner oval in this
galaxy, but there is no P/N crossing associated with it apart from weak
indication of a potential decoupling. Instead, the
three main crossings are associated with the complex spiral arms.

\begin{figure}                                                                 
\figurenum{2.131}
\plotone{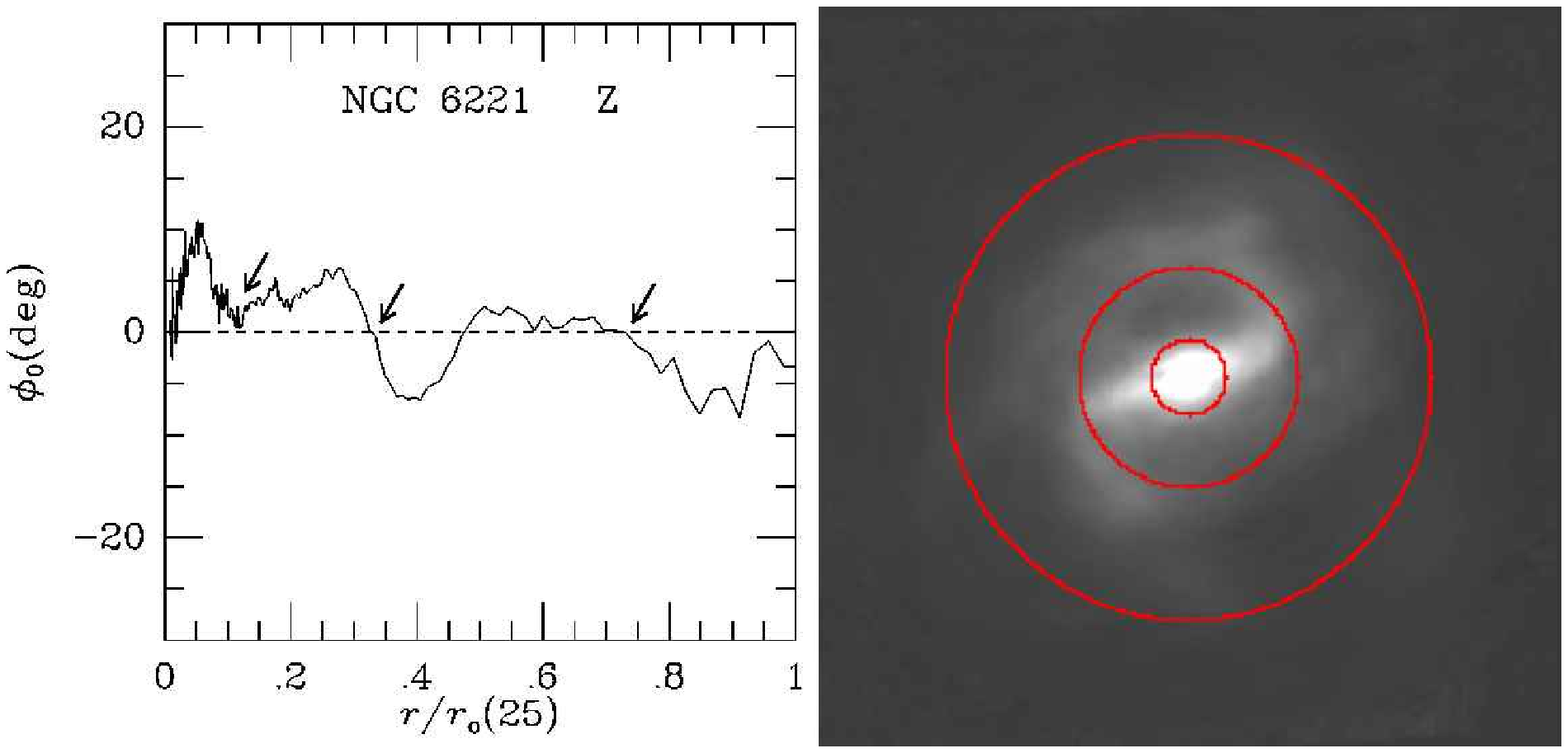}
 \vspace{2.0truecm}                                                             
\caption{Same as Figure 2.1 for NGC 6221}                                         
\label{ngc6221}                                                                 
 \end{figure}

NGC 6221 (Figure~\ref{ngc6221}) - The crossing at $r$=14\arcss 5 (0.11)
is connected to a small inner oval which is barely decoupled. 
The one at $r$=42\arcs\ (0.33) circles the
ends of the bar. The largest crossing at $r$=93\arcs\ (0.73) is probably
associated with the outer arms. In optical light this galaxy is
distorted and seems clearly interacting.

\begin{figure}                                                                 
\figurenum{2.132}
\plotone{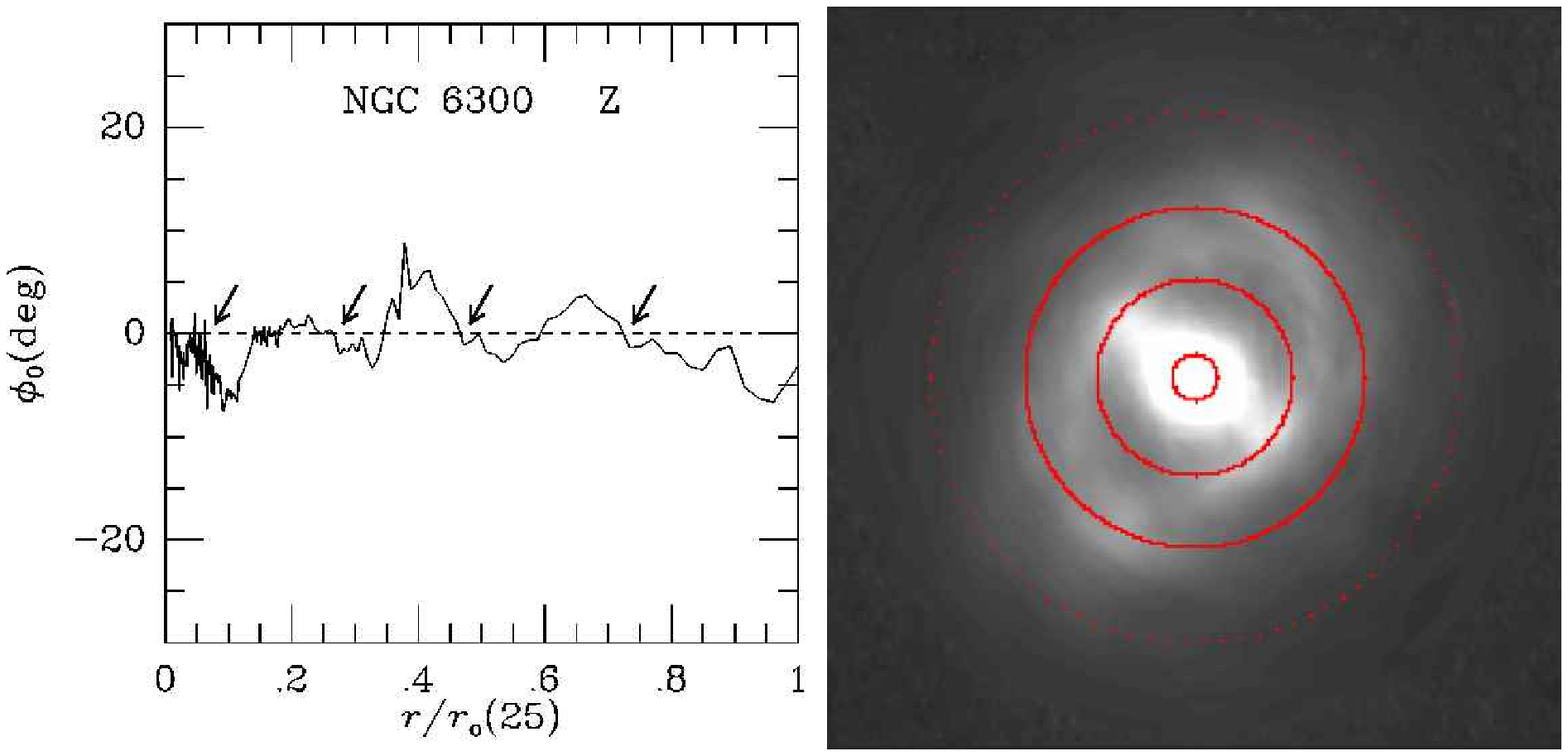}
 \vspace{2.0truecm}                                                             
\caption{Same as Figure 2.1 for NGC 6300}                                         
\label{ngc6300}                                                                 
 \end{figure}

NGC 6300 (Figure~\ref{ngc6300}) - This is a galaxy with clear multiple
patterns. The crossing at $r$=39\arcs\ (0.27) appears to encircle the ends
of the bar, which spill only a little outside. The crossing at $r$=69\arcs\
(0.47) circles just inside the ends of the prominent oval inner pseudoring. 
The innermost crossing could be associated with a small inner oval.

\begin{figure}                                                                 
\figurenum{2.133}
\plotone{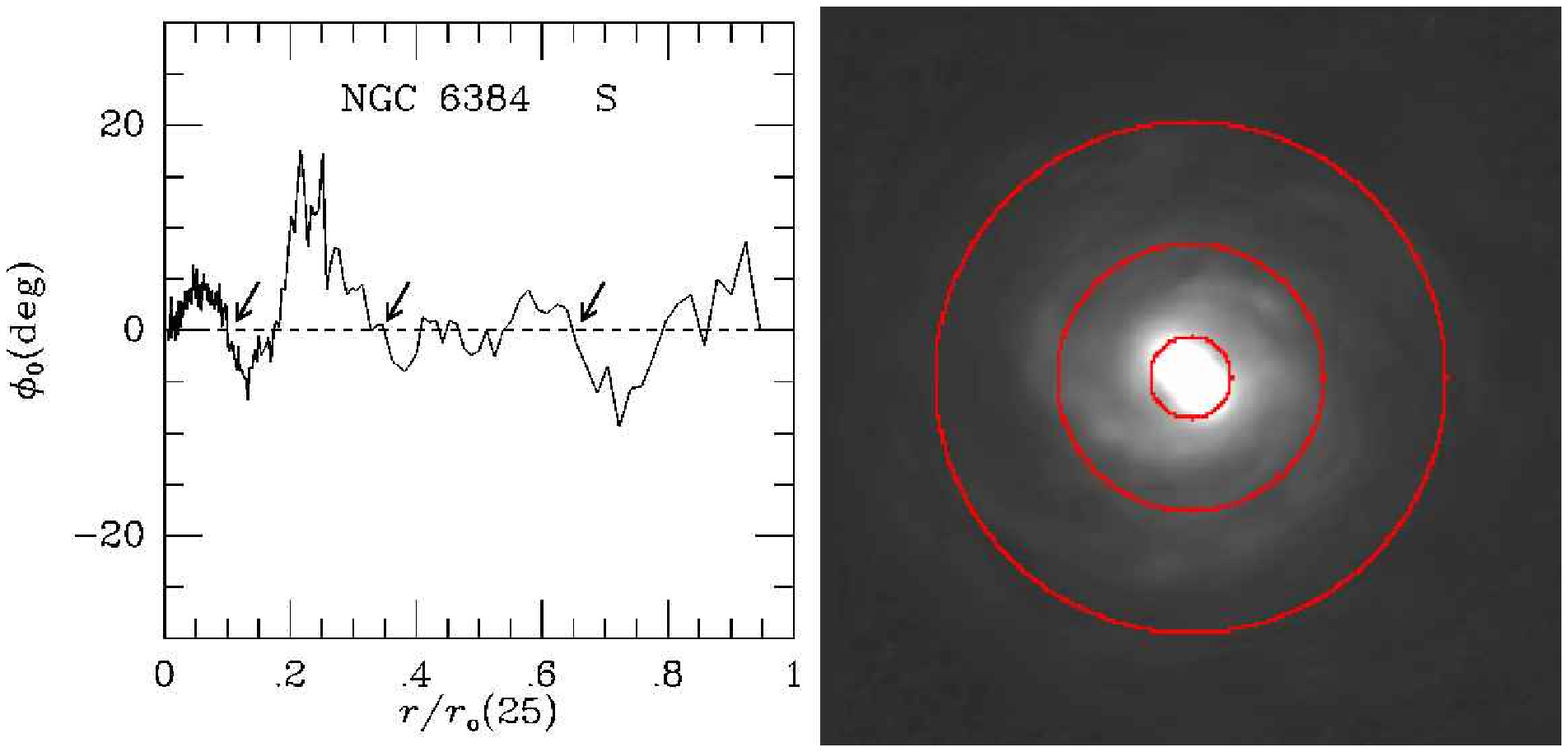}
 \vspace{2.0truecm}                                                             
\caption{Same as Figure 2.1 for NGC 6384}                                         
\label{ngc6384}                                                                 
 \end{figure}

NGC 6384 (Figure~\ref{ngc6384}) - The crossing at $r$=20\arcs\ (0.10) circles
just inside the ends of a prominent bar/oval. The other two crossings
are connected with the spiral arms. 

The best-fitting RSL08 model for this galaxy places CR at 72\arcss 5 (0.36),
very close to our second CR at 69\arcs\ (0.34).

\begin{figure}                                                                 
\figurenum{2.134}
\plotone{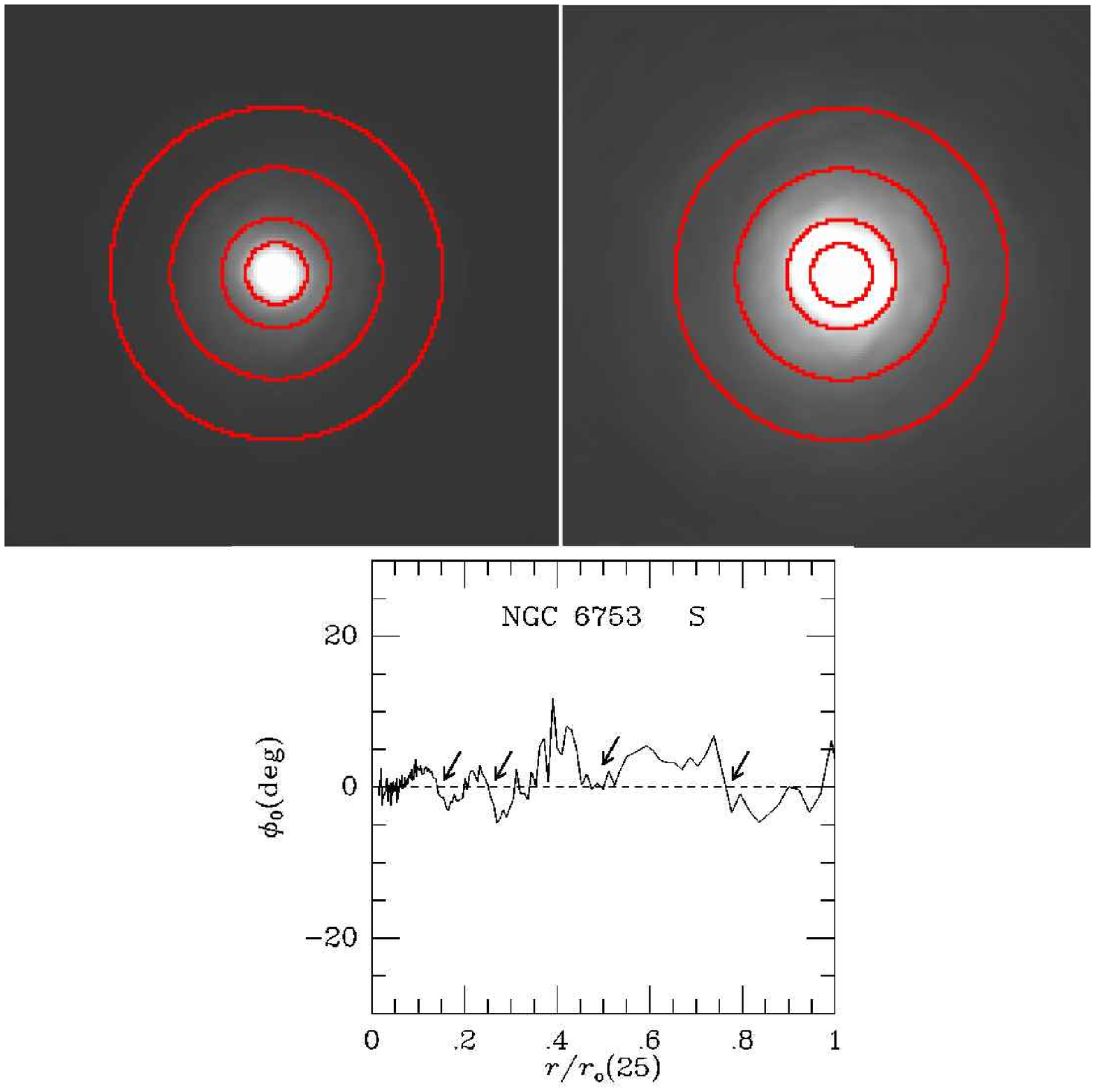}
 \vspace{2.0truecm}                                                             
\caption{Same as Figure 2.1 for NGC 6753}                                         
\label{ngc6753}                                                                 
 \end{figure}

NGC 6753 (Figure~\ref{ngc6753}) - The phase-shift distribution shows
four likely significant crossings. The one at $r$=11\arcs\ (0.14) is associated
with a small inner oval. The next one at $r$=19\arcs\ (0.25) seems associated
with a small inner spiral. The remaining crossings are associated with
intermediate spiral patterns, which show more prominently in an
optical image. 

\begin{figure}                                                                 
\figurenum{2.135}
\plotone{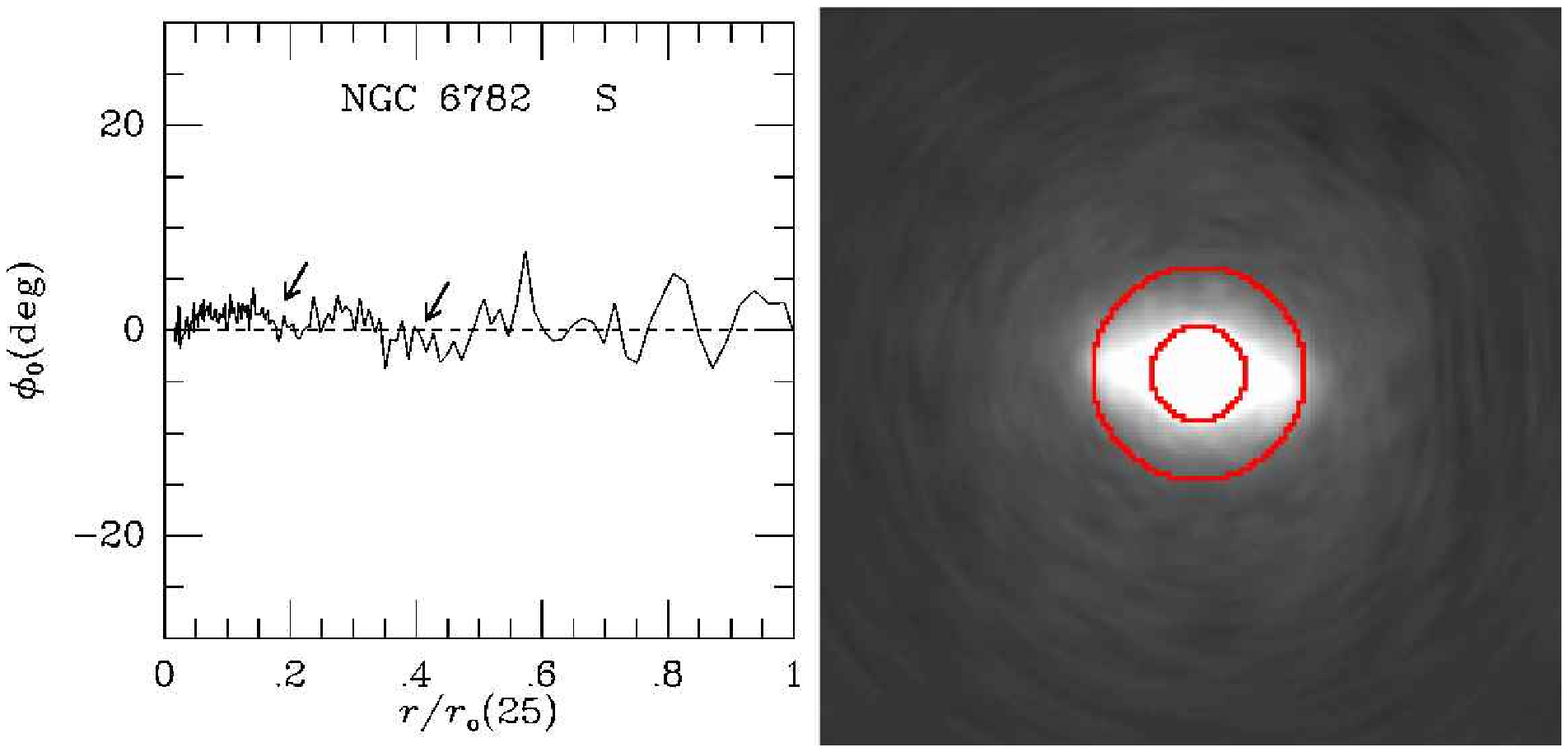}
 \vspace{2.0truecm}                                                             
\caption{Same as Figure 2.1 for NGC 6782}                                         
\label{ngc6782}                                                                 
 \end{figure}

NGC 6782 (Figure~\ref{ngc6782}) - The phase-shifts are small in this
galaxy because of the lack of skewness of the pattern.  We identify two
possibly significant but weak crossings at $r$=12\arcs\ (0.17) and
28\arcs\ (0.41).  The latter circles just inside the ends of the main
bar. The galaxy is also known to have a strong secondary bar (Buta \&
Crocker 1993). This feature lies well inside the innermost crossing.

An interesting RSL08 model places CR at 37\arcss 1 (0.54), well outside
our value which is closer to the RSL08 bar radius of 29\arcss 6 (0.43).

\begin{figure}                                                                 
\figurenum{2.136}
\plotone{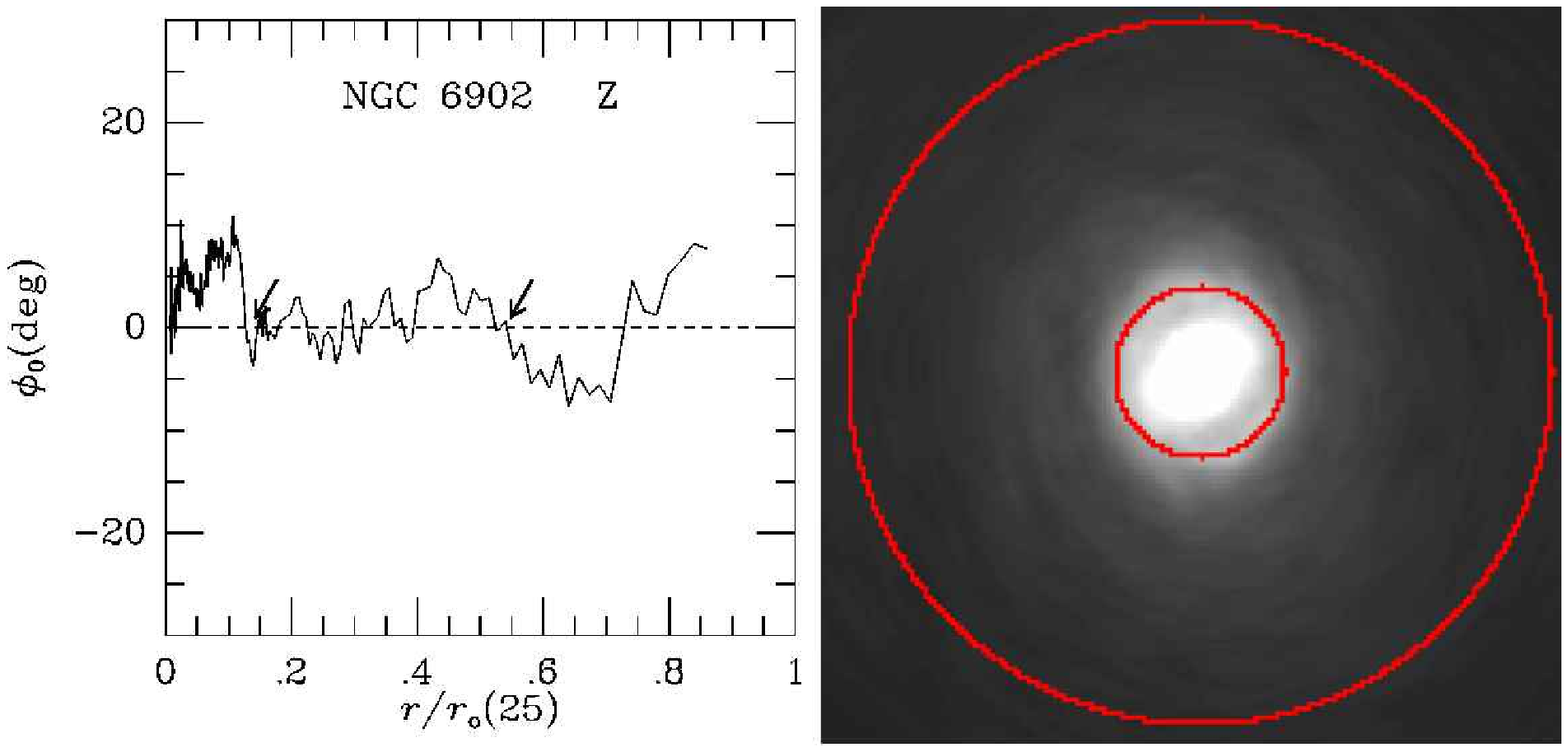}
 \vspace{2.0truecm}                                                             
\caption{Same as Figure 2.1 for NGC 6902}                                         
\label{ngc6902}                                                                 
 \end{figure}

NGC 6902 (Figure~\ref{ngc6902}) - The crossing at $r$=22\arcs\ (0.13) circles 
just outside the ends of a weak inner oval. The crossing at $r$=39\arcs\
(0.23) is weak but may be associated with the inner ring or inner spiral arms.
The spiral pattern is extensive in this galaxy, and the outer crossing
may be connected to the outer arms.  There appears to be pattern decoupling
going on in the central region.  Pattern decoupling or shearing also appears
to be going on in the intermediate radius region.

\begin{figure}                                                                 
\figurenum{2.137}
\plotone{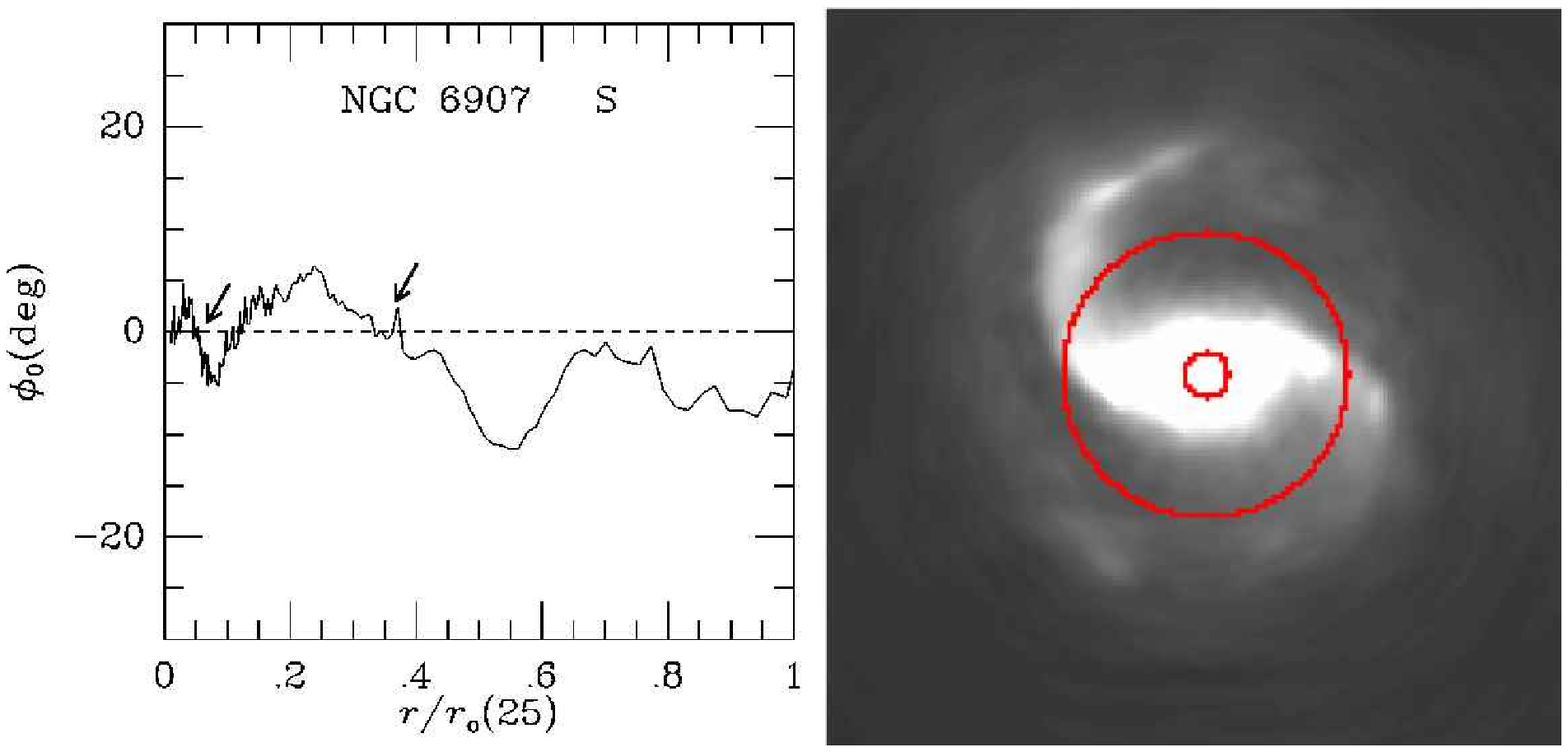}
 \vspace{2.0truecm}                                                             
\caption{Same as Figure 2.1 for NGC 6907}                                         
\label{ngc6907}                                                                 
 \end{figure}

NGC 6907 (Figure~\ref{ngc6907}) - The strong spiral and bar blend smoothly
in this galaxy. The main crossing lies just beyond the ends of the apparent
bar. The inner crossing is associated with a small inner oval.

\begin{figure}                                                                 
\figurenum{2.138}
\plotone{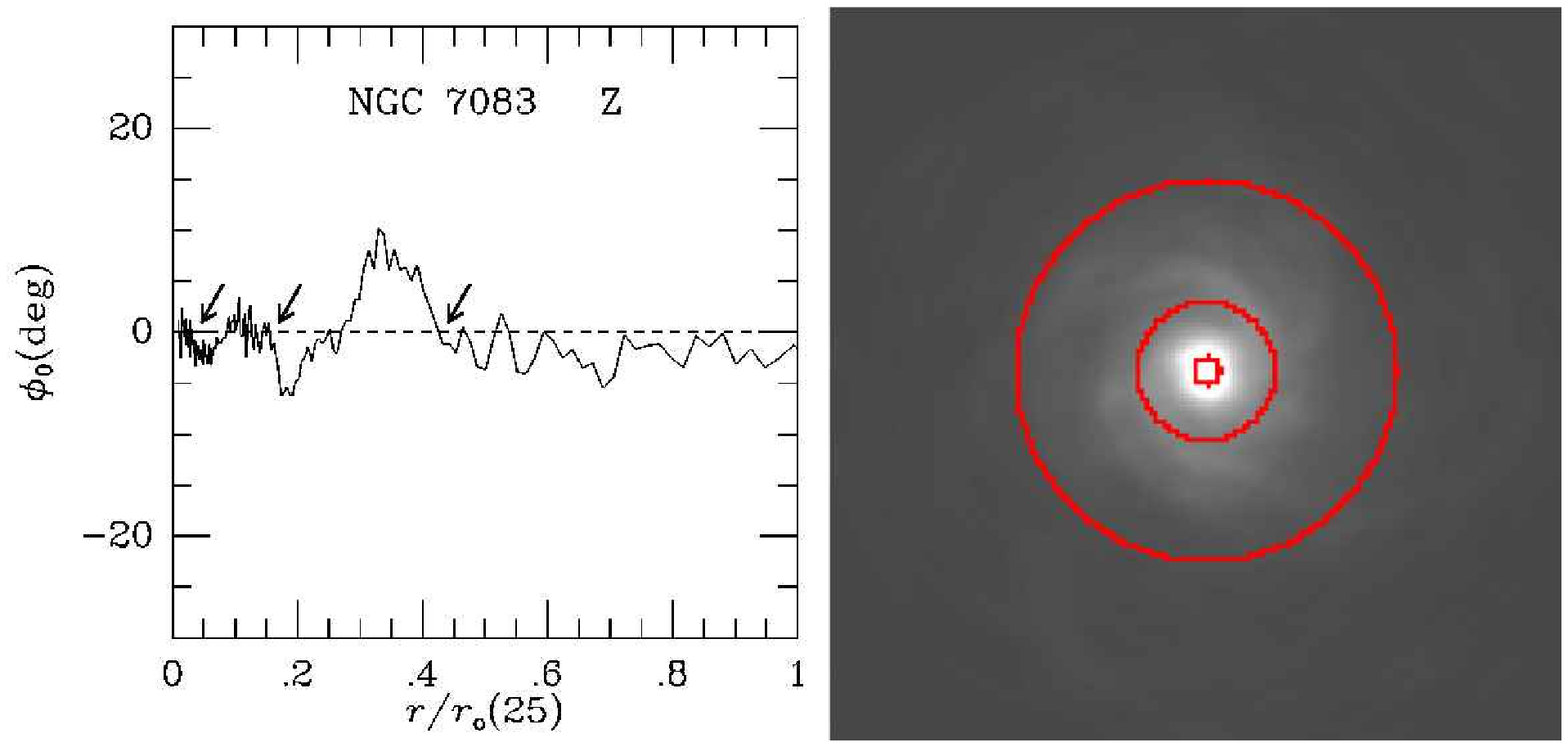}
 \vspace{2.0truecm}                                                             
\caption{Same as Figure 2.1 for NGC 7083}                                         
\label{ngc7083}                                                                 
 \end{figure}

NGC 7083 (Figure~\ref{ngc7083}) - The crossing at $r$=4\arcs\ (0.03) lies well
inside an inner oval, while that at $r$=18\arcs\ (0.15) lies well outside the
ends of this oval. The crossing at $r$=50\arcs\ (0.43) lies just outside the
inner spiral and may be associated with the outer pattern.

\begin{figure}                                                                 
\figurenum{2.139}
\plotone{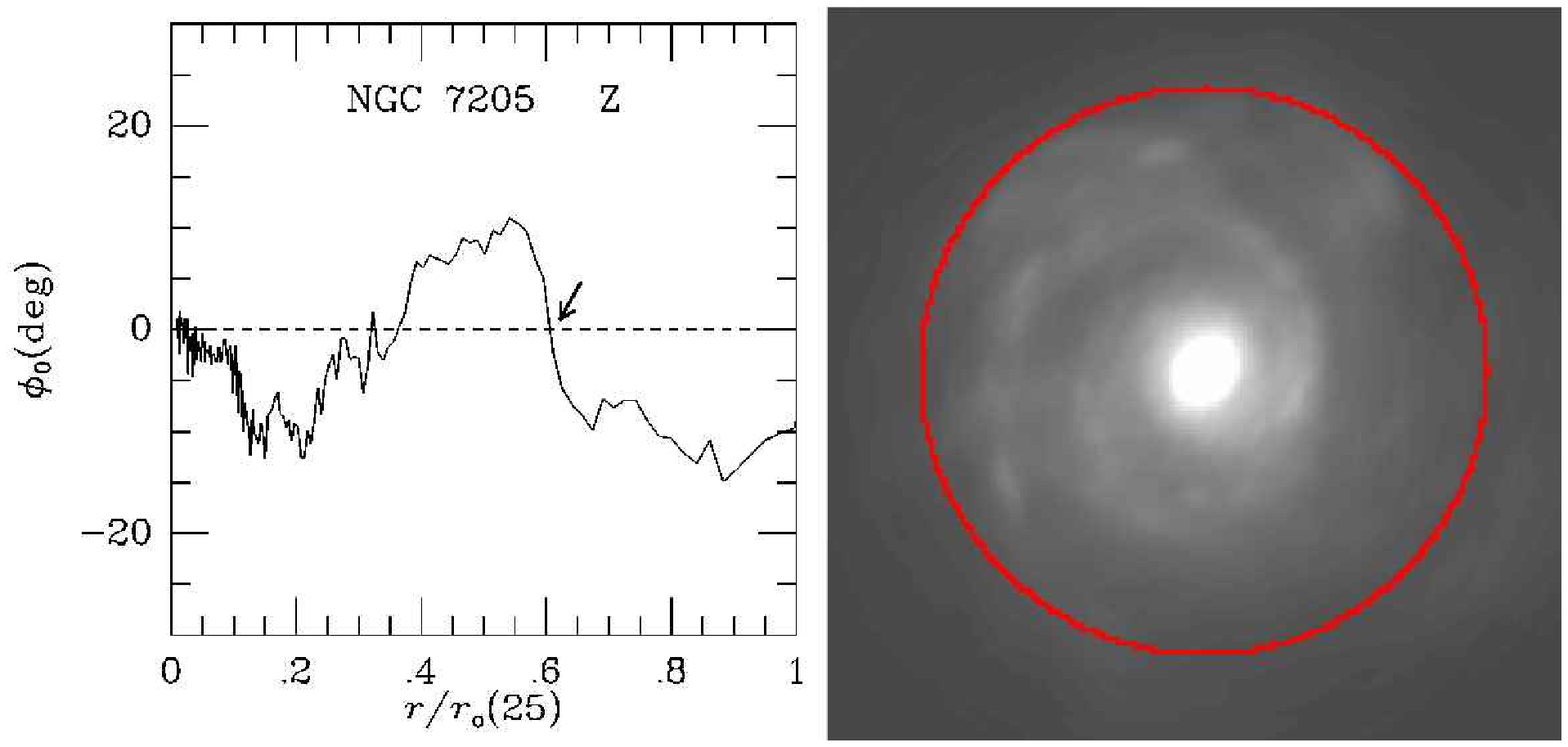}
 \vspace{2.0truecm}                                                             
\caption{Same as Figure 2.1 for NGC 7205}                                         
\label{ngc7205}                                                                 
 \end{figure}

NGC 7205 (Figure~\ref{ngc7205}) - The main crossing at $r$=74\arcs\ (0.61)
lies mainly outside the bright inner spiral pattern. The prominent
inner oval does not have an associated crossing, but we see evidence
in the phase-shift plot at $r$$\approx$12\arcs\ (0.10) for ongoing decoupling. 

\begin{figure}                                                                 
\figurenum{2.140}
\plotone{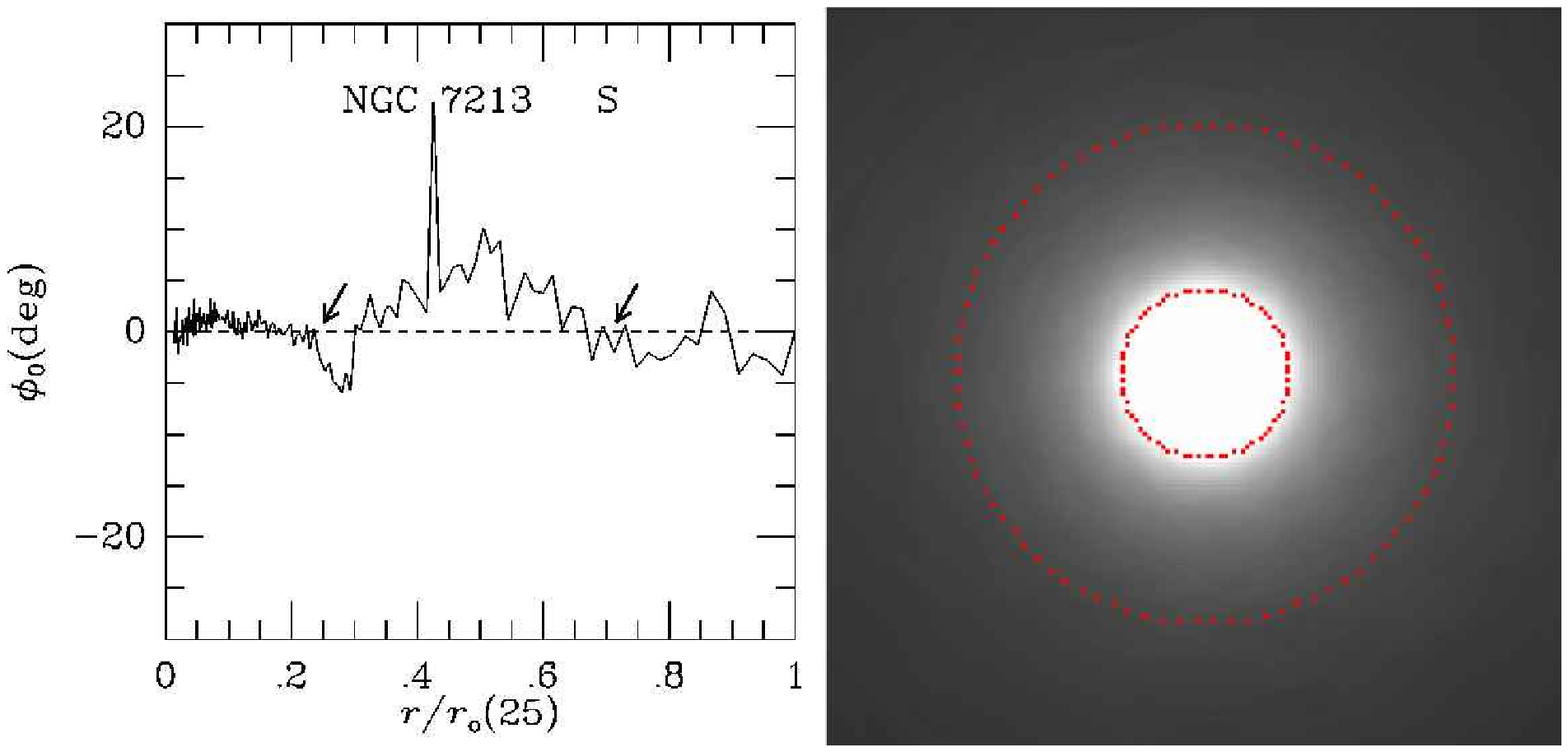}
 \vspace{2.0truecm}                                                             
\caption{Same as Figure 2.1 for NGC 7213}                                         
\label{ngc7213}                                                                 
 \end{figure}

NGC 7213 (Figure~\ref{ngc7213}) - There are no obvious density wave
patterns in the
$H$-band image, but still we see P/N crossings that might be significant.

\begin{figure}                                                                 
\figurenum{2.141}
\plotone{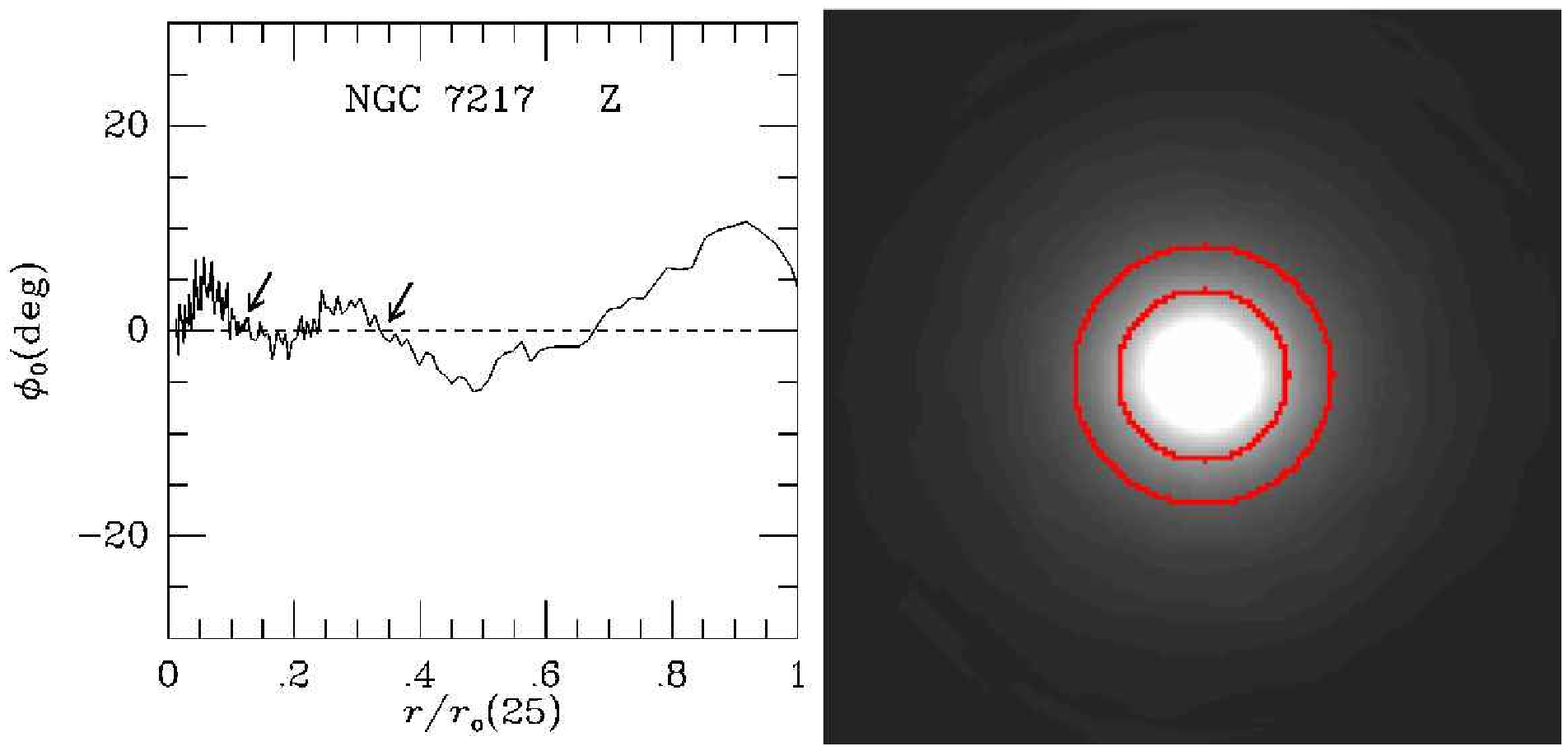}
 \vspace{2.0truecm}                                                             
\caption{Same as Figure 2.1 for NGC 7217}                                         
\label{ngc7217}                                                                 
 \end{figure}

NGC 7217 (Figure~\ref{ngc7217}) - The $H$-band image shows
a weak two-armed spiral outside the second CR at 43\arcs\
(0.34).

\begin{figure}                                                                 
\figurenum{2.142}
\plotone{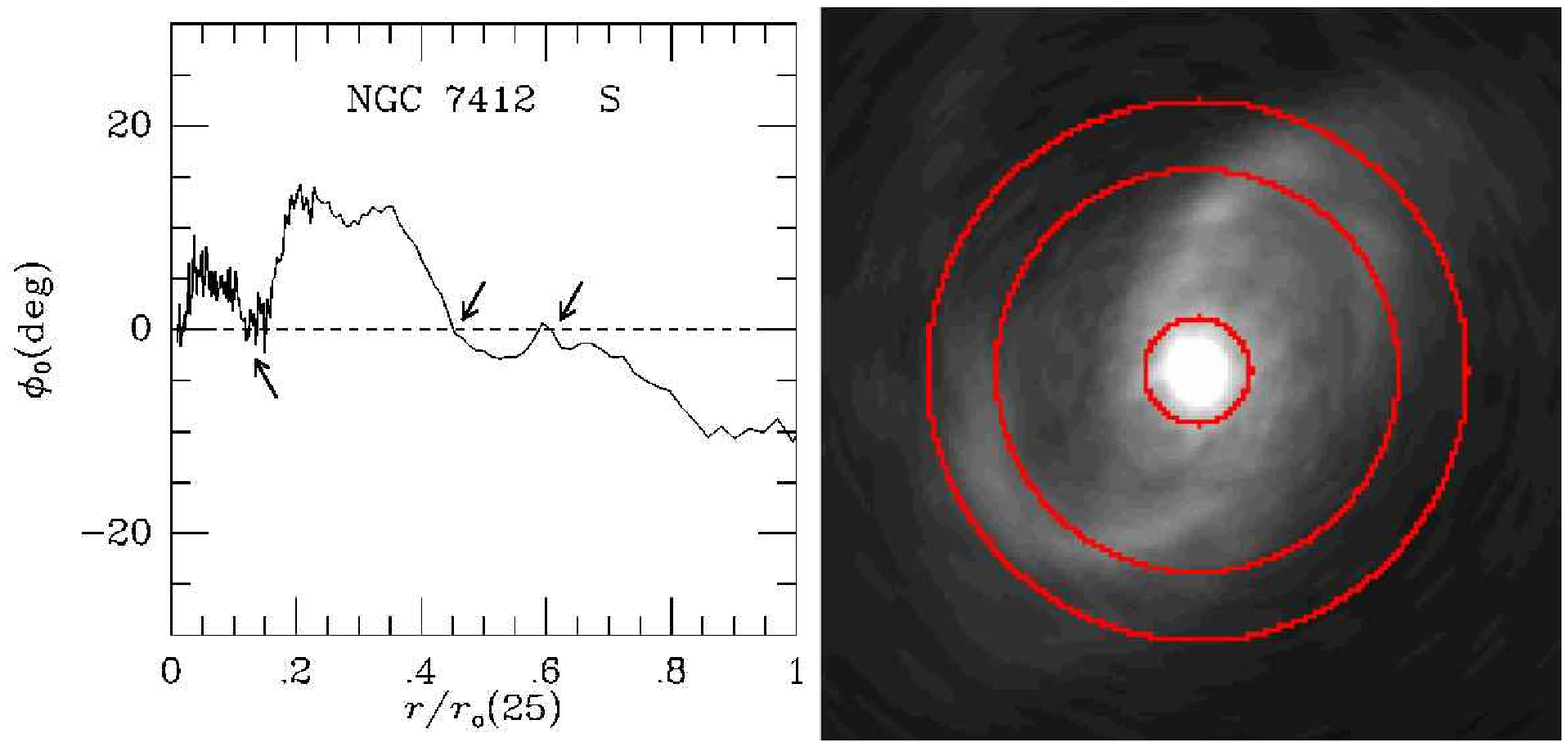}
 \vspace{2.0truecm}                                                             
\caption{Same as Figure 2.1 for NGC 7412}                                         
\label{ngc7412}                                                                 
 \end{figure}

NGC 7412 (Figure~\ref{ngc7412}) - The crossing at $r$=14\arcs\ (0.12)
corresponds to a small inner oval. The other two crossings are in the outer spiral 
arms which may be shearing.

\begin{figure}                                                                 
\figurenum{2.143}
\plotone{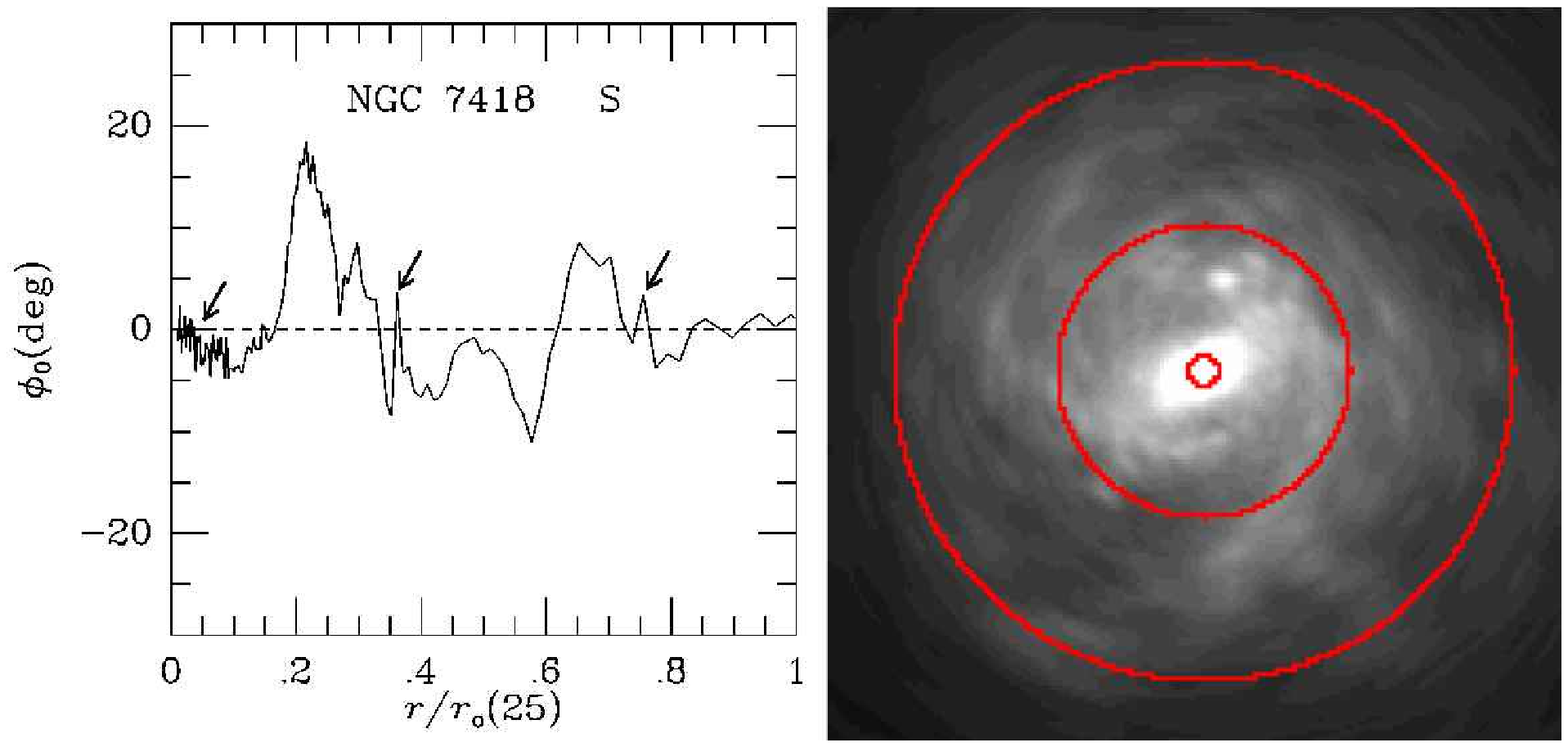}
 \vspace{2.0truecm}                                                             
\caption{Same as Figure 2.1 for NGC 7418}                                         
\label{ngc7418}                                                                 
 \end{figure}

NGC 7418 (Figure~\ref{ngc7418}) - Both of the main crossings are far
beyond the ends of the bar. The innermost crossing may indicate an
unresolved inner pattern. Although we have interpreted the
$r$=38\arcs\ (0.38) crossing as the bar CR in Table 1, the negative
phase-shifts inside this crossing suggest that the bar
may actually belong to this innermost mode.

\begin{figure}                                                                 
\figurenum{2.144}
\plotone{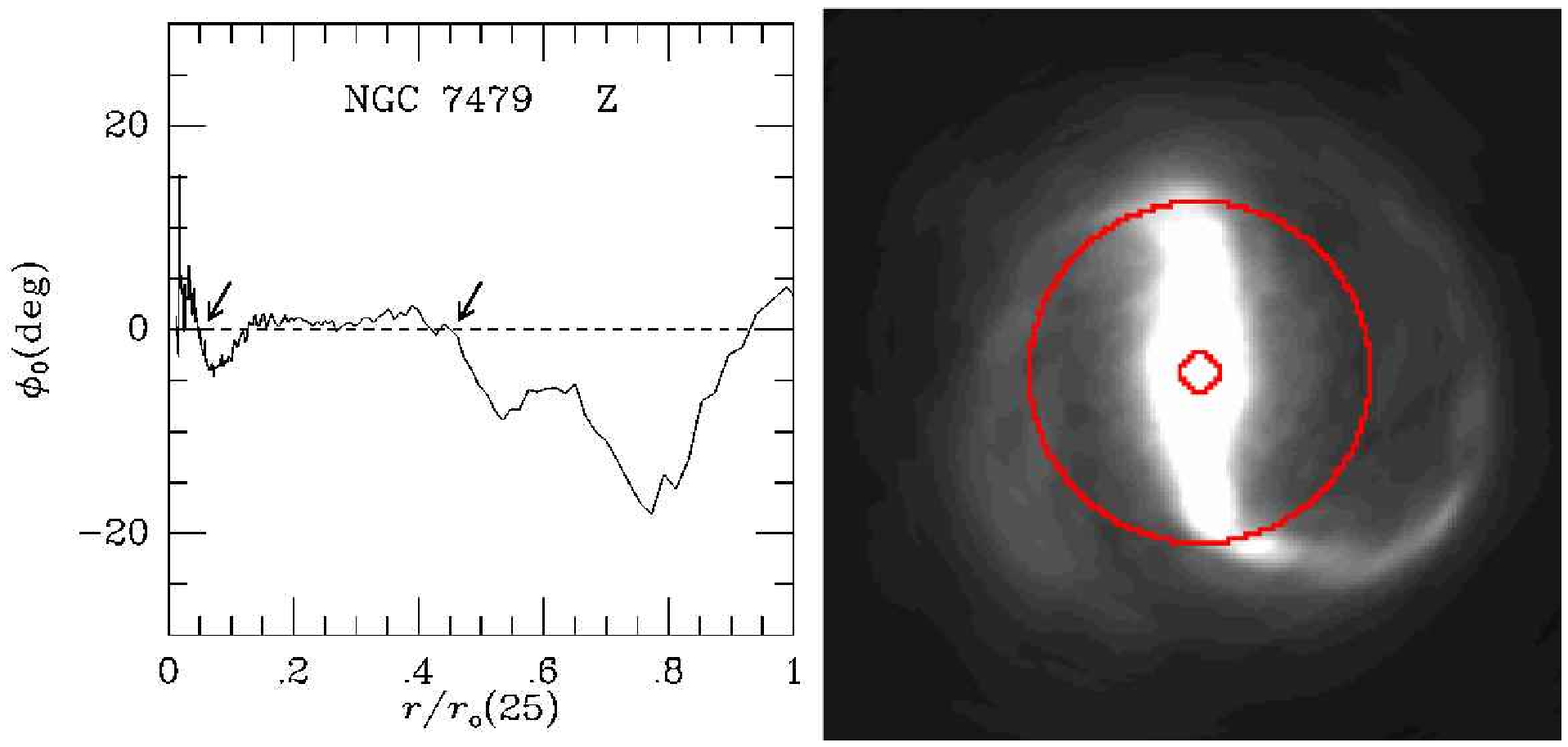}
 \vspace{2.0truecm}                                                             
\caption{Same as Figure 2.1 for NGC 7479}                                         
\label{ngc7479}                                                                 
 \end{figure}

NGC 7479 (Figure~\ref{ngc7479}) - Within $r_o$, there are only two
significant crossings in this galaxy. The crossing at $r$=58\arcs\
(0.45) encircles the ends of the bar closely. The innermost crossing
may be associated with a small inner oval. The galaxy can be thought of
as having a bar-driven spiral.  The spiral pattern outside the main CR
may be shearing.

\begin{figure}                                                                 
\figurenum{2.145}
\plotone{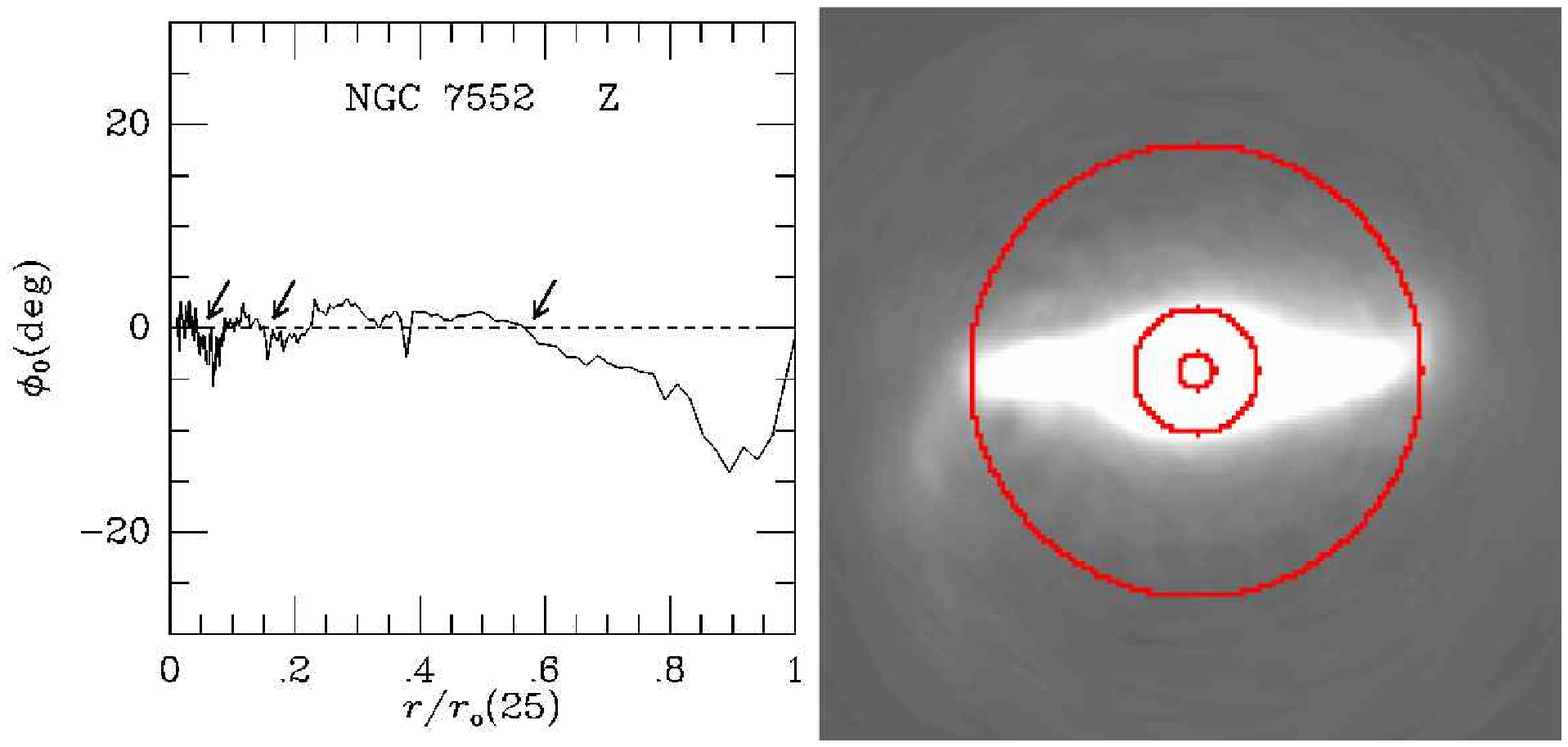}
 \vspace{2.0truecm}                                                             
\caption{Same as Figure 2.1 for NGC 7552}                                         
\label{ngc7552}                                                                 
 \end{figure}

NGC 7552 (Figure~\ref{ngc7552}) - The crossing at $r$=59\arcs\ (0.57) encircles
the bar ends. The other inner crossings may signify different sub-patterns
within the bar.

An excellent RSL08 model of this galaxy places CR at 65\arcss 0 (0.62),
close to our main CR radius.

\begin{figure}                                                                 
\figurenum{2.146}
\plotone{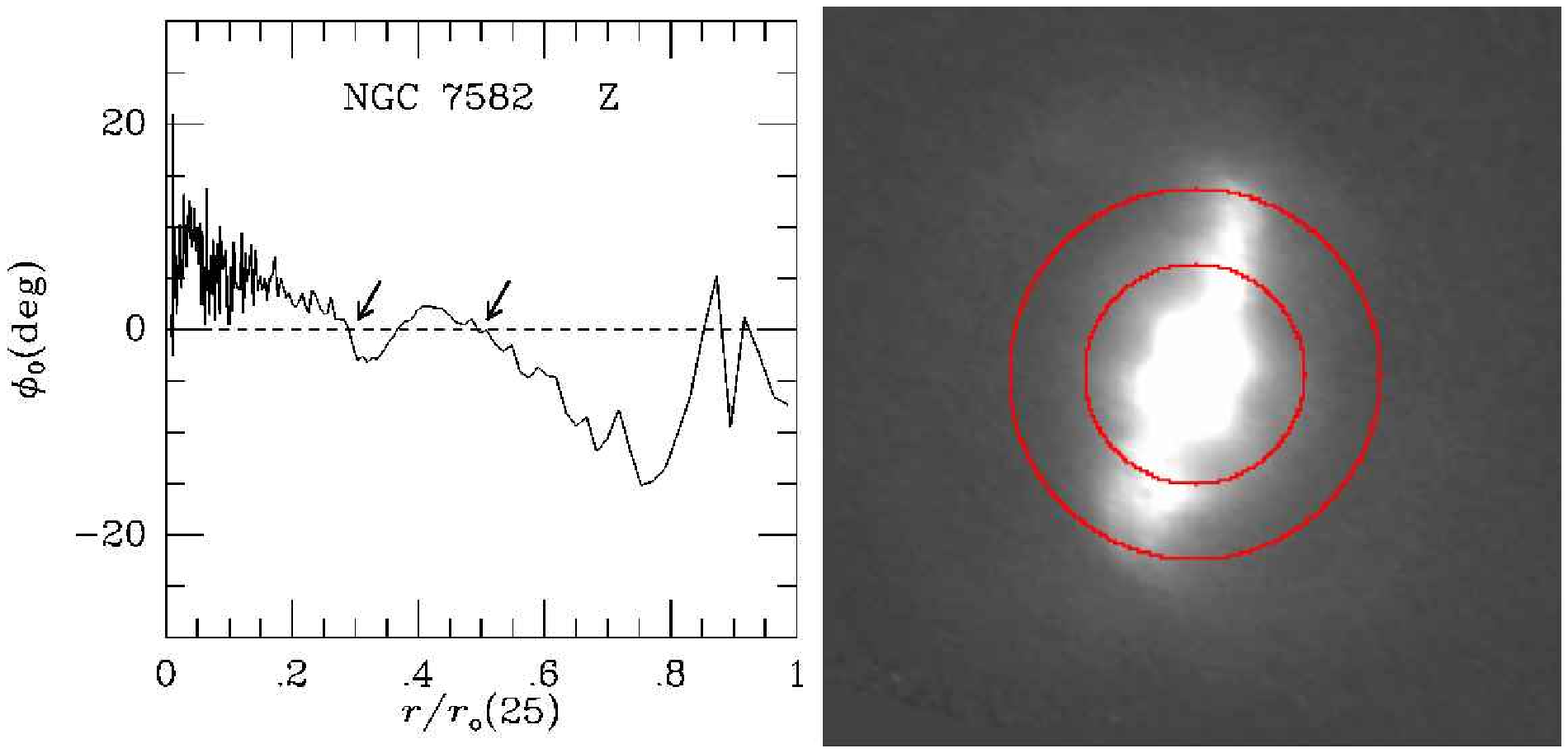}
 \vspace{2.0truecm}                                                             
\caption{Same as Figure 2.1 for NGC 7582}                                         
\label{ngc7582}                                                                 
 \end{figure}

NGC 7582 (Figure~\ref{ngc7582}) - The inner crossing at $r$=44\arcs\ (0.29)
encircles an area with strong decomposition-pinch, which may in reality correspond
to an oval feature in the galaxy center. The crossing at
$r$=74\arcs\ (0,49) encircles just inside the bar ends. Our analysis is based on an
average bulge image (section 3).

\begin{figure}                                                                 
\figurenum{2.147}
\plotone{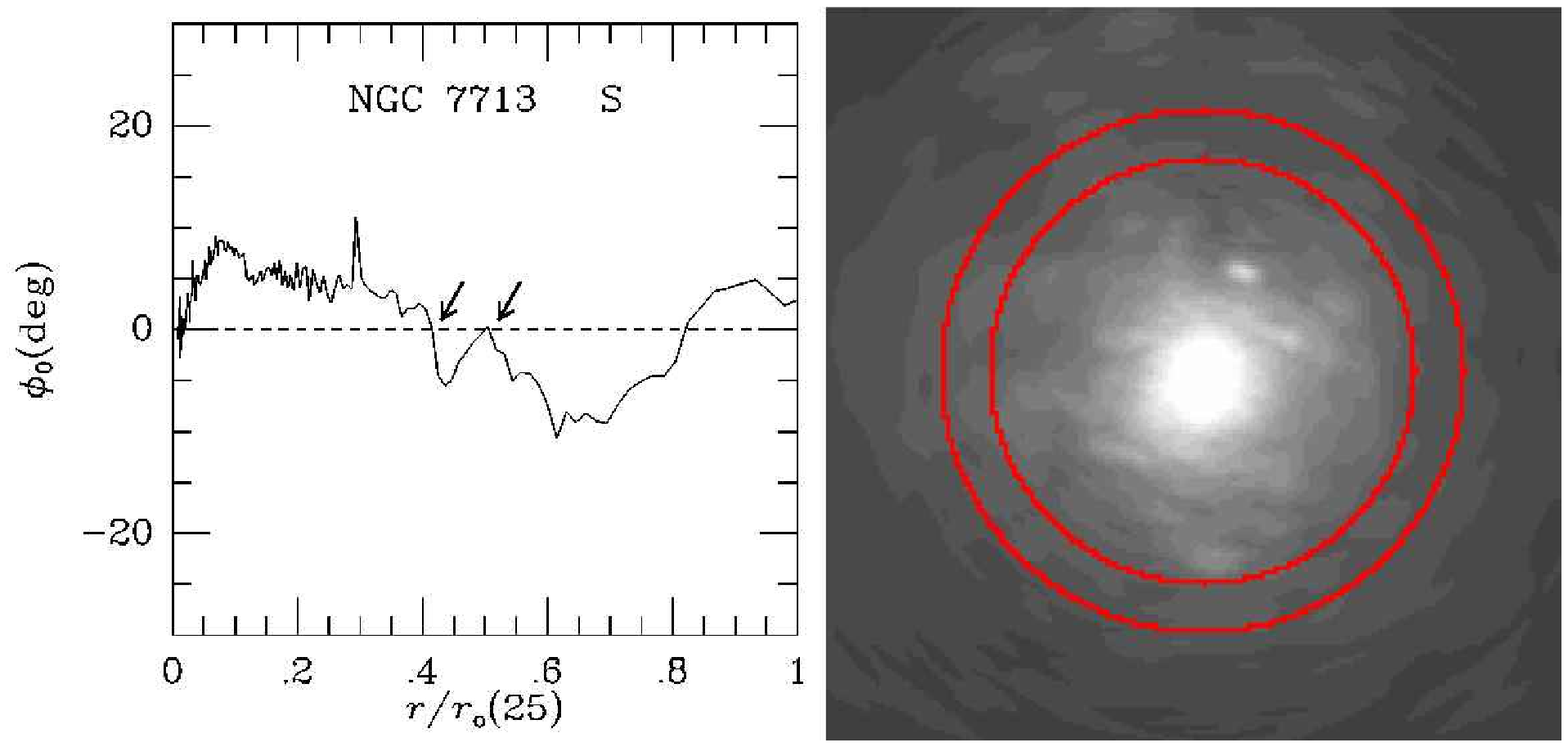}
 \vspace{2.0truecm}                                                             
\caption{Same as Figure 2.1 for NGC 7713}                                         
\label{ngc7713}                                                                 
 \end{figure}

NGC 7713 (Figure~\ref{ngc7713}) - The two closely spaced crossings are
not clearly related to any pattern.  They lie well beyond the inner oval which might
be going through a decoupling process from the outer pattern.

\begin{figure}                                                                 
\figurenum{2.148}
\plotone{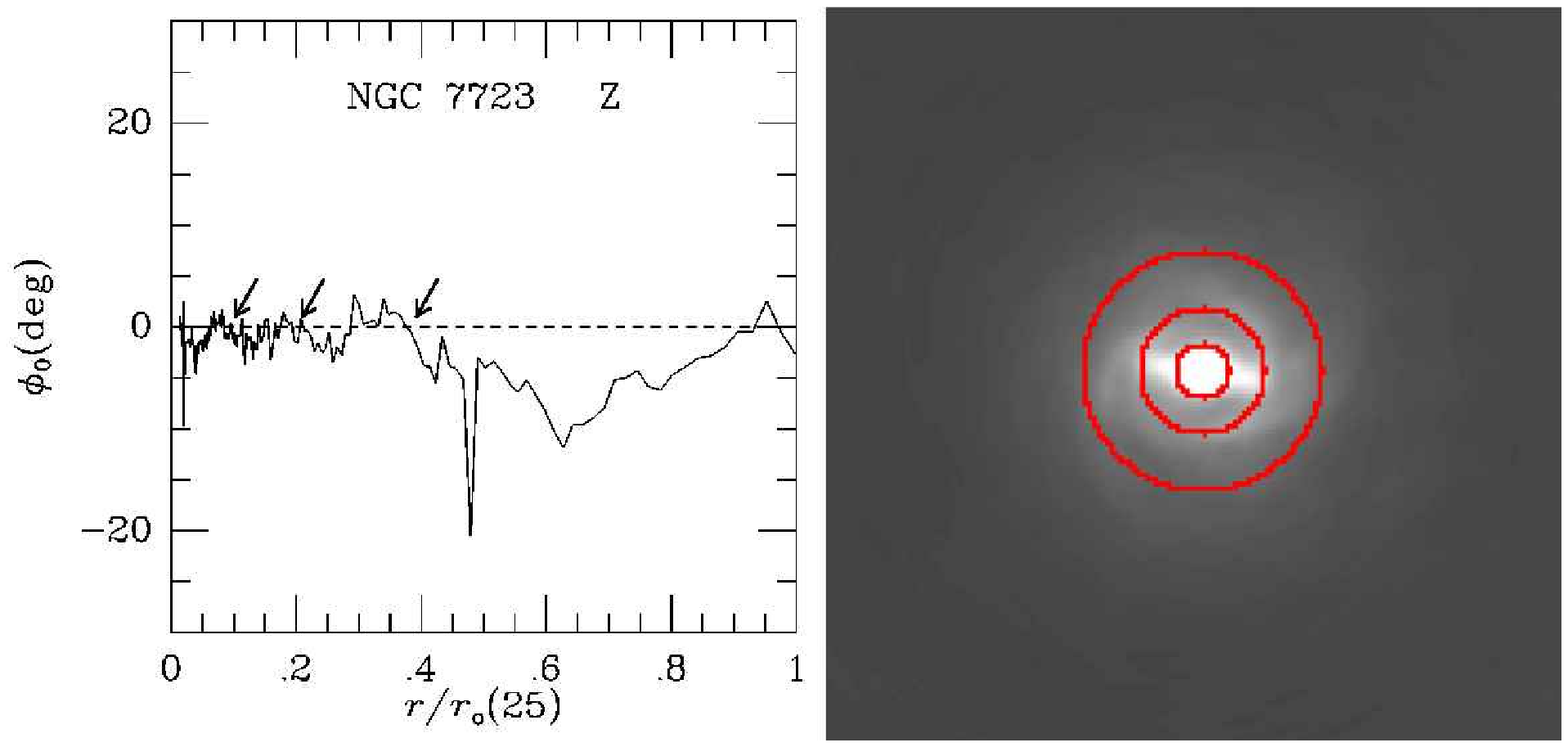}
 \vspace{2.0truecm}                                                             
\caption{Same as Figure 2.1 for NGC 7723}                                         
\label{ngc7723}                                                                 
 \end{figure}

NGC 7723 (Figure~\ref{ngc7723}) - The crossing at $r$=21\arcs\ (0.20) 
corresponds
to the ends of the bar. The crossing at $r$=40\arcs\ (0.38) corresponds to the
bright inner spiral. The innermost crossing may correspond to the small inner
oval.  There is also the indication of an unresolved inner pattern.

An RSL08 model of this galaxy places CR at 33\arcss 5 (0.31), between our two
outermost crossings.

\begin{figure}                                                                 
\figurenum{2.149}
\plotone{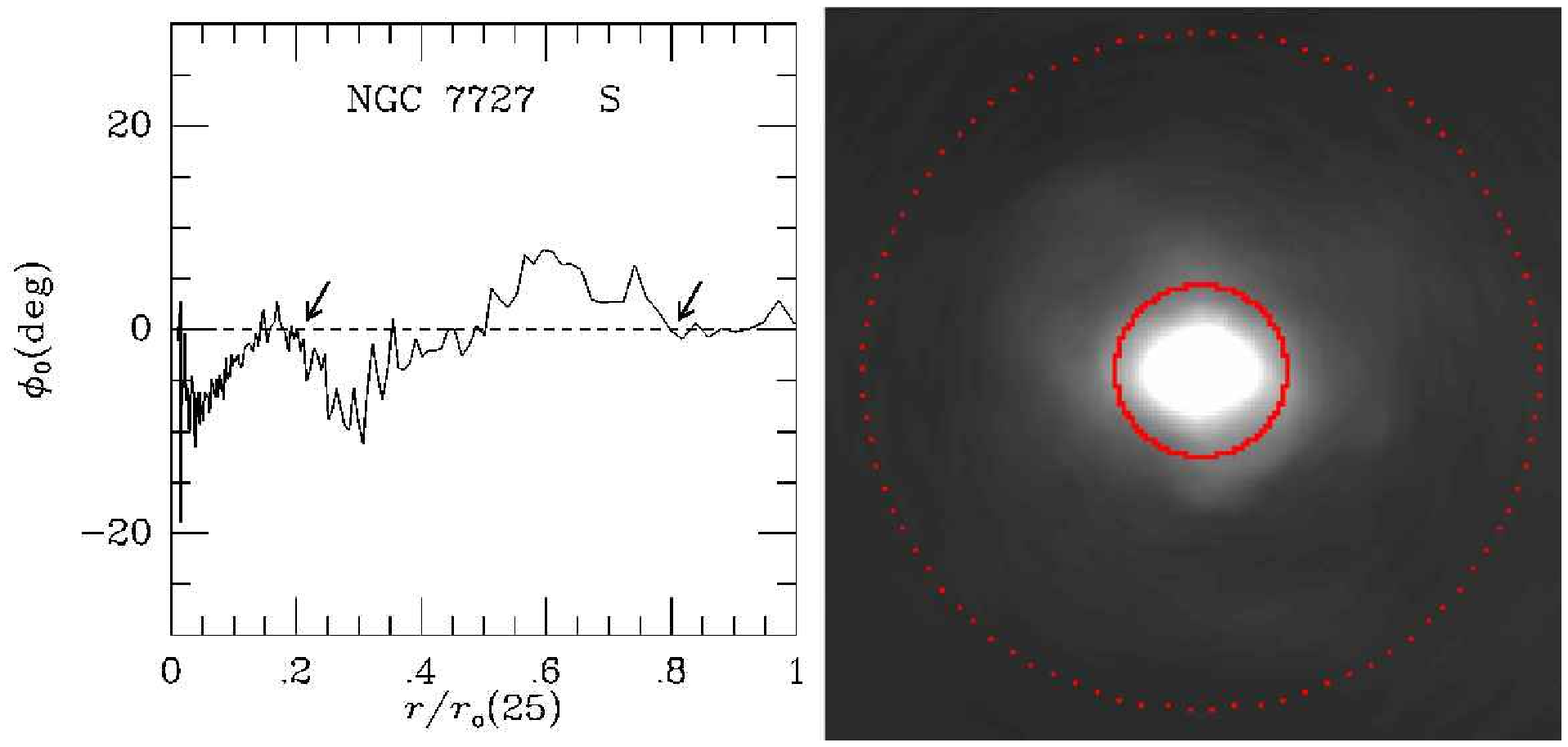}
 \vspace{2.0truecm}                                                             
\caption{Same as Figure 2.1 for NGC 7727}                                         
\label{ngc7727}                                                                 
 \end{figure}

NGC 7727 (Figure~\ref{ngc7727}) - The crossing at
$r$=29\arcs\ (0.20) encircles an inner oval. The significance of the
outer crossing is uncertain.
There is also the indication of an unresolved inner pattern.

\begin{figure}                                                                 
\figurenum{2.150}
\plotone{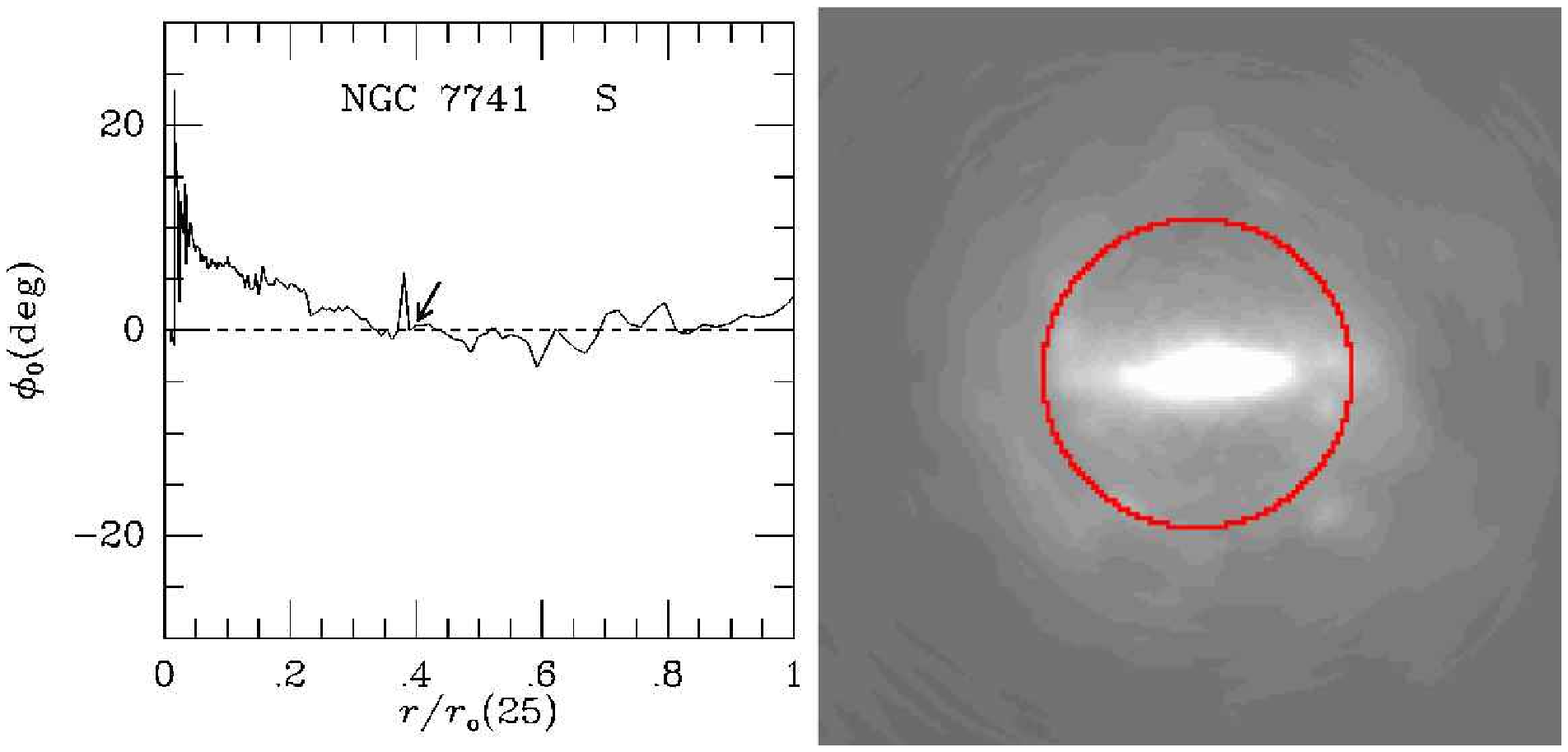}
 \vspace{2.0truecm}                                                             
\caption{Same as Figure 2.1 for NGC 7741}                                         
\label{ngc7741}                                                                 
 \end{figure}

NGC 7741 (Figure~\ref{ngc7741}) - The crossings are weak and we have
averaged crossings at 45\arcs\ (0.34) and 59\arcs\ (0.44)
as our best CR estimate.
This average circles around the ends of the bar. There is considerable
asymmetry in the bar in any case.

\begin{figure}                                                                 
\figurenum{2.151}
\plotone{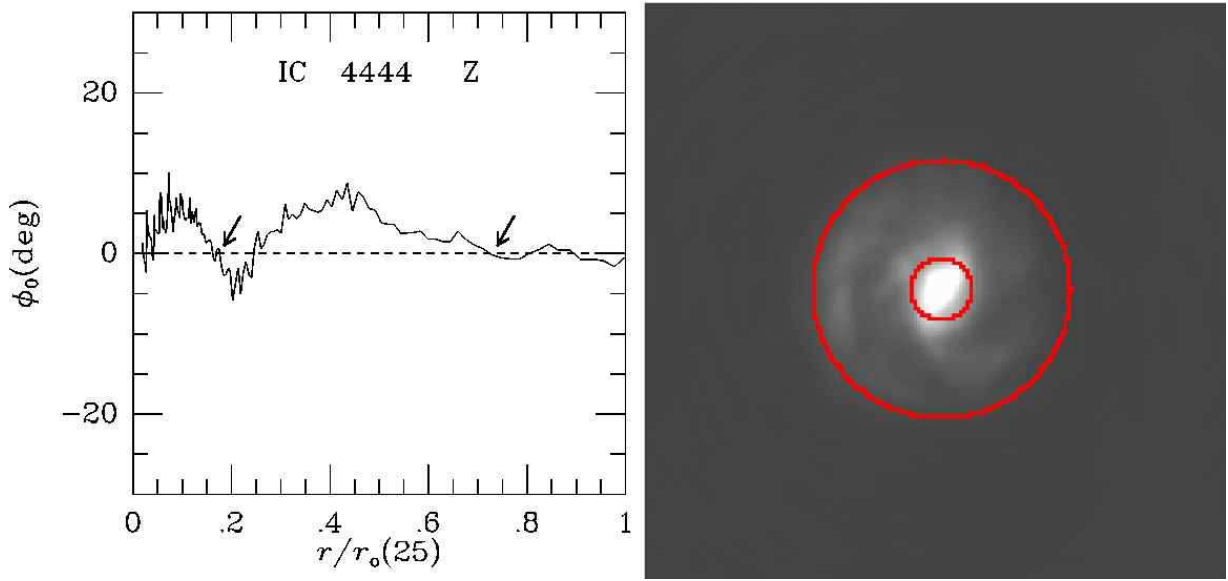}
 \vspace{2.0truecm}                                                             
\caption{Same as Figure 2.1 for IC 4444.}                                         
\label{i4444}                                                                   
 \end{figure}

IC 4444 (Figure~\ref{i4444}) - The crossing at $r$=10\arcs\ (0.17) circles
the ends of a small inner bar. The crossing at $r$=41\arcs\ (0.72) lies
just beyond the main spiral arms.

\begin{figure}                                                                 
\figurenum{2.152}
\plotone{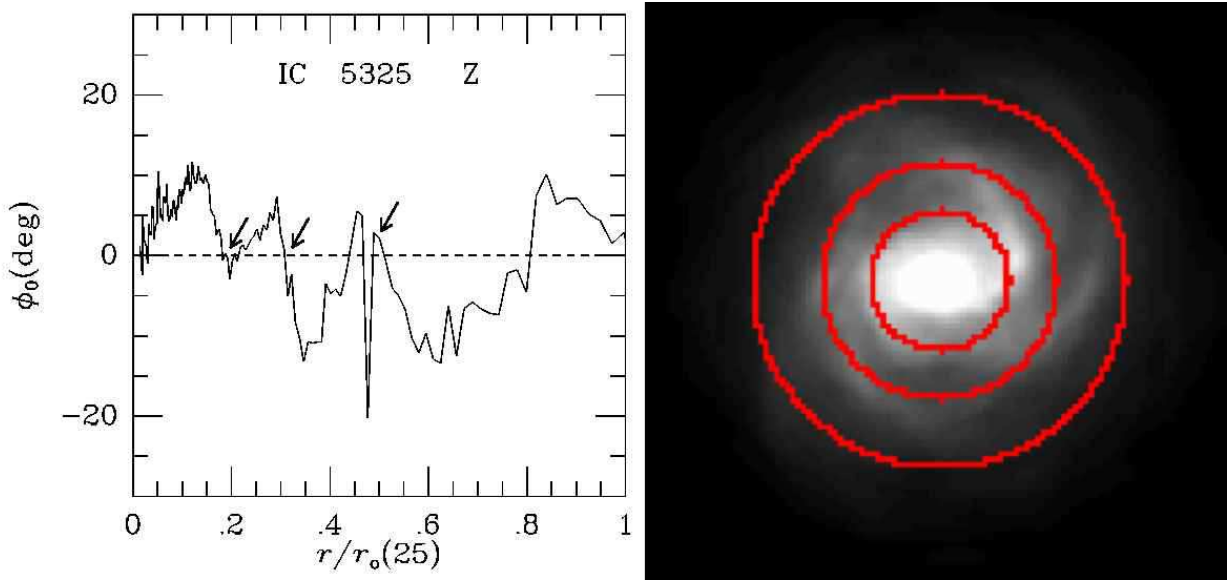}
 \vspace{2.0truecm}                                                             
\caption{Same as Figure 2.1 for IC 5325.}                                         
\label{i5325}                                                                   
 \end{figure}

IC 5325 (Figure~\ref{i5325}) - The crossing at $r$=15\arcs\ (0.18) encircles
a weak inner oval. The other two crossings seem to correspond to
the two sets of outer spiral arms.

\begin{figure}                                                                 
\figurenum{2.153}
\plotone{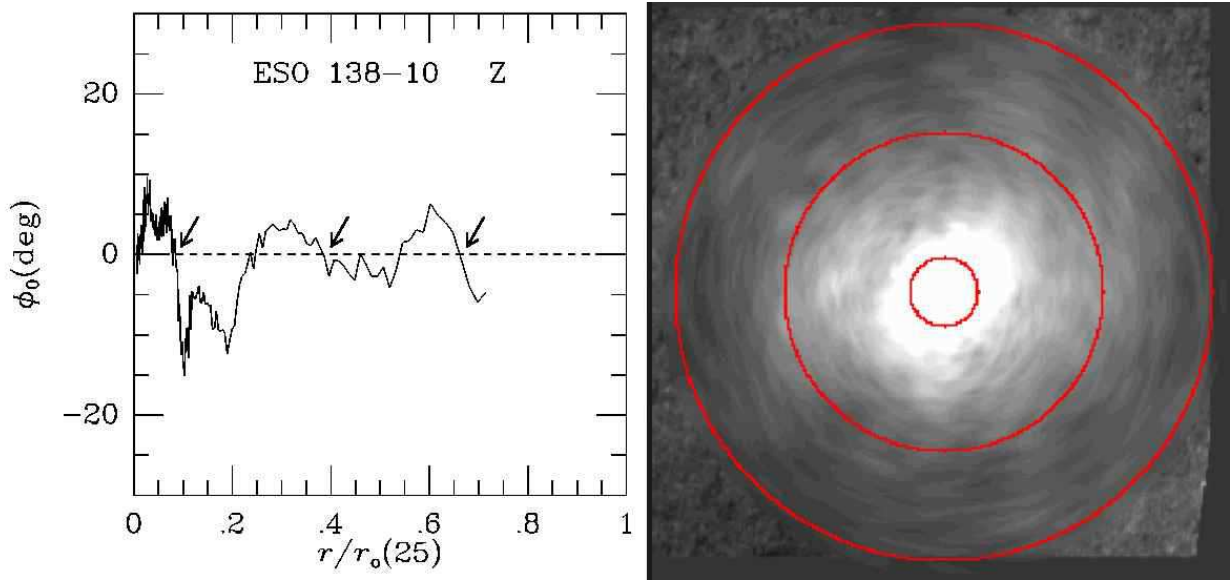}
 \vspace{2.0truecm}                                                             
\caption{Same as Figure 2.1 for ESO 138$-$10.}                                    
\label{eso138}                                                                  
 \end{figure}

ESO 138$-$10 (Figure~\ref{eso138}) - The crossing at $r$=17\arcs\ (0.08)
may be associated with a small inner oval. There is another oval at twice
the radius of this one, but it only extends about halfway to the next
crossing. There is indication of ongoing decoupling at the location of
the second oval. The outer crossings are of uncertain significance.

\section{Analysis}

Having summarized what the phase-shift distributions reveal for each
individual galaxy, we now analyze the entire galaxy sample. Table 2
provides an inventory of the features seen as well as a general
characterization of the phase-shift plot as either ``well-defined" (low
noise), or ``noisy" (having many likely noise-induced crossings), or
else ``noisy/unsteady" (implying little coherence in both the pattern
and the phase-shift plot). Some of the ovals recognized in col. 4 of
Table 2 could be artifacts of deprojection, but in general deprojection
effects should not produce false P/N crossings. Table 2 also gives an
assessment of whether a galaxy has a grand-design spiral or a
flocculent spiral. This can be difficult to judge from the near-IR
images given the low depth of exposure of some of the images. Instead,
we provide the blue-light Elmegreen Arm Class (AC; Elmegreen \&
Elmegreen 1982, 1987) as a judgment of the coherence and complexity of
the spiral pattern. $B$-band spirals that are chaotic, fragmented,
asymmetric, irregular, or dominated by only a single arm are classes AC
1-4, and are considered flocculent. Spirals with at least two symmetric
inner or outer arms, or ring-like arms, are classes 5-12 and are
considered grand-design. These classes do not perfectly translate to
what one sees in the near-IR. It is possible for a galaxy to be
``optically-flocculent" and show a grand-design spiral in the near-IR
(Thornley 1996; Buta et al. 1995). Also, some grand-design spirals are
weak in the near-IR. Nonetheless, the arm class is still an instructive measure of
the general coherence of a pattern.

The arm classes are given in col. 5 of Table 2. Most are taken from
Elmegreen \& Elmegreen (1987). In a few cases where the sample galaxy
was not included in this paper, Buta estimated the arm class from
available images in the de Vaucouleurs Atlas of Galaxies (Buta et al.
2007) and other sources. For 13 of the galaxies, the Hubble-de
Vaucouleurs type is too early to provide a reliable arm classification.
Of the remaining 140 galaxies having an arm classification, 93 would be
considered grand-design in blue light.  Of these, we independently
judged 67\% to have a ``well-defined" phase-shift plot. In the same
set, 47 galaxies would be considered flocculent, and of these we judged
the phase-shift plot to be ``well-defined" for only 28\%. This confirms
what we might have expected: a coherent grand-design pattern is likely
to be more steady and self-consistent (in a modal sense) than a flocculent, 
disorganized pattern.

Table 2 also summarizes the number of CRs we have identified. Among the
best-defined grand-design spirals (AC 9 and 12), the number of CRs
ranges from 1-5.  For the 46 AC 9 and 12 galaxies in Table 2, the mean
number of CRs we have recognized is $<n_{CR}>$ = 2.8.  For this same
subset, 78\% were judged to have well-defined phase-shift plots. The 47
AC 1-4 galaxies have $<n_{CR}>$ = 2.1.  These values largely exclude
the weak inner crossings found in many of the galaxies. These appear to
be due to largely unresolved features and we have not compiled most of them
in Table 1. We expect that higher resolution images would tell more
about the reality of such features.

One of the first conclusions we can draw from the Table 2 analysis is
that we see considerable evidence for steady patterns and genuine
density wave modes in many galaxies. If this were not the case, most
phase-shift distributions would be uninterpretable, since only for
quasi-steady modes do the potential-density phase-shift zero crossings
correspond to partially kinematic features such as corotation radii.
Nevertheless, some phase-shift distributions are complicated enough to
suggest that the patterns are unsteady. Secondly, bars and ovals in
early-type disks frequently have phase-shift crossings near or just
beyond their apparent ends, i.e., they appear as fast patterns. These
fast patterns appear to correspond to mature density wave mode features
that have completed the decoupling process and have reached
quasi-steady state, whereas density wave features in late-type disks
appear to be slower patterns (see also Elmegreen \& Elmegreen 1985;
RSL08). Thirdly, multiple pattern speeds appear to be common among
spirals, with the bar and the spiral often having different pattern
speeds and hence different corotation radii.  Nevertheless, we do find
evidence for some galaxies where the bar and the spiral have the same
pattern speed and the same corotation radius (i.e. the so-called
``bar-driven spirals''), and many such galaxies have their spirals
emanating from the end of the bar (which is often identical to the
CR).  Finally, we find that some galaxies apparently break the ``rule"
that bars cannot extend beyond their own CR radius. Especially among
intermediate-type barred galaxies, we have found cases where
the ends of the bar lie near a strong N/P crossing, which is likely to
be close to its outer Lindblad resonance radius, not CR. The strongest
grand design spirals in our sample all show major P/N crossings in the
middle of the spiral pattern, and these spiral patterns clearly must
extend well beyond their CR radii also. These issues are discussed
further in the next subsection.

\subsection{Bar Extent and Corotation}

\subsubsection{Fast and Slow Bars}

Debattista \& Sellwood (2000) defined a ``fast" bar to have a ratio of
corotation radius to bar radius $\cal R$=$r(CR)/r(bar)$ $\leq$ 1.4
while a ``slow" bar has $\cal R$ $>$ 1.4. Laurikainen et al. (2004,
their Table 3) compiled bar radii for about 100 OSUBGS galaxies which
we can use, together with the CR values derived in the current work, to
examine this ratio. [The Laurikainen et al. (2004) bar radii are
generally consistent with those of RSL08, but are smaller on average by
10\% (RSL08)]. The radii are given only for ``Fourier" bars, that is,
bars for which the relative $m$=2 Fourier amplitude maintains a
relatively constant position angle out to a certain radius. In deriving
$\cal R$ for phase-shift-derived CR radii, we are faced with a
difficulty in that many of our sample galaxies show multiple CRs. The
question is, which radius do we take for the actual CR radius of an
apparent bar? As noted in section 3, we have made such a selection in
Table 1 by placing an asterisk next to the adopted bar CR radius. In
many cases, this is a reasonable judgment: the radius lies near or just
beyond the bar ends or is the most reasonable choice because other CRs
are likely associated with other patterns. Nevertheless, there are
ambiguities, and we cannot be completely certain that all of the
asterisked choices for the bar-CRs are correct. For example, many of
the phase-shift distributions show negative values in the bar regions,
suggesting that the bar CR may in fact lie well inside the apparent
bar. Some of these are discussed in section 5.1.3. Table 1 indicates
these alternative interpretations with boldfaced values. Table 3
summarizes our values of $\cal R$, providing two values when two
interpretations are considered.

\begin{figure}
\figurenum{3}
\plotone{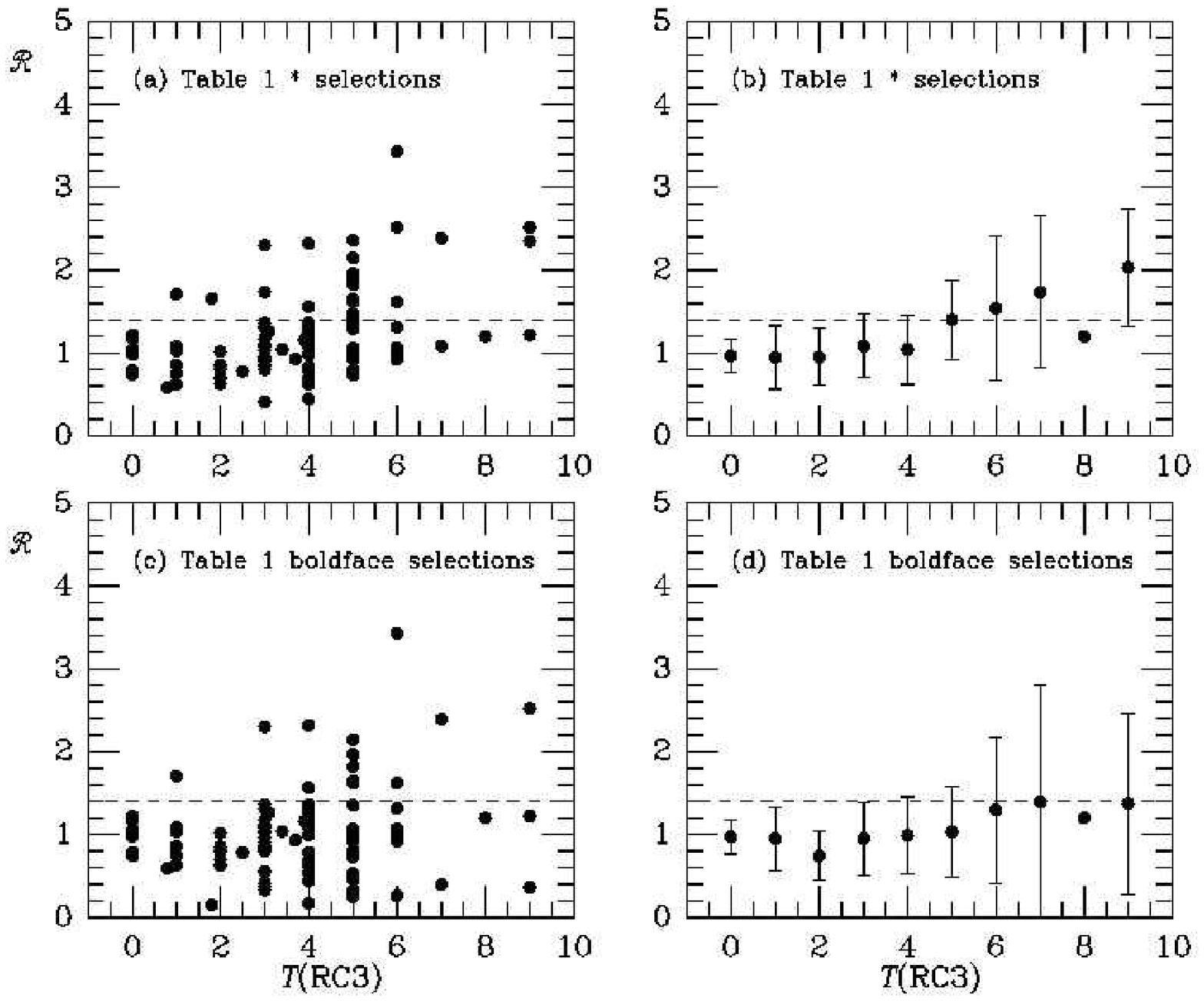}
 \vspace{2.0truecm}
\caption{Corotation to bar radius, $\cal R$ = $r(CR)/r(bar)$, for the
OSUBGS galaxies having a ``Fourier bar" (Laurikainen et al. 2004),
plotted versus numerical RC3 stage index. (a),(b) Individual points and
means with standard deviations based on the asterisked CR radii in
Table 1. (c),(d) Same as (a),(b) but using the boldfaced CR radii in
Table 1 as alternative interpretations. The dashed horizontal line is
set at the value 1.4, used to delineate a boundary between fast and
slow bars (Debattista \& Sellwood 2000).} 
\label{rcrbar}
 \end{figure}

Figure~\ref{rcrbar} shows plots of $\cal R$ versus RC3 stage index $T$,
for both the Table 1 asterisk values (upper panels) and the alternative
boldface values for some galaxies (lower panels).  The left panels show
the individual points, while the right panels show the means by type.
For the asterisk interpretations, we find for 65 galaxies of types Sbc
and earlier ($T\leq$4), the mean ratio is 1.03$\pm$0.37 (standard
deviation), indicating that the bars generally extend to CR almost
exactly.  However, for 36 galaxies of type Sc and later ($T\geq$5), the
mean ratio is 1.50$\pm$0.63(s.d.). A similar trend was first found by
RSL08 and seems consistent with Elmegreen \& Elmegreen's (1985) finding
that late-type bars are smaller than early-type bars and may fall well
short of their corotation radii. The suggestion is that fast bars are
found mainly in earlier types while slow bars are found mainly in later
types.

Figure~\ref{absmags} shows plots of $\cal R$ versus absolute blue
magnitude based on RC3 parameters, again based on the asterisked values
of the bar CR in Table 1 and based on the alternative boldfaced values
for some galaxies. For the asterisked values, the plot shows a weak
trend in the sense that $\cal R$ is smaller on average for the more
luminous galaxies in the sample. This is consistent with
Figure~\ref{rcrbar} in the sense that the later type galaxies in the
sample have the fainter absolute magnitudes.

Buta et al. (2005) estimated relative maximum bar and spiral torque
strengths for many of our sample galaxies, based on the same set of
images. If $\cal R$ has a type dependence, we might expect the ratio to
also be sensitive to bar strength because of bulge dilution in earlier
types. Figure~\ref{qbqs2} shows $\cal R$ versus $Q_b$ and $Q_s$.  Only
weak trends with bar and spiral strengths are evident, in the sense
that the ratio rises slightly with these parameters. This is consistent
with the type dependence of $Q_b$ (Buta et al.  2004) and $Q_s$ (Buta
et al. 2005). These plots are based only on the asterisked values in
Table 1.

\subsubsection{Comparison with RSL08}

The RSL08 dataset provides us with an opportunity to compare a
different CR determination method with the phase-shift method, although
the analyses are not completely independent since both approaches use
the same near-IR images to define the gravitational potential.
Nevertheless, the simulation method assumes a single pattern speed and
evolves a cloud-particle disk in the potential until the morphology of
the model matches the $B$-band morphology of the galaxy. The authors
argue that their method can narrow down the corotation radius to within
20\%.

Figure~\ref{pertti}a compares our Table 1 asterisked CR radii with
those from RSL08. (Individual comparisons were discussed in section
4). Although there is some correlation between the two datasets, our
values of the bar CR are systematically lower than those estimated by
RSL08. However, since the numerical simulation method largely focusses
on matching the spiral morphology to a model, it is worth comparing the
Rautiainen et al. CR radii with other (mostly spiral) radii in our
Table 1. Figure~\ref{pertti}b shows a comparison between the CR radii
in our Table 1 closest in absolute difference to the Rautiainen et al.
values against these same values. This naturally improves the
correlation somewhat because now our spiral CR radii are brought in.
This improvement in agreement shows that a numerical simulation using a
single pattern speed has the tendency to assign the pattern speed of
the spiral to the bar. Even after the incorporation of some spiral
pattern speeds, the plot shows that the mean difference is still such
that our radii average less than theirs.

\subsubsection{Super-fast Bars}

Probably the most important issue brought to light from this analysis
is the identification of numerous cases having or possibly having $\cal
R$ $<$ 1, confirming the initial evidence for this class found in Paper
I. Some of these are likely due to uncertainties in the estimated bar
radius and/or in the phase-shift-determined CR radius. We have
suggested in Paper I, however, that not all cases are likely to be due
to these uncertainties. If real, these cases would have to be
classified as ``super-fast" bars. The existence of bars extending
beyond their CR goes against the common wisdom of what should be the
bar orbit structure (Contopoulos 1980). In passive orbit analysis, the
stellar orbits support the bar within CR, but not outside CR.

One of the best cases of a possible super-fast bar in our sample is NGC
4902 (Figure~\ref{ngc4902}). This galaxy is almost face-on so that
deprojection uncertainties are not seriously impacting the results. The
phase-shift distribution plot shows successive P/N crossings at 9\arcss
0 and 38\arcss 6, each involving a positive phase-shift section
followed by a negative phase-shift section. The prominent N/P crossing
at 25\arcss 5 ($r/r_o(25)$=0.28) is very close to the bar radius of
22\arcs\ - 26\arcs\ estimated by Laurikainen et al. (2004) and RSL08.
As we noted in section 2.1, the negative portion of a phase-shift
distribution belongs to the inner mode, and since this covers most of
the bar of NGC 4902, we suggest that the actual bar CR in this case is
at 9\arcss 0 (boldfaced in Table 1), not at 38\arcss 6 (asterisked in
Table 1). The latter CR would therefore most likely be associated with
the prominent spiral forming a pseudoring around the bar.

Comparable cases to NGC 4902 include NGC 1187 (Figure~\ref{ngc1187}) and
NGC 5643 (Figure~\ref{ngc5643}). In each, an N/P crossing occurs at the
bar ends, indicating that the bar belongs to an inner mode. In each case
also, a second crossing farther out is most likely associated with a
bright inner spiral. Even the classical barred spiral NGC 1300 could be
in this category. 

\begin{figure}
\figurenum{4}
\plotone{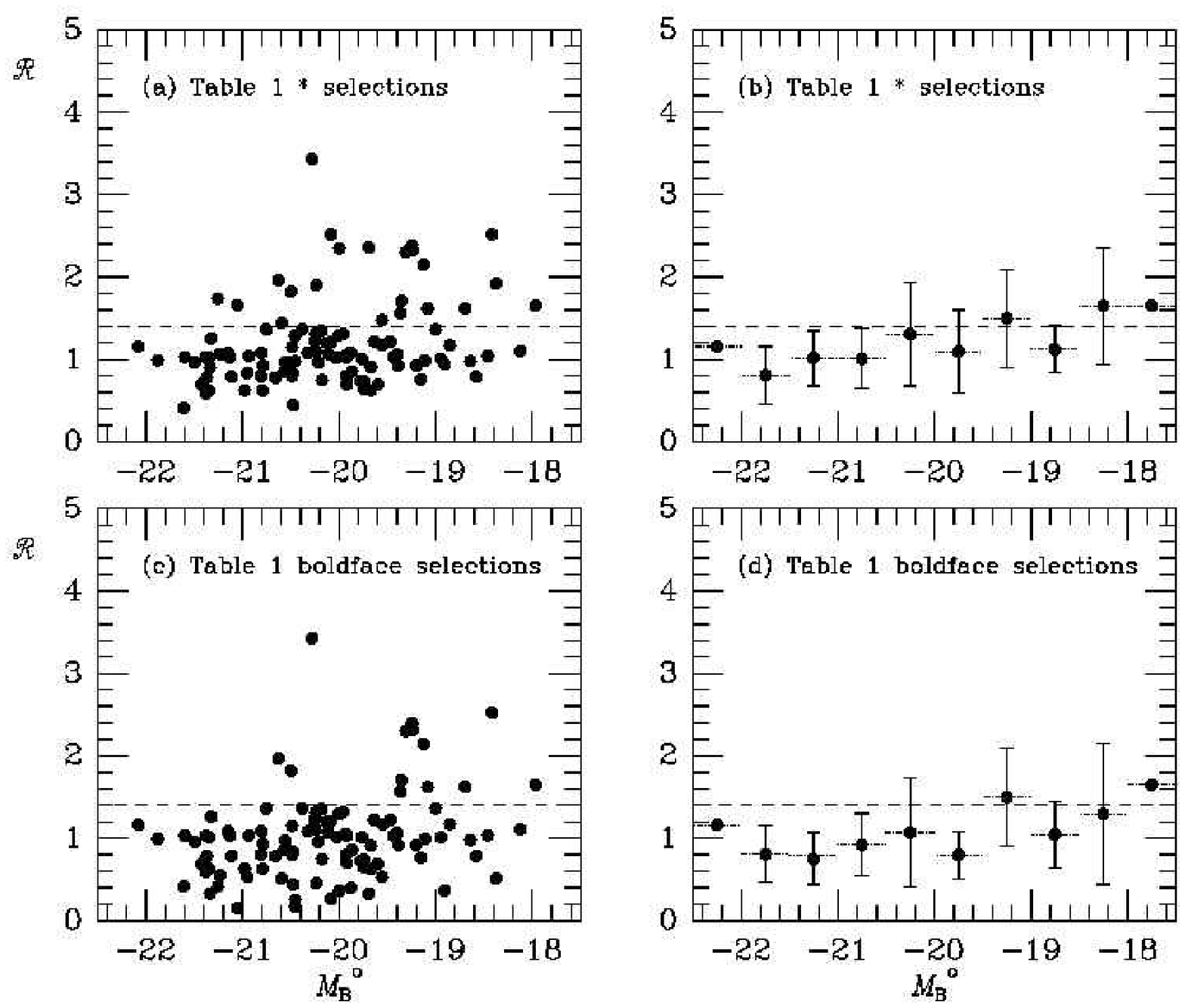}
 \vspace{2.0truecm}
\caption{Same as Figure~\ref{rcrbar} plotted versus absolute blue
magnitude, based on RC3 parameters. Left: individual points. Right:
Means in bins of 0.5 mag with standard deviations. The short dotted
horizontal lines indicate the ranges in absolute magnitude used for the
averages. The dashed horizontal line is set at the value 1.4, used to
delineate a boundary between fast and slow bars (Debattista \& Sellwood
2000).}
\label{absmags}
 \end{figure}

\begin{figure}
\figurenum{5}
\plotone{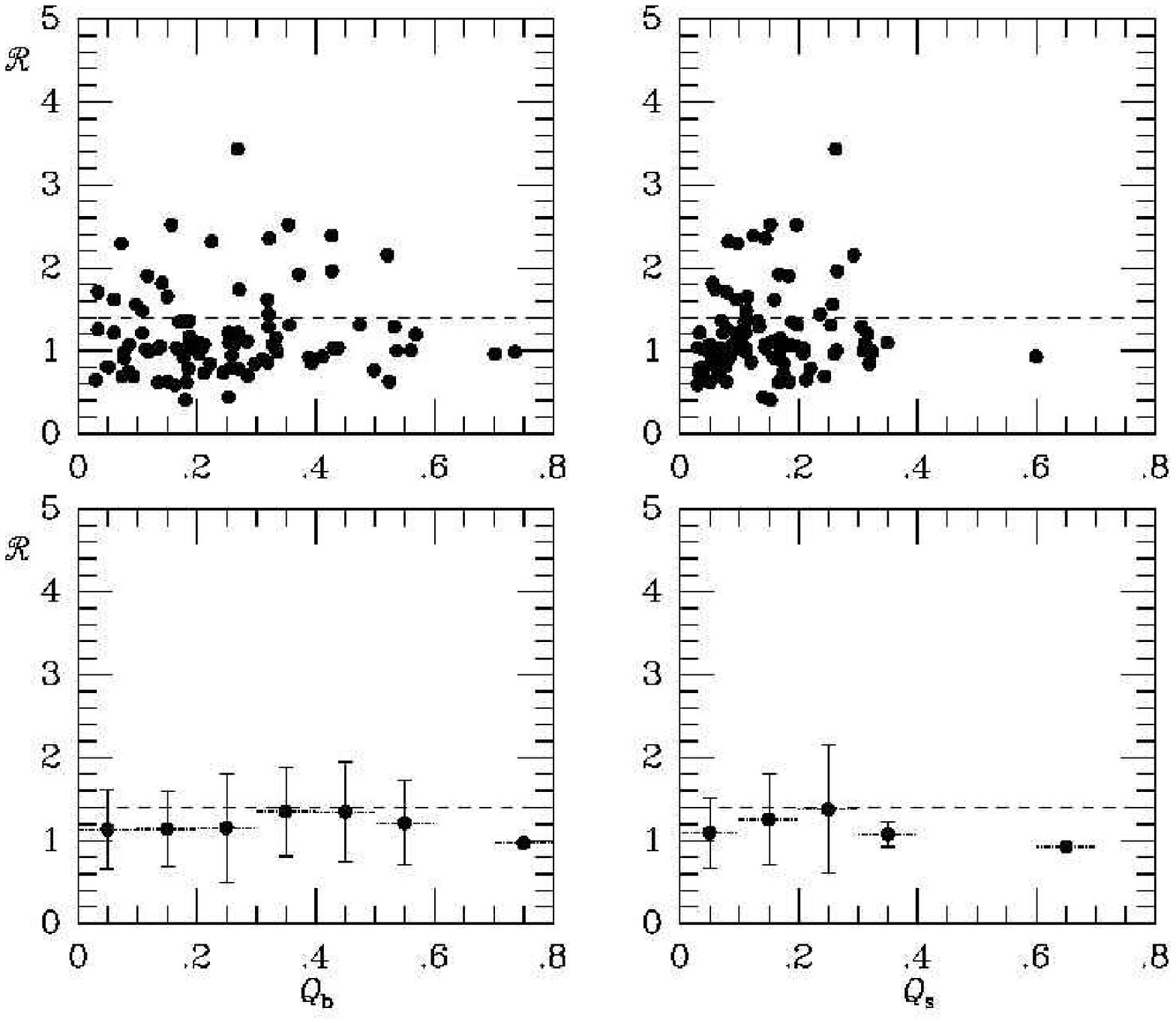}
\vspace{2.0truecm}
\caption{$\cal R$ = $r(CR)/r(bar)$ versus bar and spiral torque
strengths from Buta et al. (2005). Only the values for the asterisked
radii in Table 1 were used for these plots. In the lower panels, the
vertical lines are standard deviation error bars, while the short
dotted horizontal lines simply indicate the range in $Q_b$ and $Q_s$
used for the averages. The dashed horizontal line is set at the value
1.4, used to delineate a boundary between fast and slow bars
(Debattista \& Sellwood 2000).}
\label{qbqs2}
 \end{figure}

\begin{figure}
\figurenum{6}
\plotone{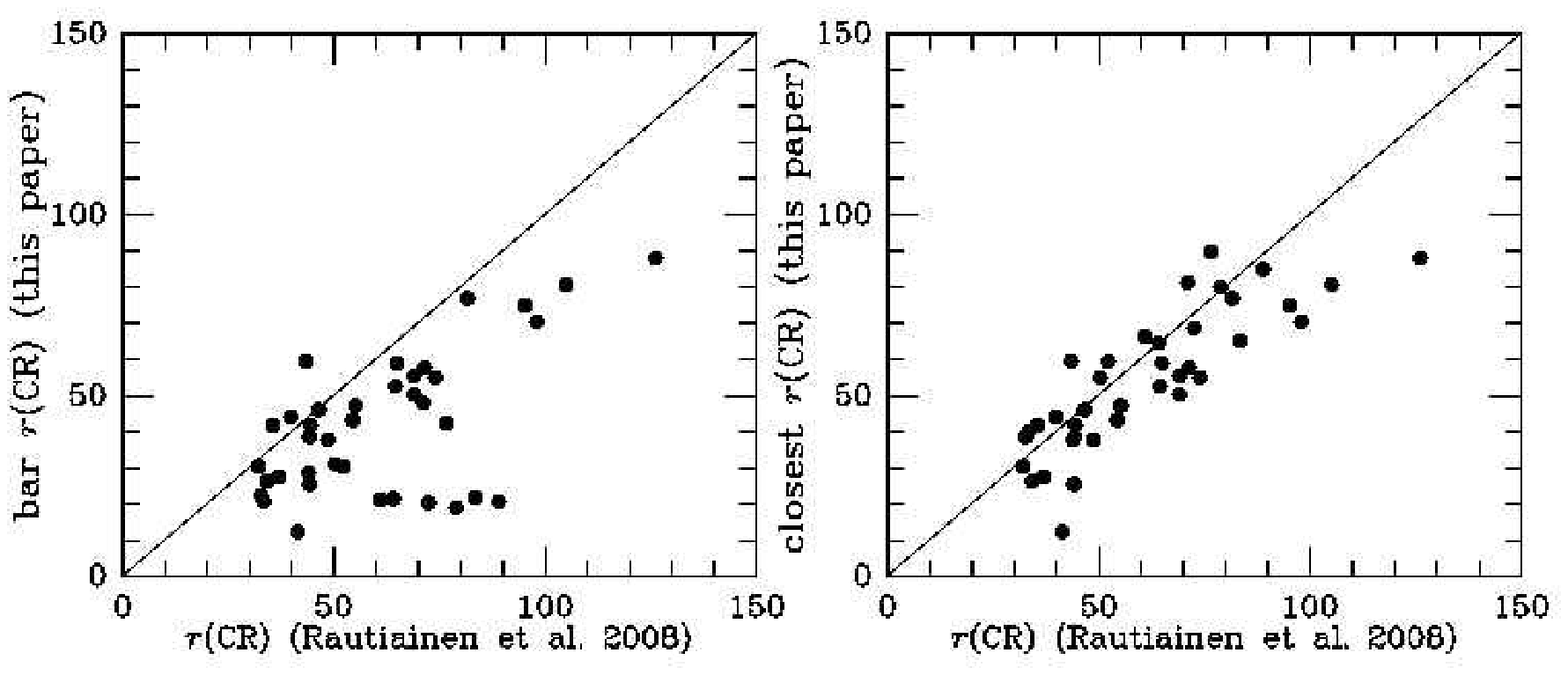}
\vspace{2.0truecm}
\caption{Comparisons of phase-shift method CR radii (in arcseconds) with
those estimated from numerical simulations by RSL05/RSL08. (a)
Comparison between our selected bar CR radii (Table 1, asterisked
values) and theirs. (b) Comparison between the phaseshift CR radii
closest in absolute value to theirs.}
\label{pertti}
\end{figure}

Figures~\ref{rcrbar}c,d show the impact of such unusual interpretations
on $\cal R$ versus type $T$, while Figures~\ref{absmags}c,d show the
impact on $\cal R$ versus absolute blue magnitude, using in each case
the bold-faced alternative bar CR values from Table 1 for 14 objects.
This considerably weakens the type dependence in $\cal R$ and changes
the means to 0.94$\pm$0.40 for types Sbc and earlier and 1.15$\pm$0.70
for types Sc and later. This suggests that ``super-fast" bars may occur
throughout the spiral sequence, not just among earlier types. Late-type
examples we have pointed out in section 3 include NGC 4496, 5334, and
7418.  Figure~\ref{absmags}c shows that with these alternative
interpretations, there are almost no slow bars having $M_B^o$ $<$
$-$19.5.

\subsection{Effect of Orientation Parameters and Bulge Shape
on Phase-Shift Distributions}

We have noted in the above discussions that the assumption of
a spherical bulge can lead to something called ``decomposition
pinch," where the inner isophotes of a galaxy show a pinched
shape in a deprojected image due to over-correction for bulge
light along the minor axis. We would like to know how this artifact
affects what we see in the phase-shift plots. We are also interested
in how uncertainties in the adopted orientation parameters for
a galaxy affect the appearance of the plots. For this purpose, we
have chosen the galaxy NGC 150, which has both a prominent bar
and a spiral. 

\begin{figure}
\figurenum{7}
\plotone{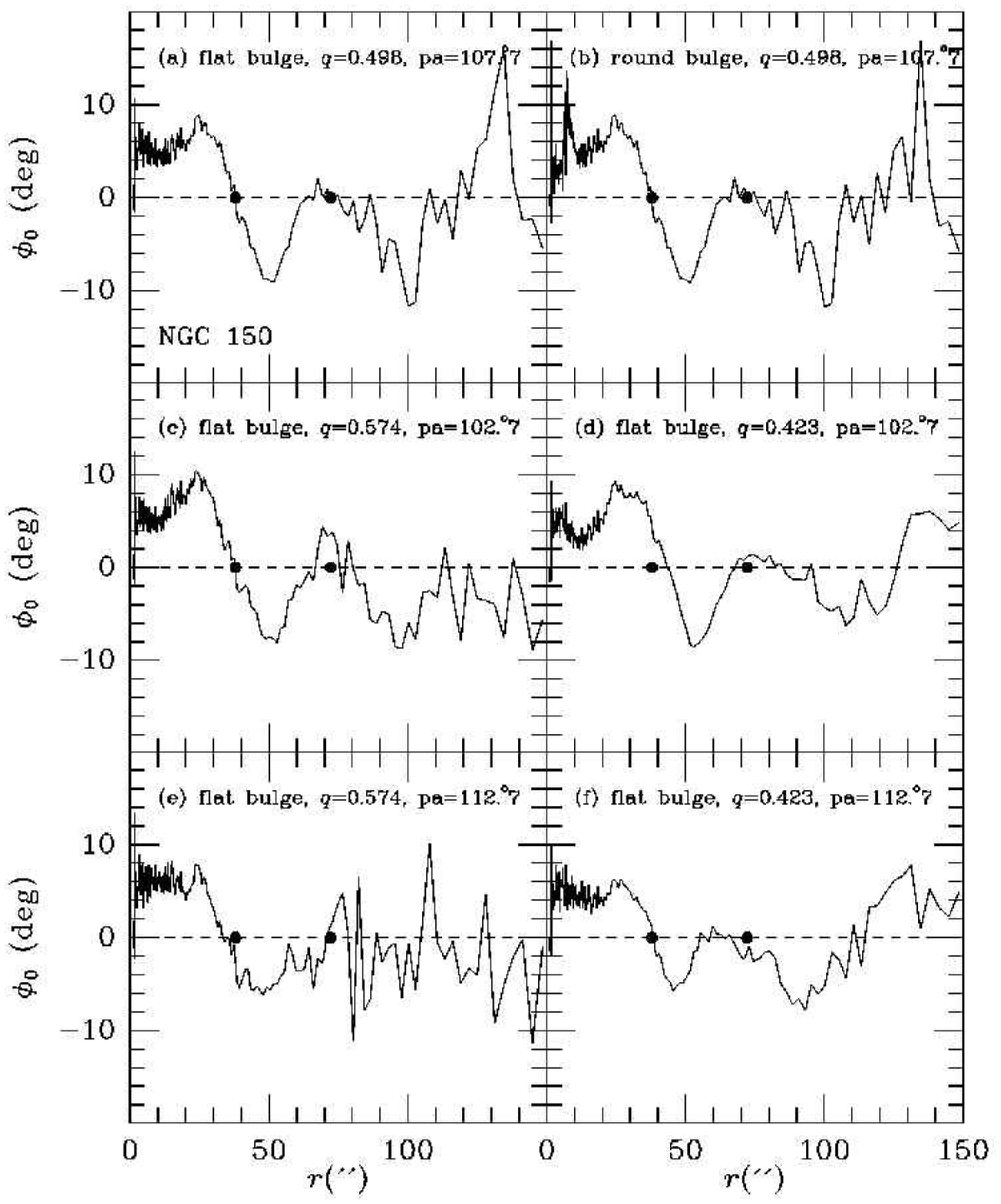}
\caption{Phase-shift distribution for NGC 150 for different orientation
parameters and assumed bulge flattenings, as indicated. For
comparison, the solid circles show the CR radii from Table 1.}
\label{testimages}
\end{figure}

\begin{figure}
\figurenum{8}
\plotone{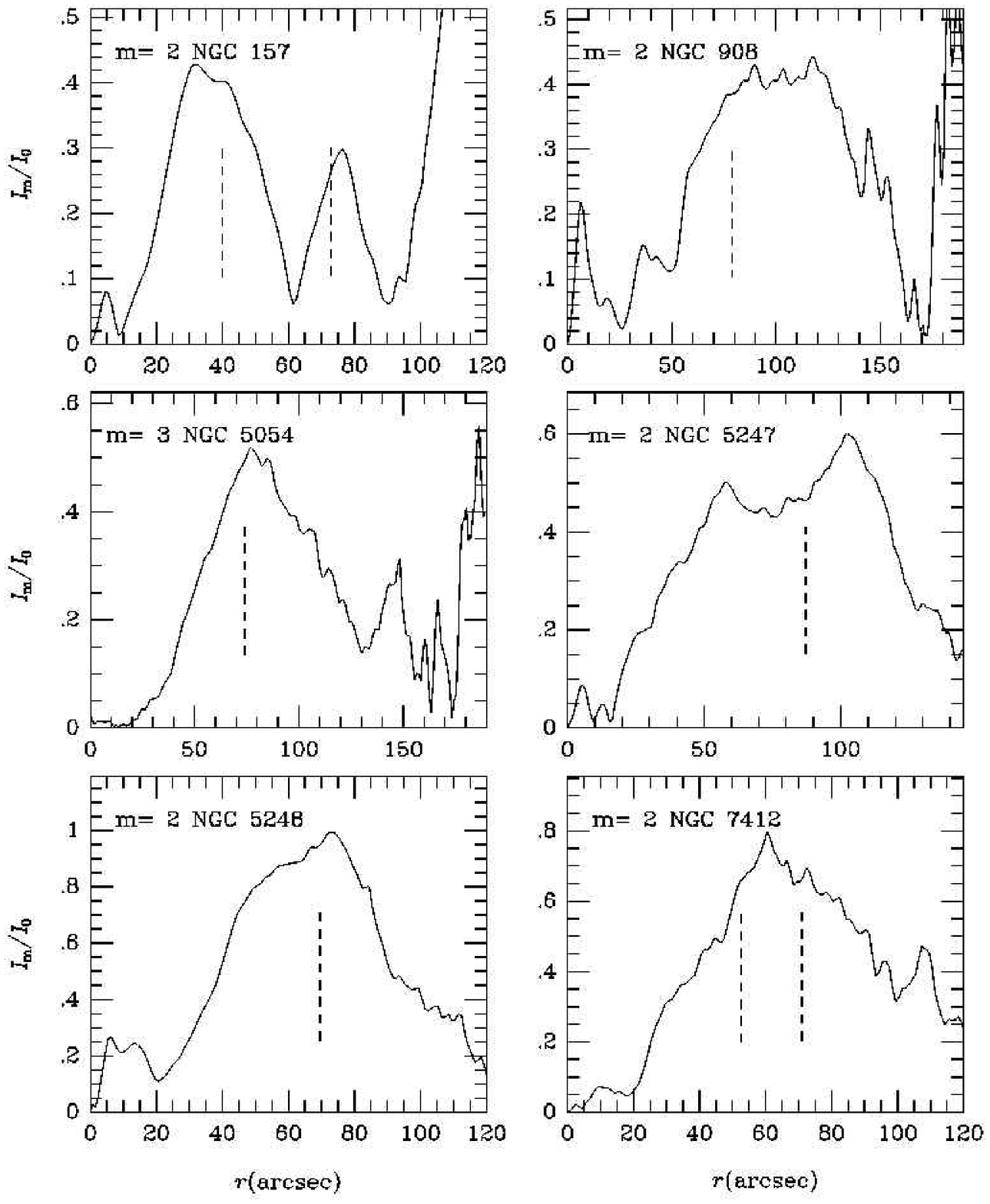}
\caption{Plots of relative Fourier $m$=2 or 3 amplitudes for six
grand-design spirals. The vertical dashed lines indicate the locations
of main corotation radii based on phase-shift crossings.}
\label{fourier_plots}
 \end{figure}

Figures~\ref{testimages}a and b show the sensitivity of the phase-shift
distribution to the assumption of bulge shape.  The test was made with
non-Fourier-smoothed images, and hence the phase-shift plot in
Figure~\ref{testimages}b is slightly noisier than that in
Figure~\ref{ngc0150}, which used the Fourier-smoothed version of the same
image. Panel (a) shows the results for an image of NGC 150 deprojected
assuming the bulge is as flat as the disk, while panel (b) shows the
results assuming the bulge is spherical.  The main difference between
these plots is the appearance of a spike in the inner regions in panel
(b). This could be a complete artifact of the decomposition pinch, but
note that it does not lead to a false crossing.  Deprojection assuming
a perfectly flat bulge causes the bulge isophotes to stretch into a
bar-like feature, but panel (a) shows that no crossing results from that
either, presumably because no additional skewness is introduced.


The remaining panels in Figure~\ref{testimages} show the effects of
different assumed inclinations (through the disk axis ratio $q$ and
major axis position angle pa). We used inclinations and position
angles departing from those used by Laurikainen et al. (2004) by 
$\pm$5$^{\circ}$. The filled circles in Figure~\ref{testimages}
show the locations of the adopted CR radii from Table 1. Naturally,
changing the orientation parameters moves the CR locations since the skewness of
the pattern is either diminished or amplified. When the
revised CR locations are plotted on the corresponding deprojected images, however,
the CR circle for the bar still lies close to or just beyond the bar ends.
We conclude that uncertainties in the assumed bulge shape and in the
orientation parameters are not seriously impacting our results. 

\subsection{High-Luminosity Grand-Design Spirals}

We have shown that in four exceptionally strong grand-design spirals,
NGC 908, 5054, 5247, 5248, the phase-shift method mostly favors a
single CR in the middle of the spiral. This is consistent with the
modal theory (Bertin et al. 1989a,b), which demonstrated that spiral modes
generally extend to their OLR.  Contopoulos \& Grosbol (1986) and
Patsis and Kaufmann (1999) argue that the strong part of a spiral mode
extends mainly to the inner 4:1 resonance, beyond which it is
insufficiently organized. While there might be galaxies where this is
the case (e.g., NGC 3627, Zhang et al.  1993), this is not likely to be
true in general. The spirals simulated by Zhang (1996, 1998, 1999) all
extend to their OLR.

Figure~\ref{fourier_plots} shows how the locations of CR radii for NGC
908, 5054, 5247, 5248, as well as NGC 157 and 7412 which show two CRs
in the spiral region, compare with the maximum in relative Fourier
amplitudes of the light distribution. These show a tendency for the CR
radii to lie near the maxima of the dominant Fourier term. For NGC 157,
each main CR is associated with a distinct $m$=2 maximum, while for NGC
7412 the two CRs share  a single $m$=2 maximum. In NGC 908, 5247, and
5248, the single main CR lies within a broad maximum distribution of
$m$=2 amplitude. For NGC 5054, the same is found but for $m$=3, which
is the dominant term in that case. These findings are also consistent
with the modal theory (Bertin et al. 1989a,b; Elmegreen and Elmegreen
1990).

\section{Discussion}

\subsection{Further comments on the Robustness of the Method and
Additional Support for the Results}

This paper is the first application of the potential-density
phase-shift method to a large sample of spiral galaxies covering almost
all Hubble types. We have shown a clear relationship between
phase-shift crossings and observed morphological features in many
galaxies, and highlighted a possibly significant trend of correlation
between the ratio $\cal R$ with the de Vaucouleurs-revised Hubble type,
which is similar to what has been found in simulation studies of the
same image database by RSL08. Despite the fact that the quantitative
value of the trend depends on which crossing we chose to use as the bar
CR radius, the gross qualitative trend appears to be rather robust
after we have experimented with the different choices of the bar CR
values in cases of potential ambiguity, as shown in
Figure~\ref{rcrbar}. We have highlighted counter-intuitive examples
where the bulk of a bar's extent may lie outside its CR radius, and
such examples may be found across the morphological sequence though
they are most concentrated among the intermediate Hubble types.

\begin{figure}
\figurenum{9}
\plotone{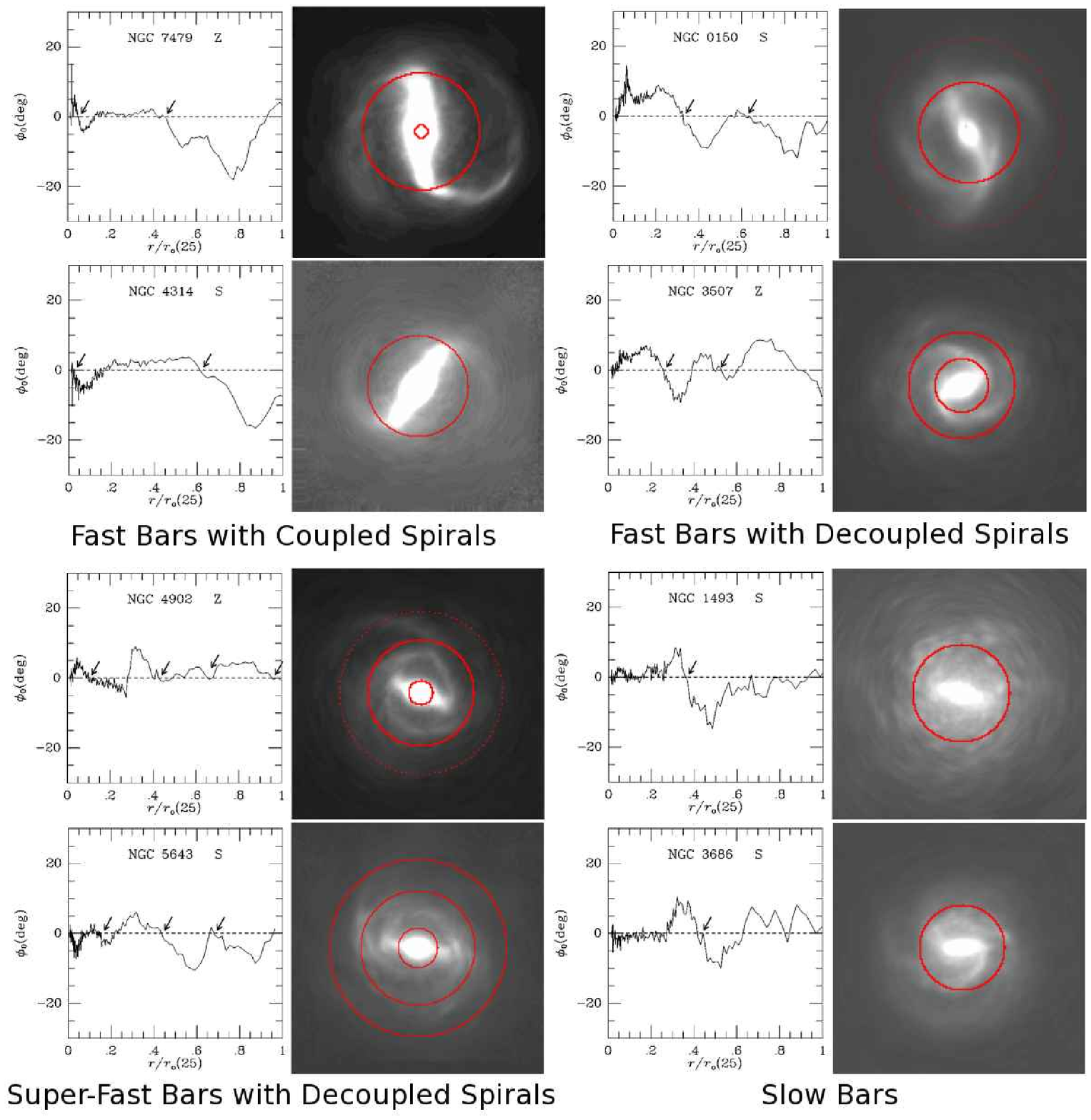}
\vspace{0.0truecm}
\caption{Montage showing examples of bars in different categories.
Fast bars with coupled spirals are cases where the bar and spiral
probably have the same pattern speed. Fast bars with decoupled
spirals are cases where the bars extend to near their CR radius,
but the spiral has a different CR radius. Super-fast bars with
decoupled spirals are very unusual cases where the CR of the bar
is likely to be well inside the ends of the bar. Slow bars are
cases where $\cal R$ $>$ 1.4.
}
\label{summary}
\end{figure}

Figure~\ref{summary} summarizes the types of bar-spiral combinations we
have highlighted in our discussion. In the upper left panels, NGC 7479
and NGC 4314 are presented as cases where the bar and the spiral have
the same pattern speed. The bars have positive phase-shifts throughout
most of their extent, and the single major P/N crossing almost exactly
encircles the bar ends. In these galaxies, the bar extends exactly to
its CR, and is a conventional fast bar. 

In the upper right panels of Figure~\ref{summary}, NGC 150 and NGC 3507
show fast bars where the spirals are likely independent patterns with
their own CR and pattern speed. Again, the bars have positive
phase-shifts throughout most of their extent, and appear to end exactly
at their CR.

The situation is noticeably different for the two galaxies in the lower
left panels of Figure~\ref{summary}. In each case, there is a well-defined
N/P crossing almost exactly at the bar ends, indicating a decoupling point.
NGC 4902, as we noted in section 5.1.3, shows the expected one positive,
followed by one negative zone for an inner mode, and taking this literally
we are forced to conclude that the bar CR corresponds to the innermost
circle, not the more comforting choice of the intermediate circle. NGC 5643 is similar
except for some negative phaseshifts near the center, which may signify
an unresolved inner mode. In both cases, the narrowest parts of the bar lie
completely outside the inner CR and extend to the N/P crossing, which
as we have argued is likely to be close to the location of the bar's OLR. Thus, NGC 4902
and 5643 are more consistently interpreted as ``super-fast" bars.

We have also found in our analysis suggestive cases of slow bars, but
clearcut examples are actually not strong. In most of the high $\cal R$
cases based on the asterisked radii in Table 1, there are negative
phase-shifts in the bar region. This suggests ongoing decoupling of the
bar from the spiral. A possible example of this is seen in NGC 3513
(Figure~\ref{ngc3513}), which has a very conspicuous bar and spiral.
Only a single P/N crossing is found, yet the bar has negative
phase-shifts throughout and a bump in the inner regions suggestive of a
decoupling pattern. Thus, although $\cal R$ = 2.15 based on the single
P/N crossing, it is not clear this is a genuine slow bar case.  The two
suggested examples of slow bars in the lower right panels of
Figure~\ref{summary} are also weak because phase-shifts are low or
nearly zero along the bar. In NGC 3686, the phase-shifts are almost
entirely negative but within the noise of being zero. It could be
argued that in both of these galaxies, the bar and spiral are still
decoupling.  Based on our analysis, we do not find a single absolutely
compelling case of a slow bar.

Even though Figure~\ref{summary} shows that we can identify systematic
trends in phase-shift distributions that have specific physical
meanings, our experience has also shown that phase-shift distributions
can be complicated and difficult to interpret. It would be a fair
criticism of the method that we do not yet have a robust rule for
deciding in general which phase-shift plots and crossings are to be
trusted, and which are not. Our approach, by combining the phase-shift
information with what we see in the image, is probably the best in
general. The ambiguity and the variety of situations we encounter in
these plots might be partly due to the fact that an existing stage of
galaxy morphology is a result of both its initial condition at the time
of galaxy formation and its subsequent interaction history.  No two
galaxies will go through exactly the same stages of evolution.  The
broad categories we use to classify the evolution stages (i.e. galaxy
morphological types) reflect only a general trend, but do not exhaust
all the real-world possibilities. Except for the best-defined bars and
spirals, this complex history of evolution, coupled with the complexity
that arises when the pattern is not yet quasi-steady, is why there are
no set rules in general for interpreting what an observed phase-shift
trend really means dynamically, apart from giving the signs of the
instantaneous torquing action between a wave and the basic state.

RSL08 have argued that the results of the phase-shift method are of
questionable validity, owing to the cases where we deduce that $\cal R$
is considerably less than 1. We have made our arguments in favor of
such cases nonetheless, but it is also important to note that the
passive orbit analysis that Contopoulos (1980) used to deduce $\cal R
\geq$ 1 as a rule is not necessarily valid when collective effects are
taken into account. These cases are also supported by several other
lines of argument. In the case of a massive bar with little spiral
component connecting to it, such as NGC 4665, the bar necessarily
extends beyond its CR because for the SWING amplifier (Toomre 1981) to
work in generating the bar mode, there needs to be density wave content
outside CR to receive the energy transmitted from the over-reflection
of the waves inside CR.

Also, as we have argued in Paper I, the $N$-body bar described by Sparke \&
Sellwood (1987) showed evidence that at least in certain stages of its
evolution the bar can extend beyond its CR, possibly all the way to its
OLR. In its earlier phases, this bar, although not well-formed and not
yet steady, extends much beyond its CR (their Figure 3, left panel),
while at later phases the bar is more developed and has a pattern speed
low enough to end near its CR.  In the N-body simulations of Sparke and
Sellwood, these different stages of bar evolution happen rather fast,
which is partly a result of the fact that the initial condition of this
N-body simulation (i.e. the basic state) is not entirely compatible
with the unstable mode that emerges out of it, so the whole system
evolves relatively fast in the course of the simulation to achieve
overall global self-consistency and equilibrium.  In physical galaxies,
the basic state and the mode are always in a co-evolution process
during the Hubble-type transformation process in a galaxy's lifetime,
so the intermediate stages of the evolution could potentially last much
longer than have been demonstrated in the past N-body simulations. A
more robust proof of the long-lasting nature of these intermediate-type
super-fast bars will likely emerge from a careful demonstration of the
morphological transformation along the Hubble sequence throughout the
lifetime of an individual galaxy using a fully-self-consistent 3D
N-body simulation, a demanding task that currently still has not yet been
achieved. 

In addition, as we have already argued, in the nested-pattern formation
scenario outlined in this paper, some inner bars decouple from a skewed
long bar, which can otherwise be regarded as a spiral.  Since the
spiral pattern has been in multiple instances shown to extend beyond
its CR, these long, skewed bars can as well, i.e., in reality there is
no hard dividing line between a bar and a spiral.

Finally, the continuity of the phase-shift curves themselves demands
that the positive-humps and negative-humps form in mutual succession,
and their relation to the direction of angular momentum exchange
(between the wave and the basic state) demands that we interpret the
N/P crossings as the locations where the successive modes decouple, and
the CRs are positioned in between at the P/N crossings. This condition
by itself dictates that a density wave mode in general extends beyond
its CR. The CR radius and bar radius can become equal only when there
is a spiral component outside the bar (and connecting to it to form the
same mode) to receive the transmitted wave.

\subsection{Support for the Modal Nature of the Spiral and Bar Patterns in
Galaxies}

We have shown that more than half of our OSUBGS sample galaxies possess
grand-design density wave patterns. Furthermore, the nested patterns
organize themselves in a clear sequence of overlapping resonance
features, which appear to develop gradually in well-coordinated
mode-decoupling processes. These and other evidences found in this work
as well as in previous works by other researchers show that the density
wave patterns in most disk galaxies are likely to be
self-organized, quasi-stationary modes which undergo slow secular
morphological transformation together with the slow evolution of the
basic state.  This picture is inconsistent with the alternative view of
the fast transient nature of density wave patterns in disk galaxies
(see, for example, Carlberg \& Sellwood 1985, hereafter CS85), which
argues that spiral patterns die out and re-emerge on the time scale
of the orbital period of the disk stars (see the first paragraph of the
Introduction section of CS85.  The orbital period of stars for the solar
neighborhood is on the order of 250 Myr).  Such hypothesized fast
change is incompatible, in a statistical sense, with the predominance
of well-organized, strong density wave patterns which appear to be the
norm among especially the intermediate and early Hubble-type galaxies
in this study and in other near-infrared samples of disk galaxies.

Furthermore, our results and the previous results of Zhang
(1996,1998,1999) could not be understood in a ``broadening of
resonances'' sense as advocated by some transient spiral supporters.
This is because in the resonance-broadening scenario,
even though the mass flow directions
predicted near the ILR and OLR are consistent with what we found using
the phase shift method, these same mass flow directions at the ILR and
OLR were found (CS85, p. 88) to be ``independent of the winding sense
of the spiral (see LBK)''.  This result of CS85 or LBK is expected since
resonant interaction is a local effect, so the winding sense of the
pattern does not matter.  In contrast, for the collective effect we are
dealing with here, the sense of the phase-shift is critically dependent
on whether the wave is leading or trailing, since the geometrical phase
shift is determined by the Poisson equation, whereas the physical sense
of whether the potential is torquing the density forward or backward
depends on which direction the pattern is rotating. This fact alone
tells us that the collective dissipation effect in spiral galaxies
cannot be viewed as a broadening-of-resonance effect.  In addition, the
resonance-broadening effect is a kind of top-down-control effect -- it
is a single-orbit's response to the applied potential (CS85, p.
81, mentioned the equivalence of their current approach with 
LBK's single-orbit approach, even though one is Eulerian and the other
is Lagrangian).  In a self-organized instability it is the
``sideways'' interactions among stars themselves that leads to pattern
coherence and collective dissipation.  This latter effect cannot be
modeled entirely as a top-down-control hierarchy due to an {\em applied}
spiral or bar potential.  An additional piece of evidence of the
disparity of the two approaches is that, near the CR region, the sense
of angular momentum exchange between the basic state and the wave, and
thus the sense of mass flow predicted by the broadening-of-resonance
approach (see Sellwood \& Binney 2002, Figure 4) are exactly opposite to that
predicted by the phase shift method (the latter being consistent with
that needed to build up the Hubble sequence), and thus the behavior at
CR of the resonance-broadening approach could not lead to the secular mass
re-distribution needed to build up the Hubble sequence.

Finally, we comment that in many of the past N-body simulations of
transient spiral formation (see, e.g. Sellwood 2008 and the references
therein), the spiral patterns' transience is a direct result of the
choice of basic state, i.e., Sellwood and coworkers had consistently
chosen the kind of disks that are {\em stable} to spiral mode
formation.  However, in the simulation of barred galaxies (e.g. Sparke
\& Sellwood 1987), Sellwood and coworkers had consistently chosen basic
states that are {\em unstable} to bar mode formation. Such disparity in
the choice of basic state, which reflected partly the authors' bias,
appears now to be untenable in accounting for the morphological
features of real galaxies: As we have seen amply in this study, the
spiral and bar modes intermingle and inter-transform throughout the
radial extent of the disk of a galaxy, and throughout the lifetime of a
galaxy. It is no longer possible for a simulator to arbitrarily choose
one kind of basic state characteristics for spirals and another for
bars, since these features are interconnected. Both kinds of density
wave patterns appear to have originated as unstable modes in galaxy
disks, and the self-organization and collective dissipation processes
go hand in hand to guarantee both the short-term quasisteady nature of
the pattern, as well as the long-term slow evolution of the parent
galaxy's mass and velocity distribution -- and the latter evolution of
the basic state of the disk in turn leads to the slow transformation of
the density-wave pattern's modal morphology so as to be compatible with
the evolving basic-state's characteristics (Zhang 1998, 1999; Bertin et
al. 1989a,b).

\section{Conclusions}

We have used the potential-density phase-shift method to locate
corotation radii in 153 OSUBGS galaxies using deprojected $H$-band
images. This is the first application of the method to a large number
of galaxies across the full range of spiral sub-types. Our main
conclusions are:

\noindent
1. Most phase-shift distributions do not appear to be simply noise. Most
OSUBGS galaxies show well-defined phase-shift crossings which correspond
to grand-design density wave resonance features.

\noindent
2. Multiple corotations are found in many galaxies. For example, NGC
4303 shows five well-defined crossings within $r_o(25)$. Nevertheless,
a single-pattern-speed model of RSL08 had also been shown to be able to account for
many aspects of this galaxy's morphology.

\noindent
3. Some phase-shift distributions (and corroborated by galaxy images)
show evidence of pattern decoupling, where a change in the phase-shift
leads mostly to a near-crossing rather than a full-fledged crossing.
This type of resonant pattern evolution and successive decoupling
favors a continuously-present and evolving density wave pattern, not a
recurrent one as is advocated in the case of bar formation by Bournaud
\& Combes (2002).  It also contradicts the rapidly-changing recurrent
spiral scenario of Carlberg \& Sellwood (1985).

\noindent
4. For grand-design spirals, the phase-shift method places corotation
in the middle of the spiral, suggesting that these spirals actually
extend to near their OLR and not the inner 4:1 resonance as suggested
by some earlier theoretical studies. We also find that the CR radii of
several grand-design spirals lie near the maximum of their dominant
Fourier term, as expected from the modal theory of spiral structure.

\noindent
5. A comparison between phase-shift bar CR radii and numerical
simulation CR radii (RSL08) shows that the phase-shift method generally
gives smaller values. Some of this disagreement is due to the tendency
for the numerical simulation method to latch on to the pattern speed of
the spiral rather than the pattern speed of the bar, as also
acknowledged by RSL08. When more outer phase-shift CR radii are
compared with RSL08 values, much better agreement is found.

\noindent
6. We confirm a likely type-dependence in the ratio $\cal R$ =
$r(CR)/r(bar)$ if we generally select the P/N crossings that lie near
or just outside the ends of the bar as the bar CR radius.  The average
$\cal R$ ranges from 1.03$\pm$0.37 for types Sbc and earlier, to
1.50$\pm$0.63 for types Sc and later. This is similar to what was
found by RSL08 in spite of the use of different CR radii. Nevertheless,
we have shown that if we follow our rule of one positive and one
negative portion of the phase-shift distribution for a single mode,
then many of the bars we observe are actually ``super-fast" and the
type dependence in $\cal R$ is considerably weakened, reducing to
average values of 0.94$\pm$0.40 for types Sbc and earlier and
1.15$\pm$0.70 for types Sc and later. Super-fast bars still appear to
favor the intermediate to late types. Such cases of super-fast bars are
{\it not} predicted by passive-orbit theory of galaxy model
construction, which clearly favors no bar extending beyond its CR
radius (Contopoulos 1980).

\noindent
7. We find compelling evidence of normal fast bars having $\cal R$
$\approx$ 1 and either a coupled spiral having the same pattern speed
(e.g., NGC 4314, 7479), or a decoupled one having a different pattern
speed (e.g., NGC 150, 3507). We do not find as much compelling evidence
for genuine slow bars in our sample. Many of those having $\cal R >$
1.4 based on our Table 1, asterisked CR selections have near-zero or negative
phase-shifts in the bar region, suggesting ongoing pattern decoupling.
An example of this is NGC 3513 which has only a single P/N crossing at
more than twice the bar radius and yet the phase-shift distribution is
negative throughout the bar's extent. Our two best slow bar cases are
NGC 1493 and 3686 and even these are not clearcut.

\noindent
8. Our confirmation of the presence of quasi-steady {\it nested wave
modes} and their close correspondence with the phase-shift distribution
is in direct contradiction with Lynden-Bell and Kalnajs (1972)'s and 
Binney \& Tremaine (2008)'s conclusion of a constant angular momentum flux 
throughout the galactic disk: such a flux pattern does not deposit angular 
momentum locally onto the disk {\em en route} of the outward angular momentum
transport by the wave, and thus does not promote the growth of the nested 
resonance patterns that we have found in real galaxies.

We thank E. Laurikainen and H. Salo for the deprojected OSUBGS images
used for this study. RB acknowledges the support of NSF grant
AST-0507140 to the University of Alabama. Funding for the OSUBGS was
provided by grants from the NSF (grants AST 92-17716 and AST 96-17006),
with additional funding from the Ohio State University. NED is operated
by the Jet Propulsion Laboratory, California Institute of Technology,
under contract with NASA. Funding for the creation and distribution of
the SDSS Archive has been provided by the Alfred P. Sloan Foundation,
the Participating Institutions, NASA, NSF, the U.S. Department of
Energy, the Japanese Monbukagakusho, and Max Planck Society.

\newpage
\noindent
References

\noindent
Bertin, G., Lin, C.C., Lowe, S.A., \& Thurstans, R.P. 1989a,
ApJ, 338, 78

\noindent
Bertin, G., Lin, C.C., Lowe, S.A., \& Thurstans, R.P. 1989b,
ApJ, 338, 104

\noindent Binney, J., \& Tremaine, S. 2008, Galactic Dynamics
(Princeton:Princeton Univ. Press)

\noindent
Bournaud, F. \& Combes, F. 2002, \aap, 392, 83

\noindent
Buta, R. \& Crocker, D. A. 1993, \aj, 106, 1344

\noindent
Buta, R., Laurikainen, E., \& Salo, H. 2004, \aj, 127, 279 

\noindent
Buta, R., van Driel, W., Braine, J., Combes, F., Wakamatsu, K., Sofue,
Y., \& Tomita, A. 1995, ApJ, 450, 593

\noindent 
Buta, R., Vasylyev, S., Salo, H., and Laurikainen, E. 2005, \aj, 130,
506

\noindent
Buta, R. J., Corwin, H. G., \& Odewahn, S. C. 2007, The de Vaucouleurs Atlas of
Galaxies, Cambridge: Cambridge U. Press 

\noindent
Carlberg, R.G., \& Sellwood, J.A., 1985, \apj, 292, 79

\noindent
Contopoulos, G. 1980, A\&A, 31, 198

\noindent
Contopoulos, G. \& Grosbol, P. 1986, A\&A, 155, 11

\noindent
Debattista, V.P., \& Sellwood, J.A. 2000, ApJ, 543, 704

\noindent
de Vaucouleurs, G., de Vaucouleurs, A., Corwin, H. G., Buta, R. J., Paturel, G.,
\& Fouque, P. 1991, Third Reference Catalog of Bright Galaxies (New York: Springer) (RC3)

\noindent
Elmegreen, B. G. \& Elmegreen, D. M. 1985, \apj, 288, 438

\noindent
Elmegreen, B. G. \& Elmegreen, D. M. 1990, \apj, 355, 52

\noindent
Elmegreen, D. M. \& Elmegreen, B. G. 1982, \mnras, 201, 1021

\noindent
Elmegreen, D. M. \& Elmegreen, B. G. 1987, \apj, 314, 3

\noindent
Eskridge, P. B., Frogel, J. A., Pogge, R. W., et al. 2002, ApJS, 143, 73

\noindent
Garcia-Burillo, S., Fernandez-Garcia, S., Combes, F., Hunt, L. K.,
Haan, S., Schinnerer, E., Boone, F., Krips, M., \& Marquez, I. 2008,
astro-ph 0810.4892

\noindent
Gnedin, O.Y., Goodman, J., \& Frei, Z. 1995, \aj, 110, 1105

\noindent
Kalnajs, A. 1971, \apj, 166, 275

\noindent
Knapen, J. Whyte, L. F., De Blok, W. J. G., \& van der Hulst, J. M. 2004, \aap, 423, 481

\noindent
Kormendy, J. and Kennicutt, R. 2004, ARAA, 42, 603

\noindent
Laurikainen, E., Salo, H., Buta, R., \& Vasylyev, S. 2004, \mnras, 355, 1251

\noindent
Laurikainen, E., Salo, H., Buta, R., \& Knapen, J. H. 2007,
\mnras, 381, 401

\noindent
Lynden-Bell, D., \& Kalnajs, A.J., 1972, MNRAS, 157, 1

\noindent
Masset, F. \& Tagger, M. 1997, A\&A, 322, 442

\noindent
Mazzuca, L. M., Knapen, J. H., Veilleux, S., \& Regan, M. W. 2008, \apjs, 174, 337

\noindent
Mihos, C. \& Bothun, G. 1997, \apj, 481, 781

\noindent
Patsis, P. A. \& Kaufmann, D. E. 1999, \aap, 352, 469

\noindent
Quillen, A. C., Frogel, J. A., \& Gonz\'alez, R. A. 1994, \apj, 437, 162

\noindent
Rautiainen, P. \& Salo, H. 1999, A\&A. 348, 737

\noindent
Rautiainen, P., Salo, H., \& Laurikainen, E. 2005, ApJ, 631, L129

\noindent
Rautiainen, P., Salo, H., \& Laurikainen, E. 2008, \mnras, 388, 1803

\noindent
Sellwood, J.A. 2008, in Formation and Evolution of Galaxy Disks,
Eds. J.G. Funes, S.J., and E.M. Corsini (SFO: ASP), 241

\noindent
Sellwood, J.A., \& Binney, J.J. 2002, MNRAS, 336, 785

\noindent
Sempere, M. J. \& Rozas, M. 1997, \aap, 317, 405

\noindent
Snow, C. 1952, Hypergeometric and Legendre Functions with Applications
to Integral Equations of Potential Theory (NBS/AMS 19; Washington, DC:
National Bureau of Standards)

\noindent
Sparke, L.S., \& Sellwood, J.A. 1987, MNRAS, 225, 653

\noindent
Thornley, M. D. 1996, ApJ, 469, L45

\noindent
Toomre, A.  1981, in Structure and Dynamics of Normal Galaxies,
ed. S. M. Fall \& D. Lynden-Bell (Cambridge: Cambridge Univ. Press), 111

\noindent
Zhang, X. 1996, ApJ, 457, 125

\noindent
Zhang, X. 1998, ApJ, 499, 93

\noindent
Zhang, X. 1999, ApJ, 518, 613

\noindent
Zhang, X. 2003, JKAS, 36, 223

\noindent
Zhang, X. \& Buta, R. 2007, \aj, 133, 2584

\noindent
Zhang, X., Wright, M., \& Alexander, P. 1993, \apj, 418, 100

\clearpage
\include{t01}
\include{t02}
\include{t03}

\clearpage

\clearpage

\noindent
Fig. 1 (cont.) Schematics of the possible evolution of a phase-shift
distribution. In phase 1, the bar and the spiral share a single pattern
speed and there is one major P/N crossing. In phase 2, there is a weak
P/N crossing in the intermediate region of the bar, indicating
decoupling of patterns. In phase 3, this decoupling has been largely
completed, and there are now separate crossings for the bar and the
spiral. The ``model" for phase 1 is NGC 7479, that for phase 2 is NGC
613, and that for phase 3 is NGC 4593. In each case, the radii have
been scaled to place the main CR (arrows) in the same relative
position. 

\clearpage
  
\clearpage                                                                      

\clearpage                                                                      

\clearpage                                                                      
                                                                                
\clearpage                                                                      
                                                                                
\clearpage                                                                      
                                                                                
\clearpage                                                                      
                                                                                
\clearpage                                                                      
                                                                                
\clearpage                                                                      
                                                                                
\clearpage                                                                      
                                                                                
\clearpage                                                                      
                                                                                
\clearpage                                                                      
                                                                                
\clearpage                                                                      
                                                                                
\clearpage                                                                      
                                                                                
\clearpage                                                                      
                                                                                
\clearpage                                                                      
                                                                                
\clearpage                                                                      
                                                                                
\clearpage                                                                      
                                                                                
\clearpage                                                                      
                                                                                
\clearpage                                                                      
                                                                                
\clearpage                                                                      
                                                                                
\clearpage                                                                      
                                                                                
\clearpage                                                                      
                                                                                
\clearpage                                                                      
                                                                                
\clearpage                                                                      
                                                                                
\clearpage                                                                      
                                                                                
\clearpage                                                                      
                                                                                
\clearpage                                                                      
                                                                                
\clearpage                                                                      
                                                                                
\clearpage                                                                      
                                                                                
\clearpage                                                                      
                                                                                
\clearpage                                                                      
                                                                                
\clearpage                                                                      
                                                                                
\clearpage                                                                      
                                                                                
\clearpage                                                                      
                                                                                
\clearpage                                                                      
                                                                                
\clearpage                                                                      
                                                                                
\clearpage                                                                      
                                                                                
\clearpage                                                                      
                                                                                
\clearpage                                                                      
                                                                                
\clearpage                                                                      
                                                                                
\clearpage                                                                      
                                                                                
\clearpage                                                                      
                                                                                
\clearpage                                                                      
                                                                                
\clearpage                                                                      
                                                                                
\clearpage                                                                      
                                                                                
\clearpage                                                                      
                                                                                
\clearpage                                                                      
                                                                                
\clearpage                                                                      
                                                                                
\clearpage                                                                      
                                                                                
\clearpage                                                                      
                                                                                
\clearpage                                                                      
                                                                                
\clearpage                                                                      
                                                                                
\clearpage                                                                      
                                                                                
\clearpage                                                                      
                                                                                
\clearpage                                                                      
                                                                                
\clearpage                                                                      
                                                                                
\clearpage                                                                      
                                                                                
\clearpage                                                                      
                                                                                
\clearpage                                                                      
                                                                                
\clearpage                                                                      
                                                                                
\clearpage                                                                      
                                                                                
\clearpage                                                                      
                                                                                
\clearpage                                                                      
                                                                                
\clearpage                                                                      
                                                                                
\clearpage                                                                      
                                                                                
\clearpage                                                                      
                                                                                
\clearpage                                                                      
                                                                                
\clearpage                                                                      
                                                                                
\clearpage                                                                      
                                                                                
\clearpage                                                                      
                                                                                
\clearpage                                                                      
                                                                                
\clearpage                                                                      
                                                                                
\clearpage                                                                      
                                                                                
                                                                                
\clearpage                                                                      
                                                                                
\clearpage                                                                      
                                                                                
\clearpage                                                                      
                                                                                
\clearpage                                                                      
                                                                                
\clearpage                                                                      
                                                                                
\clearpage                                                                      
                                                                                
\clearpage                                                                      
                                                                                
\clearpage                                                                      
                                                                                
\clearpage                                                                      
                                                                                
\clearpage                                                                      
                                                                                
\clearpage                                                                      
                                                                                
\clearpage                                                                      
                                                                                
\clearpage                                                                      
                                                                                
\clearpage                                                                      
                                                                                
\clearpage                                                                      
                                                                                
\clearpage                                                                      
                                                                                
\clearpage                                                                      
                                                                                
\clearpage                                                                      
                                                                                
\clearpage                                                                      
                                                                                
\clearpage                                                                      
                                                                                
\clearpage                                                                      
                                                                                
\clearpage                                                                      
                                                                                
\clearpage                                                                      
                                                                                
\clearpage                                                                      
                                                                                
\clearpage                                                                      
                                                                                
\clearpage                                                                      
                                                                                
\clearpage                                                                      
                                                                                
\clearpage                                                                      
                                                                                
\clearpage                                                                      
                                                                                
\clearpage                                                                      
                                                                                
\clearpage                                                                      
                                                                                
\clearpage                                                                      
                                                                                
\clearpage                                                                      
                                                                                
\clearpage                                                                      
                                                                                
\clearpage                                                                      
                                                                                
\clearpage                                                                      
                                                                                
\clearpage                                                                      
                                                                                
\clearpage                                                                      
                                                                                
\clearpage                                                                      
                                                                                
\clearpage                                                                      
                                                                                
\clearpage                                                                      
                                                                                
\clearpage                                                                      
                                                                                
\clearpage                                                                      
                                                                                
\clearpage                                                                      
                                                                                
\clearpage                                                                      
                                                                                
\clearpage                                                                      
                                                                                
\clearpage                                                                      
                                                                                
\clearpage                                                                      
                                                                                
\clearpage                                                                      
                                                                                
\clearpage                                                                      
                                                                                
\clearpage                                                                      
                                                                                
\clearpage                                                                      
                                                                                
\clearpage                                                                      
                                                                                
\clearpage                                                                      
                                                                                
\clearpage                                                                      
                                                                                
\clearpage                                                                      
                                                                                
\clearpage                                                                      
                                                                                
\clearpage                                                                      
                                                                                
\clearpage                                                                      
                                                                                
\clearpage                                                                      
                                                                                
\clearpage                                                                      
                                                                                
\clearpage                                                                      
                                                                                
\clearpage                                                                      
                                                                                
\clearpage                                                                      
                                                                                
\clearpage                                                                      
                                                                                
\clearpage                                                                      
                                                                                
\clearpage                                                                      
                                                                                
\clearpage                                                                      
                                                                                
\clearpage                                                                      
                                                                                
\clearpage                                                                      
                                                                                
\clearpage                                                                      
                                                                                
\clearpage                                                                      
                                                                                
\clearpage                                                                      
                                                                                
\clearpage                                                                      
                                                                                
\clearpage                                                                      
                                                                                
\clearpage                                                                      
                                                                                
\clearpage                                                                      
                                                                                
\clearpage                                                                      
                                                                                
\clearpage                                                                      
                                                                                
\clearpage                                                                      
                                                                                
\clearpage

\clearpage

\clearpage

\clearpage

\clearpage
  
\clearpage

\clearpage

\end{document}

%% file: t01.tex
\makeatletter
\def\jnl@aj{AJ}
\ifx\revtex@jnl\jnl@aj\let\tablebreak=\nl\fi
\makeatother

\begin{deluxetable}{lrccrccrccrcc}
\tablenum{1}
\tablewidth{45pc}
\tablecaption{Summary of Corotation Radii in OSUBGS Galaxies\tablenotemark{a}}
\tablehead{
\colhead{Galaxy} &
\colhead{CR$_1$} &
\colhead{err} &
\colhead{code} &
\colhead{CR$_2$} &
\colhead{err} &
\colhead{code} &
\colhead{CR$_3$} &
\colhead{err} &
\colhead{code} &
\colhead{CR$_4$} & 
\colhead{err} & 
\colhead{code} \\
\colhead{1} &
\colhead{2} &
\colhead{3} &
\colhead{4} &
\colhead{5} &
\colhead{6} &
\colhead{7} &
\colhead{8} &
\colhead{9} &
\colhead{10} &
\colhead{11} & 
\colhead{12} & 
\colhead{13} 
}
\startdata
NGC   150  &  38.0* & ... &  3 &  72.2  & ... &  1 &   ...  & ... &  . &   ...  & ... &  . \\
NGC   157  &   6.4  & ... &  3 &  40.0  & ... &  3 &  72.9  & ... &  3 &   ...  & ... &  . \\
NGC   210  &  36.7* & 3.8 &  3 &  80.8  & ... &  3 &  ....  & ... &  . &   ...  & ... &  . \\
NGC   278  &  25.3  & ... &  3 &   ...  & ... &  . &   ...  & ... &  . &   ...  & ... &  . \\
NGC   289  &   3.9  & 0.3 &  1 &  21.3* & ... &  3 &  66.2  & ... &  3 & 133.2  & ... &  1 \\
NGC   428  &  54.8* & ... &  3 &   ...  & ... &  . &   ...  & ... &  . &   ...  & ... &  . \\
NGC   488  &  13.6  & 0.8 &  1 &  34.5  & ... &  1 &  70.3  & ... &  3 &   ...  & ... &  . \\
NGC   578  &  19.1* & ... &  3 &  38.8  & ... &  3 &  79.9  & ... &  1 &   ...  & ... &  . \\
NGC   613  &   5.0  & ... &  3 &  {\bf 55.1}  & ... &  3 &  88.0* & ... &  3 &   ...  & ... &  . \\
NGC   685  &  19.3* & ... &  3 &  92.5  & ... &  1 &   ...  & ... &  . &   ...  & ... &  . \\
NGC   864  &   {\bf 6.5}  & ... &  3 &  33.0* & ... &  3 &  57.3  & ... &  1 &   ...  & ... &  . \\
NGC   908  &   7.9  & 0.6 &  3 &  21.0  & ... &  1 &  78.6  & ... &  3 &   ...  & ... &  . \\
NGC  1042  &   3.3  & ... &  3 &  34.7  & ... &  3 &  84.4  & ... &  3 &   ...  & ... &  . \\
NGC  1058  &  30.5  & ... &  3 &   ...  & ... &  . &   ...  & ... &  . &   ...  & ... &  . \\
NGC  1073  &  38.0* & ... &  3 &  89.3  & ... &  3 &   ...  & ... &  . &   ...  & ... &  . \\
NGC  1084  &  22.4  & ... &  3 &  44.9  & 1.6 &  1 &   ...  & ... &  . &   ...  & ... &  . \\
NGC  1087  &   2.1  & ... &  3 &  35.3* & ... &  3 &   ...  & ... &  . &   ...  & ... &  . \\
NGC  1187  &  {\bf 13.0}  & 1.9 &  3 &  55.0* & ... &  3 & 113.1  & ... &  1 &   ...  & ... &  . \\
NGC  1241  &  12.4* & 1.1 &  3 &   ...  & ... &  . &   ...  & ... &  . &   ...  & ... &  . \\
NGC  1300  &  54.8* & 1.4 &  3 & 142.3  & ... &  3 &   ...  & ... &  . &   ...  & ... &  . \\
NGC  1302  &  31.0* & ... &  3 &  55.0  & ... &  1 &  83.6  & ... &  1 &   ...  & ... &  . \\
NGC  1309  &   7.8  & ... &  1 &  15.3  & ... &  3 &  34.2  & 1.9 &  3 &   ...  & ... &  . \\
NGC  1317  &  18.8  & ... &  3 &  43.3* & ... &  3 & 105.4  & ... &  1 &   ...  & ... &  . \\
NGC  1350  &  {\bf 11.8}  & 1.1 &  3 & 135.0* & ... &  3 &   ...  & ... &  . &   ...  & ... &  . \\
NGC  1371  &  19.5  & ... &  3 &  63.0  & ... &  1 &  81.5  & ... &  1 &   ...  & ... &  . \\
NGC  1385  &  31.9* & ... &  3 &   ...  & ... &  . &   ...  & ... &  . &   ...  & ... &  . \\
NGC  1493  &  37.5* & ... &  3 &   ...  & ... &  . &   ...  & ... &  . &   ...  & ... &  . \\
NGC  1559  &  18.6* & ... &  3 &   ...  & ... &  . &   ...  & ... &  . &   ...  & ... &  . \\
NGC  1617  &  14.2  & ... &  3 &  38.0* & ... &  3 &   ...  & ... &  . &   ...  & ... &  . \\
NGC  1637  &   3.2  & ... &  1 &  23.3* & 1.2 &  3 &  50.5  & ... &  3 &   ...  & ... &  . \\
NGC  1703  &  25.5* & ... &  3 &  59.2  & ... &  1 &   ...  & ... &  . &   ...  & ... &  . \\
NGC  1792  &  37.0  & ... &  1 &  55.8  & ... &  1 &   ...  & ... &  . &   ...  & ... &  . \\
NGC  1808  &   7.0  & 0.2 &  3 &  74.4* & ... &  3 & 106.0  & ... &  1 &   ...  & ... &  . \\
NGC  1832  &  22.6* & ... &  3 &  38.6  & ... &  1 &   ...  & ... &  . &   ...  & ... &  . \\
NGC  2090  &  25.1  & ... &  3 &  78.8  & ... &  1 &   ...  & ... &  . &   ...  & ... &  . \\
NGC  2139  &  21.7* & ... &  3 &   ...  & ... &  . &   ...  & ... &  . &   ...  & ... &  . \\
NGC  2196  &  12.8  & 1.8 &  3 &  64.8  & ... &  1 &   ...  & ... &  . &   ...  & ... &  . \\
NGC  2442  &  11.4  & 2.6 &  1 &  86.6* & ... &  3 &   ...  & ... &  . &   ...  & ... &  . \\
NGC  2559  &  38.5* & ... &  3 &   ...  & ... &  . &   ...  & ... &  . &   ...  & ... &  . \\
NGC  2566  &  56.2* & ... &  1 &  98.6  & ... &  3 &   ...  & ... &  . &   ...  & ... &  . \\
NGC  2775  &  64.0  & ... &  1 & 104.0  & ... &  1 &   ...  & ... &  . &   ...  & ... &  . \\
NGC  2964  &   3.0  & ... &  1 &  36.7* & ... &  3 &   ...  & ... &  . &   ...  & ... &  . \\
NGC  3059  &  25.9* & 2.0 &  3 &   ...  & ... &  . &   ...  & ... &  . &   ...  & ... &  . \\
NGC  3166  &  54.6* & 4.8 &  3 &   ...  & ... &  . &   ...  & ... &  . &   ...  & ... &  . \\
NGC  3223  &  23.2  & ... &  3 &  62.0  & ... &  3 &   ...  & ... &  . &   ...  & ... &  . \\
NGC  3227  &  47.0* & 4.7 &  3 & 108.7  & ... &  3 &   ...  & ... &  . &   ...  & ... &  . \\
NGC  3261  &  28.6* & ... &  3 &  37.7  & ... &  1 &  52.3  & ... &  1 &   ...  & ... &  . \\
NGC  3275  &   6.1  & 0.2 &  3 &  25.7* & 1.3 &  3 &   ...  & ... &  . &   ...  & ... &  . \\
NGC  3319  &   9.5  & ... &  3 &  37.7* & ... &  3 & 126.4  & ... &  3 &   ...  & ... &  . \\
NGC  3338  &  18.1* & 1.4 &  3 &  50.3  & ... &  3 & 163.3  & ... &  3 &   ...  & ... &  . \\
NGC  3423  &  17.9  & 1.3 &  3 &   ...  & ... &  . &   ...  & ... &  . &   ...  & ... &  . \\
NGC  3504  &  24.3  & ... &  3 &  41.9* & ... &  1 &   ...  & ... &  . &   ...  & ... &  . \\
NGC  3507  &  26.3* & ... &  3 &  52.1  & 1.7 &  3 &   ...  & ... &  . &   ...  & ... &  . \\
NGC  3513  &   2.0  & ... &  3 &  59.6* & ... &  3 &   ...  & ... &  . &   ...  & ... &  . \\
NGC  3583  &   3.3  & ... &  1 &  30.5* & ... &  3 &  64.6  & ... &  1 &   ...  & ... &  . \\
NGC  3593  &   8.8  & 0.7 &  1 &  29.5  & ... &  1 &  51.6  & ... &  1 &   ...  & ... &  . \\
NGC  3596  &  12.3  & ... &  3 &  32.2  & ... &  3 &  49.8  & ... &  1 &  71.1  & ... &  1 \\
NGC  3646  &  15.4  & ... &  3 &  74.4  & ... &  3 &   ...  & ... &  . &   ...  & ... &  . \\
NGC  3675  &   6.7  & ... &  3 &  27.4* & ... &  1 &  83.9  & ... &  3 &   ...  & ... &  . \\
NGC  3681  &  20.4* & ... &  3 &   ...  & ... &  . &   ...  & ... &  . &   ...  & ... &  . \\
NGC  3684  &  27.7  & ... &  3 &  60.8  & ... &  3 &   ...  & ... &  . &   ...  & ... &  . \\
NGC  3686  &  41.8* & ... &  3 &   ...  & ... &  . &   ...  & ... &  . &   ...  & ... &  . \\
NGC  3726  &  21.9* & 1.0 &  3 &  65.2  & 2.1 &  3 & 114.1  & ... &  1 &   ...  & ... &  . \\
NGC  3810  &  14.2  & ... &  3 &  64.3  & ... &  1 &  83.8  & ... &  1 &   ...  & ... &  . \\
NGC  3887  &  28.0* & ... &  3 &  57.1  & ... &  3 &   ...  & ... &  . &   ...  & ... &  . \\
NGC  3893  &  20.8  & ... &  3 &  61.0  & ... &  3 &   ...  & ... &  . &   ...  & ... &  . \\
NGC  3938  &  25.0  & ... &  3 &  61.7  & ... &  3 &   ...  & ... &  . &   ...  & ... &  . \\
NGC  3949  &   7.8  & ... &  1 &  37.5  & ... &  1 &   ...  & ... &  . &   ...  & ... &  . \\
NGC  4027  &  24.0* & ... &  3 &  53.2  & ... &  1 &   ...  & ... &  . &   ...  & ... &  . \\
NGC  4030  &  10.4  & ... &  3 &  29.7  & ... &  3 &  55.7  & ... &  3 &  91.5  & ... &  1 \\
NGC  4051  &   3.3  & ... &  3 &  70.4* & ... &  3 &   ...  & ... &  . &   ...  & ... &  . \\
NGC  4123  &  18.0  & ... &  3 &  55.4* & ... &  3 & 108.9  & ... &  1 &   ...  & ... &  . \\
NGC  4136  &   6.8  & ... &  3 &  24.7* & 1.9 &  3 &  34.2  & 0.6 &  3 &  60.8  & ... &  1 \\
NGC  4145  &  46.6* & ... &  3 & 122.4  & ... &  3 &   ...  & ... &  . &   ...  & ... &  . \\
NGC  4151  &   5.7  & ... &  3 &  69.8  & ... &  3 &  99.4* & ... &  3 &   ...  & ... &  . \\
NGC  4212  &  35.6  & ... &  3 &   ...  & ... &  . &   ...  & ... &  . &   ...  & ... &  . \\
NGC  4242  &  30.4  & ... &  3 &   ...  & ... &  . &   ...  & ... &  . &   ...  & ... &  . \\
NGC  4254  &   7.5  & ... &  3 &  20.8  & ... &  3 &  55.8  & ... &  3 &  90.7  & ... &  3 \\
NGC  4293  &  15.0  & 1.5 &  1 &  79.3* & ... &  3 & 142.3  & ... &  1 &   ...  & ... &  . \\
NGC  4303\tablenotemark{b}  &  20.8* & ... &  3 &  48.8  & ... &  3 &  69.6  & ... &  1 &  84.9  & ... &  1 \\
NGC  4314  &   3.3  & ... &  3 &  76.9* & ... &  3 &   ...  & ... &  . &   ...  & ... &  . \\
NGC  4394  &  {\bf 16.7}  & ... &  3 &  42.4* & ... &  3 &  89.6  & ... &  1 &   ...  & ... &  . \\
NGC  4414  &  20.7  & ... &  3 &  49.3  & ... &  3 &  76.2  & ... &  3 &   ...  & ... &  . \\
NGC  4450  &   4.5  & ... &  1 &  30.5* & ... &  3 &  59.4  & ... &  3 & 156.5  & ... &  1 \\
NGC  4457  &  44.2* & ... &  3 &   ...  & ... &  . &   ...  & ... &  . &   ...  & ... &  . \\
NGC  4487  &  20.5* & 0.6 &  3 &  42.8  & 1.6 &  1 &   ...  & ... &  . &   ...  & ... &  . \\
NGC  4496  &   9.1  & ... &  3 &  60.0* & ... &  3 & 102.5  & ... &  1 &   ...  & ... &  . \\
NGC  4504  &  10.3  & 1.2 &  1 &  26.6  & ... &  3 &   ...  & ... &  . &   ...  & ... &  . \\
NGC  4527  &  65.2* & ... &  3 &   ...  & ... &  . &   ...  & ... &  . &   ...  & ... &  . \\
NGC  4548  &  19.0  & ... &  3 &  74.8* & ... &  3 &   ...  & ... &  . &   ...  & ... &  . \\
NGC  4571  &   5.3  & ... &  1 &  50.5  & 2.6 &  3 &   ...  & ... &  . &   ...  & ... &  . \\
NGC  4579  &   8.7  & ... &  1 &  {\bf 24.7}  & ... &  3 &  48.0* & ... &  1 &  81.3  & ... &  3 \\
NGC  4580  &  15.4  & ... &  3 &  38.7  & 0.8 &  3 &   ...  & ... &  . &   ...  & ... &  . \\
NGC  4593  &  12.3  & ... &  3 &  65.8* & ... &  3 &  97.4  & ... &  3 &   ...  & ... &  . \\
NGC  4618  &  83.0* & ... &  3 &   ...  & ... &  . &   ...  & ... &  . &   ...  & ... &  . \\
NGC  4643  &   6.2  & ... &  1 &  50.2* & ... &  1 &   ...  & ... &  . &   ...  & ... &  . \\
NGC  4647  &   {\bf 6.1}  & ... &  3 &  17.0* & ... &  3 &  50.5  & ... &  1 &   ...  & ... &  . \\
NGC  4651  &   6.1  & 0.5 &  3 &  36.5* & ... &  3 &  51.8  & ... &  3 &  88.9  & 2.5 &  3 \\
NGC  4654  &   4.3  & 0.3 &  3 &  20.1* & 0.7 &  3 &  78.6  & 1.9 &  3 &   ...  & ... &  . \\
NGC  4665  &  16.8  & ... &  3 &  47.2* & ... &  3 &   ...  & ... &  . &   ...  & ... &  . \\
NGC  4689  &  34.1  & 0.8 &  3 &  66.5  & 1.4 &  3 & 103.0  & ... &  3 &   ...  & ... &  . \\
NGC  4691  &  34.1* & ... &  3 &   ...  & ... &  . &   ...  & ... &  . &   ...  & ... &  . \\
NGC  4699  &   2.5  & ... &  1 &  14.9* & ... &  3 &   ...  & ... &  . &   ...  & ... &  . \\
NGC  4772  &  52.5  & ... &  1 &   ...  & ... &  . &   ...  & ... &  . &   ...  & ... &  . \\
NGC  4775  &  38.0  & ... &  3 &   ...  & ... &  . &   ...  & ... &  . &   ...  & ... &  . \\
NGC  4781  &   2.4  & ... &  1 &  {\bf 15.2}  & ... &  3 &  42.4* & ... &  3 &  61.3  & ... &  3 \\
NGC  4900  &   {\bf 9.2}  & ... &  3 &  34.6* & ... &  3 &   ...  & ... &  . &   ...  & ... &  . \\
NGC  4902  &   {\bf 9.0}  & ... &  3 &  38.6* & ... &  3 &  59.5  & ... &  1 &  86.6  & ... &  1 \\
NGC  4930  &   4.5  & ... &  3 &  46.2* & ... &  3 &  96.8  & 7.3 &  3 &   ...  & ... &  . \\
NGC  4939  &  17.3* & ... &  3 &  37.6  & ... &  3 &  69.3  & ... &  3 & 120.3  & 4.6 &  3 \\
NGC  4941  &  57.1  & ... &  1 &   ...  & ... &  . &   ...  & ... &  . &   ...  & ... &  . \\
NGC  4995  &  21.6* & 0.8 &  3 &  37.7  & 0.8 &  3 &  64.5  & ... &  3 &   ...  & ... &  . \\
NGC  5005  &  29.4  & 0.6 &  3 &  51.6* & ... &  3 &   ...  & ... &  . &   ...  & ... &  . \\
NGC  5054  &  74.0  & ... &  3 &   ...  & ... &  . &   ...  & ... &  . &   ...  & ... &  . \\
NGC  5085  &   7.5  & ... &  3 &  21.6  & 0.3 &  3 &  47.2  & ... &  3 &  97.3  & ... &  1 \\
NGC  5101  &  10.3  & 3.0 &  3 &  54.3* & ... &  3 &  93.5  & ... &  3 &   ...  & ... &  . \\
NGC  5121  &  37.1  & ... &  3 &   ...  & ... &  . &   ...  & ... &  . &   ...  & ... &  . \\
NGC  5247  &  14.0  & 1.0 &  3 &  87.2  & ... &  3 &   ...  & ... &  . &   ...  & ... &  . \\
NGC  5248  &  12.5  & ... &  3 &  69.6  & ... &  3 &   ...  & ... &  . &   ...  & ... &  . \\
NGC  5334  &   {\bf 6.0}  & ... &  3 &  42.5* & ... &  3 &  81.8  & ... &  3 &   ...  & ... &  . \\
NGC  5427  &   7.3  & 0.2 &  3 &  38.0  & ... &  3 &  63.3  & ... &  1 &   ...  & ... &  . \\
NGC  5483  &   2.6  & ... &  3 &  18.0* & ... &  3 &  25.6  & ... &  3 &  72.7  & ... &  1 \\
NGC  5643  &  {\bf 23.6}  & ... &  3 &  66.9* & ... &  3 & 104.7  & ... &  3 &   ...  & ... &  . \\
NGC  5676  &  23.2  & ... &  3 &  38.2  & ... &  3 &  61.2  & ... &  1 &   ...  & ... &  . \\
NGC  5701  &  15.4  & ... &  3 &  52.5* & ... &  3 &  86.9  & ... &  3 &   ...  & ... &  . \\
NGC  5713  &   {\bf 5.0}  & ... &  3 &  29.3* & ... &  3 &  71.3  & ... &  3 &   ...  & ... &  . \\
NGC  5850  &  {\bf 29.9}  & ... &  3 &  80.7* & ... &  3 &   ...  & ... &  . &   ...  & ... &  . \\
NGC  5921  &  20.6  & ... &  3 &  57.8* & ... &  3 & 137.5  & ... &  3 &   ...  & ... &  . \\
NGC  5962  &  27.3* & ... &  3 &   ...  & ... &  . &   ...  & ... &  . &   ...  & ... &  . \\
NGC  6215  &  27.3  & 4.3 &  3 &  57.0  & ... &  3 &  72.5  & ... &  3 &   ...  & ... &  . \\
NGC  6221  &  14.5  & ... &  3 &  41.9* & ... &  3 &  93.0  & ... &  3 &   ...  & ... &  . \\
NGC  6300  &   9.4  & ... &  3 &  39.1* & ... &  3 &  68.5  & ... &  3 & 106.4  & ... &  1 \\
NGC  6384  &  20.4* & ... &  3 &  68.7  & ... &  3 & 131.7  & ... &  3 &   ...  & ... &  . \\
NGC  6753  &  10.9  & ... &  3 &  19.4  & ... &  3 &  37.3  & ... &  3 &  58.9  & ... &  3 \\
NGC  6782  &  12.2  & ... &  3 &  27.6* & ... &  3 &   ...  & ... &  . &   ...  & ... &  . \\
NGC  6902  &  22.0* & ... &  3 &  92.1  & 1.5 &  3 &   ...  & ... &  . &   ...  & ... &  . \\
NGC  6907  &   5.7  & 0.1 &  3 &  37.5  & 2.3 &  3 &   ...  & ... &  . &   ...  & ... &  . \\
NGC  7083  &   3.7  & ... &  3 &  18.1  & ... &  3 &  49.7  & ... &  3 &   ...  & ... &  . \\
NGC  7205  &  74.1  & ... &  3 &   ...  & ... &  . &   ...  & ... &  . &   ...  & ... &  . \\
NGC  7213  &  22.0  & ... &  1 &  64.9  & 3.0 &  1 &   ...  & ... &  . &   ...  & ... &  . \\
NGC  7217  &  14.5  & 1.6 &  3 &  43.1  & ... &  3 &   ...  & ... &  . &   ...  & ... &  . \\
NGC  7412  &  13.9  & ... &  3 &  52.7  & ... &  3 &  71.0  & ... &  3 &   ...  & ... &  . \\
NGC  7418  &   {\bf 4.0}  & ... &  3 &  38.1* & 1.8 &  3 &  81.3  & 2.0 &  3 &   ...  & ... &  . \\
NGC  7479  &   6.3  & ... &  3 &  57.7* & ... &  3 &   ...  & ... &  . &   ...  & ... &  . \\
NGC  7552  &   4.7  & ... &  3 &  15.6  & ... &  3 &  59.0* & ... &  3 &   ...  & ... &  . \\
NGC  7582  &  43.6  & ... &  3 &  74.3* & ... &  3 &   ...  & ... &  . &   ...  & ... &  . \\
NGC  7713  &  55.7  & ... &  3 &  67.9  & ... &  3 &   ...  & ... &  . &   ...  & ... &  . \\
NGC  7723  &   9.3  & ... &  3 &  20.7* & ... &  3 &  40.1  & ... &  3 &   ...  & ... &  . \\
NGC  7727  &  29.1* & ... &  3 & 114.5  & ... &  1 &   ...  & ... &  . &   ...  & ... &  . \\
NGC  7741  &  51.9* & 6.8 &  3 &   ...  & ... &  . &   ...  & ... &  . &   ...  & ... &  . \\
IC   4444  &   9.7* & 0.4 &  3 &  41.4  & ... &  3 &   ...  & ... &  . &   ...  & ... &  . \\
IC   5325  &  15.1* & ... &  3 &  25.5  & ... &  3 &  40.4  & 1.7 &  3 &   ...  & ... &  . \\
ESO  138$-$10 &  16.8  & 0.6 &  3 &  79.9  & ... &  3 & 137.4  & ... &  3 &   ...  & ... &  . \\
\enddata
\tablenotetext{a}{Col. 1: galaxy name; cols. 2,5,8,11:
radii of inferred corotations from positive-to-negative
potential-density phase-shift crossings, in arcseconds;
cols. 3,6,9,12: a number is given in these columns only
if the phase-shift crossing is noisy. Then, the value is
a standard deviation about an average crossing radius;
cols. 4,7,10,13: a code indicating how significant we
think a crossing is. Code 3 crossings are considered more 
reliable than code 1 crossings.}
\tablenotetext{b}{CR$_5$=134\rlap{.}$^{\prime\prime}$0,
code=3}
\end{deluxetable}

%% file: t02.tex
\makeatletter
\def\jnl@aj{AJ}
\ifx\revtex@jnl\jnl@aj\let\tablebreak=\nl\fi
\makeatother

\begin{deluxetable}{llccrlllccrl}
\tablenum{2}
\tablewidth{0pc}
\tablecaption{Summary of Features\tablenotemark{a}}
\tablehead{
\colhead{Galaxy} &
\colhead{$n_{CR}$} &
\colhead{bar?} &
\colhead{oval?} &
\colhead{AC} &
\colhead{PS-plot} &
\colhead{Galaxy} &
\colhead{$n_{CR}$} &
\colhead{bar?} &
\colhead{oval?} &
\colhead{AC} &
\colhead{PS-plot} \\
\colhead{1} &
\colhead{2} &
\colhead{3} &
\colhead{4} &
\colhead{5} &
\colhead{6} &
\colhead{7} &
\colhead{8} &
\colhead{9} &
\colhead{10} &
\colhead{11} &
\colhead{12} 
}
\startdata
NGC   150    & 2  & y  & n  & 12  & well-def   & NGC  3227    & 2: & n  & y  &  7  & noisy      \\
NGC   157    & 3  & n  & n  & 12  & well-def   & NGC  3261    & 3: & y  & n  &  9B & noisy      \\
NGC   210    & 2: & n  & y  &  6  & noisy      & NGC  3275    & 2  & y  & n  &  4  & noisy      \\
NGC   278    & 1: & n  & y? &  3B & noisy      & NGC  3319    & 3  & y  & n  &  5  & well-def   \\
NGC   289    & 4  & n  & y  & 12  & well-def   & NGC  3338    & 3  & n  & y? &  9  & well-def   \\
NGC   428    & 1+ & n  & y  &  1  & well-def   & NGC  3423    & 1: & n  & y: &  2  & noisy      \\
NGC   488    & 3: & n  & y  &  3  & noisy      & NGC  3504    & 2  & y  & y  &  8  & well-def   \\
NGC   578    & 3  & y  & n  &  9  & well-def   & NGC  3507    & 2  & y  & n  &  9  & well-def   \\
NGC   613    & 3  & y  & n  &  9  & well-def   & NGC  3513    & 2  & y  & n  & 12  & well-def   \\
NGC   685    & 2  & y  & n  &  2B & well-def   & NGC  3583    & 3  & y  & n  & 12  & well-def   \\
NGC   864    & 3  & y  & n  &  5  & well-def   & NGC  3593    & 3: & n  & y  & ..  & noisy/unst \\
NGC   908    & 3  & n  & n  &  9  & well-def   & NGC  3596    & 4: & n  & y  &  5  & noisy      \\
NGC  1042    & 3  & n  & n  &  9  & well-def   & NGC  3646    & 2  & n  & y? &  2  & well-def   \\
NGC  1058    & 1  & n  & y  &  3  & peculiar   & NGC  3675    & 3: & n  & y  &  3  & noisy      \\
NGC  1073    & 2+ & y  & n  &  5  & well-def   & NGC  3681    & 1: & n  & y  &  5  & noisy      \\
NGC  1084    & 2  & n  & y  &  5  & well-def   & NGC  3684    & 2  & n  & y  &  5  & noisy      \\
NGC  1087    & 2: & y  & n  &  2  & noisy      & NGC  3686    & 1  & y  & n  &  5  & well-def   \\
NGC  1187    & 3: & y  & n  &  9  & well-def   & NGC  3726    & 3  & n  & y  &  5  & well-def   \\
NGC  1241    & 1  & y  & n  &  4  & well-def   & NGC  3810    & 3: & n  & y  &  2  & noisy      \\
NGC  1300    & 2  & y  & n  & 12  & well-def   & NGC  3887    & 2  & n  & y  &  2  & well-def   \\
NGC  1302    & 3  & n  & y  &  8  & noisy      & NGC  3893    & 2  & n  & y  & 12  & well-def   \\
NGC  1309    & 3: & n  & y  &  3  & noisy/unst & NGC  3938    & 2: & n  & n  &  9  & noisy      \\
NGC  1317    & 2+ & y  & y  & ..  & noisy      & NGC  3949    & 2: & n  & y  &  1  & noisy/unst \\
NGC  1350    & 2  & y  & n  & 12B & well-def   & NGC  4027    & 2  & y  & n  &  4  & well-def   \\
NGC  1371    & 3+ & n  & y  &  4  & noisy      & NGC  4030    & 4  & n  & y: &  9  & well-def   \\
NGC  1385    & 1  & y  & n  &  4  & well-def   & NGC  4051    & 2  & n  & y  &  5  & well-def   \\
NGC  1493    & 1: & y  & n  &  5  & well-def:  & NGC  4123    & 3: & y  & y  &  9  & noisy      \\
NGC  1559    & 1  & y  & n  &  5B & well-def   & NGC  4136    & 4: & y  & n  &  9  & noisy      \\
NGC  1617    & 2: & n  & y  &  3  & noisy      & NGC  4145    & 2  & y  & n  &  4  & well-def   \\
NGC  1637    & 3  & y  & n  &  5  & well-def   & NGC  4151    & 3  & y  & y  &  5  & well-def   \\
NGC  1703    & 2  & n  & y  &  9B & well-def   & NGC  4212    & 1  & n  & y  &  3  & noisy      \\
NGC  1792    & 2  & n  & n  &  3  & noisy      & NGC  4242    & 1: & n  & y  &  1  & noisy      \\
NGC  1808    & 3  & y  & n  &  7B & well-def   & NGC  4254    & 4: & n  & y  &  9  & noisy      \\
NGC  1832    & 2  & y  & n  &  5  & well-def   & NGC  4293    & 3  & y  & n  & ..  & well-def   \\
NGC  2090    & 2  & n  & y  &  3B & noisy      & NGC  4303    & 5  & n  & y  &  9  & well-def   \\
NGC  2139    & 1: & y  & n  &  2  & noisy      & NGC  4314    & 2  & y  & n  & 12  & well-def   \\
NGC  2196    & 2  & n  & y  &  6  & noisy      & NGC  4394    & 3: & y  & n  &  6  & noisy      \\
NGC  2442    & 2  & y  & n  &  7  & well-def   & NGC  4414    & 3: & n  & y  &  3  & noisy      \\
NGC  2559    & 1  & y  & n  &  9B & well-def   & NGC  4450    & 4  & n  & y  & 12  & well-def   \\
NGC  2566    & 2  & y  & n  &  8B & well-def   & NGC  4457    & 1: & n  & y  &  4  & noisy      \\
NGC  2775    & 2? & n  & y  &  3  & noisy/unst & NGC  4487    & 2: & n  & y  &  5  & noisy      \\
NGC  2964    & 2  & y  & n  & 12  & well-def   & NGC  4496    & 3: & y  & n  &  2  & noisy      \\
NGC  3059    & 1  & y  & n  &  5  & noisy      & NGC  4504    & 2: & n  & y  &  5  & noisy/unst \\
NGC  3166    & 1  & n  & y  & ..  & noisy      & NGC  4527    & 1  & n  & y  &  5B & noisy      \\
NGC  3223    & 2: & n  & y? &  9B & noisy      & NGC  4548    & 2  & y  & n  &  5  & well-def   \\
\enddata
\end{deluxetable}
\begin{deluxetable}{llccrlllccrl}
\tablenum{2}
\tablewidth{0pc}
\tablecaption{Summary of Features (cont.)\tablenotemark{a}}
\tablehead{
\colhead{Galaxy} &
\colhead{$n_{CR}$} &
\colhead{bar?} &
\colhead{oval?} &
\colhead{AC} &
\colhead{PS-plot} &
\colhead{Galaxy} &
\colhead{$n_{CR}$} &
\colhead{bar?} &
\colhead{oval?} &
\colhead{AC} &
\colhead{PS-plot} \\
}
\startdata
NGC  4571    & 2: & n  & n: &  3  & noisy      & NGC  5643    & 3  & y  & n  &  8B & well-def   \\
NGC  4579    & 4  & y  & n  &  9  & well=def   & NGC  5676    & 3: & n  & y  &  3  & noisy      \\
NGC  4580    & 2  & n  & y  & 12  & noisy      & NGC  5701    & 3  & y  & n  & ..  & noisy      \\
NGC  4593    & 3  & y  & n  &  5  & well-def   & NGC  5713    & 3  & y  & n  &  4  & well-def   \\
NGC  4618    & 1  & y  & n  &  4  & well-def   & NGC  5850    & 2  & y  & n  &  8  & well-def   \\
NGC  4643    & 2: & y  & n  & ..  & noisy      & NGC  5921    & 3  & y  & n  &  8  & well-def   \\
NGC  4647    & 3  & n  & y  &  3  & noisy      & NGC  5962    & 1: & n  & y  &  2  & noisy      \\
NGC  4651    & 4: & n  & y  &  9  & noisy      & NGC  6215    & 3  & n  & n  & 12  & well-def   \\
NGC  4654    & 3  & n  & y  &  4  & well-def   & NGC  6221    & 3  & y  & n  &  9B & well-def   \\
NGC  4665    & 2  & y  & n  & ..  & well-def   & NGC  6300    & 4  & y  & n  &  6  & well-def   \\
NGC  4689    & 3  & n  & n  &  3  & well-def   & NGC  6384    & 3: & n  & y  &  9  & noisy      \\
NGC  4691    & 1: & y  & n  & ..  & noisy      & NGC  6753    & 4  & n  & n  &  8  & noisy      \\
NGC  4699    & 2  & n  & y  &  3  & noisy      & NGC  6782    & 2: & y  & n  & ..  & noisy      \\
NGC  4772    & 1: & n  & y  &  8  & noisy/unst & NGC  6902    & 2: & n  & y  &  8B & noisy      \\
NGC  4775    & 1  & n  & y: &  9  & well-def   & NGC  6907    & 2  & y  & n  & 12  & well-def   \\
NGC  4781    & 4: & y: & y  &  4B & noisy      & NGC  7083    & 3  & n  & y: &  9B & noisy      \\
NGC  4900    & 2  & y  & n  &  3  & noisy      & NGC  7205    & 1  & n  & y  &  5  & well-def   \\
NGC  4902    & 4  & y  & n  &  9  & well-def   & NGC  7213    & 2: & n  & n  & ..  & noisy/unst \\
NGC  4930    & 3: & y  & n  &  8B & noisy      & NGC  7217    & 2  & n  & y? &  3  & noisy      \\
NGC  4939    & 4  & n  & y  & 12  & well-def   & NGC  7412    & 3  & n  & y  &  9B & well-def   \\
NGC  4941    & 1: & n  & y: &  3  & noisy/unst & NGC  7418    & 3  & y  & n  &  5B & noisy      \\
NGC  4995    & 3  & y  & n  &  6  & well-def   & NGC  7479    & 2  & y  & n  &  9  & well-def   \\
NGC  5005    & 2: & n  & y  &  3  & noisy      & NGC  7552    & 3  & y  & n  & 12B & well-def   \\
NGC  5054    & 1  & n  & y  &  5  & well-def   & NGC  7582    & 2: & y  & n  & 12B & well-def   \\
NGC  5085    & 4: & n  & n  &  2  & well-def   & NGC  7713    & 2  & n  & y  &  1B & noisy      \\
NGC  5101    & 3  & y  & n  & ..  & noisy      & NGC  7723    & 3: & y  & n  &  5  & noisy      \\
NGC  5121    & 1: & n  & y: & ..  & noisy/unst & NGC  7727    & 2: & n  & y  &  1  & noisy      \\
NGC  5247    & 2  & n  & n  &  9  & well-def   & NGC  7741    & 1  & y  & n  &  5  & noisy      \\
NGC  5248    & 2  & n  & y  & 12  & well-def   & IC   4444    & 2  & n  & y  &  6B & well-def   \\
NGC  5334    & 3: & y  & n  &  2  & noisy      & IC   5325    & 3: & n  & y  &  6B & noisy      \\
NGC  5427    & 3  & n  & n  &  9  & well-def   & ESO  138- 10 & 3  & n  & y  & ..  & noisy      \\
NGC  5483    & 4: & n  & y  &  5B & noisy      & ............ & .. & .. & .. & ... & .........  \\
\enddata
\tablenotetext{a}{Col. 1: galaxy name; col. 2: number of significant or possibly significant
corotations (: or ? means uncertain, + means there could be more); col. 3: indicates if a 
bar is present in the near-IR image: `y' = yes, `n' = no; col. 4: indicates if an oval (bar-like)
feature is present in the near-IR image; col. 5: Elmegreen arm class, mostly from Elmegreen
\& Elmegreen (1987) or estimated by R. Buta from images in the {\it de Vaucouleurs Atlas of Galaxies}
or other available images; if estimated by Buta, a `B' follows the arm class; col. 6: an assessment
of how coherent the phase-shift plot appears; `well-def' = well-defined and `unst' \ unsteady;
cols. 7-12: same as columns 1-6.} 
\end{deluxetable}

%% file: t03.tex
\makeatletter
\def\jnl@aj{AJ}
\ifx\revtex@jnl\jnl@aj\let\tablebreak=\nl\fi
\makeatother

\begin{deluxetable}{lclclclc}
\tablenum{3}
\tablewidth{40pc}
\tablecaption{Ratio of Corotation to Selected Bar Radii for 101 Galaxies\tablenotemark{a}}
\tablehead{
\colhead{Galaxy} &
\colhead{$\cal R$} &
\colhead{Galaxy} &
\colhead{$\cal R$} &
\colhead{Galaxy} &
\colhead{$\cal R$} &
\colhead{Galaxy} &
\colhead{$\cal R$} 
}
\startdata
NGC  150 & 1.31 & NGC 2559 & 1.15 & NGC 4303 & 0.69 & NGC 5101 & 0.78\\
NGC  210 & 0.80 & NGC 2566 & 0.78 & NGC 4314 & 1.03 & NGC 5334 & 2.36\\
NGC  289 & 1.08 & NGC 2964 & 1.22 & NGC 4394 & 0.94,0.37 & NGC 5483 & 1.35\\
NGC  428 & 1.22 & NGC 3059 & 1.29 & NGC 4450 & 1.02 & NGC 5643 & 1.44,0.51\\
NGC  578 & 0.97 & NGC 3166 & 1.21 & NGC 4457 & 0.98 & NGC 5701 & 1.05\\
NGC  613 & 0.84,0.53 & NGC 3227 & 0.63 & NGC 4487 & 0.92 & NGC 5713 & 0.98,0.17\\
NGC  685 & 0.92 & NGC 3261 & 1.03 & NGC 4496 & 2.35,0.36 & NGC 5850 & 0.90,0.33\\
NGC  864 & 1.29,0.25 & NGC 3275 & 0.62 & NGC 4527 & 0.79 & NGC 5921 & 1.10\\
NGC 1073 & 1.01 & NGC 3319 & 1.00 & NGC 4548 & 1.11 & NGC 5962 & 1.82\\
NGC 1087 & 1.96 & NGC 3338 & 0.81 & NGC 4579 & 1.07,0.55 & NGC 6221 & 1.03\\
NGC 1187 & 1.90,0.45 & NGC 3504 & 0.70 & NGC 4593 & 1.08 & NGC 6300 & 0.84\\
NGC 1241 & 0.41 & NGC 3507 & 1.17 & NGC 4618 & 2.52 & NGC 6384 & 0.62\\
NGC 1300 & 0.63 & NGC 3513 & 2.15 & NGC 4643 & 0.74 & NGC 6782 & 0.59\\
NGC 1302 & 1.22 & NGC 3583 & 1.35 & NGC 4647 & 1.48,0.53 & NGC 6902 & 1.26\\
NGC 1317 & 0.75 & NGC 3675 & 0.91 & NGC 4651 & 1.62 & NGC 7418 & 2.52,0.26\\
NGC 1350 & 1.66,0.15 & NGC 3681 & 1.36 & NGC 4654 & 1.03 & NGC 7479 & 0.96\\
NGC 1385 & 3.43 & NGC 3686 & 2.32 & NGC 4665 & 0.79 & NGC 7552 & 0.85\\
NGC 1493 & 1.61 & NGC 3726 & 0.73 & NGC 4691 & 0.76 & NGC 7582 & 0.80\\
NGC 1559 & 1.07 & NGC 3887 & 0.69 & NGC 4699 & 1.04 & NGC 7723 & 0.86\\
NGC 1617 & 1.71 & NGC 4027 & 1.20 & NGC 4781 & 1.09,0.39 & NGC 7727 & 1.08\\
NGC 1637 & 1.03 & NGC 4051 & 1.56 & NGC 4900 & 1.92,0.51 & NGC 7741 & 0.99\\
NGC 1703 & 2.29 & NGC 4123 & 1.06 & NGC 4902 & 1.74,0.41 & IC  4444 & 0.44\\
NGC 1808 & 0.86 & NGC 4136 & 1.65 & NGC 4930 & 1.04 & IC  5325 & 0.65\\
NGC 1832 & 1.36 & NGC 4145 & 2.39 & NGC 4939 & 0.99 & ........ & ....\\
NGC 2139 & 1.31 & NGC 4151 & 1.02 & NGC 4995 & 0.96 & ........ & ....\\
NGC 2442 & 0.93 & NGC 4293 & 1.18 & NGC 5005 & 1.15 & ........ & ....\\
\enddata
\tablenotetext{a}{$\cal R$=$r(CR)/r(bar)$. Proposed CR radii for the
bar are indicated by asterisks in Table 1. If two values are listed,
the second is for an alternative proposed bar CR radius (indicated
by bold-face) in Table 1. Bar
radii are from Table 3 of Laurikainen et al. (2004).}
\end{deluxetable}